\documentclass[12pt]{book}
\usepackage[utf8]{inputenc}
\usepackage{hyperref} 
\hypersetup { pdfpagemode = {UseNone}, pdftitle = {Higher Spin Gravity: Quantization and Algebraic Structures}, pdfauthor = {Tung Vuong Tran}, pdflang = {en-GB} }

\usepackage[dvipsnames]{xcolor}
\usepackage{caption}
\usepackage{framed}
\usepackage{a4wide}
\usepackage{fancyhdr}
\usepackage{stmaryrd}
\usepackage[pdftex]{graphicx}
\usepackage{amsthm,amsmath,latexsym,amssymb,amsfonts,amscd}
\usepackage{graphics,lscape,fancyhdr,array,stmaryrd,euscript}
\usepackage{empheq}
\usepackage{sidecap}
\usepackage{ifpdf}
\usepackage{verbatim}
\usepackage{color}
\usepackage{relsize}
\usepackage{tikz-cd}
\usepackage{skak}
\usepackage{slashed}
\usepackage{youngtab}

\newcommand{\bep}{\begin{picture}}
\newcommand{\eep}{\end{picture}}

\newcounter{YoungHeight}\newcounter{YoungWidth}

\newcounter{Mul1}\newcounter{Mul2}\newcounter{Mul3}\newcounter{Mul4}
\newcounter{A0}\newcounter{A1}\newcounter{A2}
\newcounter{B3}
\newcounter{C3}\newcounter{C4}
\newcounter{D1}\newcounter{D2}\newcounter{D3}
\newcounter{T0}\newcounter{T1}

\newlength{\txtHShift}

\newlength{\txtWidth}

\newcommand{\HalfLength}[2]{\setcounter{Mul1}{#1}\setcounter{Mul2}{#1}\addtocounter{Mul1}{\value{Mul2}}\addtocounter{Mul1}{\value{Mul2}}%
\addtocounter{Mul1}{\value{Mul2}}\addtocounter{Mul1}{\value{Mul2}}\setcounter{#2}{\value{Mul1}}}

\newcommand{\Add}[3]{\setcounter{#1}{#2}\addtocounter{#1}{#3}}

\newcommand{\Length}[1]{#10}
\newcommand{\YoungScale}{}

\newcommand{\shiftedText}[2]{{\hspace{#1}#2}}
\newcommand{\calcHShift}[1]{\settowidth{\txtWidth}{#1}\setlength{\txtHShift}{-0.5\txtWidth}}

\newcommand{\TextCenter}[3]{{\HalfLength{#2}{T0}%
\HalfLength{#3}{T1}\addtocounter{T1}{-3}\calcHShift{#1}%
\put(\value{T0},\value{T1}){\shiftedText{\txtHShift}{#1}}}}

\newcommand{\TextCenterB}[3]{{\calcHShift{#1}\HalfLength{#2}{T0}\Add{T1}{\Length{#3}}{-7}\put(\value{T0},\value{T1}){\shiftedText{\txtHShift}{#1}}}}

\newcommand{\TextTop}[3]{{\calcHShift{#1}\HalfLength{#2}{T0}\Add{T1}{\Length{#3}}{-9}\put(\value{T0},\value{T1}){\shiftedText{\txtHShift}{#1}}}}

\newcommand{\BlockA}[2]{{\YoungScale\bep(\Length{#1},\Length{#2}){\Add{A1}{#1}{1}\Add{A2}{#2}{1}}%
\multiput(0,0)(10,0){\value{A1}}{\line(0,1){\Length{#2}}}\multiput(0,0)(0,10){\value{A2}}{\line(1,0){\Length{#1}}}%
\setcounter{YoungHeight}{\Length{#2}}\setcounter{YoungWidth}{\Length{#1}}\eep}}

\newcommand{\BlockB}[4]{{\YoungScale\Add{B3}{\Length{#2}}{\Length{#4}}%
\bep(\Length{#1},\value{B3})\put(0,\Length{#4}){\BlockA{#1}{#2}}%
\put(0,0){\BlockA{#3}{#4}}\setcounter{YoungHeight}{\value{B3}}\setcounter{YoungWidth}{\Length{#1}}\eep}}

\newcommand{\RectT}[3]{\bep(\Length{#1},\Length{#2})\put(0,0){\line(1,0){\Length{#1}}}\put(0,0){\line(0,1){\Length{#2}}}%
\put(\Length{#1},\Length{#2}){\line(-1,0){\Length{#1}}}\put(\Length{#1},\Length{#2}){\line(0,-1){\Length{#2}}}#3{#1}{#2}\eep}

\newcommand{\RectARowUp}[2]{{\bep(\Length{#1},10)\put(0,0){\RectT{#1}{1}{\TextCenterB{#2}}}\eep}}

\newcommand{\RectBRowUp}[4]{{\bep(\Length{#1},20)\put(0,0){\RectT{#2}{1}{\TextCenterB{#4}}}%
\put(0,10){\RectT{#1}{1}{\TextCenterB{#3}}}\eep}}

\newcommand{\RectCRowUp}[6]{{\bep(\Length{#1},30)\put(0,0){\RectT{#3}{1}{\TextCenterB{#6}}}%
\put(0,10){\RectT{#2}{1}{\TextCenterB{#5}}}\put(0,20){\RectT{#1}{1}{\TextCenterB{#4}}}\eep}}

\newcommand{\YoungA}{\BlockA{1}{1}}
\newcommand{\YoungB}{\BlockA{2}{1}}

\newcommand{\YoungAA}{\BlockA{1}{2}}

\newcommand{\YoungBB}{\BlockA{2}{2}}

\newcommand{\YoungCC}{\BlockA{3}{2}}

\newcommand{\YoungAAAA}{\BlockA{1}{4}}

\newcommand{\YoungBBAA}{\BlockB{2}{2}{1}{2}}

\usepackage{hyperref,setspace}

\renewcommand{\chaptermark}[1]%
         {\markboth{\thechapter.\ #1}{}}
\renewcommand{\sectionmark}[1]%
         {\markright{\thesection\ #1}}
\newcommand{\leading}{\text{lead}}
\newcommand{\subleading}{\text{sub}}
\newcommand{\be}{\begin{equation}}
\newcommand{\ee}{\end{equation}}
\newcommand{\self}{\text{self}}
\newcommand{\tri}{\text{ver}}

\newcommand{\pvec}{{\boldsymbol{p}}}
\newcommand{\qvec}{{\boldsymbol{q}}}
\newcommand{\kvec}{{\boldsymbol{k}}}
\newcommand{\pb}{{\bar{p}}}

\DeclareMathOperator*{\intergrallambda}{%
\mathchoice%
  {\ooalign{$\displaystyle\lambda$\cr$\,\displaystyle\int$\hidewidth\cr}}
  {\ooalign{\raisebox{.14\height}{\scalebox{.6}{$\lambda$}}\cr\hidewidth$\int$\hidewidth\cr}}
  {\ooalign{\raisebox{.2\height}{\scalebox{.6}{$\lambda$}}\cr$\int$\cr}}
  {\ooalign{\raisebox{.2\height}{\scalebox{.6}{$\lambda$}}\cr$\int$\cr}}
}
\newcommand{\kb}{\bar{k}}
\newcommand{\lvec}{\boldsymbol{\ell}}
\newcommand{\NN}{\mathbb{N}}
\newcommand{\MM}{\mathbb{M}}
\newcommand{\Mcal}{\mathcal{M}}
\newcommand{\delbeta}{\Delta_{\beta}}
\newcommand{\Pcal}{\mathcal{P}}
\newcommand{\Pu}{\underline{P}}
\newcommand{\Hom}{\text{Hom}}

\newcommand{\PP}{{\mathbb{P}}}

\newcommand{\PPb}{{\overline{\mathbb{P}}}}

\newcommand{\tr}{\text{Tr}}

\newcommand{\um}{\underline{m}}

\newcommand{\ads}{\text{AdS}}
\newcommand{\cft}{\text{CFT}}
\newcommand{\D}{\text{D}}
\newcommand{\YY}{\mathbb{Y}}
\newcommand{\Tr}{\text{Tr}}
\newcommand{\Vcal}{\mathcal{V}}
\newcommand{\Ncal}{\mathcal{N}}
\newcommand{\JJ}{\mathbb{J}}
\newcommand{\AdS}{\text{AdS}}
\newcommand{\CFT}{\text{CFT}}

\newcommand{\ualpha}{{\ensuremath{\underline{\alpha}}}}

\newcommand{\ua}{{\ensuremath{\underline{a}}}}
\newcommand{\ub}{{\ensuremath{\underline{b}}}}
\newcommand{\uc}{{\ensuremath{\underline{c}}}}
\newcommand{\mm}{{\ensuremath{\underline{m}}}}

\newcommand{\pl}{\partial}
\newcommand{\Ocal}{\mathcal{O}}

\newcommand{\fud}[2]{{}^{#1}{}_{#2}\,}
\newcommand{\fdu}[2]{{}_{#1}{}^{#2}\,}

\newcommand{\HS}{\text{HS}}
\newcommand{\besubeqs}{\begin{subequations}}
\newcommand{\esubeqs}{\end{subequations}}

\newcommand{\Verma}[3]{\ensuremath{\mathcal{D}\left(#1;#2,#3\right)}}
\newcommand{\Rac}{\ensuremath{\mathrm{Rac}}}
\newcommand{\Di}{\ensuremath{\mathrm{Di}}}
\newcommand{\Wi}{\ensuremath{\mathrm{Wi}}}
\newcommand{\Wib}{\ensuremath{\bar{\mathrm{Wi}}}}

\newcommand{\sWb}{\bar{\mathrm{W}}}
\newcommand{\sW}{{\mathrm{W}}}
\newcommand{\sD}{{\mathrm{D}}}
\newcommand{\Yy}[1]{\mathbb{Y}\left(#1\right)}

\newcommand{\LMUTitle}[9]{
  \thispagestyle{empty}
  \vspace*{\stretch{1}}
  {\parindent0cm
   \rule{\linewidth}{.7ex}}
  \begin{flushright}

    \vspace*{\stretch{1}}
    \sffamily\bfseries\Huge
    #1\\
    \vspace*{\stretch{1}}
    \sffamily\bfseries\large
    #2
    \vspace*{\stretch{1}}
  \end{flushright}
  \rule{\linewidth}{.7ex}
  \vspace*{\stretch{5}}
  \begin{center}
    \includegraphics[width=2in]{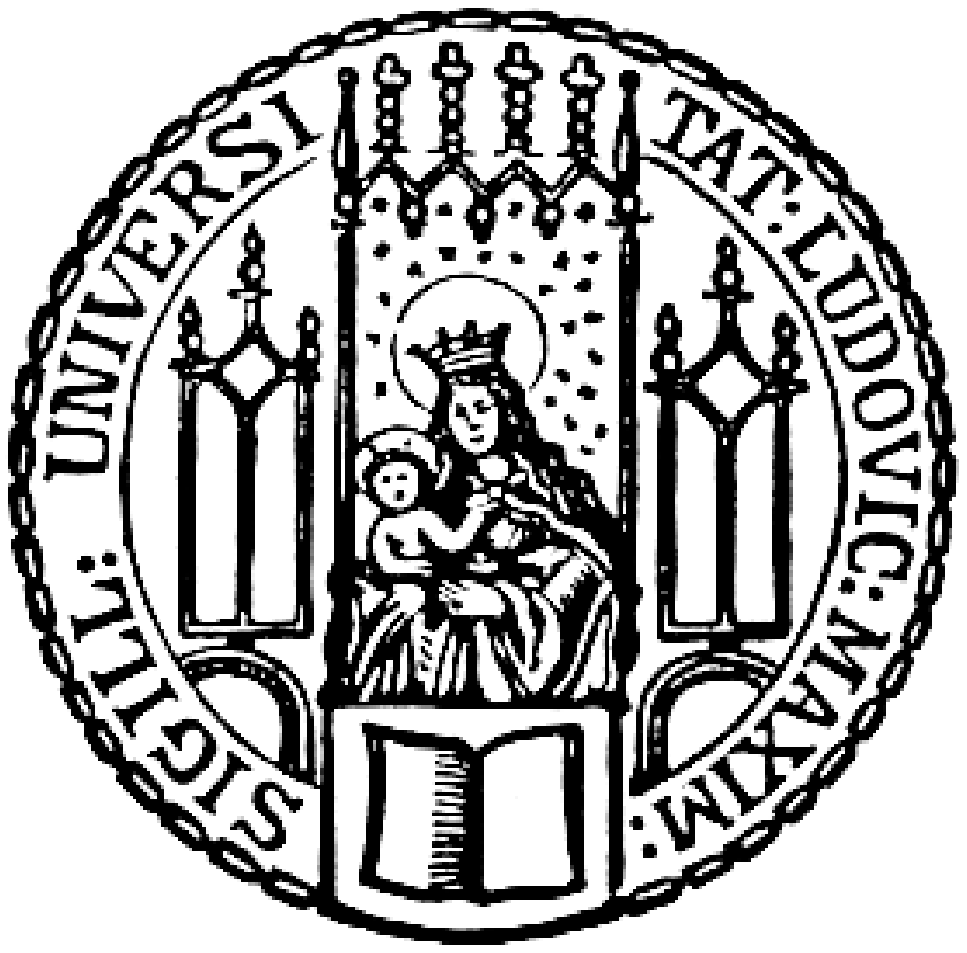}
  \end{center}
  \vspace*{\stretch{1}}
  \begin{center}\sffamily\LARGE{#5}\end{center}
  \newpage
  \thispagestyle{empty}

  \cleardoublepage
  \thispagestyle{empty}

  \vspace*{\stretch{1}}
  {\parindent0cm
  \rule{\linewidth}{.7ex}}
  \begin{flushright}
    \vspace*{\stretch{1}}
    \sffamily\bfseries\Huge
    #1\\
    \vspace*{\stretch{1}}
    \sffamily\bfseries\large
    #2
    \vspace*{\stretch{1}}
  \end{flushright}
  \rule{\linewidth}{.7ex}

  \vspace*{\stretch{3}}
  \begin{center}
    \Large Dissertation\\
    \Large at #4\\
    \Large Ludwig--Maximilians--Universitaet\\
    \Large M\"unchen\\
    \vspace*{\stretch{1}}
    \Large presented by\\
    \Large #2\\
    \Large from #3\\
    \vspace*{\stretch{2}}
    \Large Munich, #6
  \end{center}

  \newpage
  \thispagestyle{empty}

  \vspace*{\stretch{1}}
  \begin{flushleft}
  \setlength{\parskip}{\baselineskip}
    \Large \textbf{Dissertation}\\
    \noindent\hrulefill
    
    \large Submitted to the faculty of physics of the \\
    \large Ludwig-Maximilians-University Munich 
    
    \large by Tung Vuong Tran
    
    \large Supervised by Prof. Dr. Ivo Sachs \\
    \large Arnold Sommerfeld Center for Theoretical Physics , Munich, Germany 
    
    \large 1st Referee: #7 \\[1mm]
    \large 2nd Referee: #8 \\[1mm]
    \large 3rd Referee: Dr. Evgeny Skvortsov 
    
    \large Date of submission: 12/06/2020\\
    \large Date of oral examination: 29/07/2020
  \end{flushleft}

  \cleardoublepage
}

\begin{document}

  \LMUTitle
      {Higher Spin Gravity: Quantization and Algebraic Structures}               
      {Tung Vuong Tran}                       
      {Da Lat, Vietnam}                             
      {Faculty of Physics}                         
      {Munich}                          
      {12. June 2020}                            
      {Prof. Dr. Ivo Sachs}                          
      {Prof. Dr. Stefan Theisen }                         
      {Pr"ufungsdatum}                         
\clearpage

  \thispagestyle{empty}
\begin{quote}
    \textit{''It is not knowledge, but the act of learning, not possession but the act of getting there, which grants the greatest enjoyment.''}
    \\
    --- Carl Friedrich Gauss
\end{quote}

  \chapter*{Zusammenfassung}

Diese Dissertation ist den Quantenaspekten von Gravitationen höherer Spins (GRAHSs) und den ihnen zugrundeliegenden algebraischen Strukturen gewidmet. Theorien höherer Spins enthalten unendlichdimensionale Symmetrien, die m\"achtig genug sein sollten, um keine relevanten Gegenterme zuzulassen. Aus diesem Grund wird seit langem erwartet, dass GRAHSs endlich, oder zumindest renormierbar sind. Sobald gezeigt ist, dass diese Eigenschaft tats\"achlich realisiert wird, macht sie Theorien höherer Spins zu interessanten Quantengravitationsmodellen. Wenn das keine-Gegenterme-Argument funktioniert, reduziert sich das Problem, eine quantenkonsistente Theorie höherer Spins zu konstruieren, bemerkenswerterweise auf das Problem, ein konsistentes klassisches Modell von GRAHS zu finden.

Eine der interessantesten Klassen von GRAHSs ist die chirale GRAHS, die sowohl in der Minkowski- als auch in der AdS-Raumzeit existiert. Sie ist momentan die einzige Theorie mit propagierenden Feldern höherer Spins und einer recht einfachen Wirkung. Die Theorie ist auf perturbativer Ebene lokal. Die Wirkung der chiralen GRAHS ist in der Lichtkegel-Eichung bekannt und vermeidet alle Theoreme, welche die Existenz einer Theorie höherer Spins im flachen Raum verbieten. Wir studieren die Struktur der Quantenkorrekturen in der chiralen GRAHS im Minkowskiraum im Detail. Wir zeigen, dass, aufgrund einer nichttrivialen Kürzung unter den Feynmandiagrammen dank einer spezifischen Form der Wechselwirkungen (dem Kopplungs-Verschwörungs-Mechanismus), alle Baumniveau-Amplituden verschwinden; wir analysieren im Detail zwei-, drei- und vier-Punkt Einschleifenamplituden und zeigen, dass diese UV-konvergent sind. Mit Hilfe von Unitaritätsschnitten berechnen wir die komplette $n$-Punkt Einschleifenamplitude und zeigen, dass sie aus drei Faktoren besteht: (i) der Einschleifenamplitude in QCD oder SDYM mit allen Helizitäten plus; (ii) einem bestimmten kinematischen Verzierungsfaktor für höhere Spins; (iii) einem rein numerischen Faktor der Gesamtanzahl der Freiheitsgrade.

Im Kontext von AdS/KFT wird vermutet, dass GRAHSs dual zu recht einfachen konformen Feldtheorien (KFTs) sind: zu freien und kritischen Vektormodellen (Typ-A), freien Fermionen und Gross--Neveu-Modellen (Typ-B) und, allgemeiner, zu Chern--Simons-Materie-Theorien. Wir studieren im Detail die Vakuum-Einschleifenkorrekturen in verschiedenen Theorien höherer Spins in der anti-de Sitter (AdS) Raumzeit. Für die Typ-A-Theorie in $\mathrm{AdS}_{d+1}$ beweisen wir die Vermutung, dass die freie Energie für alle ganzzahligen Spins verschwindet und der freien Energie einer Kugel eines freien Skalarfeldes für alle geraden Spins gleicht. Wir erweitern dieses Resultat auf alle nicht-ganzzahligen Dimensionen und reproduzieren insbesondere die freie-Energie-Korrektur zur $4-\epsilon$ Wilson--Fisher KFT als einen Einschleifeneffekt in der Typ-A-Theorie auf $\mathrm{AdS}_{5-\epsilon}$. Wir berechnen ebenfalls die Beiträge fermionischer Felder höherer Spins, die für supersymmetrische GRAHS relevant sind. Es wird gezeigt, dass diese exakt mit der Vorhersage der KFT übereinstimmen. Der Beitrag bestimmter Felder gemischter Symmetrie, die in Typ-B GRAHS vorkommen, wird ebenfalls berechnet. Der letztere Beitrag führt (in geraden Raumzeitdimensionen) auf eine Frage, die zu beantworten bleibt.

Freie KFTs haben unendlichdimensionale globale Symmetrien, die in Algebras höherer Spins manifestiert sind. Die holographisch dualen GRAHSs sollten im Prinzip komplett durch diese Symmetrie bestimmt sein. Deshalb ist die einzige Information, die wir benötigen, um eine Theorie höherer Spins in AdS zu konstruieren, eine Algebra höherer Spins, die aus ihrer dualen freien KFT extrahiert werden kann. In dieser Dissertation rekonstruieren wir die Typ-A GRAHS in $\mathrm{AdS}_5$ auf der Ebene der formal konsistenten klassischen Bewegungsgleichungen (formale GRAHS).
  \chapter*{Summary}
This dissertation is dedicated to the quantum aspects of higher spin gravities (HSGRAs) and to their underlining algebraic structures. Higher-spin theories are governed by infinite-dimensional symmetries called higher-spin symmetries. Higher-spin symmetry should be powerful enough to leave no room for any relevant counterterms. Therefore, higher spin gravities have long been expected to be finite or at least renormalizable. This feature, once shown to be realized, makes higher-spin theories interesting toy models of Quantum Gravity. Remarkably, if the no-counterterm argument works, the problem of constructing a quantum consistent higher-spin theory downgrades to a problem of finding a consistent classical model of higher-spin gravity. 

One of the most interesting classes of HSGRAs is chiral HSGRA, which exists both in Minkowski and AdS spacetime. It is the only theory at present with propagating massless higher spin fields and a rather simple action. The theory is perturbatively local. The action of the chiral theory is known in the light-cone gauge and and avoids all No-Go theorems that forbid the existence of higher-spin theories in flat space. We study in detail the structure of quantum corrections in the Minkowski Chiral HSGRA. We show that all tree-level amplitudes vanish, which is due to a nontrivial cancellation among all Feynman diagrams thanks to the specific form of the interactions (coupling conspiracy mechanism); we analyze in detail two-, three- and four-point one-loop amplitudes and show that they are UV-convergent. Using unitarity cuts we compute the complete one-loop $n$-point amplitude and show that it consists of three factors: (i) all-plus helicity one-loop amplitude in QCD or SDYM; (ii) a certain kinematical higher spin dressing factor; (iii) a purely numerical factor of the total number of degrees of freedom.

In the context of AdS/CFT, HSGRAs are conjectured to be dual to rather simple conformal field theories (CFT): free and critical vector models (Type-A), free fermion and Gross-Neveu models (Type-B) and, more generally, to Chern-Simons Matter theories. We study in detail vacuum one-loop corrections in various higher-spin theories in anti-de Sitter (AdS) spacetime. For the Type-A theory in $\ads_{d+1}$ we prove the conjecture that the free energy vanishes for all integer spins and is equal to the sphere free energy of one free scalar field for all even spins. We extend this result to non-integer dimension and, in particular, reproduce the free energy correction to the $4-\epsilon$ Wilson-Fisher CFT as a one-loop effect in the Type-A theory on $\ads_{5-\epsilon}$. We also compute the contribution of fermionic higher spin fields that are relevant for supersymmetric HSGRA. These are shown to match precisely with the prediction of the CFT. The contribution of certain mixed-symmetry fields that appear in Type-B HSGRA is also computed. The latter leads to a puzzle (in even spacetime dimension) that remains to be resolved.

Free CFTs have infinite-dimensional global symmetries manifested in higher spin algebras. The holographic dual HSGRAs should, in principle, be completely determined by this higher spin symmetry. Therefore, to construct a higher-spin theory in AdS, the only initial data we need is a higher spin algebra extracted from its free CFT dual. In this thesis, we reconstructed the Type-A HSGRA in $\ads_5$ at the level of formally consistent classical equations of motion (Formal HSGRA).
  \chapter*{Publications}

This thesis is based on some of the author’s work during the period of November 2016 - August 2020 as a Ph.D. student at Ludwig Maximilian University in Munich and Max-Planck Institute for Gravitational Physics (Albert Einstein Institute) in Golm, Potsdam. The corresponding publications are:\\
{}\\
\cite{Gunaydin:2016amv}: \textbf{Exceptional $F(4)$ higher-spin theory in $AdS_6$ at one-loop and other tests of duality.}\\
Murat Günaydin, E.D. Skvortsov, Tung Tran.\\
\underline{Published in JHEP}: \textbf{\textcolor{RoyalBlue}{JHEP 1611 (2016) 168}}\\
\cite{Skvortsov:2017ldz}: \textbf{AdS/CFT in Fractional Dimension and Higher Spin Gravity at One Loop.}\\
E.D. Skvortsov, Tung Tran.\\
\underline{Published in Universe}: \textbf{\textcolor{RoyalBlue}{Universe 3 (2017) no.3, 61}}\\

\cite{Skvortsov:2018jea}: \textbf{Quantum Chiral Higher Spin Gravity 
}\\
Evgeny D. Skvortsov, Tung Tran, Mirian Tsulaia\\
\underline{Published in Physical Review Letters}: \textbf{\textcolor{RoyalBlue}{Phys.Rev.Lett. 121 (2018) no.3, 031601 }}\\
\cite{Skvortsov:2020wtf}: \textbf{More on Quantum Chiral Higher Spin Gravity}\\
Evgeny D. Skvortsov, Tung Tran, Mirian Tsulaia\\
\underline{Published in Physical Review D}: \textbf{\textcolor{RoyalBlue}{Phys.Rev.D 101 (2020) 10, 106001}}\\
\cite{Skvortsov:2020gpn}: \textbf{One-loop Finiteness of Chiral Higher Spin Gravity}\\
Evgeny D. Skvortsov, Tung Tran\\
\underline{Published in JHEP}: \textbf{\textcolor{RoyalBlue}{JHEP 07 (2020) 021}}\\

\cite{Sharapov:2019pdu}: \textbf{Towards Massless Sector of Tensionless Strings on AdS${}_5$}\\
Alexey Sharapov, Evgeny Skvortsov, Tung Tran\\
\underline{Published in Physics Letter B}: \textbf{\textcolor{RoyalBlue}{Phys.Lett. B800 (2020) 135094}}\\
${}$\\
The author also contributed to the following work:
{}\\
\cite{Sachs:2019wrk}: \textbf{On Non-Perturbative Unitarity in Gravitational Scattering}\\
Ivo Sachs, Tung Tran\\
\underline{Published in European Physical Journal C}: \textbf{\textcolor{RoyalBlue}{Eur.Phys.J. C79 (2019) no.11, 914}}

  \tableofcontents
  \markboth{Contents}{Contents}

  \markboth{Conclusion}{Conclusion}

  \mainmatter\setcounter{page}{1}
  \chapter{Introduction}

\section{Motivations}
Unquestionably, Quantum Field Theory (QFT) and General Relativity (GR) form the backbones of our theoretical frameworks to understand the universe. The triumph of QFT is that it successfully describes the dynamics of elementary particles with spin-$s\leq 1$ within a \textit{small zoo} of particles maybe known as the Standard Model \cite{Glashow:1961tr,Weinberg:1967tq,ASalam}, which has been verified to a remarkable level of precision through experiments down to subatomic distances of $\sim 10^{-19}$ m (or energy scale $\sim 10^4$ GeV). The Standard Model consists of matter fields and gauge bosons, which are the mediators of three out of four known fundamental interactions (electromagnetic, weak, and strong interactions) between visible matter. On the other hand, we have Einstein's gravity that describes the remaining force, gravity, which governs large scale physics where the corresponding mediator is graviton --- a spin-2 gauge boson. The most recent detection of gravitational waves by the LIGO collaboration \cite{Abbott:2016blz} and the image of Black Hole by the Event Horizon Telescopes \cite{Akiyama:2019cqa} showed that GR still endures as one of the most successful theories of all times, after one hundred-years from the original formulation \cite{Einstein:1916}. 

Despite these successes, the two pivotal realms of QFT and GR still resist
unification into an ultimate framework known as Quantum Gravity (QG) that should describe Nature all at once. Roughly speaking, the scale where features of Quantum Gravity becomes relevant is the Planck scale $l_p\sim$ $10^{-35}$ m (or $10^{19}$ GeV), where gravity should become strongly coupled. Therefore, GR should lose its predictive power the moment we approach the Planck scale. The very first evidence of the objection for unification was that the perturbative approach in QFT led to non-renormalizable UV divergences of GR starting from two-loops \cite{Goroff:1985th,vandeVen:1991gw}. Several notable attempts to soften this UV behaviors are superstring/M-theory, supergravity (SUGRA), higher spin gravity (HSGRA), etc. In these examples, the field content and symmetries are extended by considering supersymmetry, extra dimensions and higher spin states. Opposite to the general expectation that supersymmetry should be the main factor to make a gravitational theory UV-finite \cite{deWit:1977fk,Cremmer:1978ds,deWit:1982bul,Bern:2006kd}, it turns out that it is higher spin fields that are indispensable ingredients for UV divergence cancellations \cite{Veneziano:1968yb,Amati:1987wq}. Therefore, if we want to formulate a UV-finite QG \textit{perturbatively}, it seems unavoidable to introduce higher-spin fields. 

Seeking unification of the laws of physics has shown throughout the history of modern physics to be a fruitful approach to gain a deeper level of understanding of why things are the way they are. The intuition is that a more elegant formulation for pre-existing theories can bring new insights to the pathway toward a theory of everything. The prime examples are the theory of electromagnetism \cite{Maxwell:1865} and the unification of space and time into an entity \textit{spacetime} in special relativity \cite{Einstein:1905}. A more recent achievement was the electro-weak theory \cite{Salam:1959zz,Weinberg:1967tq}, a low-energy effective theory of the Standard Model, that unified electromagnetism and the weak force with the gauge group of $SU(2)\times U(1)$ at the energy of order $250$ GeV. Under spontaneous symmetry breaking of the group $SU(2)\times U(1)$, it gives rise to the masses of the $W$-bosons and the $Z$-boson in the Standard Model. If we take unification as the guiding principle for the search of the $final$ theory, one would naively expect that, by going to higher and higher energy, we will reach the ultimate consistent quantum theory. This theory has to contain the Standard Model and Einstein's theory as its low-energy effective field theories. At the moment, it is still impossible to experimentally access the energy where we can observe gravitational quantum effects, but it may be possible in not so distant future with the help of CMB measurements and with the dawn of the gravitational wave physics. Therefore the formulation of Quantum Gravity is driven mainly by finding examples utilizing unification, UV-completeness and symmetry as guiding principles.

String theory is a strong contender on the race towards the final theory since it contains an infinite tower of massive higher spin-fields and is UV-finite (at least up to two loops). Heuristically speaking, the finiteness of string theory is because there is no $point$-$like$ interactions. In particular, the string length $l_s$ is a natural UV cutoff that manifest in the theory and is related to the universal Regge slope $\alpha'$ as $l_s^2=\alpha'$. The $\alpha'$ is the $only$ free parameter in string theory, which makes it aesthetically pleasing compared to the Standard Model, where there are many free parameters that have to be fixed by experiments. There are two typical limits of $\alpha'$ that are usually considered: (i) the point-particle limit where $\alpha'\rightarrow 0$ in which string theory reduces to SUGRA \cite{Schwarz:1982jn,Green:1982sw}; (ii) the tensionless-limit proposed by Gross where $\alpha'\rightarrow \infty$ \cite{Gross:1988ue}. The former case, i.e. the point-particle limit, corresponds to the low energy limit of string theory and has been intensively studied in various contexts, see e.g. \cite{Hewett:1988xc,Kaplunovsky:1993rd,Derendinger:1985kk}. On the other hand, the tensionless-limit should correspond to trans-Planckian energy limit of string theory. In this limit, all the massive higher spin fields become effectively massless and therefore the theory should acquire an infinite-dimensional $gauge$ $symmetry$. This symmetry of a symmetry that is associated with massless higher spin fields is precisely the starting point of \textit{higher spin gravity theories} even though they have little to do with string theories at present. The first systematic attempts to construct higher spin theories were undertaken by Fronsdal \cite{Fronsdal:1978rb};  Brink, Bengtsson, Bengtsson \cite{Bengtsson:1983pd,Bengtsson:1983pg}; Fradkin and Vasiliev \cite{Fradkin:1987ks,Fradkin:1986qy}.

In this thesis, we will study various aspects of higher spin gravity (HSGRA). HSGRA are expected to be one of the simplest models of Quantum Gravity where the graviton becomes a part of the higher spin multiplet of massless higher spin gauge fields. The infinite dimensional symmetries should render higher spin gravities UV-finite. Therefore, the study of HSGRAs should shed more light on the Quantum Gravity Problem and can even lead to new insights into string theory. Indeed, HSGRA in $AdS_5\times S^5$, is conjectured to describe the tensionless limit of type IIB string theory \cite{Sundborg:2000wp,Sezgin:2002rt}. Besides its relation to string theory, HSGRA by itself is interesting because of the AdS/CFT Correspondence. In this context, HSGRAs should be dual to many interesting CFTs \cite{Klebanov:2002ja,Sezgin:2003pt,Giombi:2011kc} that describe real physics, with the most notable examples of Wilson-Fisher O(N) Vector Model (Ising) and Chern-Simons Matter theories \cite{Giombi:2011kc}. The latter class of theories have been recently conjectured to exhibit several remarkable dualities \cite{Maldacena:2012sf,Seiberg:2016gmd,Karch:2016sxi}. HSGRA indicates that, for example, the three-dimensional bosonization duality, at least in the large-N limit, should be a consequence of higher spin symmetries. 
\section{Overview}
In this thesis, we exclusively focus on the quantum aspect of HSGRA and its elegant algebraic structure. In particular, we will study HSGRA in the context of the AdS/CFT duality \cite{Maldacena:1997re,Gubser:1998bc,Witten:1998qj}. AdS/CFT is a remarkable relation between $d$-dimensional non-gravitational conformal field theory (CFT) as an \textit{image} on the boundary of a quantum gravity theory that lives in $d+1$-dimensional asymptotically anti-de Sitter spacetime (AdS):
\begin{align}\label{ads/cft}
    \text{AdS}_{d+1}\  \text{QG}=\text{CFT}_d
\end{align}
In principle, given two independent definitions of a CFT and its hypothetical AdS dual, the AdS/CFT correspondence can be proven by matching the CFT's correlation functions with the holographic S-matrix in AdS to all orders in coupling constant(s), i.e.
\begin{align}
    \langle \Ocal(x_1)...\Ocal(x_n)\rangle_{\text{CFT}} = \text{Holographic S-matrix}\,,
\end{align}
where $\Ocal(x_i)$ are operators at the points $x_i$ on the boundary of AdS. The duality \eqref{ads/cft} has shown its versatility by allowing us to study quantum gravity via its dual CFT and vice versa. In the original proposal by Maldacena \cite{Maldacena:1997re}, Type-IIB Superstring theory on $AdS_5\times S^5$ is conjectured to be dual to $\mathcal{N}=4$ superconformal Yang-Mill (SYM) theory on the boundary of $AdS_5$. Even though an outstanding progress in computing correlators of strongly coupled SYM has been achieved, see e.g. \cite{Aharony:2016dwx,Alday:2017xua,Alday:2017vkk}, we still know very little about how to compute correlation functions of (super-)string theories on AdS background. For this reason, the original form of the AdS/CFT correspondence remains a conjecture since there is no complete proof after two decades of efforts. 

The above conjecture, however, can be relaxed if we take some particular limits of the dimensionless 't Hooft coupling $\lambda=g_{YM}^2N$ where $N$ is the number of degrees of freedom on the CFT side which are supposed to be quantized. The  relation between $\lambda$ and the string length $l_s$ is given by 
\begin{align}\label{thoofcoupling}
    \lambda \sim \Big(\frac{R^2}{l_s^2}\Big)^{\frac{d}{2}}\,,
\end{align}
where $R$ is the AdS radius. This relation reveals the strong-weak nature of AdS/CFT duality where there are two particularly interesting limits: 
\begin{enumerate}
    \item The strongly coupled limit of CFTs, which is also known as the particle limit in AdS, is the limit $\frac{l_s^2}{R^2}\rightarrow 0$.
    \item The tensionless limit (high energy limit) of string theories \cite{Sundborg:2000wp,Sezgin:2002rt} in which $\frac{l_s^2}{R^2}\rightarrow\infty$.
\end{enumerate}
In the latter limit, i.e., the tensionless limit, the dual CFT is essentially free \cite{HaggiMani:2000ru,Sundborg:2000wp,Tseytlin:2002gz,Aharony:2006th} because the 't Hooft coupling $\lambda\rightarrow 0$. Free CFTs have higher-spin symmetry due to the emergence of an infinite tower of conserved currents. As we will show later, higher-spin symmetry is a global symmetry on the CFT side, and it defines a so-called higher-spin algebra \textbf{hs}, which is a crucial ingredient of the construction of the dual bulk theory. Moreover, the study of the bulk theory should be more feasible due to the infinite-dimensional higher-spin symmetry. Up to date, the tensionless limit of superstring theory is understood only in $\ads_3$ \cite{Eberhardt:2018ouy}, and there is no known description in higher dimensions. Moreover, the tensionless limit by itself does not have to lead to any weakly-coupled field theory description.  Therefore, the emergence of higher spin gravity on $\ads_{d+1}$ for $d\geq 3$ remains a mystery. On account of AdS/CFT correspondence, we can start from the \textit{easier} side, i.e., the CFT side, where the field theories are much better understood instead of directly dealing with the bulk theory. 

In any free CFT, we can construct conserved higher spin rank-$s$ tensors, $\JJ_s\equiv \JJ_{a_1...a_s}$, $s=1,2,...,\infty$, also called higher spin currents. These conserved higher spin tensors are bilinear in matter fields (scalar, fermion or massless spin-one field) that can take values in various representation of a gauge group. Let us, for example, consider free scalars in the fundamental representation, i.e. vector models. The AdS/CFT correspondence then tells us that the same higher spin symmetry should govern the dual gravitational theories in bulk. The bulk theory should contain higher spin fields $\Phi_s\equiv\Phi_{\um_1...\um_s}$ (with $s=0$ being the scalar field that is dual to $\JJ_0=\bar{\phi}\phi$), \footnote{The $underlined$ indices live in AdS with one extra dimension compared to $a_i$.} which are massless. Based on these observations, Klebanov and Polyakov conjectured that the gravitation AdS-dual of vector models should be given by some HSGRA \cite{Klebanov:2002ja}:
\begin{align}\label{HS/Vectormodel}
    \text{AdS}_{d+1}\; \text{unbroken HSGRA} = \text{ CFT}_d \,\text{Free Vector Model}.
\end{align}
This equation indicates that unbroken HSGRAs are expected to be dual to free CFTs  \cite{Sundborg:2000wp,Sezgin:2002rt,Klebanov:2002ja,Fernando:2015tiu}. The fact that matter fields of vector models lie inside the fundamental representation rather than the adjoint \cite{Klebanov:2002ja}  simplifies the spectrum of single-trace operators and reduces the field content of the dual bulk theory as compared to string theory.\footnote{The spectrum of String theory has a finite number of massless states and an infinite set of massive states. In particular there are infinite states with the same spin. On the contrary, the spectrum of the simplest HSGRA can contain all integer spins, each in one copy. } In the case where the matter field transforms in the fundamental representation, upon changing boundary conditions of the bulk scalar field with $s=0$, the HS theory is no longer dual to free but to the critical vector model (weakly coupled CFT in the large-$N$). Astonishingly, by studying HSGRA, we can, at the same time, understand weakly coupled CFT's that describe physics of critical phenomena.  

When matter fields are in the adjoint representation, i.e. matrix valued fields, the spectrum of single-trace operators is much bigger. In particular, Sundborg conjectured that HSGRA theory in $AdS_5\times S^5$ should be dual to free $\mathcal{N}=4$ SYM \cite{Sundborg:2000wp}. 

The higher-spin theories in AdS are \textit{not} quite conventional field theories: (i) they usually contain infinitely-many fields; (ii) higher spin fields require higher derivative interactions, as a result, the number of derivatives is unbounded; (iii) their relation to the poorly understood tensionless limit of string theory. This makes generic HSGRAs hard to study and construct. In particular, the most canonical way of constructing theories, the Noether procedure is not applicable \cite{Sleight:2017pcz}. There are, however, three well-defined theories with local enough interactions that luckily avoid and evade the numerous no-go theorems that have been proven over the years:
\begin{enumerate}
    \item Three-dimensional higher spin theory \cite{Blencowe:1988gj,Bergshoeff:1989ns,Campoleoni:2010zq,Henneaux:2010xg} (a generalization of the Chern-Simons formulation of $3d$ gravity); 
    \item Four-dimensional Conformal higher spin gravity \cite{Segal:2002gd,Tseytlin:2002gz,Bekaert:2010ky}, which is an extension of conformal gravity. There is, in some sense, a combination of the previous two cases --- conformal HSGRA in three dimensions \cite{Pope:1989vj,Fradkin:1989xt,Grigoriev:2019xmp};
    \item Four dimensional Chiral HSGRA \cite{Metsaev:1991mt,Metsaev:1991nb,Ponomarev:2016lrm,Ponomarev:2017nrr,Metsaev:2018xip,Skvortsov:2018uru}.
\end{enumerate} 
Nevertheless, a general expectation is that the $\infty$-dimensional higher-spin symmetry should be substantial to fix any meaningful physical observable. An essential set of such observables are encoded in the (holographic) S-matrix, which should be equivalent to CFT correlators. At least for the case of $unbroken$ higher spin symmetries, the symmetry itself can unambiguously fix all correlation functions \cite{Maldacena:2011jn,Boulanger:2013zza,Alba:2013yda,Alba:2015upa} and imply that they are those of the same free CFT that generated the higher spin algebra. In fact, the free CFT's correlation functions are the simplest higher spin invariants \cite{Colombo:2012jx,Didenko:2013bj,Didenko:2012tv,Bonezzi:2017vha}. Therefore, to a large extent, one can avoid non-locality problems if one sticks to the higher spin invariant observables rather than to the problems of its formulation as a local field theory (see e.g. \cite{Sezgin:2011hq,Sharapov:2018kjz,Sharapov:2019vyd} for a discussion). 

The study of higher-spin theories with \textit{interactions} has a long history because there are first of all many No-Go theorems \cite{Weinberg:1964ew,Coleman:1967ad,Weinberg:1980kq} that forbid higher-spin interaction in flat space and AdS \cite{Bekaert:2015tva,Maldacena:2015iua,Sleight:2017pcz,Ponomarev:2017qab}. While these theorems restrict the $S$-matrix of Minkowski HSGRA to be trivial, they have little to say about local effects. Intriguingly, using the light-front approach \cite{Bengtsson:1983pd,Bengtsson:1983pg}, one can show that local cubic interaction for any triplet of spins does exist \cite{Metsaev:2005ar}. Moreover, the solution that respects Poincare (conformal) symmetries is called \textit{chiral} HSGRA, which exists both in flat and AdS spacetime \cite{Metsaev:1991mt,Metsaev:1991nb,Ponomarev:2016lrm,Metsaev:2018xip,Skvortsov:2018uru}. 

Just a few years after Fronsdal started the entire higher spin programme and after the first light-front results were obtained, Fradkin and Vasiliev showed that cubic vertices for higher spin fields could be written covariantly on (A)dS backgrounds \cite{Fradkin:1987ks,Fradkin:1986qy}, including the gravitational interactions. Later, Vasiliev himself constructed the higher-spin system (or Vasiliev's equations) that has the form:
\begin{subequations}\label{vasiliev}
\begin{align}
d\omega&=\omega\star \omega +\Vcal_3(\omega,\omega,C)+\Vcal_4(\omega,\omega,C,C)+... \\ 
dC&=\omega\star C-C\star \pi(\omega)+\Vcal_3(\omega,C,C)+\Vcal_4(\omega,C,C,C)+...
\end{align}
\end{subequations}
The system has several important features. It is background-independent. The fields here are the one-form connection of the higher spin algebra $\omega$ and $C$ is a zero-form taking values in the same algebra. The bilinear terms are fixed entirely by the higher spin algebra. The higher order terms, interaction vertices, were fixed by requiring the equations to be \textit{formally consistent}, i.e. they are consistent with $dd\equiv0$, where $d$ is the exterior derivative. The price to pay is that there is an infinite set of auxiliary fields which are economically packed inside $\omega$ and $C$. To date, there is no known action that can be used to derive the above system,\footnote{In a metric-like formalism, a part of the higher-spin action is known \cite{Bekaert:2015tva} from holography.} but there is a somewhat non-standard action \cite{Boulanger:2011dd}. This is in contrast with the examples of HSGRAs above, where not only do the theories exist, but they also have rather simple actions.

There is however a big difference between being formally consistent and being actually consistent in the sense of giving concrete and well-defined predictions for higher spin interactions \cite{Boulanger:2015ova}. The problem is that $C$ encodes an unbounded number of derivatives of the physical fields. Therefore, expressions that are nonlinear in $C$ can easily form infinite sums over all derivatives that are problematic and disagree with the known interactions of higher spin fields.\footnote{The problem has little to do with HSGRA. Any field theory's equations, e.g. pure gravity, can be written in a form similar to \eqref{vasiliev} and will lead to some zero-form $C$ that encodes an unbounded number of derivatives of the physical fields. Therefore, the vertices have to be constrained more than just by the formal consistency. } As a result, it is not known how to systematically extract correct interactions out of \eqref{vasiliev}. Nevertheless, \eqref{vasiliev} captures certain algebraic structures of interactions that are hardly accessible via perturbative methods like the Noether procedure.

Let us summarize some of the proposals and methods to study higher-spin theories. These include:
\begin{itemize}
    \item Noether procedure: a canonical perturbative method to introduce interactions with the requirement that the full action $S$ has to be gauge invariant, namely
    \begin{align}
        0=\delta S=\delta_0S_2+\delta_0S_3+\delta_1S_2+...\,.
    \end{align}
    Here, $S_2$ is the free action and $\delta_0$ is a linearized gauge transformation. In addition, we also have $S_{3,4,...}$ --- higher order interactions, and $\delta_{1,2,...}$ --- the field-dependent deformation of the gauge transformations. Due to the conceptual difficulty of understanding locality in higher-spin theories, a complete example of HSGRA constructed via the Noether procedure is not known. It seems that higher spin theories can not be conventional field theories due to non-local interactions \cite{Sleight:2017pcz}. 
    
    \item Light-cone approach: the main idea is to construct the charges of the Poincare (or any other spacetime symmetry) algebra directly in terms of physical degrees of freedom. Whenever a covariant formulation is available, one can simply impose the light-cone gauge. Then, all unphysical degrees of freedom are gone and the stress-tensor generates all the required charges. The power of the light-cone approach is that it helps to study the problem of interactions in full generality without having to use one or another covariant realization of a given set of physical degrees of freedom (there can be many such realizations that are not equivalent as far as the problem of interactions is concerned). More about the light-cone approach can be found in Chapter \ref{chapter4}.
    
    \item Reconstruction: an approach that reconstructs the bulk theory through information obtained from a given conjectural CFT dual. For type-A HSGRA, the cubic action and some part of the quartic action were reconstructed in \cite{Sleight:2016dba,Bekaert:2015tva}. In this approach, AdS/CFT is automatically being proved (or better say, trivialized) at classical level since the reconstructed interactions give exactly the correlation functions we started with. The main issues here are whether the reconstructed action is local enough for it to be taken seriously (for free CFT's it is not and the non-localities are yet to be tamed) and what happens at the quantum level.
    
    \item Collective dipole pushes the idea of reconstruction till the end. It was so far applied only for $\ads_4/\cft^3$. The fields on the CFT side are bi-local fields $\Psi(x,y)=\sum_i\phi^i(x)\phi^i(y)$. Here $\phi^i(x)$ are spacetime scalars that are $O(N)$-vectors, $i=1,...,N$. The idea is to take the free/critical vector model path integral and change the integration variables from $\phi^i$ to $\Psi(x,y)$. The latter can be interpreted as higher spin fields in $\ads_4$. Due to the change of variables, there is a non-trivial Jacobian $\log J =\frac{N}{2}\Tr \log \Psi$ in the partition function. It is conjectured that the action in terms of the bi-local fields $\Psi$ is equivalent to the action of a higher spin theory in $\ads_4$ \cite{Koch:2010cy}. It reproduces all correlation functions by construction. 
    \item IKKT matrix model for the fuzzy sphere: an idea that treats space-time as a dynamical physical system with intrinsic quantum structure by studying the IKKT matrix model \cite{Steinacker:2019fcb}. There is a specific solution whose internal structure leads to a consistent and ghost-free higher-spin gauge theory. 
    
    \item Formal HSGRA: this is essentially an approach suggested by Vasiliev's equations \eqref{vasiliev}, i.e. to try to understand the deformation of higher-spin symmetries caused by \eqref{vasiliev}-like equations. It turns out that the interaction vertices can be derived from a strong homotopy algebra that can be constructed in a simple way for any given higher spin algebra \cite{Sharapov:2018kjz,Sharapov:2019vyd}. We will apply this in Chapter \ref{chapter5} to HSGRA in $\ads_5$.
\end{itemize}

\section{Summary of this thesis}
We aim to \textit{unfold} some of the key features of higher spin theories in this thesis. In particular, we will focus on UV-finiteness and algebraic structures of HSGRA both in flat and AdS spacetime.
\subsection{Main results}
As is already mentioned, higher-spin theories share some features with string theory like an infinite tower of fields with ever increasing spin. In some cases the interactions in HSGRA are also known to be non-local. Generically, the spectrum of \textit{non-minimal} higher spin model involves massless fields with spin-$s=0,1,2,...,\infty$, and there exists a truncation to $minimal$ model which has only infinitely many $even$ spins. It is important to note that the minimal model is so far the only consistent truncation from the non-minimal one. Any attempt to consider a finite subset of higher spin fields (or drop some spins from the even ones) will lead to an inconsistency. Higher spin fields are gauge fields and they have linearized gauge symmetries of the form $\delta \Phi_{\um_1...\um_s}=\nabla_{(\um_1}\xi_{\um_2...\um_s)}$. Here $\xi_{\um_1...\um_{s-1}}$ are the corresponding gauge parameters. Then, for any massless gauge field, there will be a dual higher spin conserved tensors, $\JJ_s$, on the CFT side and vice versa, i.e.
\begin{align}\label{current-field}
    \pl^b\JJ_{ba_2...a_s}=0 \quad \Longleftrightarrow \quad \delta \Phi_{\um_1...\um_s}=\nabla_{(\um_1}\xi_{\um_2...\um_s)}\,.
\end{align}
We will assume the HS/vector model duality conjecture only at the level of the basic dictionary \eqref{current-field}, and audaciously try to construct the bulk theory from the CFT side. This construction can be referred as a \textit{reconstruction} approach. 
Together with \eqref{current-field}, if we further assume that AdS/CFT holds at the classical level, the cubic action and part of the quartic action can be completely determined \cite{Bekaert:2015tva}. For free CFTs, the computation of correlation functions is straightforward by utilizing Wick's contraction. We can then use the knowledge gained from these correlators to infer the form of interactions of the bulk theory since they should be equivalent to Witten-diagrams in AdS. Schematically, the action for a spin-$s$ field reads
\begin{align}\label{Fronsdal}
    S=\sum_s\int \frac{1}{2}\Phi_s\big(\square-M^2_s+...\big)\Phi_{s}+\int V_3(s_1,s_2,s_3)[\Phi]+\int V_4(s_1,s_2,s_3,s_4)[\Phi]+...\,.
\end{align}
In principle, to show that HSGRA is UV-finite, we should compute loop-diagrams in AdS to see whether there is any divergence. However, since we do not know the full action, it is not yet possible to compute the full one-loop self-energy or the beta function that should provide access to the quantitative quantum properties of HSGRA. Moreover, direct loop calculation in AdS is very challenging, see, e.g. \cite{Giombi:2017hpr,Bertan:2018khc,Ponomarev:2019ofr}. Fortunately, knowledge of kinetic terms is sufficient to calculate the one-loop determinant of HSGRAs in AdS that, in turn, can tell us a little bit about UV behavior of HSGRAs. This idea was first suggested and applied in \cite{Giombi:2013fka}, where it was shown the one-loop determinant in Type-A HSGRA can be regularized and computed. Moreover, the computation gives the results that are consistent with $\ads/\cft$ under certain assumptions.

The essential ingredient for HSGRA UV-finiteness is precisely the infinite-dimensional higher spin symmetries that alleviate UV-divergence. 
For the case of chiral HSGRA in flat space, we show that the theory is consistent at both classical and quantum levels. Moreover, it is not in contradiction with the No-Go theorems since the full $S$-matrix is 1, thanks to higher spin symmetries \cite{Skvortsov:2018jea,Skvortsov:2020wtf}. The theory is constructed in light-cone gauge \cite{Metsaev:2005ar,Ponomarev:2016lrm} and has a complete action of the following schematic form
\begin{align}
    S=-\sum_{\lambda}\int \Phi_{\lambda}^{\dagger}\pvec^2\Phi^{\lambda}+\int V_3(\lambda_1,\lambda_2,\lambda_3)[\Phi]\,.
\end{align}
Chiral HSGRA in flat space is the first known example of a quantum HSGRA. We expect that the cousin of chiral HSGRA in AdS \cite{Metsaev:2018xip,Skvortsov:2018uru} should also exhibit the same features even though the loop computation in AdS can be a challenge. 

Beside the ability of rendering HSGRA UV-finite, higher spin symmetry is also useful to \textit{formally} construct HSGRA via formally consistent equations of motion \cite{Sharapov:2018kjz,Sharapov:2019vyd}. Indeed, using higher spin algebra (HSA) as the \textit{only} input from the free CFT, we show that the equations of motion \eqref{vasiliev} of bulk theory can be constructed by deforming the (HSA) with an explicit example of higher spin theory in AdS${}_5$. Generically, the HSA is nothing but the quotient of the universal enveloping algebra of $so(d,2)$ (the conformal algebra for AdS${}_{d+1}$ by the two-sided Joseph ideal $\mathcal{I}$. For the case of AdS${}_5$ the \textit{seed} that generates the universal enveloping algebra is $su(2,2)\sim so(4,2)$. Following the procedure in \cite{Sharapov:2018kjz,Sharapov:2019vyd}, we can obtain the equations of motion of the bosonic HSGRA by deforming the following commutation relation
\begin{align}\label{deformPP}
    [P_{AB},P_{CD}]=(1+\nu \kappa)\big(L_{AD}C_{BC}-L_{BD}C_{AB}-L_{AC}C_{BD}+L_{BC}C_{AD}\big)
\end{align}
by a formal deformation parameter $\nu$, while keeping all other relations of the conformal algebra intact to preserve local Lorentz algebra and its action on tensors. In fact, beside being a seed that drives the whole deformation, the above deformed $[P,P]$-commutator leads to the vacuum Einstein’s equations. Supersymmetric extension should also be possible using the same procedure even though the algebra may look a bit more sophisticated.

\underline{The main results in this thesis are}: 
\begin{enumerate}
    \item In \cite{Gunaydin:2016amv,Skvortsov:2017ldz}, we computed the vacuum one-loop effect in the dual HSGRAs in integer dimensions and fractional dimensions which opened up possibilities to study AdS/CFT in non-integer dimension. The results for type-A HSGRA and fermionic higher fields match precisely with the AdS/CFT's predictions. We discovered that the results of type-B HSGRA in even dimensions lead to a puzzle that calls for a better understanding of the duality. 
    
    \item In \cite{Skvortsov:2018jea,Skvortsov:2020wtf}, we explicitly showed how the cancellation of UV-divergences happens for the case of chiral HSGRA in flat space. The same pattern for UV-cancellation in chiral HSGRA should extend to any classes of higher spin theories. 
    
    \item In \cite{Sharapov:2019pdu}, we constructed formally consistent equations of motion of bosonic HSGRA in AdS${}_5$. The supersymmetric version of this theory, with the gauge symmetry $psu(2,2|4)$, should describe the tensionless limit of type-IIB superstring theory in $AdS_5\times S^5$.
\end{enumerate}  
\subsection{Outline}
We outline the thesis in the following:\\
${}$\\
In chapter \ref{chapter2}, we review some standard knowledge for HSGRA in metric-like, light-front and frame-like formalisms. Moreover, we also provide some fundamental concepts of AdS/CFT that are relevant in this thesis.

In chapter \ref{chapter3}, we prove the conjecture by Giombi, Klebanov et al. \cite{Giombi:2013fka,Giombi:2014iua,Giombi:2014yra} that the free energy of both free and critical Vector Models can be reproduced as a one-loop effect in the dual HSGRAs. In particular, we perform many one-loop tests for various HSGRAs on different backgrounds. Moreover, following an earlier idea by Klebanov and Polyakov, that HSGRAs/Vector Models duality \cite{Klebanov:2002ja} may also be extended to fractional dimensions, we recover the free energy of the $4-\epsilon$ Wilson-Fisher CFT \cite{Giombi:2014xxa} from the dual HSGRA in $\ads_{5-\epsilon}$, \cite{Skvortsov:2017ldz}. 

In chapter \ref{chapter4}, which is based on the original work \cite{Skvortsov:2018jea}, we show how UV divergences of HSGRAs get canceled due to interactions fine-tuned by higher spin symmetries. This effect of cancellation of UV-divergences is the most important feature of HSGRAs, which makes them models of Quantum Gravity. We observe this phenomenon in the chiral HSGRA, a class of HSGRAs \cite{Ponomarev:2016lrm} that has an action in the light-cone gauge and exists in flat and AdS backgrounds \cite{Metsaev:2018xip,Skvortsov:2018uru}.

We dedicate chapter \ref{chapter5} to the discussion of the algebraic structure of HSGRAs. Using the $A_{\infty}$-algebra, which can be constructed from a higher spin algebra \cite{Sharapov:2018kjz}, we obtain formally consistent equations of HSGRA on AdS${}_5$ \cite{Sharapov:2019pdu}. Although HSGRAs are self-contained models of Quantum Gravity, they may emerge in the tensionless limit of string theories. Therefore, supersymmetric extension of our result with the gauge symmetry $psu(2,2|4)$ should describe the massless sector of tensionless type-IIB strings on
$AdS_5\times S^5$ \cite{Sundborg:2000wp,Sezgin:2002rt}.

In chapter \ref{chapter6}, we summarize the main results of this thesis and discuss HSGRA's current state of the art.

We collect various technical details in the Appendices.
  \chapter{Review of Higher Spin Theories}\label{chapter2}

The study of higher-spin fields has a long history (see \cite{Bekaert:2010hw,Boulanger:2008tg} for a summary). The most relevant starting point for us is the work of Fronsdal \cite{Fronsdal:1978rb}. In this chapter, we review free HSGRA in the metric-like, light-front and frame-like formalisms. We also discuss some basic concepts of AdS/CFT paying attention to the case of HSGRA/Vector Model duality. The metric will come with the convention of mostly \textit{plus} components.
\section{Metric-like Formalism for HSGRA}
Since the birth of QFT, there have been many No-Go results, see e.g.  \cite{Weinberg:1964ew,Coleman:1967ad,Weinberg:1980kq,Boulanger:2008tg},  that constrain interactions between massless higher-spin fields. In other words, these theorems may rule out all QFTs with interactions whenever there are gauge fields with spin-$s>2$. However, the equation and action for free higher spin fields are known thanks to Fronsdal and Fang \cite{Fronsdal:1978rb,Fang:1978wz}. The presentation in this section follows \cite{Sorokin:2004ie,Didenko:2014dwa,Kessel:2016hld}. For simplicity, we will discuss higher spin fields with integer spins.
\subsection{Flat Space}
We adopt the following convention: a totally symmetric rank-$s$ tensor $T_{(\mu_1...\mu_s)}$ will be denoted as $T_{\mu(s)}$ for short. When indices are on the same level and denoted by the same letter, it means symmetrization is already applied, e.g. $\partial_{\mu}T_{\mu}\equiv \partial_{(\mu_1}T_{\mu_2)}$. Then, the higher spin equation for a free spin-$s$ field (this is also known as Fronsdal's equation) reads \cite{Fronsdal:1978rb}
\begin{align}\label{eq:Fronsdal}
    \square \Phi_{\mu(s)}-\partial_{\mu}\partial^{\nu}\Phi_{\nu\mu(s-1)}+\partial_{\mu}\partial_{\mu}\Phi_{\mu(s-2)\nu}^{\qquad \ \ \nu}=0. 
\end{align}
The above equation is invariant under the following gauge transformation
\begin{align}\label{eq:Fronsdalgauge}
    \delta \Phi_{\mu(s)}=\partial_{\mu}\xi_{\mu(s-1)}, \qquad \qquad  \xi_{\mu(s-2)\nu}^{\qquad \ \  \nu}=0.
\end{align}
It is not hard to see that the equation \eqref{eq:Fronsdal} is a generalization of the free equation of motion for massless fields of spin-$s=0,1,2$. The trace constraint on the gauge parameter is crucial for gauge invariance of the field $\Phi_{\mu(s)}$. We also need a somewhat \textit{unusual} constraint that $\Phi_{\mu(s)}$ should be \textit{double-traceless}, i.e.
\begin{align}
    \Phi_{\mu(s-4)\nu\sigma}^{\qquad \quad \  \nu\sigma}\equiv0.
\end{align} 
Note that the trace constraint of $\xi_{\mu(s-1)}$ in \eqref{eq:Fronsdalgauge} does not exist for the case of lower spin. In order to see that the solutions to the equation \eqref{eq:Fronsdal} is unique and carry a spin-$s$ representation of the Poincare group, we can impose \textit{transverse-traceless} (TT) gauge: $\pl^{\nu}\Phi_{\nu\mu(s-1)}=0$, $\Phi_{\mu(s-2)\nu}^{\qquad \ \  \nu}=0$, then the gauge-fixed equations and constraints read
\begin{subequations}\label{eq:flatgauge-fixed}
\begin{align}
    \square \Phi_{\mu(s)}&=0, && &\square \xi_{\mu(s-1)}&=0,\\
    \partial^{\nu}\Phi_{\nu\mu(s-1)}&=0, && &\partial^{\nu}\xi_{\nu \mu(s-2)}&=0,\\
    \Phi_{\mu(s-2)\nu}^{\qquad \ \ \nu}&=0, && &\xi_{\mu(s-3)\nu}^{\qquad \ \ \nu}&=0,\\
    \delta\Phi_{\mu(s)}&=\partial_{\mu}\xi_{\mu(s-1)}.
\end{align}
\end{subequations}
In Fourier space, the gauge-fixed equation $\square\Phi_{\mu(s)}=0$ implies $p^2=p_{\mu}p^{\mu}=0$, i.e. masslessness. To make the discussion transparent, we can go to light-cone coordinates with the metric $\eta_{+-}=\eta_{-+}=1,\  \eta_{ij}=\delta_{ij}$. We can take $p^{\mu}=a\delta_+^{\mu}$ with $a$ being some constant, where $\mu=+,-,i$  and $i=1,...,d-2$. Then, the constraint $\partial^{\nu}\Phi_{\nu\mu(s-1)}=0$ tells us that $\Phi_{+\mu(s-1)}=0$, i.e. all components of $\Phi_{\mu(s)}$ that carry at least one $+$-direction vanish. Next, the gauge symmetry $\delta \Phi_{\mu(s)}=\partial_{\mu}\xi_{\mu(s-1)}$ implements $\Phi_{-...-i...i}=0$. Therefore, the non-vanishing components for a totally symmetric spin-$s$ field are $\Phi^{i(s)}(p)$. One can check that $\Phi^{i(s)}$ is $so(d-2)$ traceless ($\Phi^{i(s-2)jk}\delta_{jk}=0$) and therefore describes a spin-$s$ particle according to Wigner's classification (see e.g. \cite{Bekaert:2006py,Didenko:2014dwa, QFTbookofWeinbergVolume1} for detailed discussion). The gauge-fixed action has a simple form
\begin{align}
    S_{\text{gf.}}=\frac{1}{2}\int_{\text{M}^d} \Phi^{\mu(s)}\square\Phi_{\mu(s)}\,.
\end{align}
\subsection{Anti-de Sitter Space}
Similarly, with flat space, we can analyze a free massless spin-$s$ gauge field on AdS${}_d$ background with the metric $g_{\mu\mu}$ --- the maximally symmetric solutions of Einstein equations with cosmological constant $\Lambda<0$. To do so, we replace partial derivatives,  $\partial$, with covariant derivatives, $\nabla$. The commutator in our convention reads
\begin{align}\label{eq:covariantderivatives}
    [\nabla_{\mu},\nabla_{\nu}]V_{\rho}=\Lambda(g_{\mu\rho}V_{\nu}-g_{\nu\rho}V_{\mu})\,.
\end{align}
The double-traceless condition becomes
\begin{align}
    \Phi_{\mu(s-4)\nu\nu\rho\rho}g^{\nu\nu}g^{\rho\rho}=0\,.
\end{align}
The gauge transformation is now
\begin{align}
    \delta\Phi_{\mu(s)}=\nabla_{\mu}\xi_{\mu(s-1)}, \qquad \qquad \xi_{\mu(s-3)\nu\nu}g^{\nu\nu}=0.
\end{align}
The Fronsdal equation gets lifted to
\begin{align}\label{eq:FronsdalAdS}
    &\square\Phi_{\mu(s)}-\nabla_{\mu}\nabla^{\nu}\Phi_{\nu\mu(s-1)}+\frac{1}{2}\nabla_{\mu}\nabla_{\mu}\Phi_{\mu(s-2)\nu}^{\qquad \ \  \nu}-M_s^2\Phi_{\mu(s)}+2\Lambda g_{\mu\mu}\Phi_{\mu(s-2)\nu}^{\qquad \ \ \nu}=0,\\
    & \text{where}\quad M_s^2=-\Lambda\Big[(s-2)(d+s-3)-s\Big]\,.
\end{align}
The factor $\frac{1}{2}$ appears in the third term because of non-commutativity of covariant derivatives \eqref{eq:covariantderivatives}. One needs $s(s-1)$ terms to symmetrize over $\mu(s)$ in AdS, while in flat space it only needs $s(s-1)/2$ terms. The \textit{mass-like} terms, the last two terms in \eqref{eq:FronsdalAdS}, appear due to gauge invariance requirement. Next, we impose TT gauge where fields are traceless and are $\nabla$-transverse ($\nabla\cdot \Phi=0$) as in flat space. In this gauge, the equations of motion reduce to
\begin{subequations}
\begin{align}
    (\square-M_s^2) \Phi_{\mu(s)}&=0, && &(\square-m_{s-1}^2) \xi_{\mu(s-1)}&=0,\\
    \nabla^{\nu}\Phi_{\nu\mu(s-1)}&=0, && &\nabla^{\nu}\xi_{\nu \mu(s-2)}&=0,\\
    \Phi_{\mu(s-2)\nu}^{\qquad \ \ \nu}&=0, && &\xi_{\mu(s-3)\nu}^{\qquad \ \ \nu}&=0,\\
    \delta\Phi_{\mu(s)}&=\nabla_{\mu}\xi_{\mu(s-1)}.
\end{align}
\end{subequations}
The gauge-fixed action in AdS reads
\begin{align}
    S_{\text{gf.}}=\frac{1}{2}\int_{\text{AdS}_d}\Phi^{\mu(s)}(\square-M_s^2)\Phi_{\mu(s)}\,. 
\end{align}
\section{Light-front Formalism for HSGRA}
It is sometimes more convenient to describe massless fields in the light-cone gauge \cite{Bengtsson:1983pd,Bengtsson:1983pg,Metsaev:2005ar}. The reason are that 
\begin{itemize}
    \item We can work directly with physical degrees of freedom and this is the most general approach to local dynamics. Therefore, unitarity is manifest;
    \item One avoids ambiguities that arise in manifestly covariant formulations, e.g. the same degrees of freedom can be embedded into different tensor fields; 
    \item Doing computation in light-cone gauge is rather simple compared to some other approaches. 
\end{itemize}
For practical purposes, we will only review free massless higher spin fields in four dimensional Minkowski and AdS.
\subsection{Flat Space}
In four dimensions, the metric in light-cone gauge reads
\begin{align}
    ds^2=2dx^+dx^-+2dzd\bar{z}\,,
\end{align}
where 
\begin{align}
    x^{\pm}=\frac{x^4\pm x^0}{\sqrt{2}}, \quad z=\frac{x^1\pm i x^2}{\sqrt{2}}\,.
\end{align}
We can make a Fourier transformation
\begin{align}
    \Phi^{\mu(s)}(\pvec)=\frac{1}{(2\pi)^{3/2}}\int \, d^{4}xe^{-ix\cdot p}\Phi^{\mu(s)}(x)\,.
\end{align}
Upon imposing the light-cone gauge $\Phi_{+\mu(s-1)}=0$, the components that describe physical d.o.f of $\Phi_{\mu(s)}$ are $\Phi_{i(s)}$, where $\Phi^{i(s-2)jk}\delta_{jk}=0$. These are irreducible rank-$s$ tensors that transform under the little group $SO(2)\sim U(1)$. The number of  independent components of a traceless symmetric rank-$s$ tensor in two dimensions is two. Therefore, effectively, we can present any massless spin-$s$ field by two scalar fields. In particular
\begin{align}
    \Phi^{i(s)}(\pvec)=((\Phi_{\pvec}^{\lambda})^{\dagger},\Phi_{\pvec}^{\lambda}), \qquad \Phi^{\lambda}(\pvec)\equiv\Phi^{\lambda}_{\pvec},\quad  (\Phi_{\pvec}^{\lambda})^{\dagger}=\Phi^{-\lambda}_{-\pvec}\,,
\end{align}
where $\pvec=(p^+,p^-,p^i)$ is the four momentum. Then, the free action for a massless field simply reads
\begin{align}
    S=-\frac{1}{2}\int d^4\pvec \,(\Phi^{\lambda}_{\pvec})^{\dagger}\pvec^2\,\Phi^{\lambda}_{\pvec}\,.
\end{align}
\subsection{Anti-de Sitter Space}
It is also possible to describe massless higher spin fields in AdS using light-cone gauge \cite{Metsaev:2018xip}. The metric of the Poincare patch reads
\begin{align}
    ds^2=\frac{1}{z^2}\big[2dx^+dx^-+dx_1^2+dz^2\big]\,.
\end{align}
Once again, we can work directly in momentum space:
\begin{align}
    \Phi(p|z)=\frac{1}{(2\pi)^{3/2}}\int dx^-dx^1dx^+\,e^{-ix\cdot p}\Phi(x^+,x^-,x^1|z)\,.
\end{align}
The two scalar fields that describe a massless spin-$s$ gauge field obey the conjugation rules as
\begin{align}
    \Phi^{\lambda}_{p,z}\equiv \Phi^{\lambda}(p|z),\qquad (\Phi^{\lambda}_{p,z})^{\dagger}=\Phi^{-\lambda}_{-p,z}\,.
\end{align}
Finally, the free action for a massless higher spin field in $\ads_4$ takes the form
\begin{align}
    S=-\frac{1}{2}\int_{z\geq 0} dz\,d^3p (\Phi^{\lambda}_{p,z})^{\dagger}(p^2-\pl_z^2)\Phi^{\lambda}_{p,z}\,.
\end{align}
The simple form of the free action is due to masslessness and light-cone gauge in $\ads_4$. In particular, this is because massless HS fields are conformally invariant in four dimensions. Those include the Maxwell field strength $F_{\mu\nu}$, Weyl tensor $W_{\mu\nu,\rho\sigma}$ and higher-spin generalization thereof. Note that $\ads_4$ looks like a \textit{half} Minkowski space in the light-cone gauge. 
\section{Frame-like Formalism for HSGRA}\label{sec:frame-like}
Despite its clarity, the metric-like formalism can sometimes be cumbersome in doing calculation. On the other hand, while light-front formalism is handy, it does not have a manifestly covariant form. There is a way to avoid all of this by adopting the frame-like formalism. That is, we will introduce auxiliary variables in terms of vielbeins and spin-connections that carry \textit{flat} indices. The idea is to treat General Relativity (GR) as a gauge theory in the new locally flat frame --- the tangent space. From here, a generalization to fields of all spins is amenable. Another reason that the introduction of the vielbeins and spin-connections is essential is to couple matter fields, e.g., fermions, nicely to gravity. Since spinors are irreducible representation of $so(d-1,1)$, to couple spinors to gravity, we need some objects with flat indices. For more details, interested readers are referred to \cite{Didenko:2014dwa,Blagojevic:2002du,Ortin:2004ms} and references therein.

In practice, vielbeins are non-degenerate matrices that transfer indices from one basis to another that we prefer more. If we prefer to work with a flat metric, i.e., $\eta_{ab}$, then its connection to the original metric $g_{\mu\nu}$ reads
\begin{align}
    g_{\mu\nu}=\eta_{ab}e^a_{\mu}e^b_{\nu}\,.
\end{align}
For this reason, $a,b$ are sometimes referred as \textit{flat} (tangent space) indices and $\mu,\nu$ are referred as the \textit{world} (target space) indices. The metric is preserved under the following local Lorentz transformation
\begin{align}
    \delta e^a=\epsilon^a_{\ b}e^b\,\qquad \text{with}\qquad \epsilon^{ab}=-\epsilon^{ba}\,.
\end{align}
In general, the vielbein can have $d^2$ components while the metric has only $d(d+1)/2$. The remaining $d(d-1)/2$ components are accounted for by the
freedom of local Lorentz rotations, which act as gauge symmetries. The corresponding gauge field $\varpi^{ab}_{\mu}$ is the spin-connection, and is anti-symmetric in $a$ and $b$. The spin-connection has the following gauge transformations
\begin{align}
    \delta \varpi^{ab}=d\epsilon^{ab}-\varpi^a_{\ c}\epsilon^{cb}-\varpi^{b}_{\ c}\epsilon^{ac}\equiv \nabla \epsilon^{ab}\,.
\end{align}
Tangent space tensor fields are defined as
\begin{align}\label{eq:fiberrules}
   T_{a_1...a_s}=e_{a_1}^{\mu_1}...e_{a_s}^{\mu_s} T_{\mu_1...\mu_s}\,.
\end{align}
The vielbein postulate  $\nabla_{\mu}e^a_{\nu}=\partial_{\mu}e^a_{\nu}-\Gamma^{\rho}_{\mu\nu}e^a_{\rho}-\varpi^a_{\mu b}e^b_{\nu}=0$ leads to $e^{a}_{\nu}\nabla_{\mu}V_a=\nabla_{\mu}V_{\nu}\,$. Here, $\Gamma$, is the Christoffel symbol that is symmetric in $\mu, \nu$. The anti-symmetrization of the vielbein postulate in $\mu,\nu$ gives 
\begin{align}\label{eq:mixtorsion}
    T^a_{[\mu\nu]}=\partial_{[\mu}e^a_{\nu]}-\varpi_{[\mu b}^{a}e^b_{\nu]}=0\,.
\end{align}
From here, it is more convenient to work with differential forms by hiding all the world indices. We introduce degree-one differential forms, $e^a=dx^{\mu}e^a_{\mu}$ and $\varpi^{ab}=dx^{\mu}\varpi^{ab}_{\mu}$. In terms of these two new variables, one can write down the Cartan structure equations:
\begin{subequations}
\begin{align}
    T^a&=de^a-\varpi^a_{\ b}\wedge e^b=0\,,\\
    R^{ab}&=d\varpi^{ab}-\varpi^a_{\ c}\wedge \varpi^{cb}\,,
\end{align}
\end{subequations}
where the two-form $T^a=\frac{1}{2}T^a_{\mu\nu}dx^{\mu}\wedge dx^{\nu}=0$ is called the torsion two-form, which follows directly from \eqref{eq:mixtorsion}. Moreover, we have $R^{ab}=\frac{1}{2}R^{ab}_{\mu\nu}dx^{\mu}\wedge dx^{\nu}$ which is the Riemann two-form that has connection to the usual Riemann tensor as
\begin{align}
    R_{\mu\nu\rho\sigma}=R^{ab}_{\mu\nu}e_{\rho a}e_{\sigma b}\,.
\end{align}
Since $e^a$ and $\varpi^{ab}$ are gauge fields, they should account for local symmetries. The total dimension of the one-form $\omega$ when we try to combine $e^a$ and $\varpi^{ab}$ together is $d^2+d(d-1)/2=d(d+1)/2$. Hence, we may try to find a Lie algebra with this dimension. The algebra should contain Lorentz generators $L_{ab}$, which go hand in hand with $\varpi^{ab}$. In addition, we should have some generator that comes with $e^a$, call it $P_a$. Both $e^a$ and $P_a$ must transform as vectors under Lorentz rotations. This fixes commutation relations of $P_a$ with $L_{ab}$. Therefore, the Lie algebra must have the following form
\begin{subequations}
\begin{align}\label{eq:AdSalgebra}
    [L_{ab},L_{cd}]&=L_{ad}\eta_{bc}-L_{bd}\eta_{ac}-L_{ac}\eta_{bd}+L_{bc}\eta_{ad}\,,\\
    [L_{ab},P_c]&=P_a\eta_{bc}-P_b\eta_{ac}\,,\\
    [P_a,P_b]&=-\Lambda L_{ab}\,,
\end{align}
\end{subequations}
where $\Lambda$ is cosmological constant. For $\Lambda>0$, we have $so(d,1)$ which is de Sitter algebra while $\Lambda<0$ accounts for the anti-de Sitter algebra $so(d-1,2)$. Finally, when $\Lambda=0$, we return to $iso(d-1,1)$ which is Poincare algebra. We can now interpret the generators $P_a$ as local translations.

By defining, $L_{ab}=T_{ab}$ and $P_a=\sqrt{|\Lambda|}T_{a5}$, we can write the above Lie algebra as
\begin{align}
    [T_{AB},T_{CD}]=T_{AD}\eta_{BC}-T_{BD}\eta_{AC}-T_{AC}\eta_{BD}+T_{BC}\eta_{AD}\,,
\end{align}
where $A=\{a,5\}$ with $5$ being an additional direction, and $T_{AB}=-T_{BA}$. Let us pack $e^a$ and $\varpi^{ab}$ into a single one-form $\Omega=e^aP_a+\frac{1}{2}\varpi^{ab}L_{ab}$ which has the following curvature (field strength)
\begin{align}
    d\Omega-\Omega\wedge\Omega=T^aP_a+\frac{1}{2}\Big(R^{ab}-\Lambda e^a\wedge e^b\Big)L_{ab}\,.
\end{align}
If we further demand that $\Omega$ is a flat connection, i.e. $d\Omega-\Omega\wedge\Omega=0$, we get
\begin{align}
   T^a=0,\qquad \text{and} \qquad  R^{ab}=\Lambda e^a\wedge e^b\,.
\end{align}
The second equation corresponds to a maximally symmetric background
\begin{align}
    R_{\mu\nu\rho\sigma}=\Lambda\Big(g_{\mu\rho}g_{\nu\sigma}-g_{\mu\sigma}g_{\nu\rho}\Big)\,.
\end{align}
The one-form flat connection $\Omega$ has the following gauge transformations
\begin{subequations}
\begin{align}
    \delta \Omega&=d\epsilon-[\Omega,\epsilon]\,, &
    \epsilon&=\epsilon^aP_a+\frac{1}{2}\epsilon^{ab}L_{ab}\,,
\end{align}
\end{subequations}
where $\epsilon^a$ and $\epsilon^{ab}$ are the gauge parameters for $P_a$ and $L_{ab}$ respectively. One can check that these equations indeed reduce to the diffeomorphism $\delta g_{\mu\nu}=\nabla_{(\mu}\epsilon_{\nu)}$ in the metric-like formalism. Now, we are ready to see how to get higher spin fields using the frame-like approach where the field equations are of first order as shown above. Let us denote the background vielbein and spin-connection as $\bar{e}^a$ and $\bar{\varpi}^{ab}$. Then, the fluctuation of the vielbein, denoted $e^a$, and spin-connection, denoted $\varpi^{ab}$, have the following gauge transformation
\begin{align}
    \delta e^a&=\nabla \epsilon^a+\bar{e}_b\xi^{a,b}\,, &
    \delta \varpi^{ab}&=\nabla \xi^{a,b}+\bar{e}_c\xi^{a,b}\,,
\end{align}
where $\xi^{a,b}=-\xi^{b,a}$. To linear order in the fluctuation of the vielbein, we have \footnote{See \cite{Didenko:2014dwa} for a nice introduction to Young diagrams and their application in higher spin theories.}
\begin{align}
    e_{a,b}=e_a^{\mu}\bar{e}_{\mu b}\sim \parbox{10pt}{\YoungA}\otimes \parbox{10pt}{\YoungA}=\parbox{10pt}{\YoungAA}\oplus\Big(\parbox{20pt}{\YoungB}\oplus \bullet\Big)\,.
\end{align}
We see that $e_{a,b}$ contains both symmetric and anti-symmetric components. They transform as
\begin{align}
    \delta e_{(a,b)}=\nabla_{(a}\epsilon_{b)}\,, \qquad \qquad  \delta e_{[a,b]}=\nabla_{[a}\epsilon_{b]}-\xi_{a,b}\,.
\end{align}
Therefore, if we want to get rid of the anti-symmetric component, we can make an appropriate choice for $\xi_{a,b}$. The remaining symmetric component transforms like a spin-2 Fronsdal field, hence we identify it with $\Phi_{ab}$. \\
${}$\\
To get higher-spin field in the frame-like approach, we introduce a generalized vielbein that is traceless \cite{Aragone:1979hx,Aragone:1980rk,Vasiliev:1980as}
\begin{align}
    e^{a(s-1)}=e^{a(s-1)}_{\mu}dx^{\mu}, \qquad \qquad  e_{a(s-3)b}^{\qquad\ \  b}=0\,.
\end{align}
This generalized vielbein transforms as 
\begin{align}
    \delta e_{a(s-1)}=\nabla \epsilon_{a(s-1)}+\bar{e}^b\xi_{a(s-1),b}\,,\qquad \epsilon_{a(s-3)b}^{\qquad \ \ b}=0, \qquad \xi_{a(s-3)b,c}^{\qquad \ \ \ b}=0\,.
\end{align}
Repeat the same treatment above, we see that the fully symmetric part of the generalized vielbein is nothing but the Fronsdal field
\begin{align}
    \Phi_{a(s)}&=e_{\mu a(s-1)}\bar{e}^{\mu}_a\,&& \delta \Phi_{a(s)}=\nabla_{a}\epsilon_{a(s-1)}\,.
\end{align}
Once again, we can gauge away other components of $e^{a(s-1),b}$ by appropriate choice of $\xi^{a(s-1),b}$. Therefore, a gauge field $\varpi^{a(s-1),b}$ as a generalized spin-connection should be introduced to \textit{host} $\xi^{a(s-1),b}$
\begin{align}
    \varpi^{a(s-1),b}=\varpi^{a(s-1),b}_{\mu}dx^{\mu}\,,\qquad  \varpi^{a(s-3)b,c}_{\qquad \ \ \ b}=0\,,\qquad \varpi^{a(s-1),a}=0\,.
\end{align}
It turns out that $\varpi^{a(s-1),b}$ also has its own gauge redundancy \cite{Lopatin:1987hz} and requires a new spin-connection $\varpi^{a(s-1),bb}$, and so on. As a consequence, in the frame-like formalism we need generalized vielbein and a tower of generalized spin-connections 
\begin{align}
    e^{a(s-1)}&: \qquad  \parbox{90pt}{\bep(\Length{8},10)\put(0,0){\RectT{8}{1}{\TextTop{$s-1$}}}\eep}\\
    \varpi^{a(s-1),b(t)}&: \qquad  \parbox{90pt}{\bep(\Length{8},20)\put(0,0){\RectT{4}{1}{\TextTop{$t$}}}%
\put(0,10){\RectT{8}{1}{\TextTop{$s-1$}}}\eep}, \qquad t=1,2,...,s-1\,.
\end{align}
Eventually, we have the following set of generalized curvatures
\begin{equation}
\begin{split}
    R^{a(s-1)}&=\nabla e^{a(s-1)}+\bar{e}_b\wedge \varpi^{a(s-1),b}\,,\\
    R^{a(s-1),b(t)}&=\nabla \varpi^{a(s-1),b(t)}+\bar{e}_c\wedge \varpi^{a(s-1),b(t)c}+f(\Lambda,\bar{e},\varpi),\qquad t=1,2,...,s-2\,\\
    R^{a(s-1),b(s-1)}&=\nabla\varpi^{a(s-1),b(s-1)}\,,
\end{split}    
\end{equation}
where $f(\Lambda,\bar{e},\varpi)$ are certain terms that depend on the cosmological constant $\Lambda$, the background vielbein $\bar{e}$ and the generalized spin-connection $\varpi^{a(s-1),b(t)}$ \cite{Lopatin:1987hz,Skvortsov:2006at}. They start as $\Lambda \bar{e}^b \omega^{a(s-1),b(t-1)}+...$, where $...$ denotes a number of terms that impose the Young symmetry and tracelessness constraints. In Minkowski limit $\Lambda\rightarrow0$ we have $f=0$. The $R^{a(s-1),b(t)}$ are invariant under
\begin{align}
    \delta \varpi^{a(s-1),b(t)}=\nabla \xi^{a(s-1),b(t)}+\bar{e}_c\wedge \xi^{a(s-1),b(t)c}+\Lambda \bar{e}^b\wedge \xi^{a(s-1),b(t-1)}+...\,.
\end{align}
We can further impose for the system above that
\begin{equation}\label{eq:generalizedR}
\begin{split}
    R^{a(s-1)}&=0\,,\\ R^{a(s-1),b(t)}&=0\,,\qquad \qquad  \text{for}\qquad t=1,2,...,s-2\,,\\
    R^{a(s-1),b(s-1)}&=\bar{e}_c\wedge \bar{e}_d\mathcal{W}^{a(s-1)c,b(s-1)d}\,.
    \end{split}
\end{equation}
These equations are torsion-like constraints that can be used to solve for the generalized spin-connections. In the last equation, instead of zero on the r.h.s we have the 0-form $\mathcal{W}^{a(s),b(s)}$ which is the HS generalization of the Weyl tensor built out of order-s curl of the Fronsdal field
\begin{align}
    \mathcal{W}^{a(s),b(s)}\sim \nabla^{b_1}...\nabla^{b_s}\Phi^{a_1...a_s}-\text{traces}\,\qquad (\text{anti-symmetrized in $b$ and $a$})\,.
\end{align}
Given a set of connections relevant for the description of free higher spin fields, a natural question to ask is whether there is an algebra whose connection contains all these frame-like fields. It turns out that we can pack everything into a \textit{master} one-form
\begin{align}
    \omega=\Big(\varpi^{AB}T_{AB}+\sum_{s>2} \varpi^{A(s-1),B(s-1)}T_{A(s-1),B(s-1)}\Big)\,,
\end{align}
where
\begin{align}
    \varpi^{AB}&=(e^a,\varpi^{ab}),\qquad \varpi^{A(s-1),B(s-1)}=\{ e^{a(s-1)},\varpi^{a(s-1),b(t)}, t=1,...,s-1\}\\
    T_{AB}&=(P_a,L_{ab}),\qquad T_{A(s-1),B(s-1)}=\parbox{70pt}{\bep(\Length{7},20)\put(0,0){\RectT{7}{1}{\TextTop{$s-1$}}}%
\put(0,10){\RectT{7}{1}{\TextTop{$s-1$}}}\eep}=f(P_a,L_{ab})\label{eq:generators}\,.
\end{align}
Note that $T_{A(s-1),B(s-1)}$ are the generators of HS algebra that can be written in terms of polynomials in $P_a$ and $L_{ab}$ (see discussion below). Then, the generalized curvature 2-form can be written as
\begin{align}\label{eq:hsbackgroundform}
    R=d\omega-\omega\wedge \omega\,.
\end{align}
with the following natural gauge transformations for $\omega$ 
\begin{align}
    \delta \omega=d\xi-[\omega,\xi]\,,
\end{align}
where
\begin{align}
    \xi=\sum_s\xi^{A(s-1),B(s-1)}T_{A(s-1),B(s-1)}\,.
\end{align}
Here, $\xi^{A(s-1),B(s-1)}$ are gauge parameters associated to $\varpi^{A(s-1),B(s-1)}$ gauge fields.

\paragraph{Going back to Fronsdal:} Note that to return to the Fronsdal equation from \eqref{eq:generalizedR}, we do not need all the auxiliary fields $\varpi^{a(s-1),b(t)}$ with $t\geq 2$. For a Minkowski background, the equations of motion for free higher-spin fields are given by
\begin{align}\label{eq:backtoFronsdal1}
    R^{a(s-1)}=0\,,\qquad \qquad  \bar{e}^{a\mu}\bar{e}^{\nu}_bR^{a(s-1),b}_{\mu\nu}=0\,.
\end{align}
In flat space, $\nabla=\pl$ and $\bar{e}^{\mu}_a=\delta^{\mu}_a$ and the curvatures can be written as
\begin{align}\label{eq:backtoFronsdal2}
    R^{a(s-1),b(t)}_{\mu\nu}\bar{e}^{\mu c} \bar{e}^{\nu d}&=\pl^c\varpi^{a(s-1),b(t)}_{\nu} \bar{e}^{d\nu}+\varpi^{a(s-1),b(t)c}_{\nu}\bar{e}^{d\nu}-(c\leftrightarrow d)\,,\quad (t\leq s-2)\,.
\end{align}
From \eqref{eq:backtoFronsdal2}, we get
\begin{subequations}\label{eq:backtoFronsdal3}
\begin{align}
    t&=0:\qquad \pl^{c}e^{a(s-1)}_{\mu}\bar{e}^{d\mu}+\varpi^{a(s-1),c}_{\mu} \bar{e}^{d\mu}-(c\leftrightarrow d)=0\,,\label{eq:backtoFronsdal3a}\\
    t&=1: \qquad \pl^c\varpi^{a(s-1),b}_{\mu} \bar{e}^{d\mu}+\varpi^{a(s-1),bc}_{\mu} \bar{e}^{d\mu} -(c\leftrightarrow d)=0\,.\label{eq:backtoFronsdal3b}
\end{align}
\end{subequations}
If we symmetrize \eqref{eq:backtoFronsdal3} with respect to $a\leftrightarrow c$, then contract the resulting equation of \eqref{eq:backtoFronsdal3b} with $\eta_{bd}$, we get
\begin{subequations}
\begin{align}
  t&=0:\qquad   \varpi^{a(s-1),b}_{\mu} \bar{e}^{a\mu}=\pl^a e^{a(s-1)}_{\mu} \bar{e}^{b\mu}-\pl^b e^{a(s-1)}_{\mu} \bar{e}^{a\mu}\,,\label{eq:backtoFronsdal4}\\
  t&=1:\qquad \pl_b\varpi^{a(s-1),b}_{\mu}\bar{e}^{a\mu}-\pl^a\varpi^{a(s-1),b}_{\mu} \bar{e}_b^{\mu}=0\,.\label{eq:backtoFronsdal5}
\end{align}
\end{subequations}
The Fronsdal equations of motion for free higher-spin fields can be obtained by plugging \eqref{eq:backtoFronsdal4} into \eqref{eq:backtoFronsdal5} and making the identification that $\Phi^{a(s)}=e^{a(s-1)}_{\mu}\bar{e}^{a\mu}$. We simply get
\begin{align}
    \square \Phi^{a(s)}-\pl^a\pl_b\Phi^{ba(s-1)}+\pl^a\pl^a\Phi^{a(s-2)b}_{\qquad \ \ b}=0\,.
\end{align}
It is clear that to get Fronsdal equations we just need auxiliary fields with depth-$t=0,1$ only. Therefore, one may wonder why we need other extra fields. It turns out that those extra fields should be present for the consistency of HSGRA's system in the frame-like formulation. Moreover, those extra auxiliary fields are the gauge fields associated to elements of higher-spin algebra that will be explained below.
\section{AdS/CFT and Higher Spin/Vector model duality}
The AdS/CFT correspondence is one of the most celebrated discoveries of string theory \cite{Maldacena:1997re,Witten:1998qj,Gubser:1998bc}. In the original proposal, $\mathcal{N}=4$ super Yang-Mill theory in four dimensions is conjectured to be dual to Type-IIB super string theory in $AdS_5\times S^5$. At present, the conjecture is extended to more general cases. The idea is the following. 
In the most general context, AdS/CFT correspondence implies an (almost) one-to-one relation between conformal field theories, CFT's, in $d$-dimensions, and theories of quantum gravity in AdS space of one dimension higher. In principle, any CFT's correlation functions can be rewritten as Witten diagrams for a theory in AdS. However, the resulting gravitational theory may be very non-local. Therefore, the adverb 'almost' above stands for the fact that CFT's that are dual to perturbatively local weakly-coupled theories of quantum gravity have to have very special properties (for example, the large-$N$ SYM theory has these properties at strong coupling, but not at weak or intermediate coupling). In fact, CFT's that are dual to higher spin gravities do not share many of the required properties. Nevertheless, they are simple and well-defined CFT's, which gives us a hope of better understanding higher spin gravities.

\subsection{Formulation of AdS/CFT}
To visualize better, take $\{\Psi_i(x_i)\}$ to be the set of sources $\Psi_i$ of the bulk fields $\Phi_i(x_i,z_i)$ (the corresponding set of $\Phi_i$ is $\{\Phi_i\}$) when $z_i\rightarrow 0$, and $\Ocal_i$ to be some operators cooked up by the matter fields $\Upsilon$ that constitute the CFTs. The partion function on the CFT reads (we use Euclidean signature)
\begin{align}
    Z_{\text{CFT}}[\{\Psi_i\}]=\int \D \Upsilon\exp\Big[-S_{\text{CFT}}[\Upsilon]+\int d^d x \,\sum_i\Psi_i(x_i)\Ocal_i(x_i)\Big]. 
\end{align}
The duality is established whenever
\begin{align}
    Z_{\text{AdS}}[\{\Psi_i\}]=Z_{\text{CFT}}[\{\Psi_i\}]\,,
\end{align}
where
\begin{align}
    Z_{\text{AdS}}[\{\Psi_i\}]=\int_{\Phi_i\rightarrow \Psi_i} \D\Phi_i\,\exp\Big[-\frac{1}{G}S_{\text{AdS}}[\{\Phi_i\}]\Big].
\end{align}
In terms of CFT correlation functions and perturbative holographical scattering amplitude, the following relation should hold
\begin{align}\label{AdS/CFTassumption}
    \langle \Ocal_1(x_1)...\Ocal_n(x_n)\rangle = (-)^{n}\frac{\delta}{\delta \Psi_1(x_1)}...\frac{\delta}{\delta \Psi_n(x_n)}S_{\text{AdS}}[\{\Psi_i\}]\Big|_{\Psi_i=0}\,.
\end{align}
\subsection{HSGRA/Vector Model Duality}\label{HS/Vec-HSalgebra}
The AdS/CFT opens up a possibility to study HSGRAs via their dual free CFTs \cite{Sundborg:2000wp,Sezgin:2002rt,Klebanov:2002ja,Leigh:2003gk}. Recall that the coupling constant on the CFT side is the number of degrees of freedom $N\sim (R/l_p)^{d-1}$. When $N$ is large, we can make the following expansion
\begin{align}
    -\ln Z_{\cft}=F_{\cft}=N F^0_{\cft}+F^1_{\cft}+\frac{1}{N}F^2_{\cft}+...\,.
\end{align}
On the other hand, the weak coupling expansion in the dimensionless coupling $g=\frac{G}{R^{d-1}}$ gives us 
\begin{align}
    -\ln Z_{\ads}=F_{\ads}=\frac{1}{g}F^0_{\ads}+F^1_{\ads}+gF^2_{\ads}+....\,.
\end{align}
At least at large $N$, we should have $N^{-1}\sim g$. Hence, to prove Higher Spin/Vector Model duality perturbatively, we must show at each order in the coupling constants (or better non-perturbatively) that  $F^i_{\cft}=F^i_{\ads}$. \\
${}$\\
Higher-spin gauge fields are dual to conserved tensors of rank greater than two, i.e higher-spin conserved tensors \footnote{We use $a,b,c,...=0,...,d-1$ to denote $\cft^d$ Lorentz indices and $\ua,\ub,\uc,...=0,...,d$ for $\ads_{d+1}$ bulk Lorentz indices.}, $\JJ$,
\begin{align}
    \pl^m\JJ_{ma(s-1)}=0 \Longleftrightarrow \delta \Phi_{\ua(s)}=\nabla_\ua\xi_{\ua(s-1)}\,.
\end{align}
The above equation is essential in HSGRA/Vector Model duality. Since, bulk theory and CFT are governed by the same symmetry group $SO(d,2)$, fields and operators must live in the same representation of $SO(d,2)$. The presence of (at least one) higher-spin conserved tensors in a $\cft^d$, with $d\geq3$, makes this CFT a free one (possibly in disguise in the sense that the correlation functions of single-trace operators have to be those of a free CFT, but we do not claim the existence of the fundamental free fields). In particular, the moment a conserved tensor with rank higher than two appears in a CFT, it follows that a tower of infinitely many higher-spin conserved tensors must emerge to make the CFT consistent \cite{Maldacena:2011jn,Boulanger:2013zza,Alba:2013yda,Stanev:2013qra,Alba:2015upa}.

\subsubsection{Higher-spin Symmetry from Free CFT}
From conserved higher-spin rank-$s$ tensors, we can construct higher-spin conserved currents $\mathbf{J}_m^{s}$ and the corresponding charges $Q^{s}$ by contracting $\JJ_{a(s)}$ with conformal Killing tensors $t^{a(s-1)}$:\footnote{Conformal Killing tensors $t^{a(s-1)}$ obey: $\pl^a t^{a(s-2)}-\text{traces}=0\,$.}
\begin{align}
    \mathbf{J}^{s}_m(t)=\JJ_{ma(s-1)}t^{a(s-1)}\,,\qquad \quad Q^{s}(t)=\int d^{d-1}x \, \mathbf{J}_0^{s}\,.
\end{align}
These higher-spin Noether charges generate higher-spin symmetries which defines higher-spin algebras, \textbf{hs} \cite{Fradkin:1986ka,Eastwood:2002su}. Note that higher-spin algebras contain the $so(d,2)$ conformal algebra as a subalgebra. 

Being realized by the charges $Q^{s}$ above, the action of $Q^s$ on various operators in the corresponding CFT should be constrained by the Ward identities \cite{Maldacena:2012sf}. Assuming that we have $\JJ_{a(2)}$ and some other conserved higher-spin tensors with $s>2$, we have at least two charges $Q^2$ and $Q^s$. By the CFT axioms, the algebra that $Q^2$ and $Q^s$ form should contain some non-vanishing structure constants, i.e.
\begin{align}
    [Q^2,Q^s]=Q^s+...\,,\qquad [Q^s,Q^s]=Q^2+...\,.
\end{align}
The Ward/Jacobi identities then imply that there should be some other non-vanishing structure constants as well in the above ellipses. As a result one can prove that in order to satisfy the Ward identities we need $Q^s$ of all spins. In other words, the result in \cite{Maldacena:2012sf,Boulanger:2013zza,Alba:2013yda,Stanev:2013qra,Alba:2015upa} implies that the presence of at least one $\JJ_{a(s)}$ with $s>2$ leads to the presence of infinitely many of them. That is to say, CFT with exact higher-spin symmetry is essentially a free theory.

To understand algebra {\bf hs}, let us take the simplest example of a free CFT constituted by a scalar field where $\square \phi=0$. The conserved higher-spin tensors take the form
\begin{align}
    \JJ_s=\JJ_{a(s)}=\phi\partial_{a_1}...\partial_{a_s}\phi+...\,,\qquad \partial^m\JJ_{ma(s-1)}=0\quad (\text{on-shell})\,.
\end{align}
According to Eastwood \cite{Eastwood:2002su}, each element of the higher-spin algebra is in one-to-one correspondence with conformal Killing tensors $t^{a_1...a_{s-1}}$. Consider the following linear differential operators
\begin{align}
    \mathcal{D}(t,s)=t^{a_1...a_{s-1}}\partial_{a_1}...\partial_{a_{s-1}}+\text{lower orders}\,.
\end{align}
We can show that $\square \mathcal{D}\phi=\mathcal{D}'\square\phi=0$. Therefore, $\mathcal{D}(t,s)$ is a symmetry of $\square \phi=0$ since it maps solution to solution. Moreover, we can prove that $\mathcal{D}_1(t_1,s_1)...\mathcal{D}_n(t_n,s_n)$ is also a symmetry of $\square\phi=0$ (keeping the order of $\mathcal{D}_i$). As a consequence, it is easy to see that {\bf hs} is associative.\\ 
${}$\\
We can also define {\bf hs} as a quotient of universal enveloping algebra $\mathcal{U}(so(d,2))$ by a two-sided Joseph ideal $\mathcal{I}$. Recall that the generators $T^{AB}$ of $so(d,2)$ are anti-symmetric in $A,B$, which corresponds to the adjoint representation and can be depicted by  Young diagram $\parbox{10pt}{\YoungAA}\,$. From here, we can construct each elements of the universal enveloping algebra $\mathcal{U}(so(d,2))$ as polynomials in $T$'s
\begin{equation}
\begin{split}
    \mathcal{U}(so(d,2))&=\bullet\oplus\parbox{10pt}{\YoungAA}\oplus\Big(\parbox{10pt}{\YoungAA}\otimes \parbox{10pt}{\YoungAA}\Big)_S\oplus\Big(\parbox{10pt}{\YoungAA}\otimes \parbox{10pt}{\YoungAA}\otimes \parbox{10pt}{\YoungAA}\Big)_S\oplus...\\
    &=\bullet\oplus \parbox{10pt}{\YoungAA}\oplus \Bigg(\bullet\oplus\,\parbox{20pt}{\YoungB}\oplus\parbox{20pt}{\YoungBB}\oplus \parbox{10pt}{\YoungAAAA}\,\Bigg)\oplus\Bigg(\,\parbox{30pt}{\YoungCC}\oplus \parbox{20pt}{\YoungBBAA}\oplus ...\Bigg)\oplus ...\,,
\end{split}
\end{equation}
where $()_S$ denotes the symmetrized tensor product of the adjoint representation of $so(d,2)$. The first bullet $\bullet$ indicates the first singlet of $\mathcal{U}(so(d,2))$ which is the unit of $\mathcal{U}(so(d,2))$, while the second is the quadratic Casimir operator
\begin{align}
    C_2=-\frac{1}{2}T_{AB}T^{AB}\,.
\end{align}
Consider the following two-sided Joseph ideal \footnote{To get non-trivial quotient, one should be careful in choosing the elements from $\mathcal{U}(so(d,2))$ to generate the ideal. Some choice can make the ideal coincides with the full $\mathcal{U}(so(d,2))$, which results in a trivial quotient.}
\begin{align}
    \mathcal{I}=\mathcal{U}(so(d,2))\Bigg(\big[C_2-\lambda\big]\oplus \parbox{20pt}{\YoungB}\oplus\parbox{10pt}{\YoungAAAA}\,\Bigg)\mathcal{U}(so(d,2))\,,\qquad \lambda=-\frac{1}{4}(d^2-4)\,,
\end{align}
Then, one can show that
\begin{align}
    \textbf{hs}=\mathcal{U}(so(d,2))/\mathcal{I}=\bullet\oplus \bigoplus_{s\geq 1}\parbox{70pt}{\bep(\Length{7},20)\put(0,0){\RectT{7}{1}{\TextTop{$s$}}}%
\put(0,10){\RectT{7}{1}{\TextTop{$s$}}}\eep}\,.
\end{align}
Since $\mathcal{U}(so(d,2))$ is also associative, {\bf hs} is associative. Note that the generators with $s\geq 1$ in {\bf hs} are nothing but the $T_{A(s-1),B(s-1)}$ in \eqref{eq:generators}. The spectrum of HS theories is then determined by {\bf hs}. For more details see e.g. \cite{Vasiliev:1990en,Vasiliev:2003ev,Konshtein:1988yg,Vasiliev:2004cm, Bekaert:2009fg, Bekaert:2013zya, Alkalaev:2014nsa}. 
\subsubsection{Higher-spin Symmetry as Gauge Symmetry in AdS}
The higher-spin symmetry is a global symmetry of the dual CFT  and therefore it needs to be a gauge symmetry of the corresponding higher spin gravity. A natural object that can take this task is the one-form $\omega$, which takes valued in the higher-spin algebra. The simplest equation we can write down for the one-form is 
\begin{align}\label{eq:hsemptyspace}
    d\omega=\omega\star \omega\,,
\end{align}
where $\star$ is the product in {\bf hs}. Since $T^{AB}=(P^a,L^{ab})$, we can write any elements of {\bf hs} as polynomials in terms of $P^a$ and $L^{ab}$. Therefore, it is easy to notice that the equation \eqref{eq:hsemptyspace} is sets the previously defined curvature \eqref{eq:hsbackgroundform} to zero and describes an empty space with free higher-spin fields. The problem is now to add interactions to \eqref{eq:hsemptyspace}, which, as will be shown later, requires 0-form $C$. We will discuss this matter in chapter \ref{chapter5}.

\subsubsection{A Brief Summary of Chapter 2}
We reviewed free higher spin fields in metric-like, light-front and frame-like formalisms. We summarize the three approaches by the following diagram.
\begin{figure}[h!]
    \centering
    \includegraphics[scale=0.56]{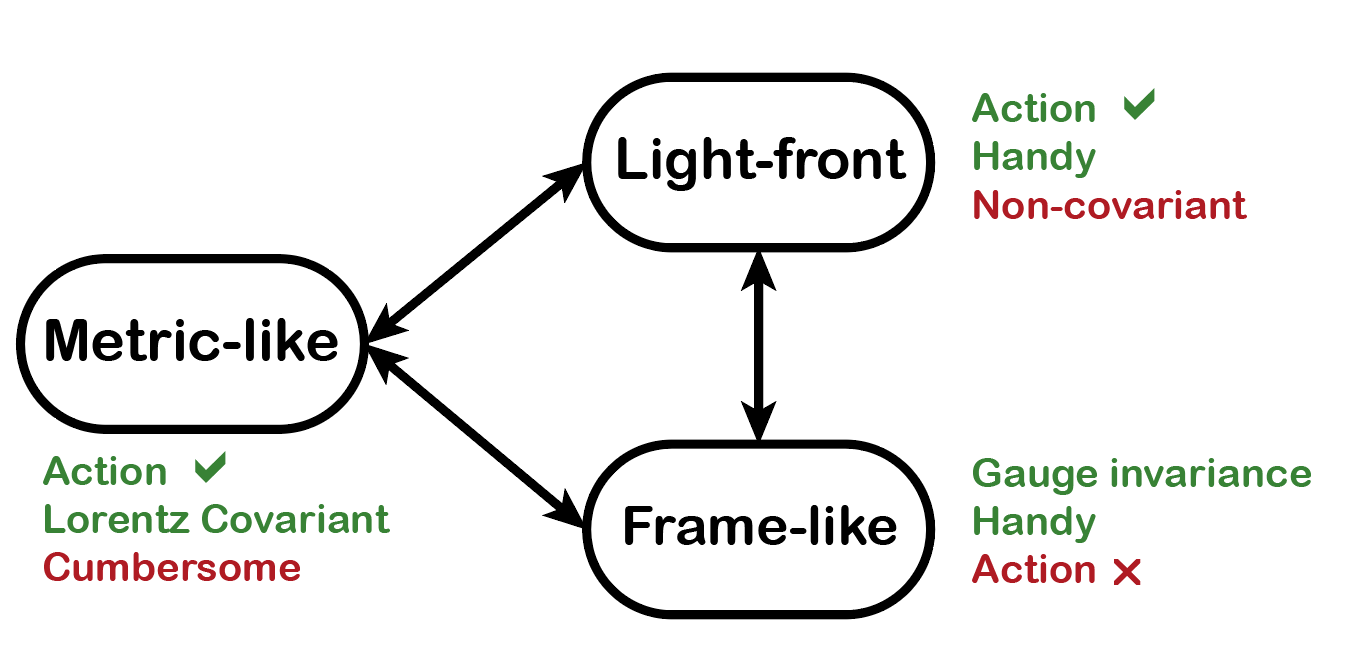}
\end{figure}

We also briefly studied the AdS/CFT correspondence paying attention to HSGRA/Vector Model duality.

\subsubsection{Interactions and Quantum Aspect of HSGRAs}
Constructing a consistent HS theory with interactions is one of the main problems of the HS program. The second problem that we want to address is that whether HSGRA can be a toy model for Quantum Gravity. The criteria we are paying attention to are: (i) there should be a graviton inside the spectrum of the theory; (ii) the theory should be UV-finite.

$\triangleright$ \underline{{\bf In flat space:}} The interaction vertices were found by studying the consistency of the deformed Poincare algebra in the light-front approach \cite{Bengtsson:1983pd,Bengtsson:1983pg,Metsaev:2005ar}. Using the light-front approach, a special class of HSGRA was found under the name chiral HSGRA \cite{Ponomarev:2016lrm}. In this theory we can show that all quantum corrections vanish \cite{Skvortsov:2018jea,Skvortsov:2020wtf}. Another class of theories that have flat space as a background are conformal theories in $3d$ and $4d$. 

$\triangleright$ \underline{{\bf In AdS:}} The list of HSGRAs with interactions is quite short: Chern-Simmons-type theories in $3d$; conformal higher spin theories in $3d$ and $4d$; chiral theory can also be extended to $\ads_4$. Another approach is to construct formally consistent classical equations of motion, a program that was pioneered by Vasiliev \cite{Vasiliev:1990en,Vasiliev:1999ba}. In the latter case there is an important conceptual problem of how to extract meaningful interaction vertices \cite{Boulanger:2015ova}. Fortunately, when HSGRA/Vector model duality conjecture was formulated, it opened up the possibility to understand the rather complicated bulk theory by studying free (or critical) Vector Models \cite{Klebanov:2002ja}. A number of non-trivial checks to test the validity of the HSGRA/Vector model duality conjecture has been performed at tree-level \cite{Giombi:2009wh,Giombi:2010vg,Didenko:2012tv}, and at one-loop see e.g. \cite{Giombi:2013fka,Giombi:2014iua,Giombi:2014yra,Beccaria:2014xda,Beccaria:2015vaa,Bae:2016rgm,Giombi:2016pvg,Gunaydin:2016amv,Skvortsov:2017ldz,Pang:2016ofv,Basile:2018zoy,Basile:2018acb,Ponomarev:2019ltz}. The general take is that since the (holographic) S-matrix is fixed by higher-spin symmetry on the CFT side \cite{Maldacena:2011jn,Maldacena:2012sf,Boulanger:2013zza,Sharapov:2018kjz}, HSGRAs should be UV-finite.



\chapter{HSGRA at One-Loop in AdS}\label{chapter3}
The main message of this thesis to a large extent is that higher spin theories should be UV finite theories due to the large amount of symmetries and the simplicity of their dual partners, vector models, on the boundary of AdS. This chapter is dedicated to the tests of several types of higher spin theories at one-loop in AdS using the spectral zeta function approach pioneered by Dowker and Hawking \cite{Dowker:1975tf,Hawking:1976ja}. For self-contained overview purposes, we briefly review some technicalities in the first few sections while the details are covered in Appendices \ref{app:chap3} and \ref{app:chap35}. 

\section{Motivation}
Computing loop diagrams in AdS to compare with results from CFT is the next step to confirm the validity of AdS/CFT conjecture. Some progress has been made in this direction, see for examples \cite{Giombi:2017hpr,Aprile:2017bgs,Alday:2017xua,Alday:2017vkk,Ponomarev:2019ltz}.\footnote{See also, \cite{Bertan:2018afl} for a direct bulk computation for $\phi^4$ theory.} These results opened a direct access to the quantum properties of the bulk theories and also a link to the anomalous dimensions of some CFTs \cite{Aharony:2016dwx}. 

At present, we do not have the full action for type-A,B and SUSY HSGRAs etc., which somewhat limits our access to quantum properties of HSGRAs in AdS.\footnote{Some part of the action for type-A HS is now understood via holographic reconstruction \cite{Sleight:2016dba,Bekaert:2014cea,Bekaert:2015tva} and in $3d$ the current interaction between the matter sector and Chern-Simons sector was added in \cite{Kessel:2015kna}.} Fortunately, with the knowledge of kinetic terms of various classes of HSGRA's that are discussed in the following section, it is sufficient to perform many nontrivial consistency checks to confirm HSGRA/Vector Model duality at the quantum level. Many one-loop tests have already been performed in a series of papers \cite{Giombi:2013fka,Giombi:2014iua,Giombi:2014yra,Beccaria:2014xda,Beccaria:2014jxa,Beccaria:2014zma,Beccaria:2014qea,Beccaria:2015vaa,Beccaria:2016tqy,Bae:2016rgm,Bae:2016hfy,Giombi:2016pvg}, see also \cite{Gupta:2012he,Gaberdiel:2011zw} for the $3d$ case. The main lessons are as follows. Each of the fields in the spectrum of HS theories contributes a certain amount to one of the computable quantities: sphere free energy, Casimir Energy, $a$- and $c$-anomaly coefficients. The sum over all spins is formally divergent and requires a regularization. Refined in this way the sum over spins becomes finite and matches the corresponding quantity on the CFT side, which in many cases leads to nontrivial tests rather than $0=0$ equalities.

Vacuum one-loop corrections in higher-spin (HS) theories in AdS require one simple ingredient as an input data: a CFT with infinitely many conserved higher-rank tensors conventionally called higher-spin currents $\JJ_s$. These type of CFTs are free or behave like free theories in the strict $N\rightarrow\infty$ limit. The algebra of HS currents determines the field content of the dual HS theory and allows one to perform many one-loop tests. The simplest free CFTs provide the basic examples of HSGRA/Vector Models dualities: the free scalar field is dual to Type-A HS theory with spectrum made of totally-symmetric HS fields and the free fermion is dual to Type-B whose spectrum contains specific mixed-symmetry fields that include totally-symmetric HS fields too. There are also SUSY extensions of HS, see for example \cite{Fernando:2009fq,Fernando:2010dp,Fernando:2014pya,Fernando:2015tiu,Sezgin:2012ag}. This implies that the HSGRA/Vector Model duality should be, in principle, extendable to all unbroken higher-spin theories and their supersymmetric extensions.

Then, our contributions are the following:
\begin{enumerate}
    \item We derive the spectral zeta-function for arbitrary mixed-symmetry bosonic and fermionic fields.
    \item We compute one-loop determinants for Type-A and Type-B theories.
    \item We study the contributions of fermionic HS fields in diverse dimensions, which is crucial for the consistency of SUSY HS theories.
    \item In $\ads_5$ we study Type-D,E,... HS theories that are supposed to be dual to higher-spin doubletons with spin greater than one and find that they do not pass the one-loop test.
    \item Partially-massless fields are also briefly discussed.
    \item A simple expression for the $a$-anomaly of an arbitrary-spin free field is found.
    \item We also discovered that the Type-B theories in all even dimensions lead to puzzling results that require better understanding of the duality, the bulk result, however,  still can be represented as a change of the $F$-energy.
    \item We prove analytically and also extend to all dimensions including fractional one{\color{red}s} the results for Type-A theory in section \ref{sec:fractionaltest} which supports the conjecture of Klebanov and Polyakov \cite{Klebanov:2002ja} that HSGRA/Vector Model duality may be extensible to fractional dimensions. 
\end{enumerate}

\section{Classes of Higher-Spin Theories}
\label{sec:HSTheories}
In this section, we will classify some classes of HSGRAs via their CFT duals. First of all, we select a number of distinct free fields $\Upsilon_i$ (and their conjugates $\bar{\Upsilon}^i$) that take values in some representation of some group $G_i$. Then, we impose the singlet constraint by projecting onto the invariants of some subgroup $H\in G_i$. The spectrum of the $\ads$-dual theory is then generated by all single-trace quasi-primary operators that are $H$-singlets. Schematically, the single-trace operators, which are dual to \textit{single-particle} states in AdS, have the form
\begin{align}
    \JJ_{s}=\Tr\Big[\bar{\Upsilon}^i \pl^{s}\Upsilon^i\Big]+ ...\equiv\sum_i \Tr\Big[\bar{\Upsilon}^i \pl_{(a_1}...\pl_{a_s)}\Upsilon^i\Big]+ ...\,, 
\end{align}
when $\Upsilon^i$ sit in the fundamental representation, and the form
\begin{align}\label{eq:adjsingletrace}
    \Tr\Big[\Upsilon \pl^{s_1}\Upsilon\pl^{s_2}\Upsilon ... \pl^{s_n}\Upsilon\Big]\,.
\end{align}
in the adjoint representation. The latter (\ref{eq:adjsingletrace}) corresponds to an exponentially growing number of states in AdS. It is interesting to note that the dual bulk theories of free CFT's with matter
in adjoint representations look like the dual theories of CFT's with fundamental matter coupled to certain matter multiplets \cite{Sundborg:2000wp,Skvortsov:2015pea, Bae:2016rgm,Bae:2016hfy}. 

Below, we classify some of the free CFTs that have HS duals of type-A, B, SUSY HSGRA's and certain others, which are considered in this thesis. To proceed, it is suggestive to use the language and pictures of Young diagrams which refer to $so(d)$ representations to describe HS currents/fields. We denote the Young diagrams as $\YY(s_1,...,s_n)$ where $s_i$ are the length of each rows and $s_1\geq s_2 \geq ... \geq s_n$. 
\begin{align}
   \YY(s_1,s_2,...,s_n)= \parbox{85pt}{{\bep(80,50)\unitlength=0.38mm%
\put(0,0){\RectBRowUp{5}{4}{$...$}{$s_n$}}%
\put(0,20){\RectCRowUp{12}{8}{7}{$s_1$}{$s_2$}{$s_3$}}\eep}}
\end{align}

\paragraph{Type-A.} A free scalar field $\square \phi=0$ as a representation of the conformal algebra is usually called $\Rac$. With one complex scalar one can construct conserved higher-spin currents, which are totally-symmetric tensors:
\begin{align}
    \JJ_{s}&= \bar{\phi} \pl^s \phi +...\,, && \Delta_s=d+s-2\,,\\
    \JJ_0&= \bar{\phi}\phi\,, && \Delta_0=d-2\,.
\end{align}
Here, we add the 'spin-zero current' $\JJ_0=\bar{\phi}\phi $. These currents can be described by $
    \parbox{60pt}{{\bep(\Length{6},6)\put(0,0){\RectT{6}{1}{\TextTop{$s$}}}\eep}}\,$. If the scalar is real then the currents of odd ranks vanish. According to Flato-Fronsdal theorem, \cite{Flato:1978qz,Craigie:1983fb,Vasiliev:2004cm,Dolan:2005wy}:
\begin{align}
    \Rac\otimes \Rac&= \sum_s \JJ_s\,.\label{typeaope}
\end{align}
We can, further, make $\phi$ take values in some fundamental representation $V$ of some Lie group
$G$, so that $\phi$ belong to the space $S=\Rac\otimes V$. It is clear that the spectrum of $\JJ_s$ will correspond to the $G$-invariant part of the tensor product $S\otimes S$. For example, if $\phi^i$ is an $SO(N)$-vector and $N$ is large, then the relevant invariant tensor is $\delta_{ij}$, which is symmetric. By swapping two scalar fields, we observe that all HS currents with odd spins are projected out and the $SO(N)$-invariant single-trace operators belong to $(\Rac\otimes\Rac)_S$, i.e. have even spins,
\begin{align}
    (\Rac\otimes \Rac)_S&= \sum_k \JJ_{2k}\,.
\end{align}
Type-A HSGRA contains bosonic totally-symmetric HS fields that are duals of $\JJ_s$, known as Fronsdal fields \cite{Fronsdal:1978rb}, and an additional scalar field $\Phi_0$ that is dual to $\bar{\phi}\phi$. At the free level Fronsdal fields $s=0,1,2,3,...$ obey
\begin{align}
    (-\square+M_s^2) \Phi^{\ua(s)}&=0, && &M_s^2=(d+s-2)(s-2)-s\,,
\end{align}
where we imposed the transverse traceless (TT) gauge as in Section 2. HSGRA that has totally-symmetric HS fields, $s=0,1,2,...$ is called the \textit{non-minimal} Type-A, which is the $U(N)$-singlet projection, and the one with even spins only, $s=0,2,4,...$ is the \textit{minimal} Type-A, which is the $O(N)$-singlet projection. One can also define the $usp(N)$-singlet theory whose spectrum is made of three copies of odd spins and one copy of even spins \cite{Giombi:2013fka}.

\paragraph{Type-B.} In this case, one can take a free fermion $\slashed {\pl}\psi=0$ called $\Di$. The spectrum of single-trace operators is more complicated \cite{Flato:1978qz,Vasiliev:2004cm,Dolan:2005wy,Alkalaev:2012rg,Alkalaev:2012ic}. They have the symmetry of all \textit{hook} Young diagrams $\mathbb{Y}(s,1^p)$:\footnote{Notation $1^p$ means $p$ rows of length one.}
\begin{align}
    \JJ_{s,p}&=\JJ_{a(s),m[p]}= \bar{\psi} \gamma...\gamma \pl^{s-1}\psi+...\,.
\end{align}
The conserved currents $\JJ_{a(s),m[p]}$,\footnote{Note that the conservation is not simply $\pl\cdot \JJ=0$ due to the Young symmetry. One has to project onto the right irreducible component, otherwise there are no solutions or unitarity is lost. The projection is done by anti-symmetrizing over all $m$ indices in the second line.} which obey Young condition, have now \textit{mixed-symmetry} and vanishing traces. In particular, $\JJ_{a(s),m[p]}$ are symmetric in $a_1...a_s$ and anti-symmetric in $m_1...m_p$. In summary,
\begin{align}
\JJ_{a_1...a_s,m_1...m_p}&= \bar{\psi} \gamma_{a_s m_1...m_p} \pl_{a_1...a_{s-1}} \psi+...\,,\label{typebcurrent}\\ \notag
    \begin{aligned}
    &\text{conservation:} & \pl^n \JJ_{a(s-2)mn,m[p]}&=0\,,  \\
    &\text{Young:}& \JJ_{a(s),am[p-1]}&=0\,,\\
    &\text{tracelessness:}& \JJ\fud{b}{a(s-2)b,m[k]}&=0\,.
    \end{aligned} &&
    \parbox{60pt}{{\bep(60,50)\unitlength=0.38mm%
    \put(0,40){\RectARowUp{6}{$s$}}%
    \put(0,0){\RectT{1}{4}{\TextCenter{$p$}}}\eep}}
\end{align}
Conserved currents correspond to $s\geq2,\forall p$ and we also have the usual usual conserved current $\bar{\psi}\gamma_a\psi$ when $s=1,p=0$.
Note that the totally-symmetric HS currents, i.e. when $p=0$, are still there. In addition, there are anti-symmetric tensors that are anomalous, i.e. not obeying any conservation law, of the form:
\begin{align}\label{masssivepforms}
    \JJ_{m[p]}&= \bar{\psi} \gamma_{m_1}...\gamma_{m_p} \psi\,, && p=0,2,3,4,...\,,
\end{align}
which are degenerate cases of the same expression \eqref{typebcurrent}. The spectrum of single-trace operators can equivalently be read off from $\Di\otimes \Di$ \cite{Flato:1978qz,Vasiliev:2004cm,Dolan:2005wy} as
\begin{align}
    \Di\otimes \Di &= \sum_{s,p} \JJ_{s,p}\,,\qquad  \text{with}\quad  p\leq \frac{d-2}{2}\,.
\end{align}
The corresponding spectrum of the Type-B theory is made of bosonic mixed-symmetry gauge fields with spin $\mathbb{Y}(s,1^p)$, $s>1,\forall p$ or $s=1,p=0$:\footnote{The ellipses hide all $\nabla \xi$-terms with different permutations due to the requirement of Young symmetry. }
\begin{align}
    &\left(-\square +M^2_{s,1^p}\right)\Phi^{\ua(s),\mm[p]}=0\,, && &M^2_{s,1^p}&=(d+s-2)(s-2)-s-p\,.
\end{align}
We call such fields {\it hooks} due to the shape of Young diagrams $\mathbb{Y}(s,1^p)$. The general formula for the mass-like term was found in \cite{Metsaev:1995re, Metsaev:1997nj}. Note that type-A HSGRA is \textit{not} a sub-theory of type-B's due to the differences in the cubic couplings \cite{Bekaert:2014cea,Skvortsov:2015pea}. Even stronger, the Type-A HS algebra is not a subalgebra of the Type-B algebra. However, in $d=3$, there is an exception where there are no mixed-symmetry fields ($p$ has to be $0$ in this case) and the HS algebras generated by free boson and free fermion are the same. In other words, the currents have the same form $\YY(s,0,0,...)$, but $\JJ_0^{A}=\phi^2$ has weight $\Delta=1$ while $\JJ_0^{B}=\bar{\psi}\psi$ has $\Delta=2$, which corresponds to the same mass-like term $M^2=-2$ in AdS.  

\paragraph{SUSY HS.} We consider super-symmetric HSGRAs that are dual to CFT's made of free scalars and fermions. Together with pure bosonic currents, there are also super-currents:\footnote{As primaries the currents must be traceless in $a(s)$ and $\gamma$-traceless in $a(s);\alpha$, the former being a consequence of the latter.}
\begin{align}
    \JJ_{s=m+\tfrac12}&= \phi \pl^m\psi+... && \Longleftrightarrow&&\JJ_{a(m);\alpha}=\phi \pl_{a_1}...\pl_{a_m}\psi_\alpha+...\,.
\end{align}
The super-currents can be expressed via  $\Di\otimes \Rac$ \cite{Flato:1978qz,Vasiliev:2004cm,Dolan:2005wy}:
\begin{align}
   \Di\otimes \Rac&=\sum_{m=0} \JJ_{s=m+\tfrac12}\,.
\end{align}
The super-currents are dual to totally-symmetric fermionic HS fields in AdS \cite{Fang:1978wz,Fang:1979hq}:
\begin{align}
    (\slashed{\nabla}+m)\Phi^{\ua(s);\ualpha}=0\,, \qquad \qquad  m^2=-\left(s+\tfrac{d-4}{2}\right)^2\,.
\end{align}
The square of the HS Dirac operators read
\begin{align}
    (-\slashed{\nabla}+m)(+\slashed{\nabla}+m)&=\left(-\square +M^2_{s}\right)\,,&    M^2_{s}&=m^2+s+\frac{d(d+1)}4\,,
\end{align}
where the mass-like terms were found in \cite{Metsaev:1998xg} for fermionic fields of any symmetry type. We can, therefore, present the simplest SUSY HSGRA through the following super-matrices
\begin{align}
    \begin{pmatrix}
    \text{Type-A}=\Rac\otimes \Rac & \Rac\times \Di\\
    \Di\times \Rac & \text{Type-B}=\Di\otimes \Di
    \end{pmatrix}=
    \sum
    \begin{pmatrix}
    \Phi^{\ua(s)} & \Psi^{\ua(s-\tfrac12);\ualpha}\\
    \Psi^{\ua(s-\tfrac12);\ualpha} & \Phi^{\ua(s),\mm[p]}
    \end{pmatrix}
\end{align}
Again, one can take a number of $\phi$'s and $\psi$'s and impose the singlet constraint with respect to some global symmetry group $G$.

\paragraph{More general HS theories. } Given some dimensions $d$ there is a list $L$ of free conformal fields that can be in $\cft^d$. Generically, $L$ can contain free scalar and free fermion and other fields which depend on the dimension $d$:
\begin{align}
    L&=\{\Di,\Rac,...\}\,.
\end{align}
The list $L$ does not exclude free conformal fields\footnote{For a comprehensive list of conformally-invariant equations we refer to \cite{Shaynkman:2004vu}.} $\phi^{\mathbb{S}}$ with any spin $\mathbb{S}$ obeying $\square^k \phi^{\mathbb{S}}+...=0$ equations of motion, where $k=1,2,...\,$. These theories, however, are usually non-unitary. In even dimension, i.e. $d=2n$, beside the singletons ($\phi$ and $\psi$), we can also have doubletons $S_j$ with spin-$j$ \cite{Gunaydin:1984wc,Gunaydin:1984fk,Metsaev:1995jp,Bekaert:2009fg,Fernando:2015tiu}, where $j=0,\tfrac12$ are the usual $\Rac$ and $\Di$. The $j=1$ case corresponds to $\tfrac{d}{2}$-forms, e.g. the Maxwell field-strength $F_{ab}$ in $d=4$. 

For any given $L$s, we can construct $\cft$s that have higher-spin conserved tensors. For example, consider $L_1=\Rac$, $L_2=\Di$ that take values in the $Nn$ and $Nm$ dimensional representations of $u(N)\times u(n)$ and $u(N)\times u(m)$, respectively. By imposing $F=u(N)$-singlet, the spectrum has HS fields of $\Rac\otimes \Rac$ with values in $u(n)$, fields of $\Di\otimes \Di$ with values in $u(m)$ and $2nm$ fermionic HS fields (see \cite{Konstein:1989ij} for $d=3$). 

For the case of Type-C HSGRA as the dual of the spin-$j=1$ doubleton $S_1$ in $\ads_5/\cft^4$ and $\ads_7/\cft^6$, one finds that the spectrum of Type-C HSGRA contains complicated mixed-symmetry fields \cite{Beccaria:2014zma}. Moreover, we can cook up some extended multiplets of type $n_b\Rac+n_f \Di+n_v S_1$ for more interesting cases. One of the most notable example is $\ads_7/\cft^6$ with $(2,0)$ tensor supermultiplet that contains $\Rac,\, \Di$ and an $S_1$ rank-3 tensor. 

For more details on how to define a general HSGRA via CFT's content, we refer the interested readers to \cite{Gunaydin:2016amv} and references therein.

\section{Higher-Spin Theories at One-Loop}
\label{sec:HSTheoriesatOneLoop}


\subsection{Overview of The One-Loop Tests}
\label{sec:onelooptests}
The idea of the one-loop tests of HS AdS/CFT was explained  in \cite{Giombi:2013fka,Giombi:2014iua}. The AdS partition function
\begin{align}
    Z_{\AdS}&=\int \prod_i \D\Phi_i\, e^{-S[\Phi_i]}\,,
\end{align}
as a function of the bulk coupling $G$ should lead to the following expansion of the free energy $F_{\AdS}$:
\begin{align}
    -\ln Z_{\AdS}&=F_{\AdS}=\frac{1}{G} F^0_{\AdS} +F^1_{\AdS} +G F^2_{\AdS}+...\,,
\end{align}
where the first term is the classical action evaluated at an extremum. $F^1$ stands for one-loop corrections, etc. On the dual CFT side there should be a similar expansion for the CFT free energy $F_{\cft}$:
\begin{align}
    -\ln Z_{\CFT}&=F_{\CFT}=N F^0_{\CFT}+F^{1}_{\CFT}+\frac1{N}F^2_{\CFT}+...\,,
\end{align}
where the large-$N$ counting suggests that $G^{-1}\sim N$. The number of dof. $N$ is expected to be quantized \cite{Maldacena:2011jn}, which is not yet seen in the bulk. In a free CFT's, all but the first term are zero, which should match $F^0_{\AdS}$. To compute $F^0_{\ads}$ is, however, still an impossible task since the classical action is not known. Nevertheless, with the knowledge from the kinetic term of the action, we can check whether $F^1_{\AdS}$ vanishes identically or produces a contribution proportional to $F^0_{\CFT}$, which can be compensated by modifying the simplest relation $G^{-1}=N$ to  $G^{-1}=a(N+\text{integer})$ \cite{Giombi:2013fka,Giombi:2014iua}. This basic idea allows to perform several non-trivial tests thanks to the fact $F^1$ can be computed on different backgrounds. The simplest ones include ${S}^d$, $\mathbb{R}\times {S}^{d-1}$ and ${S}^1\times {S}^{d-1}$ that are the boundaries of Euclidean $\ads_{d+1}=\mathbb{H}^{d+1}$, global $\ads_{d+1}$ and thermal $\ads_{d+1}$, respectively.\footnote{Note that on more complicated backgrounds one encounters the problem of light states \cite{Gaberdiel:2012uj,Banerjee:2012gh}.} In addition, due to the appearance of $\log$-divergences on both sides of AdS/CFT more numbers should agree.

\paragraph{CFT Side.} The free energy computed on ${S}^d$ of radius $R$ is a well-defined number in odd $d$ provided the power divergences are regularized away and is $a_d\log R$ when $d=2n$, where $a$ is the Weyl anomaly coefficient, see e.g. \cite{Casini:2010kt} for conformal scalar.

The free energy on ${S}_\beta^1\times {S}^{d-1}$ with the radius of the circle playing the role of inverse temperature $\beta$ should have the form
\begin{align}\label{FCasimir}
    F&= a_d \log R\Lambda +\beta E_c +F_\beta\,,
\end{align}
where $a_d$ is the anomaly and it vanishes for odd $d$ and also for $\phi$ and $\psi$ on $\mathbb{R}\times {S}^{d-1}$ and ${S}^1\times {S}^{d-1}$.
The last term $F_\beta$ goes to zero when $\beta\rightarrow\infty$, i.e. for $\mathbb{R}\times {S}^{d-1}$, and can be easily computed in a free CFT:
\begin{align}
    F_\beta&=\mathrm{tr}\, \log[1\mp e^{-H \beta}]^{\mp1}=\mp\sum_m \frac{(\pm)^m}{m}Z_0(m\beta)\,.
\end{align}
Here, $H$ is the Hamiltonian of the free CFT and $Z_0(\beta)$ is one-particle partition function
\begin{align}
    Z_0&= \mathrm{tr}\, e^{-\beta H}=\sum_n d_n e^{-\beta \omega_n}\,,
\end{align}
where $d_n$ and $\omega_n$ are degeneracies and eigen values of $H$. The second term in (\ref{FCasimir}), which is proportional to $\beta$, is the Casimir Energy. It is given by a formally divergent sum
\begin{align}
    E_c&=(-)^F\frac12 \sum_n d_n \omega_n =(-)^F \frac12 \zeta_0(-1)\,,  & \zeta_0(z)&=\sum_n \frac{d_n}{\omega_n^z}\,,
\end{align}
which is usually regularized via $\zeta$-function. For free fields it vanishes for odd $d$. The Mellin transform maps $Z_0$ into $\zeta_0$. See Appendix \ref{app:MrCasimir} for many explicit values.

It is crucial to impose the singlet constraint on the CFT side. In a free CFT, e.g. free scalar, $F_\beta$ is constructed from the character $Z_0$ of $\Rac$. After the singlet constraint is imposed, one finds, see e.g. \cite{Giombi:2014yra}, that
$F_\beta$ is built from the character $Z$ of the singlet sector instead of the $\Rac$-character $Z_0$, i.e. from the character of $\Rac\otimes \Rac$ if the CFT is just $\Rac$.
Also, the Casimir Energy is $E^{\text{sing}}_c=NN_f \beta E_c$,
where $N_fN$ is the total number of free fields with the factor of $N$ removed by the singlet constraint.

\paragraph{Quadratic action.} Using the Transverse-Traceless (TT) gauge discussed in Chapter \ref{chapter2}, we have the following free actions for that are building blocks of the simplest bosonic and SUSY HSGRA's that are cooked up from Rac's and Di's
{\allowdisplaybreaks\begin{align}
    S_0&=\frac{1}{G}\int\left[N_AS_A+N_BS_B+N_FS_{F}\right]\,,\\
    S_{A}&=\frac12 \sum_s\int \Phi_{\ua(s)}\left(-\square +M^2_{s}\right)\Phi^{\ua(s)}\,,\\
    S_{B}&=\frac12 \sum_{s,p}\int \Phi_{\ua(s),\mm[p]}\left(-\square +M^2_{s,1^p}\right)\Phi^{\ua(s),\mm[p]}\,,\\
    S_{F}&= \sum_{s}\int \bar{\Psi}_{\ua(s-\tfrac12)}\left(\slashed{\nabla}+m_s\right)\Psi^{\ua(s-\tfrac12)}\,,
\end{align}}\noindent
where the multiplicities $N_A$, $N_B$, $N_F$ depend on the multiplet chosen. 
\paragraph{AdS Side.} The one-loop free energy for a number of (massless) fields in $\ads_{d+1}$ is given by determinant of the bulk kinetic terms
\begin{align}\label{eq:gaugeghost}
    (-)^FF^1_{\AdS}&=\frac12\sum_s \mathrm{tr}\,\log|-\square+M^2_\Phi|-\frac12\sum_s\mathrm{tr}\,\log|-\square+M^2_\xi|\,,
\end{align}
where the sum is over all fields $\Phi_s$. The second term in \eqref{eq:gaugeghost} corresponds to the ghost contribution if $\Phi_s$ is a gauge field and needs to be subtracted.\footnote{See \cite{Buchbinder:1995ez} for an earlier discussion of quantization of higher-spin fields in $\ads_4$.} There is an additional minus $(-)^F$, if fields are fermions. It can be computed by the standard zeta-function regularization \cite{Dowker:1975tf,Hawking:1976ja} of one-loop determinants and leads to
\begin{align}
    (-)^FF_{\ads}^1&=-\frac12 \zeta'(0)- \zeta(0)\log R\Lambda_{UV}\,,
\end{align}
where $R$ is the $\ads$ radius, $\Lambda_{UV}$ is a UV cutoff.

In Euclidean $\ads_{d+1}$, which is also known as Lobachevsky space $\mathbb{H}^{d+1}$, the $\zeta$-function is proportional to the regularized volume of $\ads_{d+1}$ space, which is a well-defined number for $\ads_{d=2n+2}$ and contains $\log R$ for $\ads_{d=2n+1}$. In $\ads_{d=2n+2}$, we have conformal anomaly whose appearance is present by $\log\,R\Lambda_{UV}$. The one-loop free energy on the thermal $\ads_{d+1}$ with boundary $S^1_\beta\times S^{d-1}$ is expected to be
\begin{align}
    F&=\beta[a_{d+1}\log R\Lambda_{UV} +E_c]+F_\beta\,,
\end{align}
where $F_\beta$ vanishes in the high temperature $\beta\rightarrow0$ limit. In odd dimension, i.e $d=2n+1$, the $a_{d+1}$-anomaly is zero, while in even dimension it should be the same as in Euclidean $\ads$ \cite{Giombi:2014yra}. Therefore, it can be computed from the free energy in Euclidean $\ads_{d+1}$ with boundary $S^d$, i.e. $\mathbb{H}^{d+1}$. In the latter case only the total anomaly coefficient can vanish, as was shown in \cite{Giombi:2013fka,Giombi:2014iua}. Therefore, once $a_{d+1}=0$, the rest of the one-loop contribution should be feasible. 

The $N^0$ part of the free energy, $F_\beta$, counts the spectrum of states and should be automatically the same on both sides of the duality. Indeed, the spectrum of HS theories is determined by higher-spin algebra {\bf hs}, which are given by free CFTs. The spectrum of single-trace operators is the same as the spectrum of HS fields and is given by the tensor product of appropriate (multiplets of) singletons/doubletons. Therefore, the $F_\beta$ part can be ignored on both sides for a moment: it can be attributed to generalized Flato-Fronsdal theorems, see e.g. \cite{Giombi:2014yra} for some checks. While the representation theory guarantees that the spectra should match, a direct path-integral proof is needed.

As will be shown, the Casimir Energy $E_c$ does not vanish for the case of minimal theories and Type-C theory \cite{Beccaria:2014zma}, which requires to modify $G^{-1}\sim N$. Moreover, the computation we perform below depends heavily on the dimension $d$.
\paragraph{ $\boldsymbol{\ads_{2n+2}/\cft^{2n+1}}$ cases.} The CFT partition function on a sphere is a number, while $F^1_{\ads}$ in $\mathbb{H}^{2n+2}$ contains $\log R\Lambda_{UV}$-divergences for individual fields. Therefore, in other to cancel the log-divergence, we need to pick the right multiplet, otherwise the finite part of $F$ is ill-defined. Then the finite part, $-\tfrac12 \zeta'(0)$, should be compared to $F_{\CFT}^1$, which is zero in free CFT's. If $F^1_{\AdS}$ is found to be non-zero, then one can try to adjust the relation between $N$ and bulk coupling $G$ as to make the two sides agree, assuming that $F^0_{\AdS}=F^0_{\CFT}$ and $F^1_{\AdS}=m F^0_{\CFT}$ where $m\in \mathbb{Z}$. This requirement is due to the quantization of the bulk coupling. It was found \cite{Giombi:2013fka} that this is the case for the minimal models with even spins and $F^1_{\AdS}$ is equal to $F^0_{\CFT}$ for a free scalar field  \cite{Klebanov:2011gs}.

Another test is for Casimir Energy $E_c$. It vanishes on the CFT side, while every field contributes a finite amount on the $\ads$-side. Therefore, only appropriately regularized sum over spins can vanish.

\paragraph{ $\boldsymbol{\ads_{2n+1}/\cft^{2n}}$ cases.} The regularized volume of AdS-space contains $\log R$, while the sphere free energy $F_{\cft}=a_d\log R$ is given by the $a$-coefficient of the Weyl anomaly ($\log R\Lambda_{UV}$-term vanishes for every field individually). Again, $F^1_{\ads}$ either vanishes or should be equal to an integer multiple of the $a$-anomaly of the dual free CFT, $F^0_{\cft}$, and can be compensated by modifying $G^{-1}\sim N$. The same computation then gives the anomaly for the conformal HS fields --- Fradkin-Tseytlin fields, $-2a_{HS}=a_{CHS}$, \cite{Fradkin:1985am,Tseytlin:2013jya,Giombi:2013yva,Giombi:2014iua}.

Since the Casimir Energy does not have to vanish on the CFT side, we expect AdS results to be some interesting numbers. $F^1_{\AdS}$ corresponds to the order-$N^0$ corrections in CFT, which are absent for free CFT's.

For mutual consistency, if a modification of $G^{-1}=N$ is needed, it must be the same for all the tests in a given theory.

\section{One-Loop Tests}
\label{sec:tests}
In this section we perform the one-loop tests reviewed in Section \ref{sec:HSTheoriesatOneLoop}. Our new results include: computations in even dimensions, spectral zeta-function for fermionic and mixed-symmetry HS fields. Less conventional cases of partially-massless fields and higher-spin doubletons are discussed in Appendix \ref{app:strangeHS}. 

The spectrum of SUSY HSGRA is made of bosonic and fermionic HS fields. The simplest case is when the dual free CFT made of $n$ scalars and $m$ fermions, so $S=n\Rac\oplus m\Di$. By imposing different singlet constraints the spectrum of bosonic HS fields can be truncated to minimal theories. The spin of fermionic HS fields, if there are any, runs over all half-integer values $s=\tfrac{1}{2},\tfrac{3}{2},\tfrac{5}{2},...\,$. In the minimal theories the order $N^0$ one-loop corrections usually do not vanish. It is important for the consistency of SUSY HS theories that the modifications of $G^{-1}\sim N$ is necessary for consistency of Type-A and Type-B are the same, which was observed for $a$, $c$, $E_c$ in $\ads_{5,7}$  \cite{Beccaria:2014qea,Beccaria:2014xda} and for $E_c$ in all $\ads_{2n+1}$ \cite{Giombi:2014yra}. 

\subsection{Casimir Energy Test}
\label{sec:CasimirTest}
The Casimir Energy tests are the simplest since the computation of $E_c$ is not difficult and we refer to Appendix \ref{app:MrCasimir} for technicalities. Each field contributes some finite amount to the Casimir Energy. It is important to use the same regularization that has been already applied for Type-A and Type-B models. 

We will discuss HS fermions only, since the pure Type-A and Type-B contributions are discussed below. Vanishing of the Casimir energy can be seen after summation over spins with the exponential regulator $\exp[-\epsilon(s+(d-3)/2)]$. For example, in $\ads_6$ the summation of $E_c$ over all totally-symmetric HS fermionic fields reads
\begin{align}
     -\sum_{m=0}^{}\textstyle \frac{(m+1) (m+2) \left(1344 m^6+12096 m^5+39760 m^4+57120 m^3+31388 m^2+420 m-2449\right)}{967680}e^{-\epsilon(m+(d-2)/2)}\Big|_{\text{fin.}}&=0\,,
\end{align}
where $|_{\text{fin.}}$ means to take the finite $\epsilon$-part of the sum evaluated with the exponential regulator. From the character perspective, consider $\Di\otimes\Rac$ in any dimension, we have
\begin{align}
    \chi(\Di)\chi(\Rac)&=\cosh \left(\frac{\beta }{2}\right) \sinh ^{2-2 d}\left(\frac{\beta }{2}\right) 2^{\left[\frac{d}{2}\right]-2 d+3}\,,
\end{align}
which is manifestly even in $\beta$ and therefore the Casimir Energy vanishes. For the same reason $E_c$ vanishes for non-minimal Type-A,B models and is equal to that of Rac and Di for minimal cases was also applied in \cite{Giombi:2014yra}. The Casimir Energy for the fermionic subsector is bounded to always vanish, which is what we observed. The tests for more complicated mixed-symmetry fields and partially-massless fields are discussed in Appendix \ref{app:strangeHS}. 

We have the following observation that makes the computation for individual fields easier. First, the character of a conformal scalar weight $\Delta$ has the form $q^\Delta(1-q)^{-d}$. It is easy to see that the number of physical d.o.f factorizes out in the character for any $\Delta$. Second, the Casimir Energy/its first derivative can be shown to vanish for $\Delta=d/2$ for $d$ even/odd:
\begin{align}
    E_c(\Delta=\tfrac{d}2)&=0\,, && d=2k\,,\\
    \frac{\pl}{\pl \Delta}E_c(\Delta=\tfrac{d}2)&=0\,, && d=2k+1\,.
\end{align}
Moreover, the second derivative of $E_c$ with respect to the conformal weight has a very simple form:
\begin{align}
    \frac{\pl^2}{\pl \Delta^2} E_c (\Delta)&=\frac{(-)^d \Gamma (\Delta )}{2 \Gamma (d) \Gamma (\Delta-d +1)}\,.
\end{align}
\subsection{Laplace Equation and Zeta Function}
The eigenvalue problem of the Laplace operator is closely related to construction of zeta-function{\color{red}s}. We first discuss how to compute the eigenvalues and degeneracies for the Laplace operator on a sphere and then proceed to zeta-function on $\mathbb{H}^{d+1}$ (Euclidean AdS), which can be obtained from that on a sphere, see \cite{Camporesi:1994ga}.
\subsubsection{Laplace Eigenvalue Problem}
We are interested in the spectrum of the Laplacian on $S^N=SO(N+1)/SO(N)$:
\begin{align}
    (-\square+M^2) \Phi^{\mathbb{S}}_{n}=\lambda_n^{\mathbb{S}} \Phi^{\mathbb{S}}_n\,,
\end{align}
where $M^2$ is the mass-like term and $\Phi^{\mathbb{S}}$ is a transverse, traceless field with Lorentz spin $\mathbb{S}$, where $\mathbb{S}$ can be any representation which we label by a Young diagram, $\mathbb{S}=\mathbb{Y}(s_1,...,s_k)$. As is well-known, the eigenvalues $\lambda_n$ are given by the difference of two Casimir operators with a trivial shift by $M^2$:
\begin{align}
    -\lambda_n&=C^{so(N+1)}_2(\mathbb{S}_n)-C^{so(N)}_2(\mathbb{S})+M^2\,,\\
    d_n&=\mathrm{\dim}\, \mathbb{S}_n\,,
\end{align}
 Here the Young diagrams $\mathbb{S}_n$ of representations that contribute are obtained from $\mathbb{S}$ by adding a row of extra length $n$ as the first row:\footnote{In general, there are many more representations that contain $\mathbb{S}$ upon reduction to $so(N)$. The restriction to transverse and traceless fields reduces this freedom to one number, which is $n$. The TT-fields result from imposing gauges on the off-shell fields.}
\begin{align}
\mathbb{S}&=\parbox{100pt}{{\bep(80,40)\unitlength=0.38mm%
\put(0,0){\RectBRowUp{5}{4}{$...$}{$s_n$}}%
\put(0,20){\RectBRowUp{8}{7}{$s_1$}{$s_2$}}\eep}}
&
\mathbb{S}_n&=\parbox{85pt}{{\bep(80,50)\unitlength=0.38mm%
\put(0,0){\RectBRowUp{5}{4}{$...$}{$s_n$}}%
\put(0,20){\RectCRowUp{12}{8}{7}{$s_1+n$}{$s_1$}{$s_2$}}\eep}}
\end{align}
The degeneracy $d_n$ is just the dimension of $\mathbb{S}_n$. For example, for the scalar Laplacian with $M^2=0$ we have
\begin{align}
    \lambda_n&=n(N+n-1)\,, & d_n&=\mathrm{dim}^{so(N+1)}\, \mathbb{Y}(n)\,,
\end{align}
where $d_n$ is the number of components of the totally-symmetric
rank-$n$ tensor of $so(N+1)$. Analogously, for totally-symmetric rank-$s$ tensor fields we find
\begin{align}
    \lambda_n&=M^2+E(E-N+1)-s\,, \qquad \qquad E=N+s+n-1\,, \\
    d_n&=\mathrm{dim}^{so(N+1)}\, \mathbb{Y}(s+n,s)\,.
\end{align}

\subsubsection{Spectral Zeta-function}
Having eigenvalues $\lambda_n$ and degeneracy $d_n$, we can compute the spectral $\zeta$-function on $S^{d+1}$:
\begin{align}
    \zeta(z)&= \mathrm{vol}\, S^{d+1} \times \sum_n \frac{d_n}{(\lambda_n)^z}\,.
\end{align}
Extension to hyperbolic space $\mathbb{H}^{d+1}$ requires some work, see e.g. \cite{Camporesi:1992wn,Camporesi:1992tm,Camporesi:1993mz,CAMPORESI199457,Camporesi:1994ga,Camporesi:1995fb,Camporesi:1990wm,Lal:2012ax,Gupta:2012he,Gopakumar:2011qs}. The cases of $\mathbb{H}^{2n+1}$ and $\mathbb{H}^{2n}$ are very different.  Here $\zeta(z)$ is the spectral $\zeta$-function, which is the Mellin transform of the traced heat kernel at coincident points:
\begin{align}
    \zeta(z)&=\frac{1}{\Gamma[z]}\int_0^{\infty} dt \, t^{z-1}\, K(x,x;t)\,.
\end{align}
In homogeneous spaces the heat kernel at coincident points $K(x,x;t)$ does not depend on coordinates and the volume of the space factorizes out. The volume factor is a source of additional divergences and needs to be regularized properly \cite{Diaz:2007an}. 

The eigenvalues can be computed in a rather simple way for any irreducible representation of weight $\Delta$. The rule established on many examples, see e.g. \cite{Camporesi:1993mz,CAMPORESI199457} is to replace $s_1+n$, which is the length of the first row, by $i\lambda-\tfrac{d}2$ where $\lambda$ is non-negative and real:
\begin{align}
    -\lambda_n&=C^{d+2}(i\lambda-\tfrac{d}2,s_1,s_2,...)-C^{d+1}(s_1,s_2,...)+M^2=\frac{1}{4} (d-2 \Delta )^2+\lambda ^2+m^2\,,\\
    M^2&=m^2+\Delta(\Delta-d)-s_1-s_2-...\,,
\end{align}
where we took the standard normalization of the mass-like term, see e.g. \cite{Metsaev:1995re}: for $\Delta$ corresponding to gauge fields, both unitary \cite{Metsaev:1995re} and non-unitary \cite{Metsaev:1995re,Skvortsov:2009zu}, we have $m^2=0$.

The heat kernel contains only a contribution of the principal series in the odd dimensional case $\mathbb{H}^{2k+1}$. In the even dimensional case $\mathbb{H}^{2k}$ a discrete series can contribute \cite{CAMPORESI199457} too, depending on the type of representation. In what follows we will ignore the contribution of discrete series, but it would be interesting to understand if they play any role in HS AdS/CFT in $d>2$. 

Zeta-function naturally has several different factors and the general expression is usually written in the following form:
\begin{align}
    \zeta&= \frac{\mathrm{vol}(\mathbb{H}^{d+1})}{\mathrm{vol}(S^d)} v_d g(s)\int_0^\infty d\lambda\, \frac{\mu(\lambda)}{\left[\frac{1}{4} (d-2 \Delta )^2+\lambda ^2\right]^z}\,,\label{zetageneral}
\end{align}
where $\mu(\lambda)$ is the spectral density that is normalized to its flat-space value:
\begin{align}
    \mu(\lambda)|_{\lambda\rightarrow\infty}&=w_d \lambda^{d}\,, &w_d&=\frac{\pi}{[2^{d-1}\Gamma(\tfrac{d+1}2)]^2}\,.
\end{align}
$g(s)$ is the number of components of the irreducible transverse traceless tensor that corresponds to the spin of the field. The volume factors are self-evident. Lastly
\begin{align}
    v_d&= \frac{2^{d-1}}{\pi}\,, && u_d=v_dw_d=\frac{(\mathrm{vol}(S^d))^2}{(2 \pi )^{d+1}}\,.
\end{align}
We discuss separately the cases of odd and even dimensions below.
\paragraph{Odd dimensions.} In the case of odd dimensions, $\mathbb{H}^{2k+1}$, $d=2k$, the $\zeta$-function is obtained by a simple replacement $s_1+n\rightarrow i \lambda-\tfrac{d}2$:
\begin{align}
    \boldsymbol{\mu}(\lambda)&=\frac{1}{\mathrm{vol} ({S}^{2k+1})}\mathrm{dim}^{so(d+1,1)}\,\left[i\lambda-\tfrac{d}2,\mathbb{S}\right]\,,
\end{align}
where the boldface $\boldsymbol{\mu}(\lambda)$ contains all the factors from \eqref{zetageneral} except for the ratio of volumes. We then extract $g(s)$, $v_d$ and $w_d$ factors. For example, for any even $d$ we find for type-A with totally-symmetric spin-$s$ bosonic fields, SUSY HS with spin $s=m+\tfrac12$ fermionic fields and type-B for bosonic hook fields  $\mathbb{Y}(s,1^p)$:
\begin{align}
    \text{Type-A}&: && \mu^A(\lambda)=w_d \left(\left(\frac{d-2}{2}+s\right)^2+\lambda ^2\right)\prod_{j=0}^{\frac{d-4}{2}} \left(j^2+\lambda ^2\right)\label{evendim}\,,\\
    \text{Fermions}&: && \mu^{ferm}(\lambda)=w_d \left(\left(\frac{d-1}{2}+m\right)^2+\lambda ^2\right)\prod _{j=0}^{\frac{d-4}{2}} \left(\left(j+\frac{1}{2}\right)^2+\lambda ^2\right)\,,\\
    \text{Type-B}&: && \mu^B(\lambda)=w_d\frac{\left(\left(\frac{d-2}{2}+s\right)^2+\lambda ^2\right)}{\left(\lambda ^2+\left(\frac{d}2-p-1\right)^2\right)}\prod _{j=0}^{\frac{d-2}{2}} \left(j^2+\lambda ^2\right)\,,
\end{align}
where the spin factors are: 
\begin{align}
    g^A(s)&=\frac{(d+2 s-2) \Gamma (d+s-2)}{\Gamma (d-1) \Gamma (s+1)}=\mathrm{dim}^{so(d)}\, \mathbb{Y}(s)\label{eq:spinfactorA}\,,\\
    g^{ferm}(m)&=\frac{\Gamma (d+m-1) 2^{\left[\frac{d}{2}\right]}}{\Gamma (d-1) \Gamma (m+1)}=\mathrm{dim}^{so(d)}\, \mathbb{Y}_{\tfrac12}(m)\,, \\
    g^B(s,p)&=\frac{(d+2 s-2) \Gamma (d+s-1)}{(p+s) \Gamma (p+1) \Gamma (s) (d-p+s-2) \Gamma (d-p-1)}=\mathrm{dim}^{so(d)}\, \mathbb{Y}(s,1^p)\,.
\end{align}
The $s=1$ case of hooks corresponds to $(p+1)$-forms studied in \cite{CAMPORESI199457}; spin-$s$ bosons were investigated in \cite{Camporesi:1994ga}. The most general case in $\ads_5$ and $\ads_7$ was studied in \cite{Beccaria:2014xda,Beccaria:2014qea}.

\paragraph{Even dimensions.} In the case of even dimensions, $\mathbb{H}^{2k+2}$, $d=2k+1$, there are two complications: there can be additional discrete modes and the Plancherel measure is not a polynomial. If we ignore the discrete modes, the spectral density is a product of a formally continued dimension $d_n$ and a hyperbolic function
\begin{align}\label{schemeofzeta}
    \boldsymbol{\mu}(\lambda)&=\frac{i}{\mathrm{vol} ({S}^{2k+2})}\,\mathrm{dim}^{so(d+1,1)}\,\left[i\lambda-\tfrac{d}2,\mathbb{S}\right]h(\lambda)\,,\\
    h(\lambda)&=\begin{cases}
                    \tanh{\pi\lambda}\,, & \text{bosons}\,,\\
                    \coth{\pi \lambda}\,, & \text{fermions}\,.
                \end{cases}
\end{align}
For example, for any even $d$ we find for totally-symmetric spin-$s$ bosonic fields, spin $s=m+\tfrac12$ fermionic fields and for bosonic fields with the shape of $\mathbb{Y}(s,1^p)$-hook:
\small
\begin{align}
    \text{Type-A}&: && \mu^A(\lambda)=w_d\lambda  \tanh (\pi  \lambda ) \left(\left(\frac{d-2}{2}+s\right)^2+\lambda ^2\right)\prod_{j=1/2}^{\frac{d-4}{2}} \left(j^2+\lambda ^2\right)\label{oddim}\,,\\
    \text{SUSY}&: && \mu^{SUSY}(\lambda)=w_d\lambda  \coth (\pi  \lambda ) \left(\left(\frac{d-1}{2}+m\right)^2+\lambda ^2\right)\prod _{j=1/2}^{\frac{d-4}{2}} \left(\left(j+\frac{1}{2}\right)^2+\lambda ^2\right)\,,\\
    \text{Type-B}&: && \mu^B(\lambda)=w_d\lambda  \tanh (\pi  \lambda )\frac{\left(\left(\frac{d-2}{2}+s\right)^2+\lambda ^2\right)}{\left(\lambda ^2+\left(\frac{d}2-p-1\right)^2\right)}\prod _{j=1/2}^{\frac{d-2}{2}} \left(j^2+\lambda ^2\right)\,,
\end{align}
\normalsize
where the spin factors are the same. Degenerate hooks with $s=1$ again correspond to $(p+1)$-forms studied in \cite{CAMPORESI199457}. For symmetric bosonic fields we refer to \cite{Camporesi:1994ga}.

\paragraph{Mixed-Symmetry Fields.} As one more example of interest let us take a mixed-symmetry field of shape $\mathbb{Y}(s_1,s_2)$:
\begin{align}
   \mu^M(\lambda)&=w_d \left(\left(\frac{d-2}{2}+s_1\right)^2+\lambda ^2\right) \left(\left(\frac{d-4}{2}+s_2\right)^2+\lambda ^2\right)\times f_{E/O}\,,\\
   g^M(s_1,s_2)&=\mathrm{dim}^{so(d)}\, \mathbb{Y}(s_1,s_2)\,,\\
   f_O&=\prod _{j=0}^{\frac{d-6}{2}} \left(j^2+\lambda ^2\right)\,, \qquad\qquad \qquad \qquad \text{odd dimensions}\,,\\
   f_E&=\prod _{j=1/2}^{\frac{d-6}{2}} \left(j^2+\lambda ^2\right)\lambda  \tanh (\pi  \lambda )\,,\qquad\quad \text{even dimensions}\,.
\end{align}
The expression for the most general mixed-symmetry field with spin defined by $so(d)$ Young diagram $\mathbb{Y}(s_1,s_2,...,s_k)$ with $k$ rows follows the same pattern:
{\allowdisplaybreaks\begin{align}
   \mu^M(\lambda)&=w_d \prod_{i=1}^{i=k}\left(\left(\frac{d-2i}{2}+s_1\right)^2+\lambda ^2\right) \times f_{E/O}\,,\\
   g^M(s_1,s_2,...,s_k)&=\mathrm{dim}^{so(d)}\, \mathbb{Y}(s_1,s_2,...,s_k)\,,\\
   f_O&=\prod _{j=0}^{\frac{d-2k-2}{2}} \left(j^2+\lambda ^2\right)\,, \qquad\qquad \qquad \qquad \text{odd dimensions}\,,\\
   f_E&=\prod _{j=1/2}^{\frac{d-2k-2}{2}} \left(j^2+\lambda ^2\right)\lambda  \tanh (\pi  \lambda )\,,\qquad\quad \text{even dimensions}\,.
\end{align}}

For fermionic mixed-symmetry fields one has to correct $f_{E/O}$ factors only:
{\allowdisplaybreaks\begin{align}
   f_O&=\prod _{j=0}^{\frac{d-2k-2}{2}} \left((j+\tfrac12)^2+\lambda ^2\right)\,, \qquad\qquad \qquad \qquad \text{odd dimensions}\,,\\
   f_E&=\prod _{j=1/2}^{\frac{d-2k-2}{2}} \left((j+\tfrac12)^2+\lambda ^2\right)\lambda  \coth (\pi  \lambda )\,,\qquad\qquad \text{even dimensions}\,.
\end{align}}

Let us collect the relevant formulae with all factors now added to $\mu(\lambda)$, which we call $\tilde{\mu}(\lambda)$.
The complete spectral zeta-function is
\begin{align}
    \zeta(z)&=\int_{0}^{\infty}d\lambda\,\frac{\tilde{\mu}(\lambda)}{\left[\lambda^2+\left(\Delta-\tfrac{d}{2}\right)^2\right]^z}\,.
\end{align}
It is worth stressing that these are the zeta-functions for transverse, traceless tensors. Then, the ghosts that associated for each massless fields always come with weight $\Delta+1$ and spin $s-1$ as compared to $(\Delta,s)$ of the fields themselves and their contributions need to be substracted. Schematically, the full zeta function of HS, $\zeta_{\HS}$, is the result of the following infinity sum 
\begin{align}\label{hszetascheme}
\begin{aligned}
    \zeta^{}_{\text{HS}}(z)&=\zeta_{\Delta,0}+\sum_{s=1,2,...}\left[\zeta_{\Delta_s,s}-\zeta_{\Delta_{s+1},s-1}\right]
\end{aligned}
\end{align}
Below, we collect some simplest formulae for $\tilde{\mu}(\lambda)$ in different dimensions and for different types of HS.

\paragraph{Four Dimensions.} In four-dimensions there are no mixed-symmetry fields and bosons and fermions are described by almost the same formulae \cite{Camporesi:1993mz}
\begin{align}
    \text{bosons/fermions}&: &\tilde{\mu}(\lambda)&=\frac{\lambda  (2 s+1)  \left(\lambda ^2+\left(s+\frac{1}{2}\right)^2\right)}{6}\times \begin{cases}
                    \tanh{\pi\lambda}\,, & \text{bosons}\,,\\
                    \coth{\pi \lambda}\,, & \text{fermions}\,.
                \end{cases}
\end{align}

\paragraph{Five Dimensions.} The explicit formulae in five dimensions, i.e. $\ads_5$, are, see also \cite{Beccaria:2014xda}:
{\allowdisplaybreaks\begin{align*}
    \text{bosons}&: &\tilde{\mu}(\lambda)&=\log R \frac{\lambda ^2  (s+1)^2 \left(\lambda ^2+(s+1)^2\right)}{12 \pi }\,,\\
    \text{fermions}&: &\tilde{\mu}(\lambda)&=\log R\frac{\left(\lambda ^2+\frac{1}{4}\right)  (2 s+1) (2 s+3) \left(\lambda ^2+(s+1)^2\right)}{24 \pi }\,,\\
    \text{height-one hooks}&: &\tilde{\mu}(\lambda)&=\log R\frac{\left(\lambda ^2+1\right)  s (s+2) \left(\lambda ^2+(s+1)^2\right)}{6 \pi }\,,\\
    \text{two-row}&: &\tilde{\mu}(\lambda)&=\log R\frac{ \left(\lambda ^2+(s_1+1)^2\right) (s_1-s_2+1) (s_1+s_2+1) \left(\lambda ^2+s_2^2\right)}{6 \pi }\,.
\end{align*}}\noindent
\paragraph{Six Dimensions.} For application to HS theory based on  $F(4)$ we are also interested in six-dimensional anti-de Sitter space:{\footnotesize
{\allowdisplaybreaks\begin{align*}
    \text{bosons}&: &\tilde{\mu}(\lambda)&=-\frac{\lambda  \left(\lambda ^2+\frac{1}{4}\right) (s+1) (s+2) (2 s+3) \tanh (\pi  \lambda ) \left(\lambda ^2+\left(s+\frac{3}{2}\right)^2\right)}{720 }\,,\\
    \text{fermions}&: &\tilde{\mu}(\lambda)&=-\frac{\lambda  \left(\lambda ^2+1\right) \left(s+\frac{1}{2}\right) \left(s+\frac{3}{2}\right) \left(s+\frac{5}{2}\right) \coth (\pi  \lambda ) \left(\lambda ^2+\left(s+\frac{3}{2}\right)^2\right)}{180 }\,,\\
    \text{hooks}&: &\tilde{\mu}(\lambda)&=-\frac{\lambda  \left(\lambda ^2+\frac{9}{4}\right) s (s+3) (2 s+3) \tanh (\pi  \lambda ) \left(\lambda ^2+\left(s+\frac{3}{2}\right)^2\right)}{ 240}\,,\\
    \text{two-row}&: &\tilde{\mu}(\lambda)&=-\frac{\lambda  (2 s_1+3) (2 s_2+1) \tanh (\pi  \lambda ) (s_1-s_2+1) (s_1+s_2+2) \left(\lambda ^2+\left(s_1+\frac{3}{2}\right)^2\right) \left(\lambda ^2+\left(s_2+\frac{1}{2}\right)^2\right)}{720 }\,.
\end{align*}}}\noindent Note that for fermions we use spin $s$, rather than integer $m=s-\tfrac12$. The only hooks in Type-B theory are of shape $\mathbb{Y}(s,1)$. Also, the bosonic cases are all mutually consistent and follow from the two-row one.  We stress that fermions cannot be obtained as $s\rightarrow s+1/2$ from bosons in this case, contrary to $d=3$.

\subsection{Zeta Function Tests: Odd Dimensions}
\label{sec:zetatestsodd}
Odd dimensions are easier since evaluation of $\zeta(0)$ and $\zeta'(0)$ is of no technical difficulty. In particular, $\zeta(0)=0$ for each field individually. The new results are on mixed-symmetry fields that belong to Type-B theories and fermionic HS fields, where all the tests are successfully passed. Also, we found a general formula for the $a$-anomaly. The zeta-function for the whole multiplet of some HS theory is denoted as $\zeta_{\HS}$.

\subsubsection{Fermionic HS Fields}
\label{sec:fermionictestsO}

Firstly, $\zeta_s(0)=0$ for any $s$ and therefore the bulk result is well-defined. It is proportional to $\log R$ due to the regularized volume of $\ads_{2k+1}$. On the boundary it should be equal to the Weyl anomaly coefficient, $a\log R$, but this has been already accounted for by the contribution of bosonic HS fields. Therefore, we should check that $\zeta'_{\HS}(0)=0$. To give few examples, in $\ads_5$, see also \cite{Beccaria:2014xda}, we find that
\begin{align*}
    \frac{\zeta'_s(0)}{\log R}&=\frac{ (2 s+1)^2 (2 s (s+1) (28 s (s+1)-31)-7)}{1440}\,, && s>\frac12\,,\\
        \frac{\zeta'_s(0)}{\log R}&=-\frac{11}{180}\,,  && s=\frac12\,.
\end{align*}
Using the same exponential cut-off $\exp[-\epsilon(s+\tfrac{d-3}2)]$ we find the total $a$-coefficient to vanish
\begin{align}
    \zeta_{\HS}'(0)&=\sum_{s=\tfrac32,\tfrac52,...} \zeta'_s(0)+\zeta'_{\tfrac12}(0)=0\,.
\end{align}
In $\ads_7$ we have a more complicated formulae, but fortunately with the same result that $\zeta_{\HS}'(0)=0$, see also \cite{Beccaria:2014qea}:
\begin{align*}
    \frac{\zeta'_s(0)}{\log R}&=\frac{ (2 s+1) (2 s+3)^2 (2 s+5) (2 s (s+3) (16 s (s+3) (11 s (s+3)-1)-981)-695)}{9676800}\,, && s>\frac12\,,\\
        \frac{\zeta'_s(0)}{\log R}&=-\frac{13}{280}\,,   && s=\frac12\,.
\end{align*}
In general dimension the computation can be simplified by introducing $P_d(\lambda)=P_d(-\lambda)$:
\begin{align}
    P_d(\lambda)&=\sum_k \alpha_k \lambda^k =\prod _{j=0}^{\frac{d-4}{2}} \left(\left(j+\frac{1}{2}\right)^2+\lambda ^2\right)\,.
\end{align}
Then, with the help of the simple integration formula
\begin{align}
    a(z)=\int_0^\infty d\lambda\, \frac{\lambda ^k}{\left(b^2+\lambda ^2\right)^z}&= \frac{\Gamma \left(\frac{k+1}{2}\right) b^{k-2 z+1} \Gamma \left(-\frac{k}{2}+z-\frac{1}{2}\right)}{2 \Gamma (z)}\,,
\end{align}
where $b=\Delta-d/2$, one finds that $\zeta(0)=0$ and $\zeta'(0)$ can be obtained from (only even $k$ matters)
\begin{align}
    \pl_z a(z)\Big|_{z=0}&=\frac{-i^k(\Delta-\tfrac{d}2)^{k+1}}{4(k+1)}\,.
\end{align}
Then, it can be effortlessly checked up to any given dimension that the total $\zeta'_{\HS}(0)$ vanishes identically. In fact, it also vanishes when restricted to 'even half-integer' spins  $s=\tfrac12+2n$.

\subsubsection{Symmetric HS Fields}
\label{sec:bosonictestsO}
The case of Type-A was studied in \cite{Giombi:2014iua,Giombi:2013fka,Beccaria:2014qea,Beccaria:2014zma,Beccaria:2014xda}. Let us quote the results. As always in odd dimensions $\zeta_s(0)=0$, while  $\zeta'_s(0)$ can be computed the same way as we did for fermions. The final output is\footnote{We note that non-min. stands for non minimal which corresponds for all spins while min. stands for minimal that corresponds to even spins only.}
\begin{align}
    \zeta'_{\HS, \text{non-min.}}(0)&=0\,,\\
    \zeta'_{\HS, \text{min.}}(0)&= -2 a_\phi\log R\,,
\end{align}
where $a^d_\phi$ is the Weyl-anomaly coefficient of the free scalar field in $\cft^d$, for which one finds, see e.g. \cite{Casini:2010kt},
\begin{align}
    a_\phi^{4}&=\frac{1}{90}\,, &
    a_\phi^{6}&=-\frac{1}{756}\,, &
    a_\phi^{8}&=\frac{23}{113400}\,, &
    a_\phi^{10}&=-\frac{263}{7484400}\,. &
\end{align}

\subsubsection{Mixed-Symmetry HS Fields}
\label{sec:mixedsymmetrytestsO}
We will discuss various versions of the Type-B theory that contains mixed-symmetry fields with Young diagrams of hook shape \eqref{typebcurrent}.
The contribution of certain mixed-symmetry fields has been already studied in lower-dimensional cases of $\ads_{5,7}$ in \cite{Beccaria:2014xda,Beccaria:2014qea,Beccaria:2014zma}. With the help of the general formula for the zeta-function we can extend these results for the Type-B theory to any dimension. Here we should find that $F^1_{\ads}$ is either zero or is a multiple of the free fermion Weyl anomaly $a^d_{\psi}$, see e.g. \cite{Aros:2011iz}:
\begin{align}
       a^4_\psi&=\frac{11}{180}\,, &a^6_\psi&=-\frac{191}{7560}\,, & a^8_\psi&=\frac{2497}{226800}\,,&
       a^{10}_\psi&=-\frac{14797}{2993760}\,.
\end{align}
First of all, the spectrum of the non-minimal theory is given by the tensor product of Dirac free fermion $\Di$ that decomposes into a direct sum $\Wi\oplus\Wib$ of two Weyl fermions. With the help of Appendix \ref{app:dimschars} one finds for $\ads_{2k+1}$:
\begin{align}
    \Di\otimes\Di&=\bigoplus_{n}\Yy{n,1^{k-1}}_+\oplus\bigoplus_{n}\Yy{n,1^{k-1}}_-\oplus2\bigoplus_{n=1,i=1} \Yy{n,1^{k-i-1}}\oplus 2\bullet\,,
\end{align}
where we indicate the spin of the fields only as the conformal weight/AdS energy is obvious.\\
${}$\\
For example, in seven dimensions the contribution of the scalar field and the total contributions of hooks of height $p=0,1,2$ are:\footnote{The zeta-function for hooks with $p+1>d/2$ is the same as for the dual fields with $p+1<d/2$.}
\begin{align}
    \zeta'_0(0)&=\frac{8}{945}\,, & \zeta'_p(0)&=\left\{\frac{1}{756},-\frac{8}{945},-\frac{1}{378}\right\}\label{Adsseven}\,,
\end{align}
while in nine dimensions the contribution of the scalar field and the total contributions of hooks of height $p=0,1,2,3$ are:
\begin{align}
    \zeta'_0(0)&=\frac{9}{1400}\,, & \zeta'_p(0)&= \left\{\frac{13}{14175},-\frac{353}{56700},-\frac{13}{14175},-\frac{23}{56700}\right\}\,,\label{Adsnine}
\end{align}
the total sum being zero, as is expected.

As for the minimal theories, there are several surprises. First of all, one can take just $U(N)$-singlet sector of $\Wi$. With the help of Appendix \ref{app:dimschars} the spectrum reads
\begin{align}
    so(d=4k)&: &&
    \left\{\begin{aligned}
    (\Wi\otimes \Wi)=&\bigoplus_n\Yy{n,1^{2k-1}}_+\oplus\bigoplus_{n,i} \Yy{n,1^{2k-4i-1}}\oplus\\
    &\bigoplus_{n,i} \Yy{n,1^{2k-4i-3}}\oplus \bullet
    \end{aligned}\right.  \\
    so(d=4k+2)&: &&
    \left\{\begin{aligned}
    (\Wi\otimes \Wi)=&\bigoplus_n\Yy{n,1^{2k}}_+\oplus\bigoplus_{n,i} \Yy{n,1^{2k-4i}}\oplus\\
    &\bigoplus_{n,i} \Yy{n,1^{2k-4i-2}}
    \end{aligned} \right.
\end{align}
We see that for $d=4k$, i.e. $\ads_{4k+1}$, the spectrum does not contain symmetric higher-spin fields at all. In particular, there is no graviton. Nevertheless, the total $\zeta'_{\HS}(0)$ can be found to vanish. For example, consider $\ads_9$, for which the results on the row-by-row basis were quoted in \eqref{Adsnine}. The spectrum of $U(N)$ Weyl fermion $\Wi$ is
\begin{align}
    \Wi\otimes \Wi&=\bullet \oplus \bigoplus_{n} \Yy{n,1}\oplus \Yy{n,1,1,1}_+\,,
\end{align}
and we see that $9/1400-(353/56700)-(23/113400)=0$. The same is of course true for the $\Wi\otimes \Wib$ sub-sector: $13/14175-(13/14175)=0$. The latter sector contains symmetric HS fields, including the graviton:
\begin{align}
    \Wi\otimes \Wib&=\bigoplus_{n} \Yy{n}\oplus \Yy{n,1,1}\,.
\end{align}
For $d=4k+2$, i.e. $\ads_{4k+3}$, the $U(N)$ Weyl fermion does include totally-symmetric HS fields, so the theory looks healthy. The spectrum of the two parts is
\begin{align}
    \Wi\otimes \Wi&= \bigoplus_{n} \Yy{n}\oplus \Yy{n,1,1}_+\,,\\
    \Wi\otimes \Wib&=\bullet\oplus\bigoplus_{n} \Yy{n,1}\,.
\end{align}
Again, the two sub-sectors result in $\zeta'_{HS}(0)=0$ independently: $1/756-(1/756)=0$ and $8/945-(8/945)=0$.

As for the minimal Type-B theory there are several options. Firstly, one can take the anti-symmetric part of $\Di\otimes\Di$, which would be the minimal Type-B. Secondly, one can take the anti-symmetric part of only $\Wi\otimes\Wi$, which would be the minimalistic option. The spectrum of the minimalistic Type-B theory is even more peculiar. We refer to Appendix \ref{app:dimschars} for more detail, while giving two examples here-below. In $\ads_7$ we find, see also \cite{Beccaria:2014qea},
\begin{align}
    (\Wi\otimes \Wi)_{O(N)}&= \bigoplus_{n} \Yy{2n+1}\oplus \Yy{2n,1,1}_+\,.
\end{align}
The total $\zeta'_{\HS}(0)$ is $-(1/378)+211/7560=191/7560$, which is in accordance with the $a$-anomaly of one Weyl fermion on ${S}^6$, see also Appendix \ref{app:MrCasimir}. In $\ads_9$ the spectrum of the minimalistic Type-B is
\begin{align}
    (\Wi\otimes \Wi)_{O(N)}&=\bigoplus_{n} \Yy{2n+1,1}\oplus \Yy{2n,1,1,1}_+\,,
\end{align}
and the contribution to $\zeta'_{\HS}(0)$ is $23/5400-(3463/226800)=2497/226800$, which is again in accordance with the $a$-anomaly of the free fermion. The contribution of the symmetric part of the tensor product
\begin{align}
    (\Wi\otimes \Wi)_{S}&=\bullet\oplus\bigoplus_{n} \Yy{2n,1}\oplus \Yy{2n+1,1,1,1}_+\,,
\end{align}
which would be relevant for the $usp(N)$-singlet theory comes with the opposite sign, $-2497/226800$. The latter is obvious, of course, without any computation since the total anomaly was found to vanish.

The same pattern can be observed in other dimensions. According to the quite general law \cite{Gubser:2002vv, Diaz:2007an,Tseytlin:2013jya}, the $a$-anomaly of conformal HS fields on the boundary can be computed from the AdS side due to the fact that $a_{\text{CHS}}=-2a_{\HS}$, which is related to more general results on the ratio of determinants \cite{Barvinsky:2005ms}. Therefore, vanishing of total $a_{\HS}$ for the mixed-symmetry fields of Type-B implies the one-loop consistency of the conformal higher-spin theory with spectrum of conformal HS fields given by the sources to the single-trace operators built out of free fermion. As in the case of Type-A conformal HS theory \cite{Fradkin:1985am,Segal:2002gd}, the action is given by the $\log \Lambda$-part of the generating function of correlators of mixed-symmetry currents $\JJ_{s,p}$, \eqref{typebcurrent}:
\begin{align}
     S_{\text{CHS}}[\Psi_{s,p}]&= \log \Lambda\text{-part of} \log \int D\bar{\psi} D\psi\, e^{\int \bar{\psi}\slashed{\pl} \psi + \sum_{s,p} \JJ_{s,p}\Psi_{s,p}}\,,
\end{align}
where $\Psi_{s,p}$ are the sources for $\JJ_{s,p}$.

\subsubsection{Simplifying a-anomaly}
\label{sec:aanomaly}
We now understand that $\zeta'(0)$, which is related to the boundary $a$-anomaly, is a quite complicated expression. However, we can express $a$ through $\zeta'$ by  considering the formula \cite{Giombi:2013yva,Giombi:2014iua,Beccaria:2014qea,Beccaria:2014xda}
\begin{align}
    a'(\Delta)=\frac{1}{\log R}\frac{1}{2\Delta-d}\frac{\pl} {\pl \Delta} \zeta'_\Delta(0)\,,
\end{align}
for any $\Delta$ and any irreducible representation $\mathbb{S}$ defined by some Young diagram $\Yy{s_1,...,s_n}$ with $n$ rows. Then we find that
\begin{align}
   a'(\Delta)&=(-)^{n+1}\mathrm{dim}\, \Yy{s_1,...,s_n}\frac{\Gamma[\Delta-n]\prod_{i=1}^{n}(\Delta+s_i-i)(d+s_i-\Delta-i)}{\Gamma[\Delta-d+n+1]\Gamma[d+1]}\,.
\end{align}
$a$ does not have a nice factorized form, but it is always proportional to $\Delta-d/2$, i.e. it vanishes at $\Delta=d/2$, which is a boundary condition for the integral that allows to reconstruct $a$ from $a'$:
\begin{align}
    a(\Delta)=\frac{1}{\log R}\zeta'_\Delta(0)&=\int_{d/2}^{\Delta}dx\, (2x-d) a'(x)\,.
\end{align}

\subsection{Zeta Function Tests: Even Dimensions}
\label{sec:zetatests}
For the case of $\ads_{2n+2}$, it is much harder to compute the zeta function because of the complexity of spectral density. It is no longer a simple polynomial, but contains the functions $\tanh$ or $\coth$. Moreover, $\zeta(0)$ is generally non-zero for each field (which is due to the conformal anomaly). Below we present the main results with the technicalities devoted to Appendix \ref{app:generalzetaHS}. The most interesting case is that of mixed-symmetry fields from the Type-B theory.

\subsubsection{Fermionic HS Fields}
\label{sec:fermionictestsE}
Let us start with few examples. Computation of $\zeta(0)$ is not too difficult thanks to a handful of papers \cite{Camporesi:1993mz,Camporesi:1991nw,Giombi:2013fka}. For example, in $\ads_4$ and $\ads_6$ the sum over all fermions is zero
\begin{align*}
    \sum_{m=0}&\frac{-1200 m^4-2400 m^3-1560 m^2-360 m-47}{2880}=0\,,\\
    -\sum_{m=0}&\frac{(m+1) (m+2) \left(2016 m^6+18144 m^5+60704 m^4+92064 m^3+56462 m^2+42 m-9061\right)}{483840}=0\,.
\end{align*}
As different from odd dimensional AdS, the sum over all 'even half-integer' spins does not vanish.

The computation of $\zeta'(0)$ is an art, see Appendices for details, but it can be shown on a dimension by dimension basis that for $\ads_{4,6,8,...}$ one finds $\zeta_{\text{fermions}}'(0)=0$. Therefore, adding fermionic HS fields is consistent to a given order, which is a necessary condition for the existence of SUSY HS theories.

\subsubsection{Symmetric HS Fields}
\label{sec:bosonictestsE}
The case of symmetric HS fields was already studied in \cite{Giombi:2013fka,Giombi:2014iua}. The summary is that $\zeta_{\HS}(0)=0$ both for minimal and non-minimal Type-A theories while $\zeta'(0)$ does not vanish for the minimal Type-A and is equal to the sphere free energy of one free scalar:
\begin{align}
    \zeta_{\HS, non-min}(0)&=0\,, & \zeta_{\HS, min.}(0)&=0\,,\\
    -\tfrac12\zeta'_{\HS, non-min}(0)&=0\,, & -\tfrac12\zeta'_{\HS, min.}(0)&=F^{\phi}_{d}\,.
\end{align}
As before, the minimal Type-A requires $G^{-1}=N-1$.

\subsubsection{Mixed-Symmetry HS Fields}
\label{sec:mixedsymmetrytestsE}
This is the most interesting case. The Type-B theory in $\ads_4$ does not differ much from the Type-A --- the spectrum consists of totally-symmetric HS fields. This is not the case in $d>3$ where the spectrum of Type-B contains mixed-symmetry fields with Young diagrams of hook shape \eqref{typebcurrent} in accordance with the singlet spectrum of free fermion $\Di$. Much less is known about these theories\footnote{Some cubic interaction vertices for mixed-symmetry fields in AdS were constructed in \cite{Boulanger:2012dx,Boulanger:2011se,Boulanger:2011qt}. A part of the Type-B cubic action that contains $0-0-s$ vertices was found in \cite{Skvortsov:2015pea,Grigoriev:2018wrx,Sharapov:2019vyd}.} except that they should exist in any dimension since $\Di$ and $\Rac$ do.

\paragraph{Zeta.} First of all, it is important to check that $\zeta(0)=0$ and thus the bulk contribution is well-defined. It is convenient to present a contribution of the $\bar{\psi}\psi$ operator and of the hooks for each height $p$ separately. Here $p$ can run over $0,...,d-2$ with $p=0$ corresponding to totally-symmetric HS fields. However, one can (and should) take into account only half of the hooks since the rest can be dualized back to $p+1\leq d/2$ and the zeta function is the same. The latter is in accordance with the generalized Flato-Fronsdal theorem, which we now write for $\ads_{2k+2}$:
\begin{align}
    \Di\otimes\Di &=\bullet \oplus \bigoplus_{n,i}\Yy{n,1^{k-i-1}}\,,
\end{align}
where there is one scalar and half of the hooks. For example, in $\ads_6$ we find
\begin{align}
    \zeta_{\bar{\psi}\psi}(0)&=-\frac{37}{7560}\,, & \zeta_p(0)=\left\{-\frac{1}{1512},\frac{1}{180}\right\}\,, &&\sum\zeta_p(0) =\frac{37}{7560}\,.
\end{align}
Here one can see the contribution of the type-A fields with $s\geq1$, which is $-1/1512$. In Type-A this is canceled by the $\Delta=3$ scalar. Now, the contribution of $\bar{\psi}\psi$ is different, but there is the $p=1$ sector and $\zeta_{\HS}(0)=0$. In $\ads_8$ we find
\begin{align}
    \zeta_{\bar{\psi}\psi}(0)&=-\frac{119}{32400}\,, & \zeta_p(0)=\left\{-\frac{127}{226800},\frac{1}{280},\frac{1}{1512}\right\}\,, &&\sum\zeta_p(0) =\frac{119}{32400}\,.
\end{align}
It can be checked for higher dimensions that the total $\zeta_{\HS}(0)=0$. Now let us have a look at the minimal theories. The $O(N)$-singlet version of the Flato-Fronsdal theorem tells that
\begin{align}
    (\Di\otimes\Di)_{O(N)} =\bullet \oplus &\bigoplus_{n,i}\Yy{2n,1^{k-4i-1}}\oplus \Yy{2n,1^{k-4i-4}}\\
    &\bigoplus_{n,i}\Yy{2n+1,1^{k-4i-2}}\oplus \Yy{2n+1,1^{k-4i-3}}\,,
\end{align}
where the scalar is present whenever $(k-1)\mod4=0$ or $(k-2)\mod4=0$. Analogously to odd dimensions, simply taking anti-symmetric part of $\Di\otimes\Di$ can result in somewhat strange spectra, which may not contain graviton. Nevertheless, such spectra yield vanishing contribution to $\zeta_{\HS}(0)$. For example, in $\ads_6$ we find
\begin{align}
    (\Di\otimes\Di)_{O(N)} =\bullet \oplus &\bigoplus_{n,i}\Yy{2n,1}\oplus\bigoplus_{n,i}\Yy{2n+1}\,, \label{eventypebmin}
\end{align}
and the contribution of all odd spin fields is zero, while hooks of even spins give exactly $\frac{37}{7560}$ to cancel that of the scalar. Similar pattern is true in higher dimensions and both minimal and non-minimal Type-B have $\zeta_{\HS}(0)=0$.

\paragraph{Zeta Prime.}
The most challenging problem of computing one-loop effect is to find $\zeta'_{\HS}(0)$. Below we give the summary of our results in several dimensions, with technicalities devoted to the Appendices. While doing the calculation, we noticed that certain integrals cannot be evaluated analytically but they cancel each other at the end, also all complicated factors disappear from the final result. For the non-minimal theories the total contribution to $-\tfrac12 \zeta'_{\HS}(0)$ is:\footnote{We list here only those results that fit one line. See also a closely related paper \cite{Giombi:2016pvg}.}
\begin{align}
    \ads_4&: &&
    -\frac12 \zeta'_{\HS}(0)=-\frac{\zeta (3)}{8 \pi ^2}\,,\\
    \ads_6&: &&
    -\frac12 \zeta'_{\HS}(0)= -\frac{\zeta (3)}{96 \pi ^2}-\frac{\zeta (5)}{32 \pi ^4}\,, \\
    \ads_8&: &&
    -\frac12 \zeta'_{\HS}(0)= -\frac{\zeta (3)}{720 \pi ^2}-\frac{\zeta (5)}{192 \pi ^4}-\frac{\zeta (7)}{128 \pi ^6}\,,\\
    \ads_{10}&: &&
    -\frac12 \zeta'_{\HS}(0)=-\frac{\zeta (3)}{4480 \pi ^2}-\frac{7 \zeta (5)}{7680 \pi ^4}-\frac{\zeta (7)}{512 \pi ^6}-\frac{\zeta (9)}{512 \pi ^8}\,,     \\
    \ads_{12}&: &&
    -\frac12 \zeta'_{\HS}(0)=-\frac{\zeta (3)}{25200 \pi ^2}-\frac{41 \zeta (5)}{241920 \pi ^4}-\frac{13 \zeta (7)}{30720 \pi ^6}-\frac{\zeta (9)}{1536 \pi ^8}-\frac{\zeta (11)}{2048 \pi ^{10}}\,.
\end{align}
The case of $\ads_4$ was studied in \cite{Giombi:2013fka}. The discrepancy with the sphere free energy of free fermion, $F^d_\psi$, is systematic, see Appendix \ref{app:MrCasimir} for some explicit values. However, these numbers are not random. They can be reproduced as a difference in the free energy via RG-flow induced by a double-trace operator $O^2_\Delta$. If the operator $O_\Delta$ is bosonic the general formula for $\delta \tilde F^\phi_\Delta=\tilde F_{IR}-\tilde F_{UV}$ can be found in \cite{Klebanov:2011gs}:\footnote{Here we pass to generalized sphere free energy $\tilde F$ that is defined as $-\sin(\tfrac{\pi d}{2})F$, see e.g. \cite{Giombi:2014xxa}. }
\begin{align}
   \delta \tilde F^\phi_{\Delta}&=\frac{1}{\Gamma (d+1)}\int_0^{\Delta-d/2} u \sin (\pi  u) \Gamma \left(\frac{d}{2}+u\right) \Gamma \left(\frac{d}{2}-u\right) \, du\,.
\end{align}
The values of the free scalar $F$-energy can also be computed as $F$-difference:
\begin{align}
    \tilde F^\phi_d&= -\delta \tilde F^\phi_{\Delta=\tfrac{d-2}2}= \delta F^\phi_{\Delta=\tfrac{d+2}2}\,.
\end{align}
The numbers that resulted from the tedious computations in $\ads_{2n+2}$ arrange themselves into the following sequence:
\begin{align}
    -\frac12 \zeta'_{\HS}(0)&=\delta \tilde F^\phi_{\Delta=\tfrac{d-1}2}= -\delta \tilde F^\phi_{\Delta=\tfrac{d+1}2}\,. \label{FenergyA}
\end{align}
However, the dual of Type-B is supposed to be a fermionic theory, for which a generalization of \cite{Klebanov:2011gs} to fermionic $O_\Delta$ in any $d$ gives \cite{Giombi:2014xxa}:
\begin{equation}
     \delta \tilde F^\psi_{\Delta}= \frac{ 2}{\Gamma(d+1)}\int_0^{\Delta-d/2} \cos(\pi u) \Gamma\left(\frac{d+1}{2}+u\right)\Gamma\left(\frac{d+1}{2}-u\right)du\,.
\end{equation}
Again the free fermion $F$-energy can be computed as $F$-difference:
\begin{equation}
   \tilde F^\psi_d= \delta \tilde F^\psi_{\Delta=\frac{d+1}{2}}= -\delta \tilde F^\psi_{\Delta=\frac{d-1}{2}}\,.
\end{equation}
We observe that for $\Delta=\frac{d-2}{2}$ it will give $-\tfrac12 \zeta'_{\HS}(0)$ up to a factor of $\pm1/4$:
\begin{equation}
   -\frac12 \zeta'_{\HS}(0)=-\frac{1}{4} \delta \tilde F^\psi_{\Delta=\frac{d-2}{2}}=\frac{1}{4} \delta \tilde F^\psi_{\Delta=\frac{d+2}{2}}\,.\label{FenergyB}
\end{equation}
For the minimal theories the computations are even more involved, but the unwanted constants do cancel and we find\footnote{A word of warning is that the spectrum of the minimal Type-B is defined in \eqref{eventypebmin}. Other projections, e.g. the $usp$-constraint or various Majorana-Weyl projections, would result in a slightly different spectra, all of which yield similar numbers, i.e. the unwanted constants go away. } for the total contribution to $-\tfrac12 \zeta'_{\HS}(0)$:
\begin{align*}
    \ads_4&: &&
    -\frac12 \zeta'_{\HS}(0)=\frac{\log (2)}{8}-\frac{5 \zeta (3)}{16 \pi ^2}\,,\\
    \ads_6&: &&
    -\frac12 \zeta'_{\HS}(0)=\frac{45 \zeta (5)}{128 \pi ^4}-\frac{3 \zeta (3)}{64 \pi ^2}-\frac{3 \log(2)}{64}\,,  \\
    \ads_8&: &&
    -\frac12 \zeta'_{\HS}(0)=\frac{649 \zeta (3)}{23040 \pi ^2}-\frac{23 \zeta (5)}{1536 \pi ^4}-\frac{449 \zeta (7)}{1024 \pi ^6}+\frac{5 \log (2)}{256}\,,\\
    AdS_{10}&: &&
    -\frac12 \zeta'_{\HS}(0)=\frac{315 \zeta (7)}{4096 \pi ^6}+\frac{3825 \zeta (9)}{8192 \pi ^8}-\frac{617 \zeta (3)}{43008 \pi ^2}-\frac{85 \zeta (5)}{4096 \pi ^4}-\frac{35 \log (2)}{4096}\,,\\
    \ads_{12}&: &&
    -\frac12 \zeta'_{\HS}(0)=\frac{29 \zeta (7)}{49152 \pi ^6}+\frac{13579 \zeta (9)}{49152 \pi ^8}+\frac{31745 \zeta (11)}{32768 \pi ^{10}}-\frac{68843 \zeta (3)}{5160960 \pi ^2}-\frac{31033 \zeta (5)}{1105920 \pi ^4}-\frac{63 \log (2)}{8192}\,.
\end{align*}
Again, these numbers do not look random. Curiously enough the $\ads_6$ result equals $6F^{\phi}$.


\section{Toward HSGRAs/Vector Models Duality in Fractional Dimensions}
Inspired by the results in integer dimension, we extend the computation above to fractional dimension for some classes of HS theories. As it was mentioned already in \cite{Klebanov:2002ja}, see also \cite{Giombi:2014xxa}, the fact that the Wilson-Fisher critical point exists in $4-\epsilon$ expansion should allow one to make sense both of the dual higher-spin theory and of the duality itself in $\ads_{5-\epsilon}/\cft^{4-\epsilon}$. While there are some results in CFTs in fractional dimensions, see e.g. \cite{Giombi:2014xxa}, the bulk side's computation is difficult whenever we try to move away from integer dimensions. In \cite{Skvortsov:2017ldz}, we computed one-loop determinant of Type-A HSGRA in fractional dimensions in Lobachevsky space $\mathbb{H}^{d+1}$ and compared it with the sphere free energy $F=-\log Z_{S^d}$ of free and critical large-$N$ $O(N)$ vector models. The results on both sides of the AdS/CFT duality do match in \textit{all} dimensions, which gives an analytic proof of the results obtained for a number of fixed integer dimensions in \cite{Giombi:2013fka,Giombi:2014iua} and extends them to fractional dimension. Upon changing the boundary conditions we reproduce the difference between the sphere free energy under a double trace deformation $(\phi^2)^2$ that drives the free model at UV to the critical model in IR. 

To understand how it works, first of all, let us start from the better understood side, i.e. the CFT side. Here, there are different techniques available that allow one to make sense of at least some of the interacting CFTs in fractional dimensions. For example, the large-$N$ expansion, see e.g. \cite{Vasiliev:1981yc,Lang:1992zw,Petkou:1994ad,Moshe:2003xn}, and the $\epsilon$-expansion \cite{Wilson:1973jj}. Another technique that is useful is conformal bootstrap which can set up some computations in fractional dimensions \cite{El-Showk:2013nia}. One of the predictions that came from the conformal bootstrap technique is to show that the $2d$ Ising model smoothly turns into the $3d$ Ising model and ends up on the free theory in $4d$. Using $\epsilon$-expansion, we can access the free theory in $d=4$ starting from $d=4-\epsilon$ and take the limit $\epsilon\rightarrow 0$. The whole range $2<d<4$ is covered by the $1/N$-expansion whenever $N$ is large. There are recent studies \cite{Fei:2014yja,Mati:2016wjn} pointing out that the critical vector model can be extended to a wider range of dimension $4\leq d\leq6$.\\
${}$\\
As noted, the observable on the CFT that we will try to match with the bulk calculation is $F=-\log Z_{S^d}$. This observable should decrease along RG flow and be stationary at fixed points which are described by conformal field theories. The $d=2$ case is solved by the $c$-theorem \cite{Zamolodchikov:1986gt}, while the $4d$ case by the $a$-theorem \cite{Cardy:1988cwa,Komargodski:2011vj}. Both the central charge $c$ and the $a$ anomaly can be extracted from the sphere free energy: $F=a \log R$, where $R$ is the radius of the sphere and $c=-3a$ in $2d$. In $d=3$ there is no conformal anomaly but it was first conjectured \cite{Myers:2010xs,Jafferis:2011zi,Klebanov:2011gs} and then proved \cite{Casini:2011kv,Casini:2012ei} that $F$ works in $3d$ as well. More generally, $\tilde{F} =(-)^{({d-1})/{2}}\log Z_{S^d}$ is expected \cite{Klebanov:2011gs} to work in odd $d$, in particular in $d=1$ it gives the $g$-theorem \cite{Affleck:1991tk}. Following these results, the definition of $\tilde{F}$ is then generalized to $\tilde{F} =\sin(\tfrac{\pi d}{2})\log Z_{S^d}$ that works in \textit{all} dimensions \cite{Giombi:2014xxa}. This observable can interpolate smoothly between all dimensions but even ones. At even dimensions, there are poles that are resolved in such a way that the $a$-anomaly is captured, $\tilde F=(-1)^{d/2}\pi a/2$. $\tilde{F}$ was computed in \cite{Gubser:2002vv,Diaz:2007an,Allais:2010qq,Klebanov:2011gs,Aros:2011iz} for the cases of free CFT's and interaction ones that induced by a double-trace deformation. For the free scalar field it is 
\begin{align}\label{eq:FA}
    \tilde F^\phi=\frac{1}{\Gamma (d+1)}\int_0^1du\, u \sin (\pi  u) \Gamma \left(\frac{d}{2}+u\right) \Gamma \left(\frac{d}{2}-u\right) \,,
\end{align}
while for the change $\delta \tilde F$ induced by a double trace deformation due to an operator $O_\Delta$ of dimension $\Delta$ it is given by
\begin{align}\label{eq:deltaFA}
     \delta \tilde F_\Delta=\frac{1}{\Gamma (d+1)}\int_0^{\Delta-d/2}du\, u \sin (\pi  u) \Gamma \left(\frac{d}{2}+u\right) \Gamma \left(\frac{d}{2}-u\right) \,.
\end{align}
and we are interested in the case $\Delta=d-2$ that corresponds to $O=\phi^2$. 
\section{The One-loop Tests in Fractional Dimensions}\label{sec:fractionaltest}
In this section, we will restrict ourselves to (non)-minimal type-A HSGRA. Whether the dual CFT is free or interacting depends on the boundary conditions imposed on the scalar field, $s=0$, of the higher-spin multiplet: $\Delta=d-2$ for the free dual and $\Delta=2$ for the (large-$N$) interacting one. Therefore, altogether we have four different cases:\footnote{The second term in the brackets is to subtract the ghosts. }
\begin{align}\label{hszeta}
\begin{aligned}
    \zeta^{}_{\text{HS},\text{n.-m.}}(z)&=\zeta_{\Delta,0}+\sum_{s=1,2,...}\left[\zeta_{d+s-2,s}-\zeta_{d+s-1,s-1}\right]\,, \\
    \zeta^{}_{\text{HS},\text{min.}}(z)&=\zeta_{\Delta,0}+\sum_{s=2,4,...}\left[\zeta_{d+s-2,s}-\zeta_{d+s-1,s-1}\right]\,,
\end{aligned}
\end{align}
where $\Delta$ can be either $d-2$ or $2$. It was shown in a number of integer dimensions \cite{Giombi:2013fka,Giombi:2014iua,Giombi:2014yra} that: 
\begin{enumerate}
    \item While each term in the sum may depend on the cutoff $\Lambda$, the full one-loop vacuum energy does not depend on the cutoff $\Lambda$, i.e. $\zeta_{\text{HS}}(0)=0$ for the (non)-minimal Type-A models.
    \item The finite part vanishes for the non-minimal Type-A, $\zeta'_{\text{HS},\text{n.-m.}}(0)=0$, and equals the sphere free energy $F$ or the $a$-anomaly of the free scalar field, i.e. $a=-\tfrac12 \zeta'_{\text{HS},\text{min.}}(0)$ for $d$ even and $F=-\log Z_{S^d}=-\tfrac12 \zeta'_{\text{HS},\text{min.}}(0)$ for $d$ odd.
\end{enumerate} 
The one-loop effect we have calculated shows that there should be an integer shift in the relation between the bulk coupling constant $G$ and the number of fields $N$ on the CFT side, $G^{-1}\sim N-1$ (provided that $F^0_{\AdS}$ does match $F^0_{\CFT}$).

As discussed above, the one-loop vacuum energy in the bulk precisely match with the $a$-anomaly coefficient of the free scalar CFT in even dimensions and the sphere free energy $F$ in odd dimensions. Upon changing the boundary conditions for the scalar field it was also shown that the difference $-\tfrac12\delta \zeta'_{\text{HS}}(0)$ matches the sphere free energy of the large-$N$ interacting vector model in $d=3$ \cite{Giombi:2013fka} and $d=5$ \cite{Giombi:2014iua}.

In \cite{Skvortsov:2017ldz}, we showed that $\tfrac12 \sin(\tfrac{\pi d}{2})\zeta'_{\text{HS}}(0)$ for the minimal Type-A theory does reproduce the generalized sphere free energy \eqref{eq:FA} for all $d$. When we changed the weight of the scalar field to $\Delta=2$, the one-loop result matches the change in the sphere free energy \eqref{eq:deltaFA} due to the double-trace deformation induced by operator $(\phi^2)^2$ on the CFT side \cite{Klebanov:2011gs,Giombi:2013fka,Giombi:2014iua}. 

Let us briefly discuss the main steps that led to our result. First of all, thanks to Camporesi and Higuchi \cite{Camporesi:1994ga}, there is a representation of the spectral density that enters $\zeta_{\Delta,s}(z)$ such that it can be extended to non-integer dimensions. Next, we \textit{anchor} the \textit{runaway} branch cut $1/\big[\lambda^2+(\Delta-\frac{d}{2})\big]^z$ by applying the Laplace transform to the spectral density \cite{Bae:2016rgm,Bae:2016hfy}.\footnote{For increasing values of $\Delta$, the branch point will move away from the origin.} Effectively, this transformation also disentangles the integral over the spectral parameter and summation over spins. Then, we convert the integral into a sum over the residues. In order to handle the sum we change the regularization prescription, see also \cite{Bae:2016rgm}, but it can be checked that this does not affect the result. Finally, we arrive at the expression, which we refer to as {\it intermediate} form, whose regularized form gives \eqref{eq:FA}. The intermediate form can also be obtained directly on the CFT side from the determinant on the sphere. The interacting large-$N$ vector model requires taking into account the difference between the contributions of the scalar fields for $\Delta=d-2$ and $\Delta=2$. 

We present the computation for the one-loop tests in fractional dimension as follows. In section \ref{sec:TypeA}, we explain how to extend the computation of one-loop determinant for type-A theory to non-integer dimensions and apply the main technical tools that allow us to handle fractional dimensions: Laplace transform, contour integration and a modified regularization. In section \ref{sec:volume} we discuss the volume of the anti-de Sitter space that enters as an overall, but important, factor. The last steps on the AdS side --- summation over spins and extraction of $\zeta_{\text{HS}}(0)$ and $\zeta'_{\text{HS}}(0)$ are done in sections \ref{sec:nonminA} and \ref{sec:minA}, where we arrive at certain {\it intermediate} forms of the result that can be matched with the CFT side. The intermediate form is directly related to the free and critical vector models in Sections \ref{sec:matchingFree} and \ref{sec:matchingCritical}, which completes the proof.


\section{Higher-Spin Partition Function in Fractional Dimensions}
\label{sec:TypeA}
Coming to fractional dimensions we prefer to isolate all the factors, including the volume of the hyperbolic space, and denote the leftover as $\tilde{\mu}(\lambda)$
\begin{align}\label{mathcalN}
    \zeta(z)&= \mathcal{N} g(s)\int_{0}^{\infty}d\lambda\,\frac{\tilde{\mu}(\lambda)}{\left[\lambda^2+\left(\Delta-\tfrac{d}{2}\right)^2\right]^z}\,, && \mathcal{N}=\frac{v_d w_d \mathrm{vol}(\mathbb{H}^{d+1})}{\mathrm{vol}(S^d)}\,.
\end{align}
where 
\begin{align}\label{eq:volumeinteger}
    \text{vol}\,S^d=\frac{2\pi^{(d+1)/2}}{\Gamma\big(\frac{d+1}{2}\big)}, \qquad\qquad \text{vol}\,\mathbb{H}^{d+1}=\begin{cases}
    \frac{2(-\pi)^{d/2}}{\Gamma\big(\frac{d+2}{2}\big)}\log R, \quad d=2k\,,\\
    \pi^{d/2}\Gamma\big(-\frac{d}{2}\big),\quad d=2k+1\,.
    \end{cases}
\end{align}
It is important to stress that the $\Ncal$ in this section is the normalized constant not the number of supersymmetries. Also, the appearance of $\log R$ signals conformal anomaly. There is a representation of the spectral density that works in all dimensions \cite{Camporesi:1994ga}:
\begin{align}\label{campor}
\tilde{\mu}(\lambda)&=\left(\left(\frac{d-2}{2}+s\right)^2+\lambda ^2\right)\left|\frac{\Gamma \left(\frac{d-2}{2}+i \lambda \right)}{\Gamma (i \lambda )}\right|^2\,.
\end{align}
This is our starting point. Note that we do not have to make an assumption that $d$ is an integer in the above expression. In general, the spectral density is not a polynomial in all dimensions, including fractional ones, except for the case of even $d$. Therefore, we will treat the $zeta$-function carefully whenever $d$ approach an even number. The computation we perform below is valid for all $d$ except even (which is of measure zero on the real line). The result for even $d$ is then obtained as a continuation from non-integer $d$. 

Let us begin with the expression for the zeta-function that is obtained by collecting all the factors and expanding the gamma functions:
\begin{equation}\label{eq:zetaA}
    \zeta_{\nu,s}(z) = \mathcal{N}\frac{g(s)}{\pi}\int_0^{\infty} d\lambda \frac{\lambda \sinh(\pi \lambda)\left(\lambda^2+\left(\frac{d}{2}+s-1\right)^2\right) \Gamma\left(\frac{d}{2}+i\lambda -1\right)\Gamma\left(\frac{d}{2}-i\lambda -1\right)}{(\lambda^2+\nu^2)^z}\,,
\end{equation}
where $\nu=\Delta-\tfrac{d}{2}$. The integrand is an even function of $\lambda$ and therefore we can extend the range of integration to $(-\infty,\infty)$ at the price of $\tfrac12$. It is convenient to perform the Laplace transform, see also \cite{Bae:2016rgm,Bae:2016xmv},\footnote{One can represent the spectral zeta-function as a differential operator acting on some seed function that has enough parameters to produce $g(s)\mu(\lambda)$. Character is an example of such a function \cite{Bae:2016rgm,Bae:2016xmv}, which is also indispensable for taking tensor products. The characters are however difficult to define in non-integer dimension.} 
\begin{equation}\label{laplace}
    \frac{1}{(\lambda^2+\nu^2)^z}=\frac{\sqrt{\pi}}{\Gamma(z)}\int_0^{\infty} d\beta\, e^{-\beta \nu} \left(\frac{\beta}{2\lambda}\right)^{z-\frac{1}{2}}  J_{z-\frac{1}{2}}(\lambda \beta)\,.
\end{equation}
The main advantage is that the exponential $e^{-\beta \nu}$ times $g(s)$ can be summed over all spins in the spectrum directly. In other words, the sum over spins and the $\lambda$ integral are now decoupled. This is one of the crucial steps that allows us to calculate the full zeta function $\zeta_{\text{HS}}$, \eqref{hszeta}, in arbitrary dimension. Notice that in applying the Laplace transform we moved the branch point from $\pm i\nu$ in (\ref{eq:zetaA}) to $0$, which makes the computation feasible. Next, we split the Bessel function into 
\begin{equation}\label{Besselsplit}
    J_{\alpha}(x)=\frac{_1H_{\alpha}(x)\, +\, _2H_{\alpha}(x)}{2}\,,
\end{equation}
where $_1H_{\alpha}(x)$ and $_2H_{\alpha}(x)$ are Hankel functions of the first kind and second kind. 

\begin{figure}
    \centering
    \includegraphics[scale=0.4]{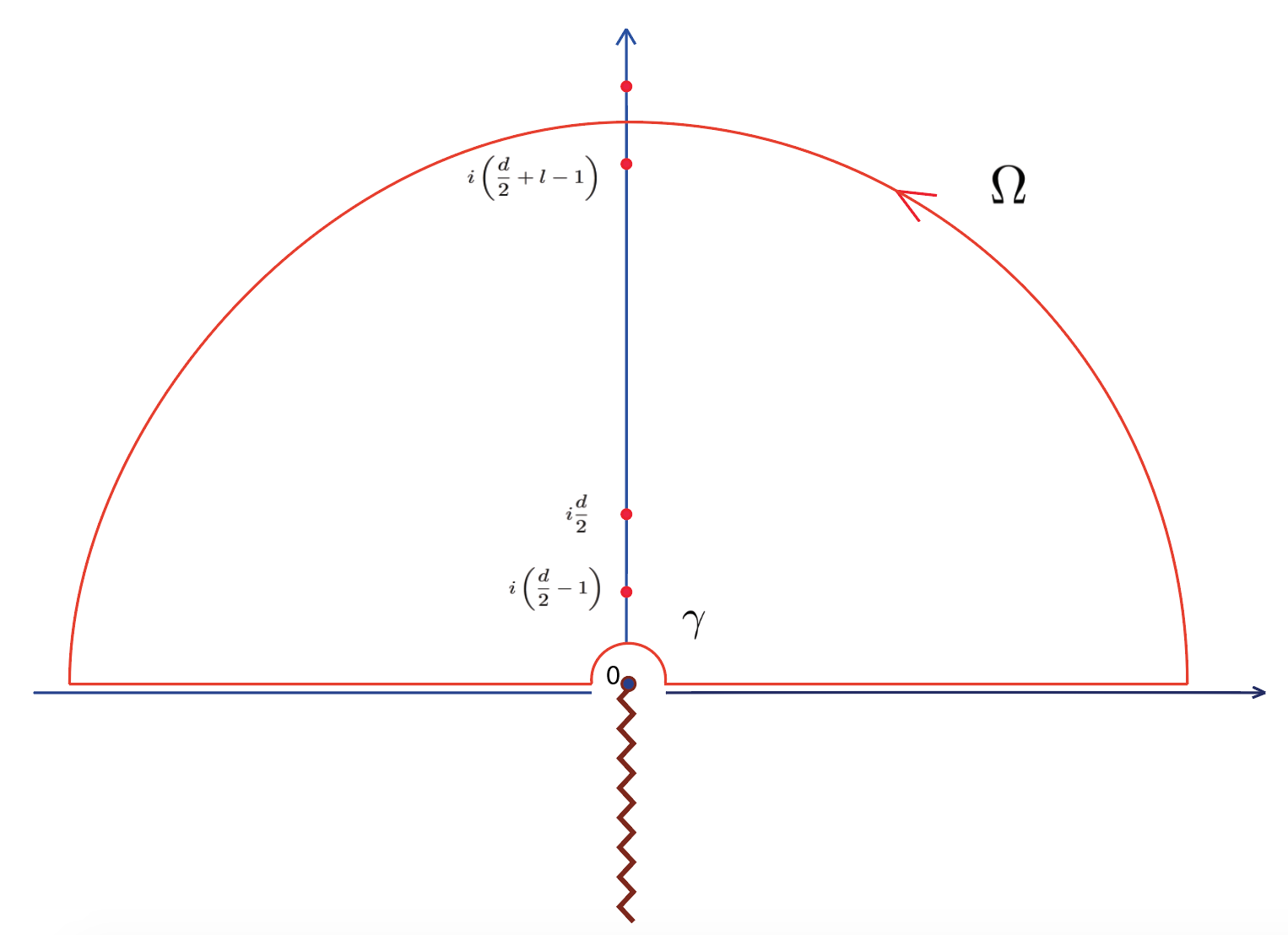}
    \caption{The contour for the part contains $_1H_{\alpha}$ lies in upper half plane where the poles are those of $\Gamma\left(\frac{d}{2}+i\lambda-1\right)$. As the $\lambda$ integral approaches $(-\infty,\infty)$, the range of $l$ also extends to infinity.}
    \label{fig:contour}
\end{figure}

Similarly to Green functions we close the contour for the part of $_1H_{\alpha}$ upward and the contour for the part of $_2H_{\alpha}$ downward. Let us show how to compute the contour integral of the part with $_1H_{\alpha}$ in (\ref{Besselsplit}) first. In order to evaluate the contribution coming from $_1H_{\alpha}$, we choose the contour as on Fig.~\ref{fig:contour}.
One needs to make sure that the upper arc of the contour does not cross any pole that comes from the $\Gamma\left(\frac{d}{2}+i\lambda-1\right)$. The residue theorem implies that
\begin{equation}\label{eq:res1}
    \oint_C f(\lambda) = 2\pi i\sum_{l=0}^{\infty} \text{Res}_{\lambda \rightarrow i\left(\frac{d}{2}+l-1\right)} \left(\lambda - i\left(\frac{d}{2}+l-1\right)\right) f(\lambda)\,,
\end{equation}
where we prefer to omit $g(s)/\pi$ for a moment:
\begin{equation}\label{eq:integrand1}
    f(\lambda)=\mathcal{N}\frac{\lambda \sinh(\pi \lambda)\left(\lambda^2+\left(\frac{d}{2}+s-1\right)^2\right) \Gamma\left(\frac{d}{2}+i\lambda -1\right)\Gamma\left(\frac{d}{2}-i\lambda -1\right) \beta^{z-\frac{1}{2}}\, _1H_{z-\frac{1}{2}}(\beta \lambda)}{2 (2\lambda)^{z-\frac{1}{2}}}
\end{equation}
We recall that the residues of $\Gamma$-function are 
\begin{equation}
 \text{Res}(\Gamma,-l) = \frac{(-1)^l}{\Gamma(l+1)}\,.
\end{equation}
We, therefore, could change the integral over $\lambda$ to an infinite sum over $l$. Before proceeding further, let us make sure that the upper arc and the contour around the branch point do not contribute to the whole contour integral. We make the change of variable $\lambda=R e^{i \theta}$:
\begin{equation}\label{eq:contour1}
    \int_{\Omega} d\lambda f(\lambda) = \lim_{R\rightarrow \infty} \int_0^{\pi} d\theta f(Re^{i\theta}) \qquad \text{and}\qquad  \int_{\gamma} d\lambda f(\lambda) = \lim_{R \rightarrow 0} \int_{\pi}^0 d\theta f(Re^{i\theta})\,.
\end{equation}
Introducing $z$ as a regulator \cite{Camporesi:1994ga,Giombi:2013fka,Giombi:2014iua,Giombi:2014yra} is useful in various ways. Let us consider the $\gamma$ contour first, if we set $z$ large enough then there is no contribution from the small contour
\begin{equation}\label{eq:bessel1}
   \lim_{\lambda \rightarrow 0} \lambda \sinh(\pi \lambda) \Gamma\left(\frac{d}{2}+i \lambda -1\right) \Gamma\left(\frac{d}{2}-i \lambda -1\right) \frac{\beta^{z-\frac{1}{2}}\, _1H_{z-\frac{1}{2}}(\beta \lambda)}{2(2\lambda)^{z-\frac{1}{2}}} = 0+ \mathcal{O}(\lambda^2) 
\end{equation}
Therefore, (\ref{eq:integrand1}) vanishes and the integral over the contour near the branch point in (\ref{eq:contour1}) also vanishes. Next, consider the large arc $\Omega$, assuming that the contour goes in between the poles of the gamma function. The integrand (\ref{eq:integrand1}) will also vanish as we make $z$ large enough in the limit where the radius $R$ goes to infinity.\footnote{For a more detailed discussion see \cite{Camporesi:1994ga}} Therefore, there is no contribution coming from $\gamma$ and $\Omega$ arcs and  (\ref{eq:res1}) is equal to
\begin{equation}
\begin{split}
    \frac{1}{2}\int_{-\infty}^{\infty}d\lambda &f(\lambda) = -\frac{\pi \mathcal{N}}{2} \sum_{l=0}^{\infty} \left(\frac{\beta^{z-\frac{1}{2}}\, _1H_{z-\frac{1}{2}}\left(i\left(\frac{d}{2}+l-1\right) \beta\right)}{\left(2 i\left(\frac{d}{2}+l-1\right)\right)^{z-\frac{1}{2}}}\right) 
    \left(\frac{d}{2}+l-1\right)\\ \times&\left(\left(\frac{d}{2}+s-1\right)^2-\left(\frac{d}{2}+l-1\right)^2\right)
    \sin\left(\pi \left(\frac{d}{2}+l-1\right)\right) \frac{\Gamma(d+l-2)}{\Gamma(l+1)} (-1)^l\,.
    \end{split}
\end{equation}
We notice that $\sin\left(\pi \left(\frac{d}{2}+l-1\right)\right) = - (-1)^l \sin\left(\frac{\pi d}{2}\right)$ and hence
\begin{equation}\label{expibeta}
\begin{split}
    \zeta(z) = & \frac{\mathcal{N}g(s)\sqrt{\pi}}{2\Gamma(z)} \int_0^{\infty} d\beta  \sum_{l=0}^{\infty} e^{-\beta \nu} \left(\frac{\beta^{z-\frac{1}{2}}\, _1H_{z-\frac{1}{2}}\left(i\left(\frac{d}{2}+l-1\right) \beta\right)}{\left(2 i\left(\frac{d}{2}+l-1\right)\right)^{z-\frac{1}{2}}}\right) 
    \\
    \times&\left(\frac{d}{2}+l-1\right)\left(\left(\frac{d}{2}+s-1\right)^2-\left(\frac{d}{2}+l-1\right)^2\right)\sin\left(\frac{\pi d}{2}\right) \frac{\Gamma(d+l-2)}{\Gamma(l+1)} \,.
    \end{split}
\end{equation}
It is difficult to say anything about the sum in general, but eventually we are interested only in few terms around $z=0$. In \cite{Bae:2016rgm}, it was argued that one can change the regularization prescription so that the $z\rightarrow0$ behaviour is not modified. Indeed, it is clear that to the leading order in $z$-expansion one can use
\begin{equation}\label{H1}
   \lim_{z\rightarrow 0}\frac{\beta^{z-\frac{1}{2}}\, _1H_{z-\frac{1}{2}}\left( \beta\lambda \right)}{2\left(2 \lambda\right)^{z-\frac{1}{2}}}= \frac{e^{i\beta\lambda}}{\sqrt{\pi}\beta}+\mathcal{O}(z)\,.
\end{equation}
This way we obtain the following contribution coming from the $_1H_{\alpha}$ function with the contour in the upper half-plane, Fig.~\ref{fig:contour}:
\begin{equation}\label{uppercontour}
\begin{split}
    \hat{\zeta}_{_1H_{\alpha}}(z) = & \frac{\mathcal{N}g(s)\sqrt{\pi}}{\Gamma(z)} \int_0^{\infty} d\beta  \sum_{l=0}^{\infty} e^{-\beta \nu} \frac{e^{-\beta\left(\frac{d}{2}+l-1\right)}}{\sqrt{\pi}\beta}
    \\
    \times&\left(\frac{d}{2}+l-1\right)\left(\left(\frac{d}{2}+s-1\right)^2-\left(\frac{d}{2}+l-1\right)^2\right)\sin\left(\frac{\pi d}{2}\right) \frac{\Gamma(d+l-2)}{\Gamma(l+1)} \,.
    \end{split}
\end{equation}
The presence of $1/\Gamma(z) \sim z$ factor in \eqref{laplace} implies that in order to get the right $\zeta(0)$ we can take only the constant term of \eqref{H1} into account. However, there should be a discrepancy between $\zeta'(0)$ computed rigorously and the one after we drop the term $\mathcal{O}(z) $ in \eqref{H1}. The difference, which we call the {\it deficit}, originates from the term of order $\mathcal{O}(z)$ in \eqref{H1}. As was noted in \cite{Bae:2016rgm} the deficit vanishes for representations that have even characters (even as a function of $\beta$, where $q=e^{-\beta}$ counts the energy via insertion of $q^E$). The deficit is discussed in Appendix \ref{app:modified}, where it is shown that it does not contribute to the full $\zeta'_{\text{HS}}(0)$.

Next, we repeat the same steps for the contribution coming from $_2H_{\alpha}$ in (\ref{Besselsplit}) where we close the contour downwards. In this case, one has to use $-2\pi i$ when applying residue theorem\footnote{The contour is the reflection image of Fig.~\ref{fig:contour} around the real axis.} for the poles of $\Gamma\left(\frac{d}{2}-i\lambda-1\right)$. 
We obtain the same structure as in (\ref{uppercontour}) since
\begin{equation}
    \lim_{\lambda \rightarrow -i(\frac{d}{2}+l-1)}   \lim_{z\rightarrow0}\, \frac{\beta^{z-1/2}\, _2H_{z-\frac{1}{2}}\left( \beta\lambda \right)}{2\left(2 \lambda)\right)^{z-\frac{1}{2}}}= \frac{e^{-\beta\left(\frac{d}{2}+l-1\right)}}{\sqrt{\pi}\beta}+\mathcal{O}(z)\,.
\end{equation}
Therefore, in order to compute the full one-loop free energy of the Type-A theory, we can write the zeta-function in a \textit{modified} form as
\begin{equation}\label{template}
\begin{split}
    \tilde{\zeta}(z) &= \frac{\mathcal{N}g(s)}{\Gamma(2z)} \int_0^{\infty} d\beta  \sum_{l=0}^{\infty} e^{-\beta \nu} \beta^{2z-1}e^{-\beta\left(\frac{d}{2}+l-1\right)} \left(\frac{d}{2}+l-1\right) \\
    & \times \left(\left(\frac{d}{2}+s-1\right)^2-\left(\frac{d}{2}+l-1\right)^2\right) \sin\left(\frac{\pi d}{2} \right) \frac{\Gamma(d+l-2)}{\Gamma(l+1)}\,.
    \end{split}
\end{equation}
Note that all but the factor $\mathcal{N}$ is usable in fractional dimensions. Below, we will regularize $\mathcal{N}$ in such a way that it allows us to work in fractional dimensions. 
\subsection{Volume of Hyperbolic Space}
\label{sec:volume}
In integer dimensions, we can use the volume form of the sphere $S^{d}$ and Hyperbolic space $\mathbb{H}^{d+1}$ as in (\ref{eq:volumeinteger}). This result arises \cite{Diaz:2007an} from the expansion of the formal volume $\pi ^{D/2} \Gamma \left(-\frac{D}{2}\right)$ in $D=d-\epsilon$:
\begin{align}\label{genvol}
  \mathrm{vol}\, \mathbb{H}^{d+1}=\frac{L_{d+1}}{\epsilon}+ V_{d+1}+ O(\epsilon)  \,,
\end{align}
where $\epsilon$-pole signals the $\log R$ divergence in $d=2k$ and $V_{d+1}$ is the finite part that makes the leading contribution for $d=2k+1$. As it was already noted in \cite{Diaz:2007an},  regularization of the volume IR divergences is not independent of regularization of the UV divergences that arise in one-loop determinants. Below, we propose an extension for the overall normalization factor which comes from the regularized volume to non-integer dimension. Note that one can write the general volume for Lobachevsky space as
\begin{equation}\label{volH}
    \text{vol} \, \mathbb{H}^{d+1}=-\frac{\pi^{\frac{d+2}{2}}}{\Gamma\left(\frac{d+2}{2}\right)\sin\left(\frac{\pi d}{2}\right)}\,,
\end{equation}
which gives the right pole as in \eqref{genvol} and reduces to $V_{d+1}$ for $d$ odd. The $\sin\left(\frac{\pi d }{2}\right)$ factor inside the modified zeta function (\ref{template}) will cancel with the one in (\ref{volH}) and gives us no poles for even dimensions. Together with the factor $\mathcal{N}$ in (\ref{mathcalN}), one arrives at the overall normalization factor in general dimensions
\begin{equation}\label{overall}
    \widetilde{\mathcal{N}}=\mathcal{N}\sin\left(\frac{\pi d }{2}\right)=-\frac{1}{\Gamma(d+1)}\,.
\end{equation}
This overall normalization factor is strikingly simple since we do not need to treat the cases of odd and even dimensions separately. Moreover, (\ref{overall}) can also be used in fractional dimension.

\section{Non-minimal Type-A in Fractional Dimensions}
\label{sec:nonminA}
Using the regularized volume, we can now write the full modified zeta-function for the Type-A as
\begin{equation}\label{eq:modzetaA}
\begin{split}
    \tilde{\zeta}(z)_{\nu,s}=&-\frac{g(s)}{\Gamma(2z)\Gamma(d+1)}\int_0^{\infty}d \beta \sum_{l=0}^{\infty} e^{-\beta \nu}\beta^{2z-1} e^{-\beta\left(\frac{d}{2}+l-1\right)}\left(\frac{d}{2}+l-1\right)\\
    \times&\left(\left(\frac{d}{2}+s-1\right)^2-\left(\frac{d}{2}+l-1\right)^2\right) \frac{\Gamma(d+l-2)}{\Gamma(l+1)} 
    \end{split}
\end{equation}
We first show that the modified zeta-function leads to $\zeta_{\text{HS}}(0)=\zeta'_{\text{HS}}(0)$ for the non-minimal Type-A theory. The total $\zeta$-function for the non-minimal Type-A is 
\begin{equation}\label{eq:nonminimalA}
    \tilde\zeta_{\text{n.-m.}}(z) = \tilde\zeta_{\frac{d}{2}-2,0}(z)+\sum_{s=1}^{\infty} \left(\tilde\zeta_{\frac{d}{2}+s-2,s}(z)-\tilde\zeta_{\frac{d}{2}+s-1,s-1}(z) \right)\,,
\end{equation}
where the labels of the zeta functions correspond to $\zeta_{\nu,s}$ as in \eqref{eq:zetaA}.
Using (\ref{eq:modzetaA}) and the spin factor in (\ref{eq:spinfactorA})
\begin{align*}
    g^A(s)=\frac{(d+2s-2)\Gamma(d+s-2)}{\Gamma(d-1)\Gamma(s+1)}
\end{align*}
we can perform the sum over all spins in \eqref{eq:nonminimalA} and get 
\begin{equation}\label{summeds}
\begin{split}
    \tilde\zeta_{\text{n.-m.}}(z)&=\sum_{l=0}^{\infty}\int_0^{\infty}\frac{d\beta\beta^{2z-1}}{\Gamma(d+1)\Gamma(2z)}\frac{e^{\frac{-\beta d}{2}}(-2+d+2l)\cosh\left(\frac{\beta}{2}\right)^2e^{-\frac{\beta}{2}(-2+d+2l)}\Gamma(-2+d+l)}{(1-e^{-\beta})^d \Gamma(l+1)}\\
    &\times (d^2+2(-2+l)l+d(-1+2l)-2l(-2+d+l)\cosh(\beta)) \\
    &=0\,.
    \end{split}
\end{equation}
It is the sum over $l$ that makes the expression  in (\ref{summeds}) vanish. Next, we need to compute $\tilde{\zeta}'_{\text{n.-m.}}(0)$ using the modified zeta-function. Remember that
\begin{equation}\label{eq:reduction1}
   \lim_{z\rightarrow 0}\frac{\beta^{2z-1}}{\Gamma(2z)}\sim \frac{2z}{\beta} + \mathcal{O}(z^2)\,.
\end{equation} 
In other words, the part of \eqref{summeds} without $1/\Gamma(2z)$ is $\tilde{\zeta}'(0)$. For the non-minimal Type-A we see that $\tilde{\zeta}'(0)$ vanishes. As a result we have proved that
\begin{equation}
    \tilde{\zeta}_{\text{n.-m.}}(0)=\tilde{\zeta}'_{\text{n.-m.}}(0)=0\,.
\end{equation}
This extends the results of \cite{Giombi:2013fka,Giombi:2014iua} to all odd dimensions as well as to fractional ones. 

\section{Minimal Type-A in Fractional Dimensions}\label{sec:minA}
The case of the minimal Type-A model is more interesting as we will not always find a $0=0$-type of equality as in the non-minimal case. The $\zeta$-function for the minimal Type-A is
\begin{equation}\label{eq:sumeven}
    \zeta_{\text{min.}}(z)=\zeta_{\frac{d}{2}-2,0}(z)+\sum_{s=2,4,...}^{\infty} \left(\zeta_{\frac{d}{2}+s-2,s}(z)-\zeta_{\frac{d}{2}+s-1,s-1}(z) \right)\,.
\end{equation}
The final result after the summation is done has a very simple form:
\begin{equation}\label{eq:important}
    \tilde{\zeta}_{\text{min.}}(z)=-\frac{1}{2\Gamma(2z)}\int_0^{\infty}d\beta \frac{\beta^{2z-1}e^{-\beta(2-d)}(1+e^{2\beta})^2}{(e^{2\beta}-1)^d}\,.
\end{equation}
To obtain (\ref{eq:important}), it is suggestive to sum over the spin-$s$ in (\ref{eq:modzetaA}) first. To do this we need to absorb all monomials in $s$ into gamma functions. For example,
\begin{equation}
    s \Gamma(d+s-2)=\Gamma(d+s-1)-(d-2)\Gamma(d+s-2)
\end{equation}
After some algebra what we obtain are several terms of the form
\begin{equation}\label{eq:block}
    \xi(\nu,p(s))=e^{-\beta \nu} \frac{\Gamma(p(s))}{\Gamma(s+1)}\,.
\end{equation}
Here $p(s)$ is of the form $s+\text{const}$ with different constants. The sums are of the usual \textit{stastistical} form. Following (\ref{eq:sumeven}) one should sum (\ref{eq:block}) according to
\begin{equation}
    \xi\left(\frac{d}{2}-2,p(0)\right)+\sum_{s=2,4,...}^{\infty} \xi\left(\frac{d}{2}+s-2,p(s)\right)-\xi\left(\frac{d}{2}+s-1,p(s-1)\right)\,,
\end{equation}
where  $\xi\left(\frac{d}{2}+s-1,p(s-1)\right)$ correspond to the ghosts. We, then, arrive at the sum over $l$:
\begin{equation*}
\begin{split}
\tilde{\zeta}_{\text{min.}}(z)&=\small\sum_{l=0}^{\infty} \int_0^{\infty}d\beta \frac{e^{-\beta\left(\frac{d}{2}+l-1\right)}(-d+2l-2)\Gamma(d+l-2)}{\Gamma(d+1)\Gamma(l+1)}\frac{e^{\beta(1-\frac{d}{2})}(-1+\coth\beta)\sinh\frac{\beta}{2}}{(1-e^{-2\beta})^d}\\
& \times \frac{\beta^{2z-1}}{\Gamma(2z)}\Bigg[-2(1+e^{-\beta})^d \left(\cosh\frac{\beta}{2}\right)^3\left((-1+d)d+2(-2+d)l+2l^2-2l(-2+d+l)\cosh\beta\right) \\
&+\cosh\beta((-1+d)d+2(-2+d)l+2l^2+2l(-2+l+1)\cosh\beta)\sinh \frac{\beta}{2}\left(1-e^{-\beta}\right)^d \Bigg]\\
    &=-\frac{1}{2\Gamma(2z)}\int_0^{\infty}d\beta \frac{\beta^{2z-1}e^{-\beta(2-d)}(1+e^{2\beta})^2}{(e^{2\beta}-1)^d}=\eqref{eq:important}\,.
    \end{split}
\end{equation*}
Formula (\ref{eq:important}) is strikingly simple. Vanishing of $\tilde{\zeta}_{\text{min.}}(0)$ is due to the fact that $\lim_{z\rightarrow0}1/\Gamma(2z)=0+\mathcal{O}(z)$. For $\tilde{\zeta}'_{\text{min.}}(0)$, using (\ref{eq:reduction1}), we arrive at 
\begin{equation}\label{eq:compactA}
    \tilde{\zeta}'_{\text{min.}}(0)=-\int_0^{\infty}d\beta \frac{e^{-\beta(2-d)}(1+e^{2\beta})^2}{\beta(e^{2\beta}-1)^d}\,. \qquad \quad 
\end{equation}
The formula above is the {\it intermediate} form.\footnote{We refer to it as intermediate as the integral is divergent and requires regularization.} After a suitable regularization it will give the correct answer for the sphere free energy as we recall in the next Section. It is worth mentioning that some of the intermediate, usually divergent, expressions on the AdS side can be directly matched with their CFT cousins, see e.g. \cite{Giombi:2014yra} for the Casimir Energy example. These facts accentuate the importance of careful adjustment of the regularization prescriptions on both sides of the duality.

\section{Matching Free Vector Model}\label{sec:matchingFree}
Having arrived at the intermediate form \eqref{eq:compactA}, we would like to show that exactly the same intermediate form emerges on the CFT side. It contains all the important information and can be directly used to derive the sphere free energy. 

Let us review the main steps in \cite{Klebanov:2011gs,Giombi:2015haa,Giombi:2014xxa} as to get the (generalized) sphere free energy $\tilde F$. The starting point is the expression for $ {F}$ for a free scalar field, which results from the sum over the eigen values of the Laplace operator on the sphere $S^d$ \cite{Diaz:2007an,Gubser:2002vv}:
\begin{equation}\label{eq:loggamma}
    F^{\phi}_{\text{min.}}= \frac{1}{2}\sum_{l=0}^{\infty} d_l \log\frac{\Gamma(\frac{d}{2}+l-1)}{\Gamma(\frac{d}{2}+l+1)}= \frac{1}{2}\sum_{l=0}^{\infty} d_l \int_0^{\infty} \frac{d\beta}{\beta}\left(-2e^{-\beta}+e^{-\beta\left(l+\frac{d}{2}\right)}+e^{-\beta\left(\frac{d}{2}+l-1\right)}\right)\,,
\end{equation}
where
\begin{equation}
    d_l=\frac{(d+2l-1)\Gamma(d+l-1)}{\Gamma(d)\Gamma(l+1)}
\end{equation}
is the degeneracy of eigen values. There is a clearly divergent part proportional to the total number of 'degrees of freedom', $\sum d_l$. This sum can be shown to vanish in a number of ways. For example, inserting cut-off $e^{-\epsilon l}$ we get
\begin{equation}
    \sum_{l=0}^{\infty} d_l e^{-\epsilon l} \sim \epsilon^{-d}\,.
\end{equation}
In order to regularize this divergence one can make $d$ negative \cite{Diaz:2007an} and then continue $d$ to the positive domain. In practice, this is equivalent to saying that the total number of degrees of freedom is zero:
\begin{equation}\label{eq:trick}
    \sum_{l=0}^{\infty}d_l=0\,.
\end{equation}
Therefore, we successfully drop the first term in \eqref{eq:loggamma}. In order to pass from $\log\Gamma$ to the intermediate form one needs to apply the integral representation of $\log \Gamma(x)$:
\begin{equation}
    \log\frac{\Gamma(\mu+\nu+1)}{\Gamma(\mu+1)}=\int_0^{\infty}\frac{d\beta}{\beta}\left(\nu e^{-\beta} -\frac{e^{-\beta \mu }-e^{-\beta(\mu+\nu)}}{e^{\beta}-1}\right)\,.
\end{equation}
As a result, (\ref{eq:loggamma}) simplifies to
\begin{equation}\label{eq:intermediatestep}
    F^{\phi}_{min}=\frac{1}{2} \int_0^{\infty}\frac{d\beta}{\beta} e^{- \frac{\beta (2 + d)}{2}} \frac{ (1 + e^{\beta})^2}{(1 - e^{-\beta})^{d}}\,.
\end{equation}
By making a change of variable, $\beta \rightarrow 2\beta$, we get exactly the intermediate form  (\ref{eq:compactA}) obtained in $AdS$ up to a factor of $(-2)$. By definition, the $AdS$ one-loop free energy is related to the sphere free energy as
\begin{equation}\label{eq:FandZeta}
    F^{\phi}_{\min}=- \frac12 \tilde{\zeta}'_{\text{min.}}(0)\,,
\end{equation}
which explains the factor $(-2)$ difference. We also note that (\ref{eq:loggamma}) leads to 
\begin{equation}\label{eq:freeF}
    F^{\phi}_{\text{min.}}= \frac{1}{2}\sum_{l=0}^{\infty} d_l \log\frac{\Gamma(\frac{d}{2}+l-1)}{\Gamma(\frac{d}{2}+l+1)}= \frac{1}{\sin\left(\frac{\pi d}{2}\right)\Gamma(d+1)}\int_0^{1} du u \sin(\pi u ) \Gamma\left(\frac{d}{2}+u\right)\Gamma\left(\frac{d}{2}-u\right)\,.
\end{equation}
In Appendix \ref{app:Proof} we show that the same result can be obtained directly from the intermediate form, i.e. the AdS result suffices to reproduce \eqref{eq:freeF} and there is no 'information loss' in going to the intermediate form. Then, the generalized sphere free energy  $\tilde{F}^{\phi}=-\sin(\tfrac{\pi d}{2})F_\phi$ is \cite{Giombi:2015haa,Giombi:2014xxa}:
\begin{equation}\label{eq:FA1}
    \tilde{F}^{\phi}_{\text{min.}}=\frac{1}{\Gamma(d+1)}\int_0^1 du \sin(\pi u) \Gamma\left(\frac{d}{2}-u\right)\Gamma\left(\frac{d}{2}+u\right)\,.
\end{equation}
Finally, we have shown that the (generalized) sphere free energy of the free scalar field results from the one-loop determinant in the minimal Type-A higher-spin theory:
\begin{equation}\label{genpoleF}
    -\sin\left(\frac{\pi d }{2}\right) F^{\phi}_{\text{min.}} = \tilde{F}^{\phi}_{\text{min.}} = \frac{1}{2}\sin\left(\frac{\pi d }{2}\right) \tilde{\zeta}'_{\text{min.}}(0) \,,
\end{equation}
which completes the proof. Despite the fact that our proof requires $d$ not to be an even integer, the final result smoothly extrapolates to $d=2k$, where there are poles that correspond to the $a$-anomaly. This extends the proof to even dimensions as well.

\section{Matching Critical Vector Model}\label{sec:matchingCritical}
Let us consider the case of the duality between the critical $O(N)$ vector model and the (non)-minimal Type-A theory where the scalar field is quantized with $\Delta=2$ ($\tilde{\nu}_{\phi}=2-\frac{d}{2}$) boundary condition. It is clear the we just need to add to $\zeta'_{\text{n.-m.}}(0)$ or $\zeta'_{\text{min.}}(0)$ the difference that is due to the change of boundary conditions for the scalar field. In this case, we see that  $\tilde{\nu}_{\phi}=-\nu_{\phi}$. As we consider the modified zeta function (\ref{eq:modzetaA}), the exponential $\exp(-\beta \nu)$ will change sign. Also, it is clear, see Appendix \ref{app:deficit}, that the deficit that can be missing from $\zeta'(0)$ due to the modified zeta-function, is absent thanks to $\tilde{\nu}_{\phi}=-\nu_{\phi}$. Repeating the procedure above, we obtain
\begin{equation}\label{eq:UV-IR}
    \delta \tilde{\zeta}'_{\phi}(0) =  \tilde{\zeta}'_{\frac{d}{2}-2,0}(0)-\tilde{\zeta}'_{2-\frac{d}{2},0}(0)  =\int_0^{\infty} \frac{ (1+e^{\beta})\left( e^{2\beta}-e^{\beta(d-2)} \right)}{\beta(e^{\beta}-1)^{d+1}}\,.
\end{equation}
This is the intermediate form that after using the same regularization as on the CFT side will give the difference between the values of the generalized sphere free energy for the free and interacting $O(N)$ vector models:
\begin{equation}
    \delta \tilde F= \tilde F_{IR}-\tilde F_{UV} = \frac{1}{\Gamma(d+1)}\int_0^{d/2-2} u \sin(\pi u)\Gamma\left(\frac{d}{2}-u\right)\Gamma\left(\frac{d}{2}+u\right)du\label{deltaF}\,.
\end{equation}
Therefore, we come to the conclusion that 
\begin{equation}\label{eq:Finteract}
    \delta F = -\frac12{\delta \tilde{\zeta}'_{\phi}}(0)
\end{equation}
Indeed, we can get (\ref{eq:Finteract}) from the CFT side through an intermediate formula which is minus one half of (\ref{eq:UV-IR}). To be more explicit,
\begin{equation}
\begin{split}
    \delta F &= \frac{1}{2}\sum_{l=0}^{\infty} d_l \log \frac{\Gamma(l+2)}{\Gamma(l+d-2)}= \frac{1}{2}\sum d_l \int_0^{\infty} \frac{d\beta}{\beta}\left((4-d)e^{-\beta}-\frac{e^{-\beta\left(l+d-3\right)}-e^{-\beta\left(l+1\right)}}{e^{\beta}-1}\right)\\
    &=-\frac{1}{2}\int_0^{\infty} \frac{ (1+e^{\beta})\left( e^{2\beta}-e^{\beta(d-2)}\right)}{\beta(e^{\beta}-1)^{d+1}}\,.
    \end{split}
\end{equation}
The same procedure as in Appendix \ref{app:Proof} allows one to relate the intermediate form to \eqref{deltaF}.

\section{Discussion and Conclusions}
\label{sec:conclusions}
In this chapter, we presented the following results:
\begin{itemize}
    \item Derivation of the spectral zeta-functions for various HSGRAs where fields are totally symmetric or mixed-symmetric.

    \item We added to the list of known one-loop results some new tests for fermions and specific mixed-symmetry fields that arise in Type-B theories. Fermionic HS fields passed both the Casimir Energy and the zeta-function tests quite easily since they are not expected to generate any one-loop corrections at all. However, it is still a non-trivial check since there is a summation over all spins which caused the cancellation at one-loop for fermionic HS fields. Type-B, which should be dual to a free fermion CFT, contains hook fields and has passed the zeta-function tests for the case of $\ads_{2n+1}/\cft^{2n}$. One finds the $a$-anomaly of free fermion in $\ads_{2n+1}$. The duality between Type-B/free fermion, however, failed naively for $\ads_{2n+2}/\cft^{2n+1}$, which was first observed for $\ads_4$ in \cite{Giombi:2013fka}. Nonetheless, we showed that the bulk one-loop results can be computed as a change in $F$-energy, \eqref{FenergyA} and \eqref{FenergyB}.

    \item After obtaining the zeta-function for a generic mixed-symmetry field, we find a very simple formula for the derivative $\pl_\Delta a(\Delta)$, which allows us to solve for $ a(\Delta)$ by a simple integration. A similar feature was observed for the second derivative of the Casimir Energy $\pl_\Delta^2 E_c$.

    \item We also tested dualities involving partially-massless fields and doubletons in the Appendices \ref{app:PMfields} and \ref{app:HSgletons}. Partially-massless fields, which belong to the spectrum of the AdS duals of the non-unitary higher-order singletons $\square^k\phi=0$, pass the tests \cite{Basile:2014wua}. On the other hand, higher-spin doubletons with $j>1$, which are unitary as representations of conformal algebra but pathological from the CFT point of view in not having a local stress tensor, do not pass the Casimir Energy test in $\ads_5/\cft^4$.
    
    \item Inspired by the results in integer dimensions, we extended the test to the fractional case for (non)-minimal Type-A HS. Our results showed that the one-loop determinants in AdS perfectly match the generalized free energy $\tilde{F}$ of a scalar on a sphere $S^d$. Upon changing boundary condition such that the scalar field is quantized with $\Delta=2$ boundary conditions, we show also that the duality between critical $O(N)$ vector model and (non)-minimal type-A theory holds for vacuum energy at one-loop.
\end{itemize}
Let us further comment on our results:

$\lozenge$ The puzzle of type-B/free fermion calls for better understanding of the duality.\footnote{It has been already noted in \cite{Giombi:2013fka} that there is a discrepancy in $\ads_4/\cft^3$ Type-B duality} Consistently with the 3d \textit{bosonization} conjecture that relates the large $N$ scalar and fermion vector models coupled to Chern-Simons theory, the free spectrum of single-trace operators built out of free fermion is identical to that of the critical boson at $N=\infty$ \cite{Leigh:2003gk}. Therefore, unless a miracle happens the two theories --- Type-A with $\Delta=2$ scalar field and Type-B --- cannot pass the one-loop test simultaneously.

$\lozenge$ The proof of the (generalized) sphere free energy $\tilde F$ of a free scalar field as a one-loop effect in the minimal Type-A higher-spin theory indicates that AdS/CFT duality may work in fractional dimensions at least for some of the models and some of the observables that are well-defined in non-integer dimensions. It would be interesting to extend the results to other models listed in Section \ref{sec:HSTheories}. For example, it should be possible to show directly in $\ads_{d+1}$ that the generalized sphere free energy of higher-spin duals of $\square^k \phi=0$ free CFT's should follow 
\begin{align}
    \tilde{F}&=\frac{1}{\Gamma (d+1)}{\int_0^{\Delta -\frac{d}{2}} u \sin (\pi  u) \Gamma \left(\frac{d}{2}-u\right) \Gamma \left(\frac{d}{2}+u\right) \, du}\,, && \Delta=\frac{1}{2} (d-2 k)\,,\quad k=1,2,...\,,
\end{align}
which is in accordance with the values for integer $d$ computed in \cite{Brust:2016gjy,Brust:2016xif}. 

For more details, we refer the readers to \cite{Gunaydin:2016amv,Skvortsov:2017ldz} and references therein. 





  \chapter{Quantum Chiral Higher Spin Gravity}\label{chapter4}

Chiral HSGRA is a special class of HSGRAs in the sense that it is the smallest higher-spin extension of gravity. The theory possesses a simple local action written in light-cone gauge in both flat and anti-de Sitter spaces \cite{Ponomarev:2016lrm,Metsaev:2018xip,Skvortsov:2018uru}, which makes it a benchmark for constructing a consistent theory of HSGRA. Numerous No-Go theorems in flat space are avoided by what we call \textit{coupling conspiracy} which is described in \cite{Skvortsov:2018jea}: local interactions conspire as to cancel each other in physical amplitudes. In this chapter, we will study quantum corrections in chiral HSGRA based on the original works \cite{Skvortsov:2018jea,Skvortsov:2020wtf,Skvortsov:2020gpn}. Due to higher-spin symmetry, we can show that the theory does not have UV-divergences in n-point amplitudes at one loop  even though the interactions are naively non-renormalizable. The same mechanism of coupling conspiracy applies to chiral HSGRA in AdS, which will improve its UV-properties. We also study Yang-Mills gaugings with $U(N)$, $SO(N)$ and $USp(N)$ groups. For $SO(N)$, see \cite{Metsaev:1991nb}, or the $USp(N)$ cases the representations that fields take values in depend on whether the spin is even or odd, which is again similar to string theory \cite{Marcus:1982fr}. Our findings indicate that higher spin fields are essential for quantization of gravity and replacing massive fields with massless ones allows us to find toy models that are much smaller and simpler than string theory, which should be helpful for understanding the quantum gravity problem. 
\section{Motivation}
HSGRAs are toy models of quantum gravity in the sense that the spin-2 graviton is a part of the spectrum that comprises massless fields of all spins and they are expected to be UV-finite due to the infinite-dimensional symmetry. This situation is very much like string theory --- a strong contender for a consistent theory of quantum gravity. String theory contains an infinite number of massive higher spin fields, which are crucial for making the theory UV-finite. Apart from having a spectrum that consists of infinitely many higher spin fields, HSGRAs also has other stringy features which make them closer to string theory rather than to conventional field theories. For example, we can have matrix-valued fields in chiral HSGRA \cite{Metsaev:1991nb}, which is reminiscent of the Chan-Paton approach \cite{Marcus:1982fr}.

Up until now, we still do not completely understand the tensionless limit of string theory, i.e. when $\alpha'\rightarrow \infty$, even in the simplest case of the bosonic string theory, (see, however, \cite{Eberhardt:2018ouy} for the tensionless limit of strings on $\ads_3$). One possible approach is as follows. We can first naively send $\alpha'\rightarrow \infty$ in the free equations \cite{Sagnotti:2003qa},\footnote{See also \cite{Sorokin:2018djm} for a recent work in this direction and \cite{Bonelli:2003kh,Lindstrom:2003mg,Bakas:2004jq,Bagchi:2016yyf} for other works on the high energy limit of string theory.} thus obtaining a consistent gauge invariant formulation of massless fields. Then, we may try to promote the original linear gauge symmetries and field equations to nonlinear ones \cite{Fotopoulos:2007nm,Sagnotti:2010at,Fotopoulos:2010ay} that result in nontrivial cubic interaction vertices \cite{Metsaev:1993ap,Metsaev:2005ar,Metsaev:2007rn,Manvelyan:2010je,Manvelyan:2010jr}. Although there is no problem at the level of cubic interactions, the quartic vertices do possess nonlocal terms which lead to failure of various consistency checks of four-point scattering amplitudes \cite{Bekaert:2010hp,Roiban:2017iqg,Bagchi:2016yyf,Dempster:2012vw}. There is not yet any consistent HSGRA that has been obtained this way.

The model we will discuss in this chapter is chiral HSGRA --- the most minimal extension of gravity with massless higher spin fields, which is constructed based on the pioneering works by Metsaev \cite{Metsaev:1991mt,Metsaev:1991nb}. At the moment, chiral HSGRA \cite{Ponomarev:2016lrm} is the only model with propagating massless higher spin fields where direct computations of quantum corrections are possible. In \cite{Skvortsov:2018jea,Skvortsov:2020wtf}, we perform the calculations for chiral HSGRA in flat space where Weinberg and Coleman-Mandula theorems dictate the S-matrix to be trivial. We show that even though the theory can avoid No-Go theorems, it does not defy the spirit of those theorems. The results in flat space hint to the expectation that other HSGRAs in AdS are UV-finite. Recall our assumption on HSGRA/Vector Model duality 
\begin{align*}
    \delta \Phi_{\ua(s)}=\nabla_{\ua}\xi_{\ua(s-1)}\qquad \Longleftrightarrow \qquad \partial^{b}\JJ_{ba(s-1)}=0\,,
\end{align*}
which means massless higher spin fields in AdS are the duals of conserved higher spin tensors. The charges associated
with the latter then form higher-spin symmetry which is an extension of the conformal symmetry. The AdS/CFT analog \cite{Maldacena:2011jn,Boulanger:2013zza,Alba:2013yda,Alba:2015upa} of the Coleman-Mandula theorem states that a CFT in $d>2$ with a higher
spin current is a free one. An immediate implication of this statement is that the holographic S-matrix (there are unique higher spin invariant holographic correlation functions \cite{Colombo:2012jx,Didenko:2013bj,Didenko:2012tv,Bonezzi:2017vha}) is also fixed by higher-spin symmetry  as in flat space.

Unlike the case of flat space where $S=1$, the holographic S-matrix of the $\ads_4$ chiral theory is shown to be nontrivial \cite{Skvortsov:2018uru} and is related to Chern-Simons Matter Theories, which should be confronted with its triviality in flat space. The reason is, when space-time is curved, higher derivative nature of the interactions becomes important and there is no perfect cancellation coming from coupling conspiracy \cite{Skvortsov:2018uru} anymore. To understand quantum consistency of AdS chiral theory, it is suggestive to first probe UV-behaviour of chiral HSGRA in flat space. If we find any UV divergence in the Minkowski space, the AdS version should suffer from the same problem. Our preliminary anticipation is that chiral HSGRA in AdS does not have UV-divergences.

One of the crucial ideas behind chiral HSGRA was to stick to the light-cone or light-front approach, which was applied to the higher spin problem in \cite{Bengtsson:1983pd,Bengtsson:1983pg} for the first time. The consistency of interactions is guaranteed by the closure of the Poincare algebra, 
\begin{align*}
    [J^{a-},J^{b-}]=0\,,\qquad \qquad [J^{a-},P^-]=0\,,
\end{align*}
much like in the light-cone quantization of string theory \cite{Goddard:1973qh}. Moreover, the light-front approach goes well with understanding gauge symmetry as just redundancy of description. An evidence for existence of higher spin theories was obtained already in 1983 \cite{Bengtsson:1983pd}: '\textit{Our conclusion is that the higher-spin theories are likely to exist, at least as classical field theories, although they may not have a manifestly covariant form}'. Due to Weinberg’s and Coleman-Mandula theorems the S-matrix approach is not applicable in flat space and we stick to the light-cone approach.  

The outline of chapter 4 is as follows. In section 2 we briefly review the analysis of deformed Poincare algebra in light-cone gauge that eventually led to the discovery of chiral HSGRA. In section 3 we give the Feynman rules, which are used in the subsequent sections to compute scattering amplitudes. In section 4 we recursively compute tree-level amplitudes by utilizing the Berends-Giele off-shell current method and show that the final amplitudes vanish on-shell. This is consistent with the Weinberg theorem. In section 5 we compute the vacuum diagrams. We shown that the vacuum loop diagrams vanish identically either due to the coupling conspiracy or due to the fact that the total number of effective degrees of freedom vanishes. In section 6 we compute the loop diagrams with external legs and demonstrate that they do not have UV-divergences and are also proportional to the total number of effective degrees of freedom, hence, can be made to vanish. Moreover, the one-loop S-matrix elements can be shown to coincide with all-plus helicity one{\color{red}-}loop amplitudes in pure QCD and SDYM, modulo a certain higher spin dressing, which is an unusual relation between the non-gravitational theories and a higher spin gravity. We conclude with section 7 that contains a summary of our results and discussion of possible future developments. We collect technicalities in Appendix \ref{app:chap4}, where  we study in detail the Chan-Paton gauging of the theory. In particular, we show that the closure of the Poincare algebra in the light-cone gauge allows for three types of gauge groups: $U(N)$, $SO(N)$ and $USp(N)$. 
\section{Chiral Higher Spin Theories}
In this section, we briefly review a recent class of HSGRA known as \textit{chiral} HSGRA \cite{Ponomarev:2016lrm} which was shown to be UV-finite up to four-point amplitude at one-loop. The theory has an action and is defined in light-cone gauge in four dimensional Minkowski and AdS spaces \cite{Metsaev:2018xip,Skvortsov:2018uru}.\footnote{We briefly discuss chiral HSGRA in AdS in Appendix \ref{app:chap4}.} 

As other theorems, the Weinberg and Coleman-Mandula theorems also have their own caveats. While the theorems restrict the impact of interactions at asymptotic region, they can not completely dictate local effects or off-shell correlators. It was shown in the past that consistent local cubic interactions of massless HS fields can exist \cite{Bengtsson:1983pd,Bengtsson:1983pg}. Later, a simple solution that ensures the closure of the Poincare algebra at the quartic order was found under the name chiral HSGRA\footnote{As the name indicates, there are more fields with positive helicities than fields with negative helicities that enter the vertices.}. 
\subsection{Flat space}
\subsubsection{Basics}
The Poincare algebra is
\begin{subequations}\label{poincarelightcone}
\begin{align}
    [L^{ab},L^{cd}]&=L^{ad}\eta^{bc}-L^{bd}\eta^{ac}-L^{ac}\eta^{bd}+L^{bc}\eta^{ad}\,\\
    [L^{ab},P^c]&=P^a\eta^{bc}-P^b\eta^{ac}\,,\\
    [P^a,P^b]&=0\,.
\end{align}
\end{subequations}
where the indices $a$ split further into $a=+,-,z,\bar{z}$ in the light-cone gauge.\footnote{Recall that the metric is $ds^2=2dx^+dx^-+2dzd\bar{z}\,$.} We will work with the \textit{light-front} approach by choosing a light-like quantization surface. The canonical choice is $x^+=0$ which makes $x^+$ behave as time and $H=P^-$ as the Hamiltonian. The \textit{dynamical} generators are generators that will receive correction when we consider interactions. There are three dynamical generators out of ten generators of $iso(3,1)$, they are
\begin{align}
    P^-=H=H_2+H_{\text{int}}\quad \text{and}\quad  J^{z-}=J_2^{z-}+J_{\text{int}}^{z-},\quad J^{\bar{z}-}=J_2^{\bar{z}-}+J_{\text{int}}^{\bar{z}-}\,,
\end{align}
where the subscript 'int' stands for interaction. For the closure of the Poincare algebra \eqref{poincarelightcone}, the equations that we need to solve are
\begin{align}\label{dynamicalPoincare}
    [H,J^{z-}]=0\,, \qquad [H,J^{\bar{z}-}]=0\,.
\end{align}
The remaining seven generators are known as \textit{kinematical} generators which are important in constraining the vertices. 
Since we work in four dimensional flat space, all massless spinning fields  have precisely two degrees of freedom. This suggests us to consider them as two scalar fields except for the case of spin-zero field --- it is just one scalar field. We can work directly in Fourier space where
\begin{align}
    \Phi^{\mu}(p,x^+)=\frac{1}{(2\pi)^{3/2}}\int e^{-i(x^-p^++p\cdot x)}\Phi^{\mu}(x,x^+)d^{3}x\,.
\end{align}
The equal time Diract bracket reads:
\begin{align}
    [\Phi^{\mu}(p,x^+),\Phi^{\lambda}(q,x^+)]=\delta^{\mu+\lambda,0}\,\frac{\delta^3(p+q)}{2p^+}\,.
\end{align}
Here, $\mu,\lambda$ are helicity labels. Denote the $p^+$ component of the four momentum $\pvec=(p^+,p^-,p,\bar{p})$ as $\beta$ from now, one finds the kinematical generators in Fourier space as \footnote{We set $x^+=0$ from now on.}
\begin{subequations}
\begin{align}
    P^+&=\beta\,,&& &P&=p\,,&& &\bar{P}&=\bar{p}\,\\
    J^{z+}&=-\beta\frac{\partial}{\partial \bar{p}}\,,&& &J^{\bar{z}+}&=-\beta\frac{\partial}{\partial p}\,, && &J^{-+}&=-\frac{\partial}{\partial \beta}\beta\,,\\
    J^{z\bar{z}}&=p\partial_p-\bar{p}\partial_{\bar{p}}-\lambda\,.
\end{align}
\end{subequations}
The dynamical generators at the free level are:
\begin{align}
    H_2&=-\frac{p\bar{p}}{\beta}\,,\\ J_2^{z-}&=\frac{\partial}{\partial \bar{p}}\frac{p\bar{p}}{\beta}+p\frac{\partial}{\partial \beta}+\lambda\frac{p}{\beta}\,,\\ J_2^{\bar{z}-}&=\frac{\partial}{\partial p}\frac{p\bar{p}}{\beta}+\bar{p}\frac{\partial}{\partial \beta}-\lambda\frac{\bar{p}}{\beta}\,.
\end{align}
The Poincare algebra is realized by charges of the form
\begin{align}
    Q_{\xi}=\int d^3p \,\beta \Phi^{-\mu}_{-p}\mathcal{O}_{\xi}(p,\pl_p)\Phi^{\mu}_p\,,\qquad \text{and}\quad  \delta_{\xi}\Phi^{\mu}(p)=[\Phi(p),Q_{\xi}]\,,
\end{align}
where $\mathcal{O}_{\xi}$ is the generator of the Poincare algebra associated with a Killing vector $\xi$\,. Due to the measure, which is $\beta$, the conjugate operators are defined as
\begin{align}
    \mathcal{O}^{\dagger}=-\frac{1}{\beta}\mathcal{O}^T(-p)\beta\,.
\end{align}
Here \textit{transpose} of $\mathcal{O}$ is defined via integration by parts. 
\subsubsection{Cubic Vertices in Flat Space}
As mentioned, the problem of a consistent HSGRA in flat space is to find $H_{\text{int}}$ and $J^{a-}_{\text{int}}$ ($a$ is refered to either $z$ or $\bar{z}$) that satisfy the Poincare algebra. The dynamical constraint \eqref{dynamicalPoincare} translates into
\begin{align}
    [H,J]=0\quad \Leftrightarrow\quad  [H_2,J_n]+[H_3,J_{n-1}]+...+[H_{n-1},J_3]+[H_n,J_2]=0\,,
\end{align}
where we write
\begin{align}
    H=H_2+H_{\text{int}}=H_2+\sum_n H_n,\qquad J=J_2+J_{\text{int}}=J_2+\sum_nJ_n\,.
\end{align}
By making an appropriate ansatz:
\begin{align}
    H_n&=\sum_n\int d^{3n}q\,\delta\Big(\sum q_i\Big)h^{q_1,...,q_n}_{\lambda_1,...,\lambda_n}\Phi^{\lambda_1}_{q_1}...\Phi^{\lambda_n}_{q_n}\,,\\
    J^{z-}_n&=\sum_n\int d^{3n}q\,\delta\Big(\sum q_i\Big)\Bigg[j_{\lambda_1...\lambda_n}^{q_1,...,q_n}-\frac{1}{n}h^{q_1,...,q_n}_{\lambda_1...\lambda_n}\Big(\sum_j\frac{\partial}{\partial \bar{q}_j}\Big)\Bigg]\Phi^{\lambda_1}_{q_1}...\Phi^{\lambda_n}_{q_n}\,,\\
     J^{\bar{z}-}_n&=\sum_n\int d^{3n}q\,\delta\Big(\sum q_i\Big)\Bigg[\bar{j}_{\lambda_1...\lambda_n}^{q_1,...,q_n}-\frac{1}{n}h^{q_1,...,q_n}_{\lambda_1...\lambda_n}\Big(\sum_j\frac{\partial}{\partial q_j}\Big)\Bigg]\Phi^{\lambda_1}_{q_1}...\Phi^{\lambda_n}_{q_n}\,,
\end{align}
with the restriction to $[H,J^{a-}]=0$ to the cubic order, i.e. we want to solve first
\begin{align}
    [H_3,J_2]=[J_3,H_2]\,.
\end{align}
The solution has the following form thanks to Metsaev \cite{Metsaev:1991mt,Metsaev:1991nb}:
\begin{align}
    h_3&=C_{+\lambda_1,+\lambda_2,+\lambda_3}\frac{\PPb^{\Lambda_3}}{\beta_1^{\lambda_1}\beta_2^{\lambda_2}\beta_3^{\lambda_3}}+\bar{C}_{-\lambda_1,-\lambda_2,-\lambda_3}\frac{\PP^{-\Lambda_3}}{\beta_1^{-\lambda_1}\beta_2^{-\lambda_2}\beta_3^{-\lambda_3}},\\
    j_3&=+\frac{2}{3}C_{+\lambda_1,+\lambda_2,+\lambda_3}\frac{\PPb^{\Lambda_3-1}}{\beta_1^{\lambda_1}\beta_2^{\lambda_2}\beta_3^{\lambda_3}}\chi^{\lambda_1,\lambda_2,\lambda_3}\,,\\
    \bar{j}_3&=-\frac{2}{3}C_{-\lambda_1,-\lambda_2,-\lambda_3}\frac{\PP^{\Lambda_3-1}}{\beta_1^{-\lambda_1}\beta_2^{-\lambda_2}\beta_3^{-\lambda_3}}\chi^{\lambda_1,\lambda_2,\lambda_3}\,,
\end{align}
where 
\begin{align}
    \Lambda_3&=\lambda_1+\lambda_2+\lambda_3,\qquad \chi=\beta_1(\lambda_2-\lambda_3)+\beta_2(\lambda_3-\lambda_1)+\beta_3(\lambda_1-\lambda_2)\\
    \PP&=\frac{1}{3}\Big[(\beta_1-\beta_2)p_3+(\beta_2-\beta_3)p_1+(\beta_3-\beta_1)p_2\Big],\qquad \text{and}\qquad \PPb=\PP(\beta,p\rightarrow \bar{p})\,.
\end{align}
Denote $\PP_{ij}=p_i\beta_j-p_j\beta_i$, one can further show that 
\begin{align}
    \PP_{12}=\PP_{23}=\PP_{31}=\PP\,,\qquad (\text{same for $\PPb$})
\end{align}
due to momentum conservation. 

The form of the cubic vertex $h_3$ is remarkably simple and can be mapped to the usual result of amplitudes with generic helicities  by the following identification
\begin{align}\label{spinor-lightcone}
    i]=\frac{2^{1/4}}{\sqrt{\beta_i}}\binom{\bar{q}_i}{-\beta_i}\qquad \Leftrightarrow\qquad [ij]=\sqrt{\frac{2}{\beta_i\beta_j}}\PPb_{ij}\,.
\end{align}
Therefore, the Hamiltonian density $h_3$ can be cast into
\begin{align}
    h_3\sim C_{\lambda_1,\lambda_2,\lambda_3}[12]^{\lambda_1+\lambda_2-\lambda_3}[23]^{\lambda_2+\lambda_3-\lambda_1}[31]^{\lambda_3+\lambda_1-\lambda_2}+c.c.\,.
\end{align}
\subsubsection{Complete Solution of HSGRA in Flat Space}
Consider the quartic level of $[H,J^{a-}]=0$ we have
\begin{align}
    [H_2,J_4^{a-}]=[J_2^{a-},H_4]+[H_3,J_3^{a-}]\,.
\end{align}
For definiteness, we consider the component of this equation with $a=z$. We note that the terms $[H_2,J_4^{a-}]$ and $[J_2^{a-},H_4]$, if non-vanishing, should be at least linear in $p$. Recall that $H_3$ contains holomorphic, denoted as $H_3(\PP)$, and anti-holomorphic parts, denoted as $H_3(\PPb)$. We see clearly that $[H_3(\PPb),J_3^{z-}]$ is $p$-independent and has to vanish by itself.\footnote{One can repeat the same analysis for $a=\bar{z}$ and see that $[H_3(\PP),J_3^{\bar{z}-}]$ is $\bar{p}$-independent.} Therefore, we have
\begin{align}
    [H_3,J_3]=0\,\qquad \Rightarrow\quad [H_3(\PPb),J_3]=0,\qquad [H_3(\PP),\bar{J}_3]=0\,.
\end{align}
Schematically, these brackets will involve the coupling constant $C,\bar{C}$ at the quadratic orders, i.e. $CC$, $\bar{C}\bar{C}$ and $C\bar{C}$. The complete solution is found by setting one of the coupling constant to zero, hence the name \textit{chiral} HSGRA. Here, we will set $\bar{C}=0$, and the cubic vertex becomes $V_3\equiv H_3=H_3(\PPb)$. Finally, we can prove that the coupling constant takes the following form \cite{Metsaev:1991mt,Metsaev:1991nb,Ponomarev:2016lrm}
\begin{align}\label{eq:magicalcoupling}
    C_{\lambda_1,\lambda_2,\lambda_3}=\frac{(l_p)^{\lambda_1+\lambda_2+\lambda_3-1}}{\Gamma[\lambda_1+\lambda_2+\lambda_3]}\equiv\frac{(l_p)^{\Lambda_3-1}}{\Gamma[\Lambda_3]}\,.
\end{align}
Here, for dimensional reason, we naturally put in by hand the Planck length $l_p$. We can see clearly that if the sum of helicities entering the vertex is less or equal to zero, the interaction will vanish, while all positive sums are allowed. Therefore, the theory violates parity. From here, one can easily write down the full Hamiltonian as
\begin{align}
    H=\int \Phi^{-\lambda}_{-p}\frac{p\bar{p}}{\beta}\Phi^{\lambda}_{p}+\int\, \delta^3\Big(\sum_{i=1}^3p_i\Big)\, h_3(\PPb)\,\Phi^{\lambda_1}_{p_1}\Phi^{\lambda_2}_{p_2}\Phi^{\lambda_3}_{p_3}\,.
\end{align}
Now, if we assume that fields take values in some matrix algebra, to be specified below, then the action reads
\begin{align}\label{eq:flatchiralaction}
    S=-\sum_{\lambda}\int d^4\pvec\,\pvec^2 \Tr\big[(\Phi^{\lambda}_{\pvec})^{\dagger}\Phi^{\lambda}_{\pvec}\big]+\sum_{\lambda_{1,2,3}}\frac{(l_p)^{\Lambda_3-1}}{\Gamma[\Lambda_3]}\int d^4\pvec_{1,2,3}\,\delta^4\big(\sum_{i=1}^3\pvec_i\big)\,\frac{\PPb^{\Lambda_3}}{\prod_{i=1}^{3}\beta_i^{\lambda_i}}\,\Tr[\prod_{i=1}^3\Phi_{\pvec_i}^{\lambda_i}]\,\,.
\end{align}
Here, as suggested, a massless gauge field with spin-$s$ is expressed by a pair of scalar fields that can carry color d.o.f, which we call Chan-Paton factors --- a terminology borrowed from string theory: 
\begin{align}
   \Phi^{\lambda}_{\pvec}\equiv (\Phi^{\lambda}_{\pvec})_aT^a\equiv (\Phi^{\lambda}_{\pvec})^A_{\ B} \quad \text{where}\quad \Phi_{\pvec}^{\pm s}\equiv \Phi^{\pm s}(\pvec)\,.
\end{align}
There are only three options for the gauge groups: 
\begin{enumerate}
    \item $U(N)$ gauging where fields are (anti)-Hermitian matrices.
    \item $SO(N)$ gauging where fields are symmetric (anti-symmetric) matrices whenever they have even (odd) spins. 
    \item $USp(N)$ gauging which is the opposite of $SO(N)$ gauging case. 
\end{enumerate}

\section{Feynman rules}
\label{sec:Feynamnrules}
Using the result in Appendices \ref{sec:u(N)color} and \ref{sec:o(N)/usp(N)color}, one can easily write down the Feynman rules for colored chiral HSGRAs. The propagator is found to be
\begin{align}
   \parbox{3.cm}{\includegraphics[scale=0.22]{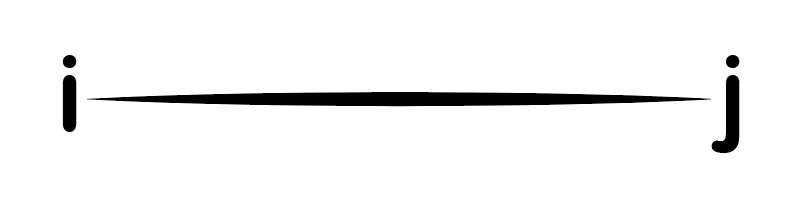}}= \parbox{3.6cm}{\includegraphics[scale=0.15]{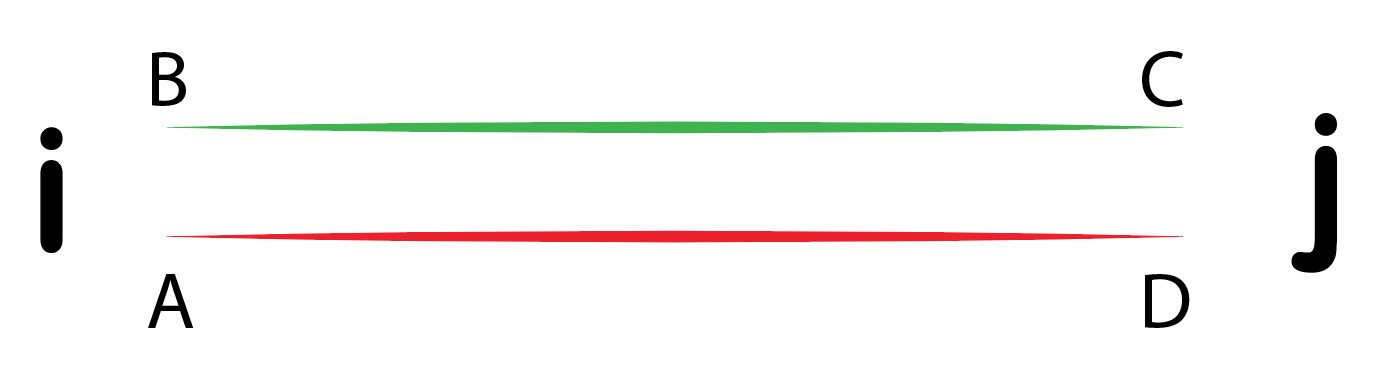}} = \frac{\delta^{\lambda_i+\lambda_j,0}\delta^4(\pvec_i+\pvec_j)}{\pvec_i^2}\,\Xi_{\text{gauge}}
\end{align}
where $\Xi_{\text{gauge}}$ is the part comes from the double line notation. For $U(N)$ gauging, which is the easiest case, we find that 
    \begin{equation}
        \Xi_{U(N)}=(-)^{\lambda_i}\delta^C_{\ B}\delta^A_{\ D}.
    \end{equation}
And, for $O(N)/USp(N)$ gauging, one finds
    \begin{align}
        \Xi_{O(N)}&=\frac{\delta_{AC}\delta_{BD}+(-)^{\lambda_i}\delta_{BC}\delta_{AD}}{2}\\
        \Xi_{USp(N)}&=\frac{C_{AC}C_{BD}+(-)^{\lambda_i+1}C_{BC}C_{AD}}{2}
    \end{align}
Note that the $\delta_{AC}\delta_{BD}$ and $C_{AC}C_{BD}$ terms corresponds to a \textit{M\"obius twist}. This makes the computation for $SO(N)/USp(N)$-valued fields a bit more subtle compare to the $U(N)$ case. 
Lastly, the vertex for all cases can be presented in the 't Hooft double lines notation as
\begin{align}
   \parbox{3.cm}{\includegraphics[scale=0.25]{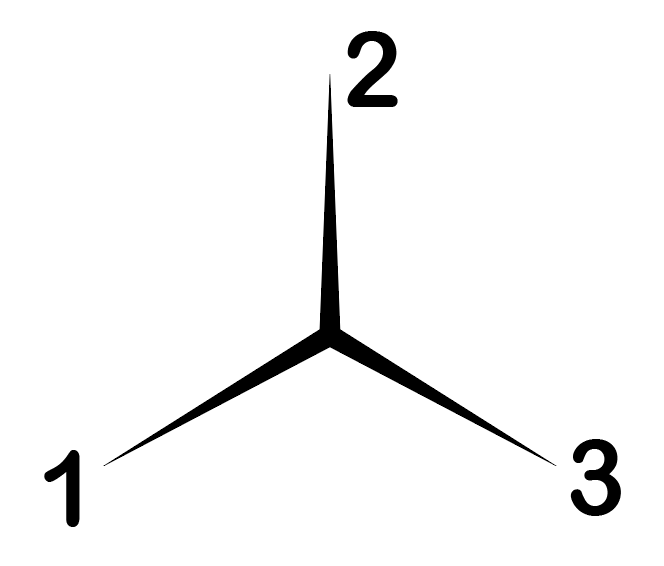}}= \parbox{3.2cm}{\includegraphics[scale=0.26]{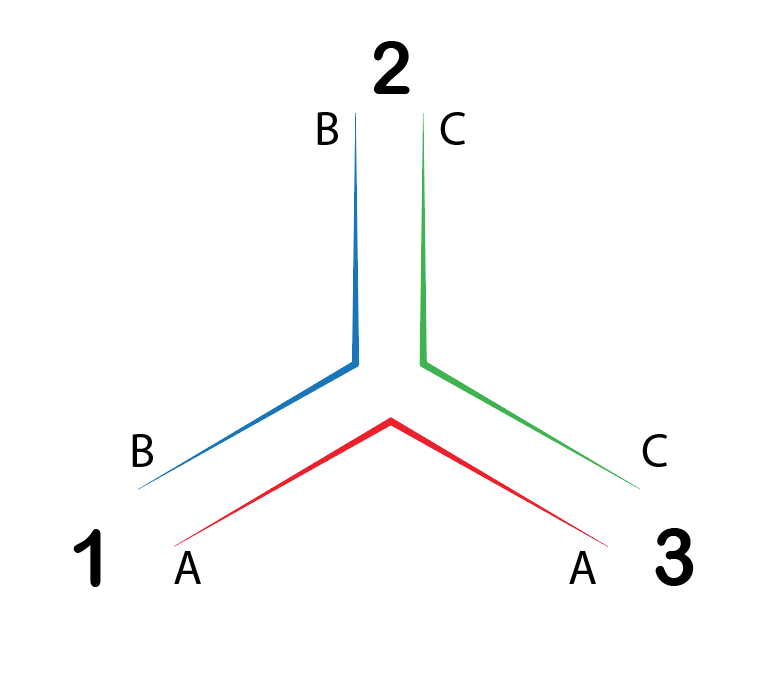}} = \delta^4(\pvec_1+\pvec_2+\pvec_3)\tr[\Phi^{\lambda_1}_{\pvec_1}\Phi^{\lambda_2}_{\pvec_2}\Phi^{\lambda_3}_{\pvec_3}]\frac{\PPb^{\lambda_1+\lambda_2+\lambda_3}}{\beta_1^{\lambda_1}\beta_2^{\lambda_2}\beta_3^{\lambda_3}}
\end{align}
where the $\tr$ is the trace over implicit $U(N)$, $O(N)/USp(N)$ indices, respectively. One should, in principles, be able to compute scattering amplitudes using all the ingredients listed here. In what follows we will compute amplitudes for the $U(N)$-case. 

\section{Tree Amplitudes}
\label{sec:trees}
In this section we compute all tree level amplitudes in chiral HSGRA. We explicitly compute $4$-, $5$- and, just for fun, $6$-point amplitudes with one off-shell leg. These amplitudes turn out to have a very simple form which leads us towards a guess for the complete $n$-point result. Then, we proceed by induction to find the $n$-point amplitude. Schematically, it can be obtained by taking one cubic vertex and attaching to two of the legs to $(n-k)$- and $k$-point amplitudes for all possible $k$, this is known as Berend-Giele off-shell current approach \cite{Berends:1987me}. This trick allows us to avoid explicit summation over all possible Feynman graph's topologies. It is crucial here to know lower order amplitudes with one off-shell leg. The result of such recursion gives us an $n+1$-point amplitude with one off-shell leg. As a matter of fact we find that all amplitudes are proportional to $\pvec^2$ of the off-shell leg and therefore vanish on-shell. We find that the S-matrix is trivial, namely $S=1$, which follows from the Weinberg soft theorem. 
\subsection{Four Point}
\label{sec:Fourpoint}
Three-point amplitudes for massless fields are identically zero dues to kinematical reasons \cite{Benincasa:2007xk}. Therefore, the simplest amplitude that may not be zero is four point. We demonstrate our work with $U(N)$ colored theory\footnote{It should be similar if one works with the case of $O(N)$ and $USp(N)$ colors. Although, as mentioned, there should be some complication due to the M\"obius twists of internal propagators.} and take advantage of the usual trick in gauge theories: to reduce everything to color-ordered amplitudes. An $n$-point amplitude can be represented as
\begin{align}
    A_n(\pvec_1,\lambda_1;...;\pvec_n,\lambda_n)&= \sum_{S_n/Z_n} \mathrm{Tr}[T_{\sigma(1)}...T_{\sigma(n)}] \hat{A}_n(\pvec_{\sigma_1},\lambda_{\sigma_1};...;\pvec_{\sigma_n},\lambda_{\sigma_n})
\end{align}
which is a sum over $(n-1)!$ permutations and $\sigma_1,...,\sigma_n$ denote various permutations of $1,...,n$. The elementary blocks, sub-amplitudes $\hat{A}_n$, should be computed using color-ordered Feynman rules. In the case of four-point the sub-amplitude consists of $s$- and $t$-channel: 
\begin{align*}
   \parbox{2.2cm}{\includegraphics[scale=0.3]{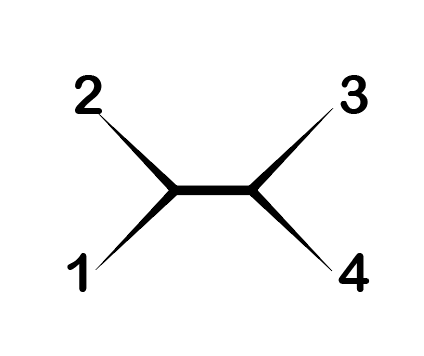}}+\parbox{2.2cm}{\includegraphics[scale=0.3]{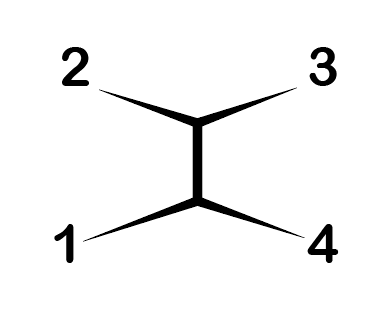}}= \parbox{2.8cm}{\includegraphics[scale=0.24]{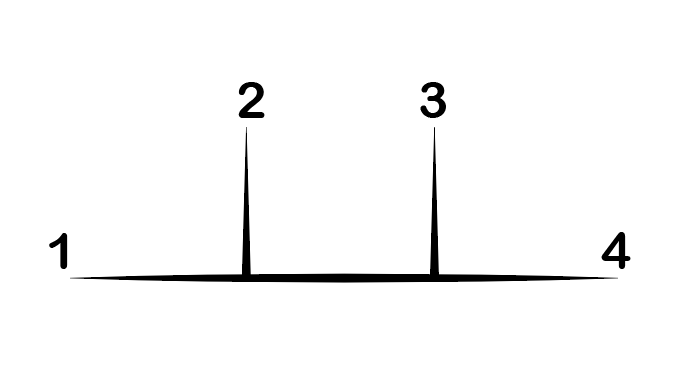}}+\parbox{2.8cm}{\includegraphics[scale=0.24]{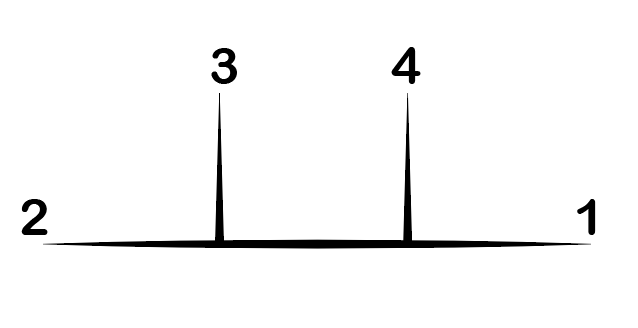}}
\end{align*}
The sum of these diagrams gives \cite{Ponomarev:2016lrm,Skvortsov:2018jea},
\begin{equation}
    A_4(1234)=\frac{\delta(\sum_i\pvec_i)}{\Gamma(\Lambda_4-1)\prod_{i=1}^4\beta_i^{\lambda_i}}\Big[\frac{\PPb_{12}\PPb_{34}(\PPb_{12}+\PPb_{34})^{\Lambda_4-2}}{(\pvec_1+\pvec_2)^2}+\frac{\PPb_{23}\PPb_{41}(\PPb_{23}+\PPb_{41})^{\Lambda_4-2}}{(\pvec_2+\pvec_3)^2}\Big]
\end{equation}
where $\Lambda_4=\lambda_1+...+\lambda_4$. In what follows we drop the overall momentum conserving $\delta$-function. 

It is important to note that the summation over helicities of the exchanged states is bounded both from above and from below due to the specific form of the magical coupling constants \eqref{eq:magicalcoupling}. If we set up an 4-pt amplitude with chiral and anti-chiral vertices, the summation is no longer bounded.

Next we use various kinematic identities from (\ref{eq:Bianchilike}) to (\ref{eq:magicidentity}) for $\PPb$ that are collected in Appendix \ref{app:kinematics}. It is easy to see that the total amplitude vanishes when all momenta are on-shell. Let us assume that the fourth momenta is off-shell, $\pvec_4^2\neq0$. Then,
\begin{equation}
    A_4(1234)=\frac{\delta(\sum_i\pvec_i)\,\alpha_4^{\Lambda_4-2}}{\Gamma(\Lambda_4-1)\prod_{i=1}^4\beta_i^{\lambda_i-1}}\frac{\beta_2\, \pvec_4^2}{2\beta_4\PP_{12}\PP_{23}}
\end{equation}
where $\alpha_4=\PPb_{12}+\PPb_{34}=\PPb_{23}+\PPb_{41}$ is cyclic invariant.
\subsection{Five Point}
\label{sec:Fivepoint}
In the case of five-point amplitude we have five diagrams, which are cyclic permutations of the comb-like diagram: 
\begin{align}
 \includegraphics[scale=0.35]{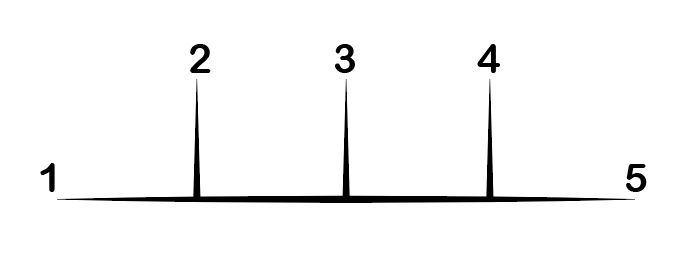}
\end{align}
After double summation over helicities, the first diagram gives 
\begin{align}\label{eq:5ptsubexample}
\hat{A}_5(12345)=\frac{1}{(\Lambda_5 -3)!\prod_{i=1}^5\beta_i^{\lambda_i}}\frac{ \PPb_{12}(\PPb_{13}+\PPb_{23})\PPb_{45}(\PPb_{45}+\PPb_{13}+\PPb_{12}+\PPb_{23})^{\Lambda_5-3}}{s_{12}\,s_{45}}
\end{align}
where $\Lambda_5=\lambda_1+...+\lambda_5$ and $s_{ij}=(\pvec_i+\pvec_j)^2$. Again, it is relatively easy to see that the full amplitude vanishes on-shell. We, however, would like to know a bit more so that we keep the fifth leg off-shell. Using the kinematic identities from Appendix \ref{app:kinematics} we can write 
\begin{align}
    \hat{A}_5(12345)+\hat{A}_5(45123)
    &=\mathcal{C}_5\frac{\PPb_{45}\beta_1\beta_2\beta_3}{2s_{45}\PP_{12}\PP_{23}}\Big[\frac{\beta_2(\beta_4+\beta_5)\pvec_5^2}{2\beta_5}-\frac{\beta_2\PPb_{45}\PP_{45}}{\beta_4\beta_5}\Big]\\
    \hat{A}_5(23451)+\hat{A}_5(51234)&=\mathcal{C}_5\frac{\PPb_{51}\beta_2\beta_3\beta_4}{2s_{51}\PP_{23}\PP_{34}}\Big[\frac{\beta_3(\beta_5+\beta_1)\pvec_5^2}{2\beta_5}-\frac{\beta_3\PPb_{51}\PP_{51}}{\beta_5\beta_1}\Big]
\end{align}
where
\begin{equation}
    \mathcal{C}_5=\frac{\alpha_5^{\Lambda_5-3}}{(\Lambda_5-3)!\prod_{i=1}^5\beta_i^{\lambda_i}}, 
\end{equation}
and $\alpha_5=\PPb_{12}+\PPb_{13}+\PPb_{23}+\PPb_{45}$ is cyclic invariant. We can simplify the above expressions further with the help of the identities
\begin{align}
   - \frac{\beta_2\PPb_{45}\PP_{45}}{\beta_4\beta_5}=\frac{1}{2}\beta_2s_{45}-\frac{\beta_2(\beta_4+\beta_5)}{2\beta_5}\pvec_5^2, \quad -\frac{\beta_3\PPb_{51}\PP_{51}}{\beta_5\beta_1}=\frac{1}{2}\beta_3s_{51}-\frac{\beta_3(\beta_1+\beta_5)}{2\beta_5}\pvec_5^2
\end{align}
where leg-5 is off-shell. Next, we collect the remnants (the parts that are not proportional to $\pvec_5^2$) and combine with the remaining comb diagram to get
\begin{equation}
\begin{split}
    \frac{1}{4}\Big(\frac{\PPb_{45}\beta_1\beta_2^2\beta_3}{\PP_{12}\PP_{23}}+\frac{\PPb_{51}\beta_2\beta_3^2\beta_4}{\PP_{23}\PP_{34}}+\frac{(\PPb_{35}+\PPb_{45})\beta_1\beta_2\beta_3\beta_4}{\PP_{12}\PP_{34}}\Big)=-\frac{\beta_1\beta_2^2\beta_3^2\beta_4\beta_5\,\pvec_5^2}{8\beta_5\PP_{12}\PP_{23}\PP_{34}}
    \end{split}
\end{equation}
The final expression of the five-point amplitude with one off-shell leg is remarkably simple:
\begin{equation}\label{eq:5ptcyclic}
    A_5(12345)=\sum_{Z_5} \hat{A}_5(12345)=-\frac{\alpha_5^{\Lambda-3}}{(\Lambda_5-3)!\prod_{i=1}^5 \beta_i^{\lambda_i-1}} \frac{\beta_2\beta_3\,\pvec_5^2}{8\beta_5\PP_{12}\PP_{23}\PP_{34}}
\end{equation}
It is quite crucial that the factor raised to power $\Lambda_5-3$ is the same for all amplitudes (even though it is not immediately obvious) and therefore we have to deal only with rather simple prefactors.
\subsection{Six Point}
\label{app:sixpoint}
Just for fun we can compute the six-point function directly. Here we have four topologies plus permutations. We will denote the topologies by the Roman numbers: $I,II,III,IV$.
\begin{figure}[h]
    \centering
    \includegraphics[scale=0.5]{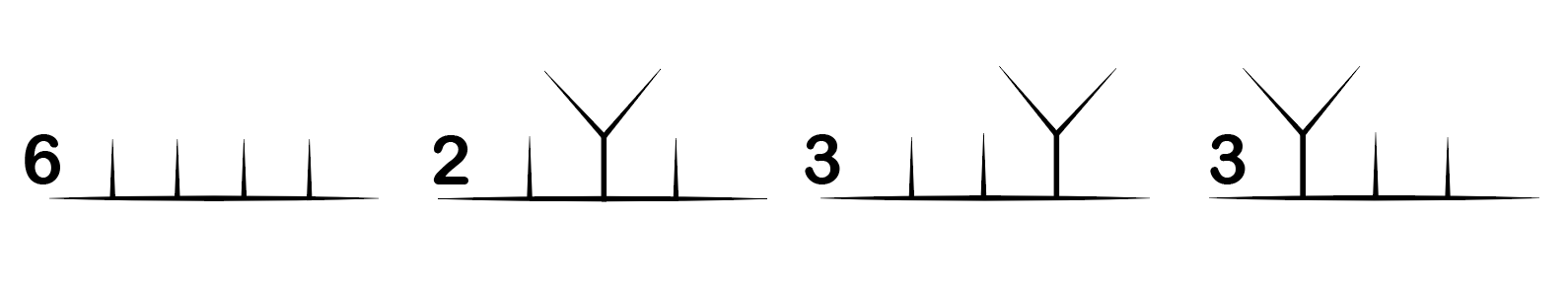}
    \caption{All possible topologies to compute 6-point amplitude. Note that the number in front of each topologies account for how many diagrams are there.}
    \label{fig:6pttopologies}
\end{figure}

The four topologies give:
\begin{align*}
     \hat{A}_I(123456)&=\frac{\PPb_{12}(\PPb_{13}+\PPb_{23})(\PPb_{14}+\PPb_{24}+\PPb_{34})\PPb_{56}}{\Gamma(\Lambda_6-3)\prod_{i=1}^6\beta_i^{\lambda_i}s_{12}s_{123}s_{56}}\alpha_6^{\Lambda_6-4}\\
     \hat{A}_{II}(123456)&=\frac{\PPb_{12}\PPb_{34}(\PPb_{61}+\PPb_{62}+\PPb_{51}+\PPb_{52})\PPb_{56}}{\Gamma(\Lambda_6-3)\prod_{i=1}^6\beta_i^{\lambda_i}s_{12}s_{34}s_{56}}\alpha_6^{\Lambda_6-4}\\
     \hat{A}_{III}(123456)&=\frac{\PPb_{12}(\PPb_{13}+\PPb_{23})(\PPb_{61}+\PPb_{62}+\PPb_{63})\PPb_{45}}{\Gamma(\Lambda_6-3)\prod_{i=1}^6\beta_i^{\lambda_i}s_{12}s_{123}s_{45}}\alpha_6^{\Lambda_6-4}\\
     \hat{A}_{IV}(456123)&=\frac{\PPb_{23}(\PPb_{13}+\PPb_{12})(\PPb_{45}+\PPb_{46})\PPb_{56}}{\Gamma(\Lambda_6-3)\prod_{i=1}^6\beta_i^{\lambda_i}s_{23}s_{456}s_{56}}\alpha_6^{\Lambda_6-4}
\end{align*}
Let us omit $\alpha_6^{\Lambda_6-4}/\Gamma(\Lambda_6-3)\prod_{i=1}^6\beta_i^{\lambda_i}$ for a moment and focus on the prefactors.\footnote{Of course, $\Lambda_6=\lambda_1+...+\lambda_6$ and $\alpha_6$ is cyclic invariant.} A short computation shows that
\begin{align*}
    A_I(123456)+A_{IV}(456123)&=\frac{\beta_1\beta_2^2\beta_3\PPb_{56}(\PPb_{14}+\PPb_{24}+\PPb_{34})}{4s_{56}\PP_{12}\PP_{23}}=\frac{\beta_1\beta_2^2\beta_3\PPb_{56}(\PPb_{45}+\PPb_{46})}{4s_{56}\PP_{12}\PP_{23}}
\end{align*}
and similarly for other permutations. Together with the contribution from  diagrams of the second topology 
\begin{align*}
    \hat{A}_{II}(123456)&=\frac{\beta_1\beta_2\beta_3\beta_4(\PPb_{61}+\PPb_{62}+\PPb_{51}+\PPb_{52})\PPb_{56}}{4\PP_{12}\PP_{34}s_{56}}=\frac{\beta_1...\beta_4(\PPb_{13}+\PPb_{14}+\PPb_{23}+\PPb_{24})\PPb_{56}}{4\PP_{12}\PP_{34}s_{56}} \\
    \hat{A}_{II}(234561)&=\frac{\beta_2\beta_3\beta_4\beta_5(\PPb_{12}+\PPb_{13}+\PPb_{62}+\PPb_{63})\PPb_{61}}{4\PP_{23}\PP_{45}s_{61}}=\frac{\beta_2...\beta_5(\PPb_{24}+\PPb_{25}+\PPb_{34}+\PPb_{35})\PPb_{61}}{4\PP_{23}\PP_{45}s_{61}} 
\end{align*}
Grouping terms proportional to $\PPb_{56}/s_{56}$, one gets
\begin{align*}
    &\beta_1...\beta_4\frac{\PPb_{56}}{4s_{56}}\Big[\frac{\beta_2(\PPb_{45}+\PPb_{46})}{\beta_4\PP_{12}\PP_{23}}+\frac{\beta_3(\PPb_{51}+\PPb_{61})}{\beta_1\PP_{23}\PP_{34}}+\frac{\PPb_{61}+\PPb_{51}+\PPb_{62}+\PPb_{52}}{\PP_{12}\PP_{34}}\Big]\\
   =& \beta_1...\beta_4\frac{\PPb_{56}}{4s_{56}}\frac{\beta_2\beta_3}{\PP_{12}\PP_{23}\PP_{34}}\Big[-\frac{(\beta_5+\beta_6)\pvec_6^2}{2\beta_6}+\frac{\PPb_{56}\PP_{56}}{\beta_5\beta_6}\Big]=-\frac{\beta_1\beta_2^2\beta_3^2\beta_4\PPb_{56}}{8\PP_{12}\PP_{23}\PP_{34}}
\end{align*}
Similarly, for terms proportional to $\PPb_{61}/s_{61}$, we get
\begin{align*}
    &\beta_2...\beta_5\frac{\PPb_{61}}{4s_{61}}\Big[\frac{\beta_3(\PPb_{51}+\PPb_{56})}{\beta_5\PP_{23}\PP_{34}}+\frac{\beta_4(\PPb_{62}+\PPb_{12})}{\beta_2\PP_{34}\PP_{45}}+\frac{\PPb_{12}+\PPb_{13}+\PPb_{62}+\PPb_{63}}{\PP_{23}\PP_{45}}\Big]\\
    =&\beta_2...\beta_5\frac{\PPb_{61}}{4s_{61}}\frac{\beta_3\beta_4}{\PP_{23}\PP_{34}\PP_{45}}\Big[-\frac{(\beta_6+\beta_1)\pvec_6^2}{2\beta_6}+\frac{\PPb_{61}\PP_{61}}{\beta_6\beta_1}\Big]=-\frac{\beta_2\beta_3^2\beta_4^2\beta_5\PPb_{61}}{8\PP_{23}\PP_{34}\PP_{45}}
\end{align*}
The remaining terms combine into
\begin{equation*}
    -\frac{\beta_1...\beta_5}{8\PP_{12}\PP_{45}}\Big[\frac{\beta_4(\PPb_{61}+\PPb_{62})}{\PP_{34}}+\frac{\beta_2(\PPb_{46}+\PPb_{56})}{\PP_{23}}\Big]=\frac{\beta_1...\beta_5}{8\PP_{12}\PP_{45}}\frac{\beta_2\beta_4}{\PP_{23}\PP_{34}}\Big[\frac{\beta_6\beta_3\pvec_6^2}{2\beta_6}+\frac{\beta_3\PPb_{61}\PP_{12}}{\beta_1\beta_2}+\frac{\beta_3\PPb_{56}\PP_{45}}{\beta_4\beta_5}\Big]
\end{equation*}
Summing all of the above partial resutls together and we get an concise expression for $6$-point amplitude:
\begin{equation}
    A(123456)=\frac{\alpha_6^{\Lambda_6-4}}{16\Gamma(\Lambda_6-3)\prod_{i=1}^6\beta_i^{\lambda_i-1}}\frac{\beta_2\beta_3\beta_4\,\pvec_6^2 }{\beta_6\PP_{12}\PP_{23}\PP_{34}\PP_{45}}
\end{equation}

\subsection{Recursive Construction for All Point}
\label{sec:npoint}
Given the results above, it is relatively easy to guess the answer for the $n$-point amplitude with one off-shell leg:
\begin{equation}\label{eq:npointrecursive}
    A_n(1...n)=\frac{(-)^n\,\alpha_n^{\Lambda_n-(n-2)}\beta_2...\beta_{n-2}\,\pvec_n^2}{2^{n-2}\Gamma(\Lambda_n-(n-3))\prod_{i=1}^{n}\beta_i^{\lambda_i-1}\beta_n\PP_{12}...\PP_{n-2,n-1}}, \quad \alpha_n=\sum_{i<j}^{n-2}\PPb_{ij}+\PPb_{n-1,n}.
\end{equation}
where $\Lambda_n=\lambda_1+...+\lambda_n$. It is easy to see that this is indeed the right answer. The $n$-point amplitude can be obtained by gluing a cubic vertex to two sub-amplitudes of $(n-k)$-point and $k$-point. It is important to know all these lower order amplitudes with one off-shell leg as to be able to attach them to the cubic vertex via propagator. The process is illustrated below 
\begin{align}\label{eq:recursivegraph}
    \parbox{10cm}{\includegraphics[scale=0.45]{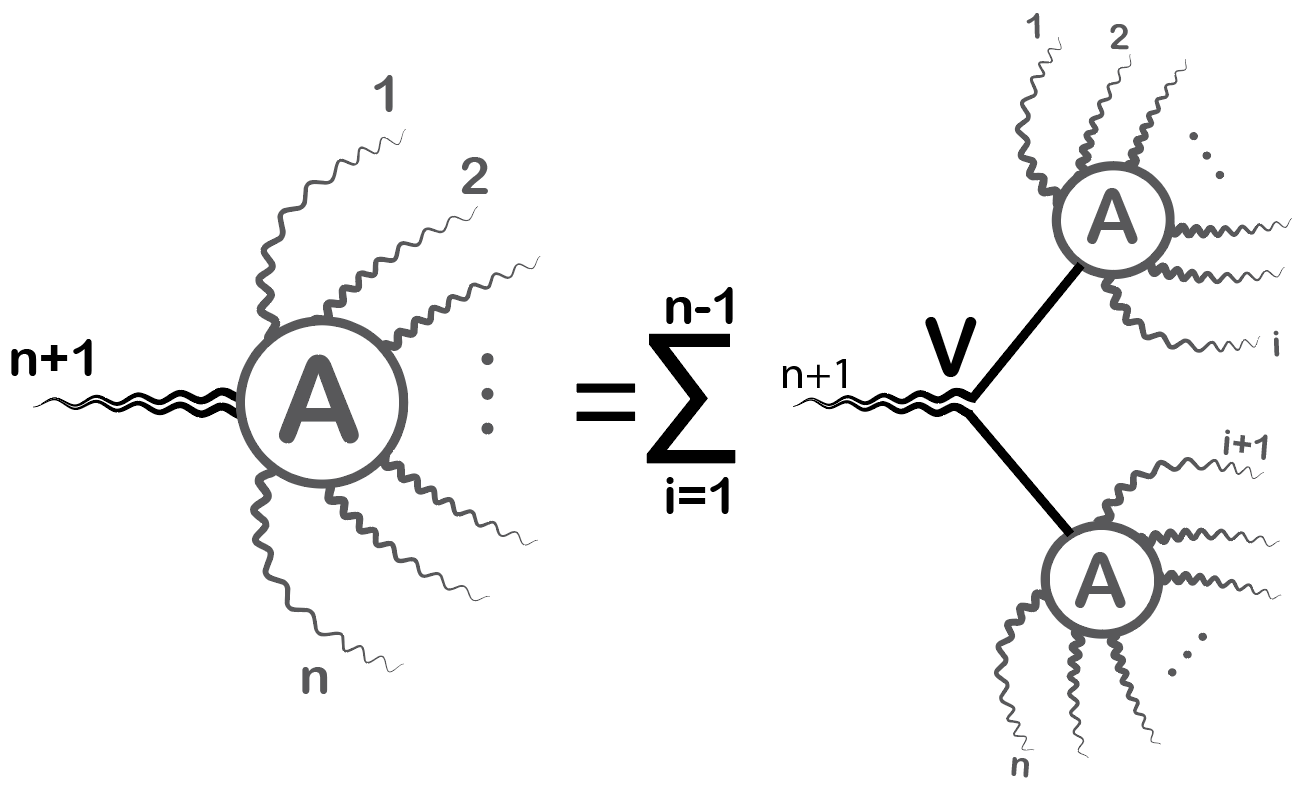}}
\end{align}
There are two position of the off-shell legs where we need to treat them specially in (\ref{eq:recursivegraph}). Note that we choose our color-ordering as (12...n+1) with clock wise order. The first diagram we want to consider is
\begin{equation}\label{eq:extendfrom1}
     \hat{A}(1...n+1)_1=\mathcal{C}_{n+1}\frac{\beta_3...\beta_{n-1}\PPb_{n+1,1}\PP_{12}}{\beta_1}
\end{equation}
where the subscript at the end of $\hat{A}(1...n+1)_1$ indicates the position of the off-shell leg that we use to glue to the cubic vertex $\boldsymbol{V}$. And, the pre-factor is
\begin{equation}
    \mathcal{C}_{n+1}=\frac{\alpha_{n+1}^{\Lambda_{n+1}-(n-1)}}{2^{n-2}\Gamma(\Lambda_{n+1}-(n-2))\prod_{i=1}^{n+1}\beta_i^{\lambda_i-1}}\frac{1}{\beta_{n+1}\PP_{12}...\PP_{n-1,n}}.
\end{equation}
It can be proven that the $\alpha_{n+1}^{\Lambda_{n+1}-(n-1)}$ factor is indeed the same for every sub-diagrams []. From leg-2 to leg-$(n-1)$, the subamplitudes after summing over helicities are
\begin{align*}
    &\hat{A}(12|3...n+1)_{2,3}=\mathcal{C}_{n+1} \underline{\beta_2}\underline{\beta_3}...\beta_{n-1}\PPb_{n+1|12}\PP_{23},\\
    &\hat{A}(123|4...n+1)_{3,4}=\mathcal{C}_{n+1} \beta_2\underline{\beta_3\beta_4}...\beta_{n-1}\PPb_{n+1|123}\PP_{34},\\
    &\cdots\\
    &\hat{A}(12...i|i+1...n+1)_{i,i+1}=\mathcal{C}_{n+1} \beta_2...\underline{\beta_i\beta_{i+1}}...\beta_{n-1}\PPb_{n+1|1...i}\PP_{i,i+1}.
\end{align*}
Here, the break $|$ between two position $i,i+1$ (also the subscript $i,i+1$) in $\hat{A}(1...i|i+1...n)_{i,i+1}$ indicates that leg-$i,i+1$ are the off-shell legs that will be glued to cubic vertex. The underlined notation means that we omit $\beta_i\beta_{i+1}$ in the above sub-amplitudes. The final piece is the sub-amplitude where we glue leg-$n$ to the cubic vertex $\boldsymbol{V}$
\begin{equation}
    \hat{A}(1...n+1)_n=\mathcal{C}_{n+1}\frac{\beta_2...\beta_{n-2}\PPb_{n+1,n}\PP_{n,n-1}}{\beta_n}.
\end{equation}
Omitting $\mathcal{C}_{n+1}$ for a moment, we have 
\begin{equation}
    \beta_3...\beta_{n-1}\frac{\PPb_{n+1|1}\PP_{12}}{\beta_1}+\sum_{i}\beta_2 ... \underline{\beta_i\beta_{i+1}}...\beta_{n-1}\PPb_{n+1|1...i}\PP_{i,i+1}+\frac{\beta_2...\beta_{n-2}\PPb_{n+1,n}\PP_{n,n-1}}{\beta_n}.
\end{equation}
Here, $\PPb_{n+1|1...i}=\PPb_{n+1,1}+\PPb_{n+1,2}+...+\PPb_{n+1,i}$ and $\PPb_{i,j}=\PPb_{ij}$. Notice that by momentum conservation
\begin{equation}
    \PPb_{n+1|1...i}\PP_{i,i+1}=\PPb_{n+1|n...i+1}\PP_{i+1,i}
\end{equation}
The proof is completed with the help of the kinematic identity:
\begin{equation}
    \beta_2\beta_4...\beta_{n-1}\sum_i\frac{\PPb_{n+1|i}\PP_{i3}}{\beta_i}=-\frac{\beta_2...\beta_{n-1}\pvec_{n+1}^2}{2\beta_{n+1}}
\end{equation}
Consequently, we have proved that
\begin{equation}
    A_{n+1}(1...n+1)=\mathcal{N}_{n+1}\frac{\beta_2...\beta_{n-1}\pvec_{n+1}^2}{\beta_{n+1}\PP_{12}...\PP_{n-1,n}}, \quad \mathcal{N}_{n+1}=\frac{(-)^{n+1}\alpha_{n+1}^{\Lambda_{n+1}-(n-1)}}{2^{n-1}\Gamma(\Lambda_{n+1}-(n-2))}
\end{equation}
The final conclusion here is that all $n$-point amplitudes with one off-shell leg have a remarkably simple form and vanish on-shell. Hence, at classical level, the chiral HSGRA is consistent with the No-Go theorems that imply $S=1$ once at least one massless higher spin particle is in the game. From the explicit calculations above it is clear that it is important to have all spins
in the spectrum without any upper/lower bounds and gaps. Moreover, the coupling constants must
have a very particular dependence on spins, $C_{\lambda_1,\lambda_2,\lambda_3}\sim 1/\Gamma(\lambda_1+\lambda_2+\lambda_3)$. This situation was
referred to as coupling conspiracy \cite{Skvortsov:2018jea}. The fact that the tree-level amplitudes vanish on-shell
indicates that there should not be any nontrivial cuts of the loop diagrams and, hence, the
loop corrections are expected to have a better UV-behaviour.
\section{Vacuum Bubbles}
\label{sec:bubbles}
Vacuum corrections stay a bit aside and could be ignored in the first approximation. Luckily, it is easy to show that all of them vanish in accordance with the naive expectation that vacuum partition function for higher-spin gravities should be one, $Z=1$, which indicates that the total regularized number of degrees of freedom is zero. This is in accordance with similar findings both in flat and AdS spaces \cite{Gopakumar:2011qs,Tseytlin:2013jya,Giombi:2013fka,Giombi:2014yra,Beccaria:2014jxa,Beccaria:2014xda,Beccaria:2015vaa,Gunaydin:2016amv,Bae:2016rgm,Skvortsov:2017ldz}.

\subsection{Determinants}
\label{sec:oneloopbubble}
The simplest vacuum corrections probe the spectrum of a theory via determinants of the kinetic operators. First, let us consider the free higher spin theory in four-dimensional flat
space \cite{Beccaria:2015vaa}. The action is the sum over all spins of the kinetic terms of massless fields:
\begin{align}
    S=\sum_s\int d^4x\Phi_{a(s)}\square \Phi^{a(s)}\,,\qquad \quad \delta \Phi_{a(s)}=\pl_{a}\xi_{a(s-1)}\,,
\end{align}
where we have choose to work in the TT-gauge. The partition function reads
\begin{align}
    \parbox{2.0cm}{\includegraphics[scale=0.34]{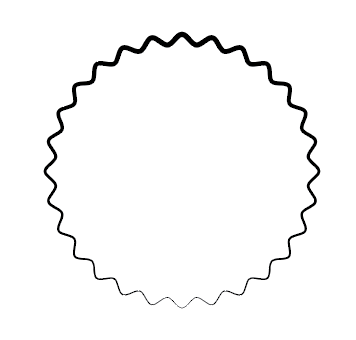}}&:  &
   Z_{\text{1-loop}}&= \frac{1}{\det^{1/2}_{0}|-\partial^2|}\prod_{s>0}\frac{\det^{1/2}_{s-1,\perp}|-\partial^2 |}{\det^{1/2}_{s,\perp}|-\partial^2|}=\frac{1}{(z_0)^{\frac{1}{2}}}\prod_{s>0}\frac{(z_{s-1})^{\frac{1}{2}}}{(z_s)^{\frac{1}{2}}}
\end{align}
where we went back to covariant description of free massless fields, which is available \cite{Tseytlin:2013jya,Beccaria:2015vaa}.  The numerator, the product of $\det_{s-1,\perp}^{1/2}$, in the formula corresponds to ghosts while the denominator, the product of $\det_{s,\perp}^{1/2}$, corresponds to massless fields with spin-$s\geq1$. The determinant of a free scalar field, $\det_0^{1/2}$ stays aside since it is not a gauge field.

At first sight, ghosts determinants seem to cancel against the rest and leave $Z_{\text{1-loop}}=1$. However, this is the same problem as determining value of the sum $1-1+1-...\,$. Indeed, for theories with infinitely many fields
a prescription of how to sum over the spectrum has to be given by hand and this is one of the instances where higher spin gravity reveals its \textit{stringy} nature. However unlike string
theory, where summation goes over relevant Riemann surfaces, we do not have any geometric understanding of how the sum over spins needs to be done.

We come up with a plausible idea as follows. The prescription of \cite{Beccaria:2015vaa} that gives $Z=1$ instructs us to count degrees
of freedom as 
\begin{align}\label{eq:dofchiralHSGRA}
    \nu_0=\sum_{\lambda} 1= 1+2 \sum_{\lambda>0}\lambda=1+2\zeta(0)=0
\end{align}
where 1 is the d.o.f for the scalar field and 2 is the total d.o.f for each massless field. Although this regularization seems to be ad hoc, the success of the zeta-function regularization in the study of determinants of higher spin theories on AdS background in chapter \ref{chapter3} provides a strong support for \eqref{eq:dofchiralHSGRA}.

Let us recall what we have learnt in Chapter 3. The kinetic operators of massless spinning fields on AdS have the form $(-\square + M_s^2)$ and the kinetic operators of the corresponding ghosts are $(-\square + m_{s-1}^2)$. The presence of spin-dependent mass-like terms does not give us naive cancellation as discussed above. However, in AdS, the determinants can be computed via spectral zeta-function \cite{Camporesi:1991nw,Camporesi:1992wn,Camporesi:1992tm,Camporesi:1993mz,Camporesi:1994ga,Camporesi:1995fb} and the spin sums can be taken with the help of zeta-function. The final
result is consistent with the AdS/CFT expectations. Therefore, the
zeta-function regularization seems to be well-tested, which justifies \eqref{eq:dofchiralHSGRA}.
\subsection{Higher Vacuum Loops}
\label{sec:Higher Vacuum Loops}
The two-loop diagram vanishes due to the chirality of interactions: assuming some helicities on the left vertex we have the opposite of those entering the vertex on the right. However, $1/\Gamma[\Lambda]$ and $1/\Gamma[-\Lambda]$ factors coming from the product of two vertices cannot both be nonzero. Hence, 
\begin{align*}
    \parbox{2.0cm}{\includegraphics[scale=0.34]{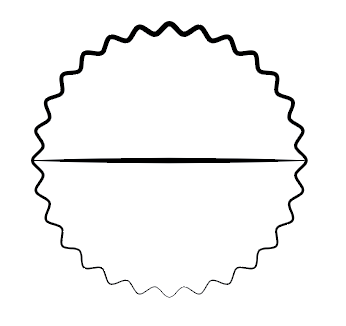}}=0
\end{align*}
The same arguments as above show that the three-loop diagrams also vanish: there is no such helicity assignment that makes all $1/\Gamma[...]$-factors nonzero.
\begin{align*}
    \parbox{2.0cm}{\includegraphics[scale=0.34]{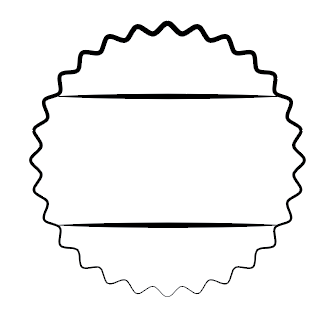}}&=0&  &
    \parbox{2.0cm}{\includegraphics[scale=0.34]{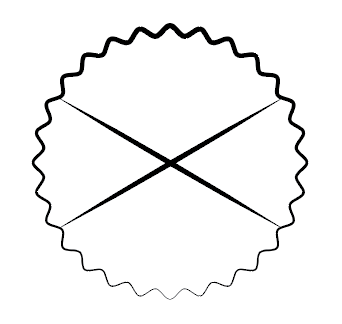}}=0
\end{align*}
It is easy to see that this is true to all loops. Indeed, the total helicity must be zero since there are no external legs and the propagator connects helicities of opposite sign. For a vacuum diagram not to vanish, the coupling constant should not be zero at each vertex. However, this is impossible due to the fact that the total helicity has to be zero. Therefore, we have to have a finite sum of positive numbers that equals zero, which shows that all vacuum diagrams vanish identically.
\section{Loops with Legs}
\label{sec:leggedloop}
We shall discuss the behaviour of legged loop diagrams by examining the tadpole, self-energy, vertex correction and 4-pt amplitude at one loop. Then, we give a general argument for multi-loop amplitudes. An important thing to remember is that vanishing of
tree-level amplitudes should eliminate all log-divergences that would lead to cuts otherwise. 
\subsection{Tadpole}
\label{sec:tadpole}
Light-cone approach is not suitable for the computation of one-point functions, like tadpole. Nevertheless, 
tadpoles for the external lines with non-zero helicity must vanish by Lorentz invariance. Indeed, at the vertex we have $\PPb_{ii}^{\Lambda}$ factor which should be zero by definition. A tadpole for the scalar field also vanishes due to the absence of the relevant vertex in the action. Lastly, if the external helicity is zero and the internal one is some $\mu$, then at the vertex we still have $\Gamma(0+\mu-\mu)^{-1}= 0$. Therefore,
\begin{align*}
    \parbox{1.5cm}{\includegraphics[scale=0.21]{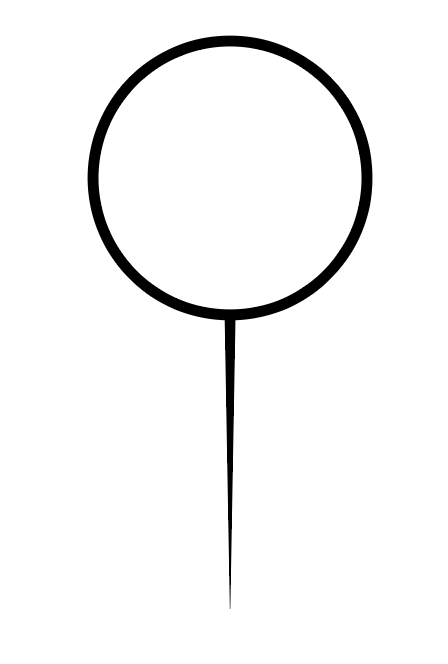}}=0
\end{align*}

\subsection{Self-energy}
\label{sec:selfenergy1loop}
Although we are studying $U(N)$-version of chiral HSGRA for concreteness, all general conclusions below are also true for the other cases ($SO(N)$ and $USp(N)$ gauging). For a given $N$ we can first have a look at the planar diagrams, which are simpler. For the
self-energy diagram, there are contributions from planar and non-planar diagrams:
\begin{align*}
   \parbox{3.7cm}{\includegraphics[scale=0.21]{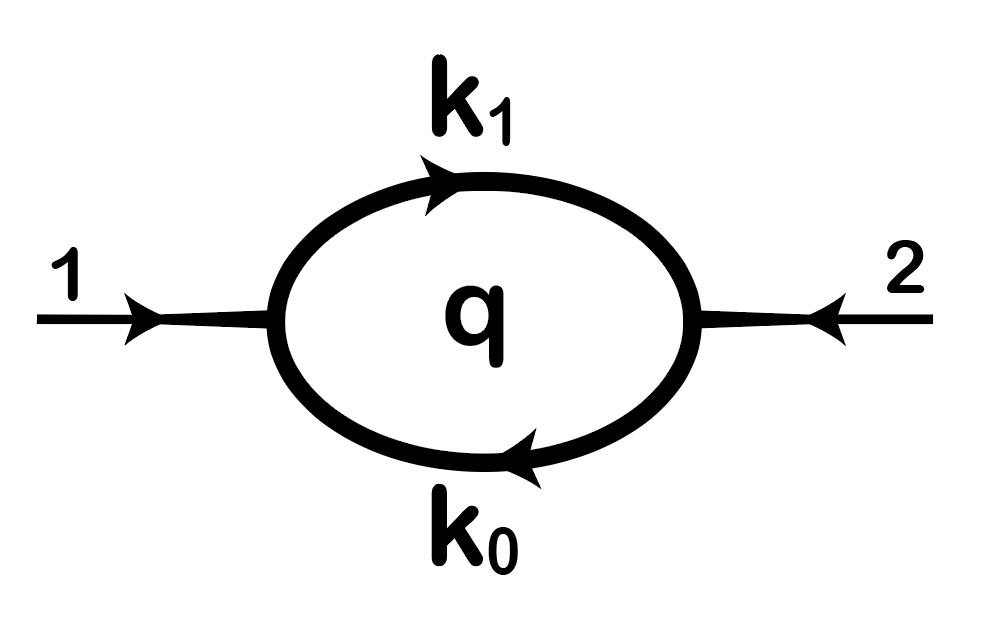}}+\parbox{3.9cm}{\includegraphics[scale=0.22]{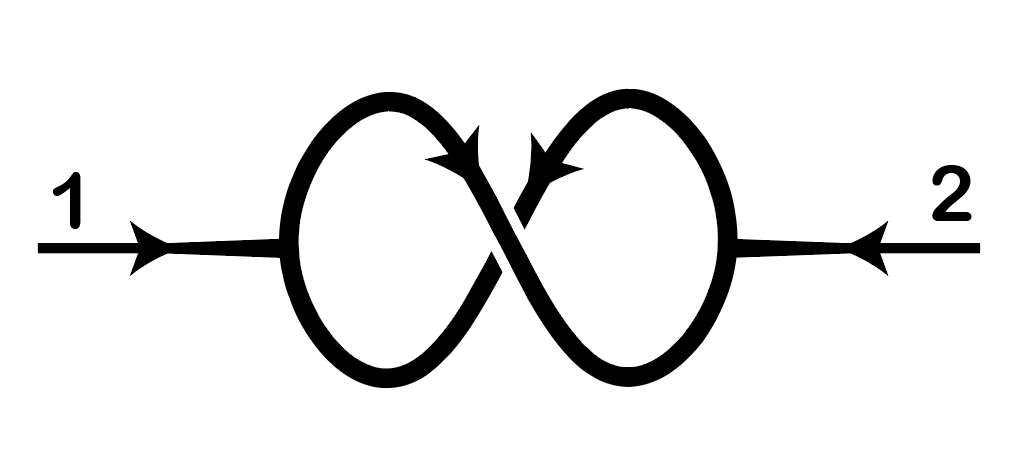}}
\end{align*}
Here, $\kvec_1,\kvec_0,\qvec$ are dual momenta and the external momentum is related to $\kvec$ as $\pvec_1=\kvec_1-\kvec_0$. The loop momentum is $\pvec=\qvec-\kvec_0$. The discussion about dual momenta can be found in \cite{Brandhuber:2007vm,Thorn:2004ie,Chakrabarti:2005ny,Chakrabarti:2006mb} (we also discuss this matter in Appendix \ref{app:Thornregulator} for completeness). 

We start our analysis by considering the simplest self-energy diagram. In order to avoid confusing and cumbersome notation, we introduce sources $h_A^{\ B}$ that can be contracted with fields. As a result each amplitude acquires factors $\Tr(hh...)$ which keeps track of the color indices. We adopt the \textit{world-sheet friendly} regularization \cite{Thorn:2004ie,Chakrabarti:2005ny,Chakrabarti:2006mb} which is used in a number of theories in light-cone gauge. The one loop self-energy reads
\begin{equation}\label{eq:selfintegrand}
\begin{split}
    \Gamma_{\text{self}}=&N\tr(h_1h_2)\sum_{\omega} \frac{(l_p)^{\Lambda_2-2}}{\beta_1^{\lambda_1}\beta_2^{\lambda_2}\Gamma(\Lambda_2-1)}\int \frac{d^4q}{(2\pi)^4} \frac{\PPb_{q-k_0,p_1}^2\delta_{\Lambda_2,2}}{(\qvec-\kvec_0)^2(\qvec-\kvec_1)^2}\\
    &-\tr(h_1)\tr(h_2)\sum_{\omega}\frac{(2l_p)^{\Lambda_2-2}}{\beta_1^{\lambda_1}\beta_2^{\lambda_2}\Gamma(\Lambda_2-1)}\int \frac{d^4q}{(2\pi)^4} \frac{\PPb_{q-k_0,p_1}^{\Lambda_2}}{(\qvec-\kvec_0)^2(\qvec-\kvec_1)^2},
    \end{split}
\end{equation}
where $d^4q=dq^-d\beta d^2q_{\perp}$ and $\Lambda_2=\lambda_1+\lambda_2$. A very important observation is that the very last sum over helicities factors out for all loop diagrams, i.e. after we sum over all but one helicities running in the loop the resulting expression does not depend on the very last helicity to be summed over. Therefore, each loop diagram has an overall factor $\nu_0=\sum_{\omega} 1$ as in the case of bubble diagrams. Let us evaluate leading contribution, namely the first term,
\begin{equation}\label{eq:selfleading}
    \Gamma_{\text{self}}^{\text{leading}}=N\tr(h_1h_2)\sum_{\omega} \frac{(l_p)^{\Lambda_2-2}}{\beta_1^{\lambda_1}\beta_2^{\lambda_2}\Gamma(\Lambda_2-1)}\int \frac{d^4q}{(2\pi)^4} \frac{\PPb_{q-k_0,p_1}^2\delta_{\Lambda_2,2}}{(\qvec-\kvec_0)^2(\qvec-\kvec_1)^2}
\end{equation}
as an example. Here, we observe that the integrand is non-vanishing only when $\Lambda_2=2$. To regulate this integral, one can introduce a cut-off $\exp[-\xi q_{\perp}^2]$, where $q_{\perp}\equiv (q,\bar{q})$ is the transverse part of $\qvec$. Then, using Schwinger parametrization and integrating out $q^-$ gives us $\delta\big(\beta(T_1+T_2)-T_1\beta_{k_0}-T_2\beta_{k_1}\big)$. Next, we replace \footnote{
Note that whenever we write $\beta_{k_i}$, it means we consider the $k^+_i$ component of the dual 4-momentum $\kvec_i$.}
\begin{equation}
    \beta=\frac{T_1\beta_{k_0}+T_2\beta_{k_1}}{T_1+T_2},
\end{equation}
and the expression (\ref{eq:selfleading}) reads (omitting the prefactor)
\begin{equation}\label{eq:intermediateself1}
    \Gamma_{\text{self}}^{\text{leading}}\sim \int \PPb_{q-k_0,p_1}^2\exp\Big[-(T+\xi)\Big(q^a-\frac{T_1k_0^a+T_2k_1^a}{T+\xi}\Big)^2-\frac{T_1T_2\pvec_1^2}{T}-\frac{\xi(T_1k_0^a+T_2k_1^a)^2}{T(T+\xi)}\Big].
\end{equation}
where we integrate over $q$ and over $T_i$ that are the Schwinger's parameters and $T=T_1+T_2$. It is now safe to send $\pvec_1^2$ on-shell and $\xi=0$ in the last two terms in the exponential in (\ref{eq:intermediateself1}). Hence, we are left with a Gaussian integral expression
\begin{equation}\label{eq:Gaussianself}
    \Gamma_{\text{self}}^{\text{leading}}\sim \int \frac{d^2q^a}{16\pi^2}\Big[(\bar{q}-\bar{k_0})\beta_1-\bar{p}_1\big(\frac{T_1\beta_{k_0}+T_2\beta_{k_1}}{T_1+T_2}-\beta_{k_0}\big)\Big]^2e^{-(T+\xi)\Big(q^a-\frac{T_1k_0^a+T_2k_1^a}{T+\xi}\Big)^2}.
\end{equation}
We can handle (\ref{eq:Gaussianself}) with the note that
\begin{equation}\label{eq:magicGuassian}
    \int d^2q_{\perp} e^{-Aq_{\perp}^2}=\frac{\pi}{A}, \quad \int d^2 q_{\perp}\,\left(\bar{q}\right)^n e^{-Aq_{\perp}^2}=0 \quad (\text{for} \, n\geq 1)
\end{equation}
then, after some manipulation we get
\begin{equation}\label{eq:selfresult}
\begin{split}
    \Gamma_{\text{self}}^{\text{leading}}&=\sum_{\omega}\frac{(l_p)^{\Lambda_2-2}\,N\tr(h_1h_2)\delta_{\Lambda_2,2}}{\beta_1^{\lambda_1-1}\beta_2^{\lambda_2-1}\Gamma[\Lambda_2-1]}\int_0^1 dx \int_0^{\infty} \frac{dT}{16\pi^2}\frac{\xi^2[x\bar{k}_0+(1-x)\bar{k}_1]^2}{(T+\xi)^3}\\
    &\xrightarrow{\xi\rightarrow 0} \nu_0\,\frac{(l_p)^{\Lambda_2-2}\,N\tr(h_1h_2)\delta_{\Lambda_2,2}}{32\pi^2\beta_1^{\lambda_1-1}\beta_2^{\lambda_2-1}\Gamma[\Lambda_2-1]}\int_0^1 dx[x\bar{k}_0+(1-x)\bar{k}_1]^2\\
    &=\nu_0\,\delta_{\Lambda_2,2}\frac{(l_p)^{\Lambda_2-2}\,N\tr(h_1h_2)\, (\bar{k}_0^2+\bar{k}_0\bar{k}_1+\bar{k}_1^2)}{96\pi^2\beta_1^{\lambda_1-1}\beta_2^{\lambda_2-1}\Gamma[\Lambda_2-1]}
    \end{split}
\end{equation}
where we made a change of variables $x=T_1/T$. Here, the $x$-integral in (\ref{eq:selfresult}) is perfectly finite and $\Gamma_{\text{self}}^{\text{leading}}$ is reminiscent of $\Pi^{++}$ amplitude in \cite{Thorn:2005ak,Chakrabarti:2005ny,Chakrabarti:2006mb}. The important feature of the computation above is that there is a factorization of $\nu_0$ which guarantee the result above vanish without the need of introducing a counter term. We note that the Lorentz invariance forbids helicity flips
for an isolated spinning particle. Therefore, if we were to find a non-vanishing contribution to $\Gamma^{\leading}_{\self}$ we would have to introduce local counterterms to cancel it.

Let us also show the result of the sub-leading term for self-energy by repeating the treatment above. The sub-leading contribution before taking the $T$-integral is
\begin{align}\label{eq:selfsub1}
    \Gamma^{\subleading}_{\self}&=\nu_0\,\frac{(2l_p)^{\Lambda_2}(-)^{\lambda_1}\tr(h_1)\tr(h_2)}{16\pi^2\Gamma[\Lambda_2-1]}\int_0^1dx \int_0^{\infty}dT \frac{\xi^{\Lambda_2}[x\bar{k}_0+(1-x)\bar{k}_1]^{\Lambda_2}}{(T+\xi)^{\Lambda_2+1}}
\end{align}
This result can be obtained using the holomorphic integral (\ref{eq:magicGuassian}). We now have a convergent integral and the result is
\begin{equation}\label{eq:selfsub2}
\begin{split}
    \Gamma^{\subleading}_{\self}&=\nu_0\,\frac{(-)^{\lambda_1}(2l_p)^{\Lambda_2}\,\tr(h_1)\tr(h_2)(\Lambda_2-1)}{16\pi^2\Gamma[\Lambda_2+1]} \int_0^1 dx[x\bar{k}_0+(1-x)\bar{k}_1]^{\Lambda_2}\\
    &=\nu_0\,\frac{(-)^{\lambda_1}(2l_p)^{\Lambda_2}\,\tr(h_1)\tr(h_2)(\Lambda_2-1)}{16\pi^2\Gamma[\Lambda_2+2]}\times\frac{\bar{k}_0^{\Lambda_2+1}-\bar{k}_1^{\Lambda_2+1}}{\bar{k}_0-\bar{k}_1}, \quad (\Lambda_2\geq 0)
    \end{split}
\end{equation} 
It is easy to see that the dangerous non-local contribution with $\Lambda_2=-1$ is zero since $\Lambda_2>1$. The kinematic part of
$\Gamma^{\subleading}_{\self}$ is finite and, hence, $\Gamma^{\subleading}_{\self}$ vanishes again due to the factorization of $\nu_0$, which takes place
regardless of the value of $\Lambda_2$.  This implies that self-energy correction of chiral HSGRA does not break Lorentz invariance.

Finally let us mention that, we can use the original momentum $\pvec_i$ and the loop momentum $\pvec$ together with the cut-off $\exp[-\xi p_{\perp}^2]$ to work with non-planar diagrams. In the case of self-energy it reads
\begin{equation}
    \begin{split}
    \Gamma^{\subleading}_{\self}&=\nu_0\,\frac{(2l_p)^{\Lambda_2}}{\beta_1^{\lambda_1}\beta_2^{\lambda_2}\Gamma[\Lambda_2-1]}\int \frac{d^4p}{(2\pi)^4}\frac{\PPb_{p1}^{\Lambda_2}}{\pvec^2(\pvec+\pvec_1)^2}\\
    &=\nu_0\,\frac{(2l_p)^{\Lambda_2}(-)^{\lambda_1}\tr(h_1)\tr(h_2)}{16\pi^2\Gamma[\Lambda_2-1]}\int_0^1dx \int_0^{\infty}dT \frac{\xi^{\Lambda_2}[x\bar{p}_1]^{\Lambda_2}}{(T+\xi)^{\Lambda_2+1}}\\
    &=\nu_0\,\frac{(2l_p)^{\Lambda_2}(-)^{\lambda_1}\tr(h_1)\tr(h_2)(\Lambda_2-1)}{16\pi^2\Gamma[\Lambda_2+1]}\int_0^1dx  [x\bar{p}_1]^{\Lambda_2}\\
    &=\nu_0\,\frac{(2l_p)^{\Lambda_2}(-)^{\lambda_1}\tr(h_1)\tr(h_2)(\Lambda_2-1)}{16\pi^2\Gamma[\Lambda_2+2]}\bar{p}_1^{\Lambda_2}, \quad (\Lambda_2\geq 0)
\end{split}
\end{equation}
Therefore, the non-planar diagram for self-energy is also UV-finite.
\subsection{Vertex correction}
\label{sec:vertexcorrection}
The next simple quantum correction we consider is the vertex correction
\begin{align*}
    \parbox{3.3cm}{\includegraphics[scale=0.19]{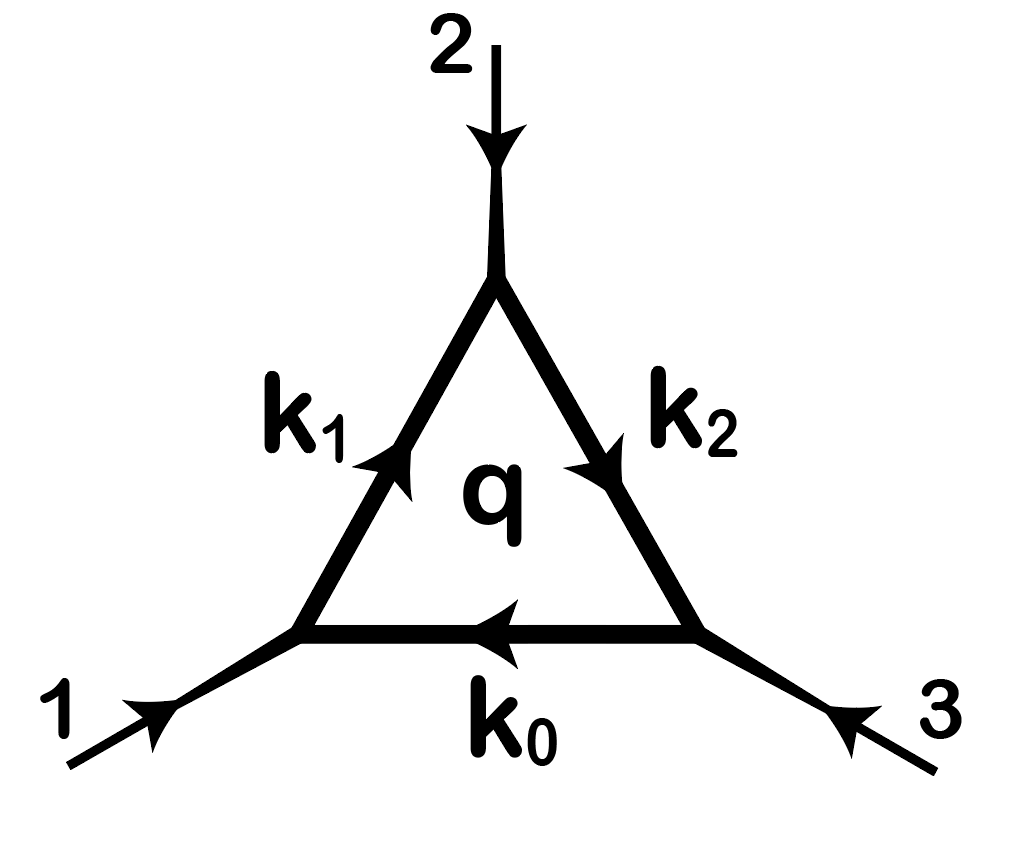}}+\parbox{2.9cm}{\includegraphics[scale=0.24]{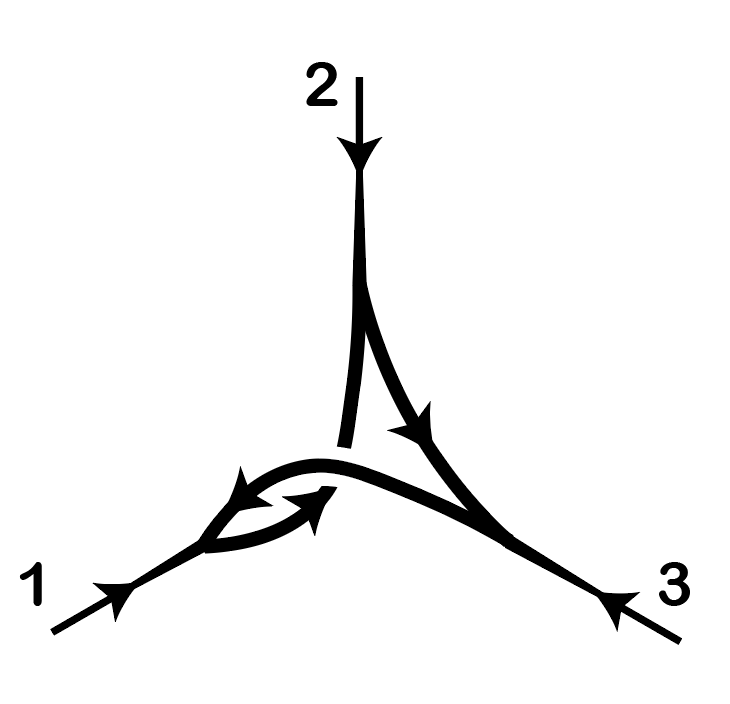}}+\parbox{3.1cm}{\includegraphics[scale=0.24]{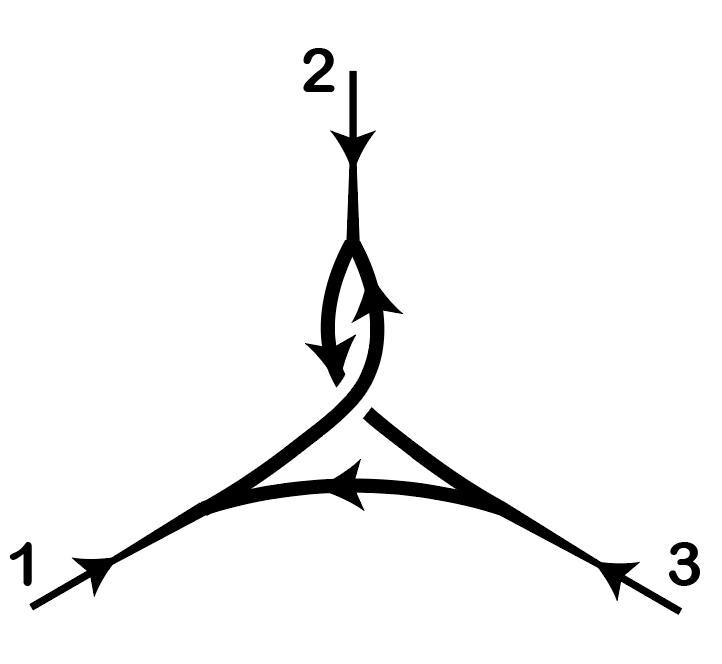}}+\parbox{3.3cm}{\includegraphics[scale=0.24]{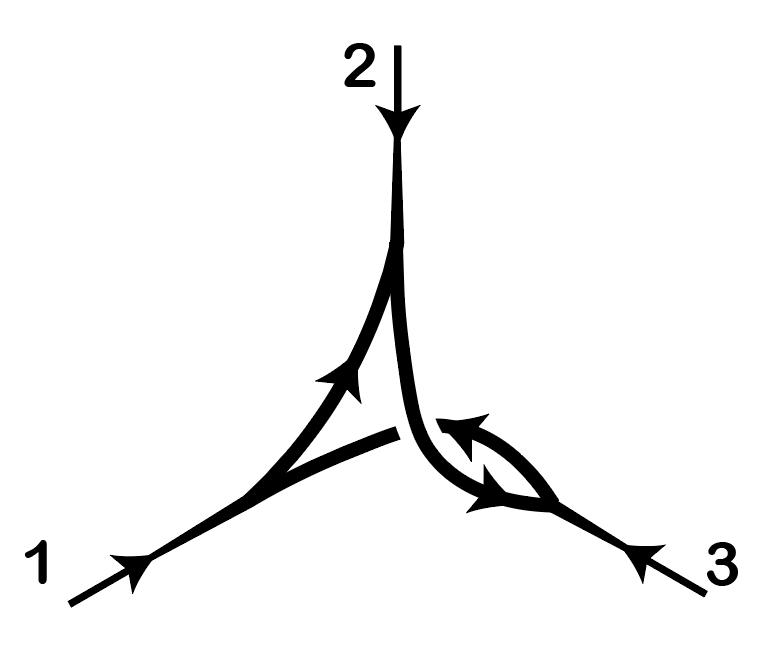}}
\end{align*}
The dual momenta in this case are $\qvec,\kvec_i$ with $i=0,1,2$. The loop momenta can be chosen to be $\pvec=\qvec-\kvec_0$ and the relation between the external momenta and dual regional momenta are $\pvec_i=\kvec_i-\kvec_{i-1}$ with $\kvec_3\equiv \kvec_0$. In other words, with clockwise order, $\pvec_i$ is the difference between the outgoing dual momenta and the ingoing dual momenta as depicted in the above figures. We keep leg-3 off-shell, i.e. $\pvec_3^2\neq 0$, and find the leading contribution to be
\small
\begin{equation}\label{eq:vertexcorrection}
\begin{split}
    \Gamma_{\tri}^{\text{lead}}=\nu_0\,\frac{\Omega_3^{\leading}(l_p \PPb_{12})^{\Lambda_3-3}}{\prod_{i=1}^3\beta_i^{\lambda_i}\Gamma[\Lambda_3-2]}\int \frac{d^4q}{(2\pi)^4} \frac{\PPb_{q-k_0,p_1}(\PPb_{q-k_1,p_2}+\PPb_{12})\PPb_{q-k_2,p_3}}{(\qvec-\kvec_0)^2(\qvec-\kvec_1)^2(\qvec-\kvec_2)^2}
    \end{split}
\end{equation}
\normalsize
The sub-leading terms come with a twist at one of the three vertices, they read
\small
\begin{equation}
\begin{split}
    \Gamma_{\tri}^{\subleading}=&-\mathcal{N}_{\tri}\tr(h_1)\tr(h_2h_3)\int\frac{d^4q}{(2\pi)^4}\frac{\PPb_{q-k_0,p_1}(\PPb_{q-k_1,p_2}+\PPb_{12})\PPb_{q-k_2,p_3}(\PPb_{12}-2\PPb_{q-k_0,p_1})^{\Lambda_3-3}}{(\qvec-\kvec_0)^2(\qvec-\kvec_1)^2(\qvec-\kvec_2)^2} \\
    &-\mathcal{N}_{\tri}\tr(h_2)\tr(h_3h_1)\int\frac{d^4q}{(2\pi)^4}\frac{\PPb_{q-k_0,p_1}(\PPb_{q-k_1,p_2}+\PPb_{12})\PPb_{q-k_2,p_3}(-2\PPb_{q-k_1,p_2}-\PPb_{12})^{\Lambda_3-3}}{(\qvec-\kvec_0)^2(\qvec-\kvec_1)^2(\qvec-\kvec_2)^2}\\
    &-\mathcal{N}_{\tri}\tr(h_3)\tr(h_1h_2)\int\frac{d^4q}{(2\pi)^4}\frac{\PPb_{q-k_0,p_1}(\PPb_{q-k_1,p_2}+\PPb_{12})\PPb_{q-k_2,p_3}(\PPb_{12}-2\PPb_{q-k_2,p_3})^{\Lambda_3-3}}{(\qvec-\kvec_0)^2(\qvec-\kvec_1)^2(\qvec-\kvec_2)^2} 
    \end{split}
\end{equation}
\normalsize
where $\mathcal{N}_{\tri}=\nu_0\,\frac{(l_p)^{\Lambda_3-3}}{\prod_{i=1}^3\beta_i^{\lambda_i}\Gamma[\Lambda_3-2]}$. Next, we show how to evaluate the integral from the leading contribution. Proceeding with the same procedure in section \ref{sec:selfenergy1loop} and appendix \ref{app:Thornregulator}, we arrive at
\small
\begin{equation}\label{eq:vertexintermediate1}
\begin{split}
   \Gamma_{\tri}^{\leading}=\nu_0\,\frac{\Omega_3^{\leading}\, (l_p \PPb_{12})^{\Lambda_3-3}}{16\pi^2\prod_{i=1}^3\beta_i^{\lambda_i}\Gamma[\Lambda_3-2]}\int \frac{\prod_{i=1}^3dT_i}{T(T+\xi)}e^{-\frac{T_1T_3\pvec_3^2}{T}}\prod_{i=1}^3\Bigg[\frac{T_{i+2}\overline{\mathbb{K}}}{T}-\xi\frac{\beta_i(\sum_{i=1}^3 T_i\bar{k}_{i-1})}{T(T+\xi)}\Bigg]
   \end{split}
\end{equation}
\normalsize
where $\Omega_3^{\leading}=N\tr(h_1h_2h_3)$. It is important to note that the integral in (\ref{eq:vertexcorrection}) is finite without the need of introducing the cut-off $\exp[-\xi q^2_{\perp}]$. In (\ref{eq:vertexintermediate1}), we identify $T_4=T_1$ and $T_5=T_2$, also
\begin{equation}
    \overline{\mathbb{K}}\equiv(\bar{k}_1-\bar{k}_0)\beta_2-(\bar{k}_2-\bar{k}_1)\beta_1=\PPb_{12}
\end{equation}
Now, it is safe to take $\xi\rightarrow 0$, we obtain
\begin{equation}\label{eq:vertexfinalresult}
\begin{split}
    \Gamma_{\tri}^{\leading}&=\nu_0\,\frac{\Omega_3^{\leading}\,(l_p)^{\Lambda_3-3}\PPb_{12}^{\Lambda_3}}{16\pi^2\Gamma(\Lambda_3-2)}\int \frac{dT_1dT_2dT_3}{\prod_{i=1}^3\beta_i^{\lambda_i}}\frac{T_1T_2T_3}{T^5}e^{-\frac{T_1T_3\pvec_3^2}{T}}\\
    &=\nu_0\,\frac{\Omega_3^{\leading}\,(l_p)^{\Lambda_3-3}\PPb_{12}^{\Lambda_3}}{16\pi^2\Gamma(\Lambda_3-2)\prod_{i=1}^3\beta_i^{\lambda_i}}\int_{x+y<1}dxdy\int_0^{\infty}dT\, xy(1-x-y)e^{-Tx(1-x-y)\pvec_3^2}\\
    &=\nu_0\,\frac{\Omega_3^{\leading}\,(l_p)^{\Lambda_3-3}\PPb_{12}^{\Lambda_3}}{96\pi^2\prod_{i=1}^{3}\beta_i^{\lambda_i}\Gamma(\Lambda_3-2)\pvec_3^2}
    \end{split}
\end{equation}
To obtain the above result, instead of using dual momentum, one can also start with the original momentums since the quantum corrections at one loop with 3 legs attached (and beyond) are perfectly finite. In terms of these variables, the vertex correction reads
\begin{equation}\label{eq:vertexalternative}
    \Gamma_{\tri}^{\leading}=\sum_{\omega}\frac{\Omega_3^{\leading}\,(l_p)^{\Lambda_3-3}\PPb_{12}^{\Lambda_3-3}}{\prod_{i=1}^3\beta_i^{\lambda_i}\Gamma(\Lambda_3-2)}\int \frac{d^4p}{(2\pi)^4}\frac{\PPb_{p1}(\PPb_{p2}+\PPb_{12})\PPb_{p3}}{\pvec^2(\pvec+\pvec_1)^2(\pvec+\pvec_1+\pvec_2)^2}
\end{equation}
Omitting the prefactor for a moment and proceed as before, we find the integral in (\ref{eq:vertexalternative}) as
\begin{equation}
\begin{split}
   \frac{\pi}{T(T+\xi)}\prod_{i=1}^3\Bigg[\frac{T_{i+2}\PPb_{12}}{T}-\xi\frac{\beta_i\big[(T_2+T_3)\bar{p}_1+T_3\bar{p}_2\big]}{T(T+\xi)}\Bigg] \xrightarrow{\xi\rightarrow 0}\frac{\pi T_1T_2T_3\PPb_{12}^3}{T^5} 
   \end{split}
\end{equation}
which is the same with (\ref{eq:vertexfinalresult}). One can immediately recognize that the final result is reminiscent of the $\Gamma^{+++}$ amplitude for QCD \cite{Thorn:2005ak,Chakrabarti:2005ny,Chakrabarti:2006mb} in the large $N$ limit. It contains the part of self-dual Yang Mill dressed with chiral HiSGRA's factor.\footnote{It would be interesting if one can find a direct relation between chiral HiSGRA and SDYM if there is any.} The overall factor $\nu_0$ makes the vertex correction vanish. 

Although we did not compute the integral for sub-leading terms of the vertex correction, they should be finite. To be more explicit, higher power of $\bar{q}$ entering the Gaussian integral of type (\ref{eq:magicGuassian}) will give zero and improve the behaviour of the cut-off $\xi$. The only place where things can diverge is at the $T$-integral. The $T$-integral have the form 
\begin{equation}
    \int_0^{\infty}dT \frac{\xi^a}{(T+\xi)^b}.
\end{equation}
It will pick up poles of the form $1/\xi^{b-a-1}$ whenever $b\geq a+2$. However, due to power counting and the magic of the holomorphic integral (\ref{eq:magicGuassian}), we should have convergent integrals. The $\nu_0$ factor again will guarantee all of the sub-leading terms to vanish due to our choice of zeta regularization. 

\subsection{Box and triangle-like diagrams}
\label{sec:boxandtri}
Next, we consider the one loop correction where we have four external legs in the large $N$ limit. This is the limit where the contribution from non-planar diagrams can be neglected since it is incomparable to the planar's contribution. Let us take a look at the box, triangle-like diagrams. We show that they are also UV finite. Consider triangle-like diagrams and take order (1234) for example:
\small
\begin{equation}
    \begin{split}
        \parbox{1.6cm}{\includegraphics[scale=0.21]{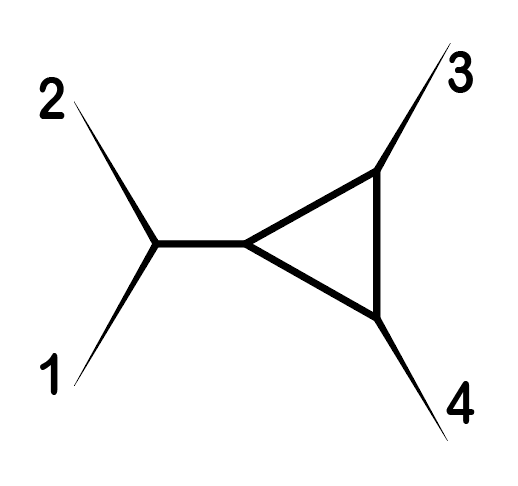}}=\Gamma_{\Delta}(1234)&=\frac{\nu_0\, (l_p)^{\Lambda_4-4}\alpha_4^{\Lambda_4-4}}{\Gamma(\Lambda_4-3)\prod_{i=1}^4\beta_i^{\lambda_i}}\frac{\PPb_{12}}{s_{12}}\int \frac{d^4p}{(2\pi)^4}\frac{(\PPb_{p1}+\PPb_{p2})(\PPb_{p3}+\PPb_{13}+\PPb_{23})\PPb_{p4}}{\pvec^2(\pvec+\pvec_1+\pvec_2)^2(\pvec+\pvec_1+\pvec_2+\pvec_3)^2}\\
        &=\frac{\nu_0\,(l_p)^{\Lambda_4-4}\alpha_4^{\Lambda_4-4}}{\Gamma(\Lambda_4-3)\prod_{i=1}^4\beta_i^{\lambda_i}}\frac{\PPb_{12}\PPb_{34}^3}{96\pi^2 s_{34}^2}\\
        &=-\frac{\nu_0\, (l_p)^{\Lambda_4-4}\alpha_4^{\Lambda_4-4}}{\Gamma(\Lambda_4-3)\prod_{i=1}^4\beta_i^{\lambda_i}}\frac{\PPb_{34}^2\PPb_{41}\PPb_{23}}{96\pi^2 s_{12}s_{23}}
    \end{split}
\end{equation}
\normalsize
Similarly,
\begin{align}
    \Gamma_{\Delta}(2341)&=\frac{\nu_0\,(l_p)^{\Lambda_4-4}\alpha_4^{\Lambda_4-4}}{\Gamma(\Lambda_4-3)\prod_{i=1}^4\beta_i^{\lambda_i}}\frac{\PPb_{23}\PPb_{41}^3}{96\pi^2 s_{23}^2}=-\frac{\nu_0\,(l_p)^{\Lambda_4-4}\alpha_4^{\Lambda_4-4}}{\Gamma(\Lambda_4-3)\prod_{i=1}^4\beta_i^{\lambda_i}}\frac{\PPb_{41}^2\PPb_{12}\PPb_{34}}{96\pi^2 s_{12}s_{23}}\\
    \Gamma_{\Delta}(3412)&=\frac{\nu_0\,(l_p)^{\Lambda_4-4}\alpha_4^{\Lambda_4-4}}{\Gamma(\Lambda_4-3)\prod_{i=1}^4\beta_i^{\lambda_i}}\frac{\PPb_{34}\PPb_{12}^3}{96\pi^2 s_{34}^2}=-\frac{\nu_0\,(l_p)^{\Lambda_4-4}\alpha_4^{\Lambda_4-4}}{\Gamma(\Lambda_4-3)\prod_{i=1}^4\beta_i^{\lambda_i}}\frac{\PPb_{12}^2\PPb_{23}\PPb_{41}}{96\pi^2 s_{12}s_{23}}\\
    \Gamma_{\Delta}(4123)&=\frac{\nu_0\,(l_p)^{\Lambda_4-4}\alpha_4^{\Lambda_4-4}}{\Gamma(\Lambda_4-3)\prod_{i=1}^4\beta_i^{\lambda_i}}\frac{\PPb_{41}\PPb_{23}^3}{96\pi^2 s_{41}^2}=-\frac{\nu_0\,(l_p)^{\Lambda_4-4}\alpha_4^{\Lambda_4-4}}{\Gamma(\Lambda_4-3)\prod_{i=1}^4\beta_i^{\lambda_i}}\frac{\PPb_{23}^2\PPb_{34}\PPb_{12}}{96\pi^2 s_{12}s_{23}}
\end{align}

As discussing in \cite{Chakrabarti:2005ny,Chakrabarti:2006mb}, one can reduce the box integral, which is in general complicated to integrate, into the triangle-like integral. The box contribution reads
\begin{equation}
\begin{split}
    \Gamma_{\square}=\sum_{\omega}\frac{\nu_0 \,(l_p)^{\Lambda_4-4}\alpha_4^{\Lambda_4-4}}{\prod_{i=1}^4\beta_i^{\lambda_i}\Gamma(\Lambda_4-3)}\int \frac{d^4p}{(2\pi)^4}\frac{\PPb_{p1}(\PPb_{p2}+\PPb_{12})(\PPb_{p3}+\PPb_{34})\PPb_{p4}}{\pvec^2(\pvec+\pvec_1)^2(\pvec+\pvec_1+\pvec_2)^2(\pvec-\pvec_4)^2}
\end{split}
\end{equation}
Notice that $\pvec$ is off-shell and we can use the following identity:

\begin{equation}\label{eq:cuttrick}
    \PPb_{pi}\PP_{pi}=-\frac{\beta_i\beta}{2}(\pvec+\pvec_i)^2+\frac{\beta_i(\beta_i+\beta)\pvec^2}{2}
\end{equation} 
to arrive at 
\begin{equation}
    \frac{\PPb_{p1}}{(\pvec+\pvec_1)^2}=-\frac{\beta\beta_1}{2\PP_{p1}}+\frac{\beta_1(\beta_1+\beta)\pvec^2}{2\PP_{p1}(\pvec+\pvec_1)^2}, \quad \frac{\PPb_{p4}}{(\pvec-\pvec_4)^2}=\frac{\beta\beta_4}{2\PP_{p4}}-\frac{\beta_4(\beta-\beta_4)\pvec^2}{2\PP_{p4}(\pvec-\pvec_4)^2}
\end{equation}
Effectively, we can reduce the box integral into triangle-like integral by canceling out one propagator in the denominator using (\ref{eq:cuttrick}). Next, we can multiply $\Gamma_{\square}$ by two for manipulation reason, then
\small
\begin{equation}
\begin{split}
 2\Gamma_{\square}=&\frac{\nu_0\,(l_p)^{\Lambda_4-4}\alpha_4^{\Lambda_4-4}}{\prod_{i=1}^4\beta_i^{\lambda_i}\Gamma(\Lambda_4-3)}\int\frac{d^4p}{(2\pi)^4}\Bigg[\frac{(\PPb_{p2}+\PPb_{12})(\PPb_{p3}+\PPb_{34})}{\pvec^2(\pvec+\pvec_1+\pvec_2)^2}\Bigg( \frac{\beta\beta_4\PPb_{p1}}{2\PP_{p4}(\pvec+\pvec_1)^2} -\frac{\beta\beta_1\PPb_{p4}}{2\PP_{p1}(\pvec-\pvec_4)^2}\Bigg)\\
&+\frac{(\PPb_{p2}+\PPb_{12})(\PPb_{p3}+\PP_{34})}{(\pvec+\pvec_1)^2(\pvec+\pvec_1+\pvec_2)^2(\pvec-\pvec_4)^2}\Bigg(\frac{\beta_1(\beta+\beta_1)\PPb_{p4}}{2\PP_{p1}}-\frac{\beta_4(\beta-\beta_4)\PPb_{p1}}{2\PP_{p4}}\Bigg)\Bigg]
 \end{split}
\end{equation}
\normalsize
Using Bianchi-like identity $\beta_{(i}\PP_{jk)}=0$, we find 
\begin{equation}
    \frac{\beta\beta_4}{2\PP_{p4}}=\frac{\PPb_{41}}{s_{41}}+\frac{\beta_4^2\PP_{p1}}{2\PP_{p4}\PP_{41}}, \quad -\frac{\beta\beta_1}{2\PP_{p1}}=\frac{\PPb_{41}}{s_{41}}+\frac{\beta_1^2\PP_{p4}}{2\PP_{p1}\PP_{41}}
\end{equation}
Then, after some straight forward algebra, the box integral becomes \footnote{Using identities listed in Appendix \ref{app:kinematics}, we can show that
\begin{align*}
    s_{ij}\PPb_{pi}\PPb_{pj}=\PPb_{ji}\Big[\PPb_{pi}(\pvec-\pvec_j)^2+\PPb_{pj}(\pvec+\pvec_i)^2-(\PPb_{pi}+\PPb_{pj}-\PPb_{ji})\pvec^2\Big]
\end{align*}}
\small
\begin{equation}
    \begin{split}
    2\Gamma_{\square}=\nu_0\,\mathcal{N}_{\square}\frac{\PPb_{41}}{s_{41}}\int \frac{d^4p}{(2\pi)^4}\frac{(\PPb_{p2}+\PPb_{12})(\PPb_{p3}+\PPb_{34})}{\pvec^2(\pvec+\pvec_1+\pvec_2)^2}\Bigg[\frac{\PPb_{p1}}{(\pvec+\pvec_1)^2}+\frac{\PPb_{p4}}{(\pvec-\pvec_4)^2}-\frac{(\PPb_{p4}+\PPb_{p1}-\PPb_{41})\pvec^2}{(\pvec+\pvec_1)^2(\pvec-\pvec_4)^2}\Bigg]+\Gamma_{\square}
\end{split}
\end{equation}
\normalsize
where $\mathcal{N}_{\square}=\frac{(l_p)^{\Lambda_4-4}\alpha_4^{\Lambda_4-4}}{\prod_{i=1}^4\beta_i^{\lambda_i}\Gamma(\Lambda_4-3)}$. Hence, 
\small
\begin{equation}\label{eq:box expression}
    \begin{split}
    \Gamma_{\square}&=\nu_0\,\mathcal{N}_{\square}\frac{\PPb_{41}}{s_{41}}\int \frac{d^4p}{(2\pi)^4}\frac{(\PPb_{p2}+\PPb_{12})(\PPb_{p3}+\PPb_{34})}{\pvec^2(\pvec+\pvec_1+\pvec_2)^2}\Bigg[\frac{\PPb_{p1}}{(\pvec+\pvec_1)^2}+\frac{\PPb_{p4}}{(\pvec-\pvec_4)^2}-\frac{(\PPb_{p4}+\PPb_{p1}-\PPb_{41})\pvec^2}{(\pvec+\pvec_1)^2(\pvec-\pvec_4)^2}\Bigg]\\
    &=\nu_0\,\mathcal{N}_{\square}\left[\frac{\PPb_{41}\big[\,\PPb_{12}^2(\PPb_{23}+\PPb_{34})+(\PPb_{12}+\PPb_{23})\PPb_{34}^2\big]}{96\pi^2 s_{12}s_{23}}+\frac{\PPb_{41}\PPb_{23}^3}{96\pi^2s_{41}^2}\right]
    \end{split}
\end{equation}
\normalsize
The last term in (\ref{eq:box expression}) cancels with the triangle $\Gamma_{\Delta}(4123)$. In the end, we obtain 
\begin{equation}
\begin{split}
   \parbox{2cm}{\includegraphics[scale=0.13]{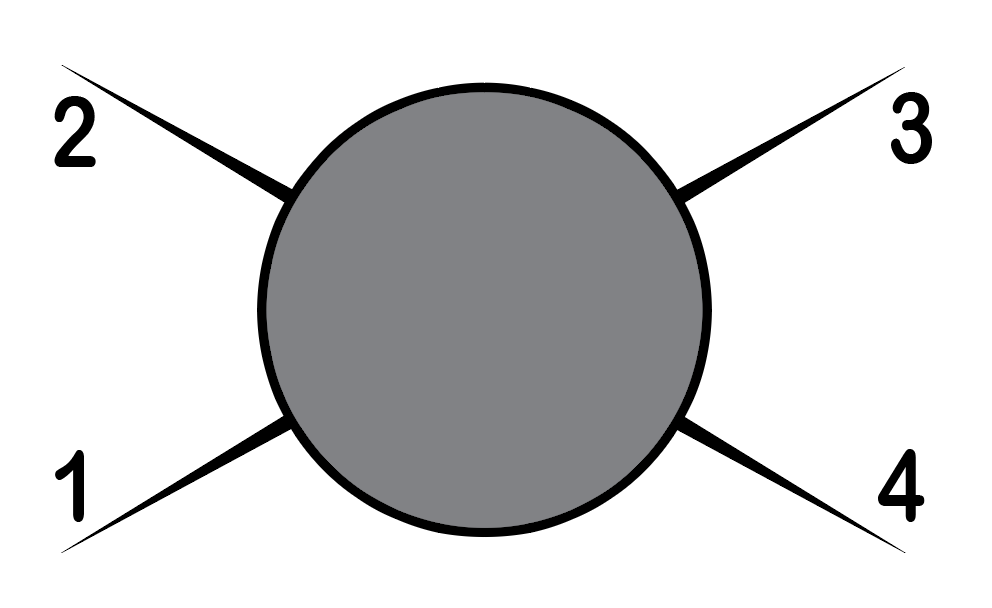}}= \Gamma_4&=\Gamma_{\square}+\big[\Gamma_{\Delta}(1234)+\text{cycl.}\big]\\
   &=\nu_0\,\frac{\mathcal{N}_{\square}}{96\pi^2}\frac{\PPb_{12}\PPb_{34}\PPb_{41}(\PPb_{12}+\PPb_{34}-\PPb_{41})}{s_{12}s_{23}}\\
   &=\nu_0\,\frac{\mathcal{N}_{\square}}{96\pi^2}\frac{\PPb_{12}\PPb_{23}\PPb_{34}\PPb_{41}}{s_{12}s_{23}}
   \end{split}
\end{equation}

which is similar to a well-know result for $\Gamma_4^{++++}$ QCD amplitude \cite{Chakrabarti:2005ny} (see also \cite{Brandhuber:2007vm}).

\subsection{The bubbles}
\label{sec:leggedbubbles}
As discussed in \cite{Chakrabarti:2005ny}, the sum over bubbles, triangle like and box diagrams should add up to zero in the case of all-plus 4pt one-loop (pure gluon) amplitude. We would like to see whether chiral HSGRA has a similar property. The last diagrams we need to compute are bubble insertions into the internal propagator, which come in two channels, $s$ and $t$, for $U(N)$ factors:
\begin{figure}[h]
    \centering
    \includegraphics[scale=0.35]{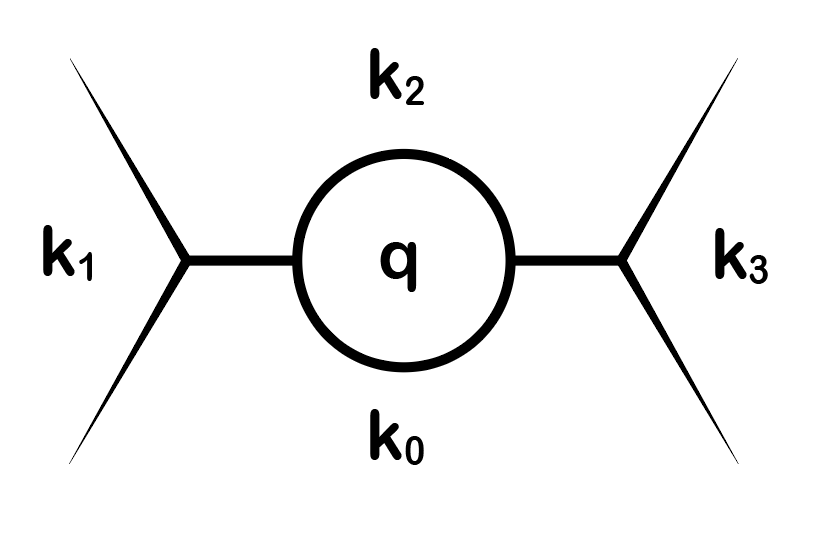}
\end{figure}\\
Here, we divided the space of dual momenta $\kvec_i$ into four regions. The external momenta $\pvec_i$ can be read off by using two adjacent regional dual momenta. For example, $\pvec_1=\kvec_1-\kvec_0$ and $\pvec_2=\kvec_2-\kvec_1$ etc. Whenever we have a close loop, we can 'put' the dual momentum $\qvec$ inside it and the loop momentum can be obtained as the difference between $\qvec$ and the nearest dual regional momentum. In the above figure, $\pvec=\qvec-\kvec_0$. Now, it is a matter of computation to show the 'internal' self-energy diagram with the four external legs labeled in clockwise order to be
\begin{equation}
\begin{split}
    \Gamma^{\text{in}}_{\bigcirc}(1234)&=-\sum_{\omega}\frac{(l_p)^{\Lambda_4-4}\alpha_4^{\Lambda_4-4}}{\prod_{i=1}^4\beta_i^{\lambda_i}\Gamma(\Lambda_4-3)}\frac{\PPb_{12}\PPb_{34}(\beta_1+\beta_2)(\beta_3+\beta_4)(\bar{k}_0^2+\bar{k}_0\bar{k}_2+\bar{k}_2^2)}{96\pi^2 s_{12}^2}\\
    &=\sum_{\omega}\frac{(l_p)^{\Lambda_4-4}\alpha_4^{\Lambda_4-4}}{\prod_{i=1}^4\beta_i^{\lambda_i}\Gamma(\Lambda_4-3)}\frac{\PPb_{41}\PPb_{23}(\beta_1+\beta_2)(\beta_3+\beta_4)(\bar{k}_0^2+\bar{k}_0\bar{k}_2+\bar{k}_2^2)}{96\pi^2 s_{12}s_{23}}\,.
    \end{split}
\end{equation}
Similarly,
\begin{equation}
\begin{split}
    \Gamma^{\text{in}}_{\bigcirc}(2341)&=-\sum_{\omega}\frac{(l_p)^{\Lambda_4-4}\alpha_4^{\Lambda_4-4}}{\prod_{i=1}^4\beta_i^{\lambda_i}\Gamma(\Lambda_4-3)}\frac{\PPb_{23}\PPb_{41}(\beta_2+\beta_3)(\beta_4+\beta_1)(\bar{k}_1^2+\bar{k}_1\bar{k}_3+\bar{k}_3^2)}{96\pi^2 s_{23}^2} \\
    &= \sum_{\omega}\frac{(l_p)^{\Lambda_4-4}\alpha_4^{\Lambda_4-4}}{\prod_{i=1}^4\beta_i^{\lambda_i}\Gamma(\Lambda_4-3)}\frac{\PPb_{12}\PPb_{34}(\beta_2+\beta_3)(\beta_4+\beta_1)(\bar{k}_1^2+\bar{k}_1\bar{k}_3+\bar{k}_3^2)}{96\pi^2 s_{12}s_{23}}\,.
    \end{split}
\end{equation}
Next, we move to the graphs where we have vacuum bubbles on the external legs. In this case, we have in total eight diagrams. Take the following diagram as an example 
\begin{figure}[h!]
    \centering
    \includegraphics[scale=0.3]{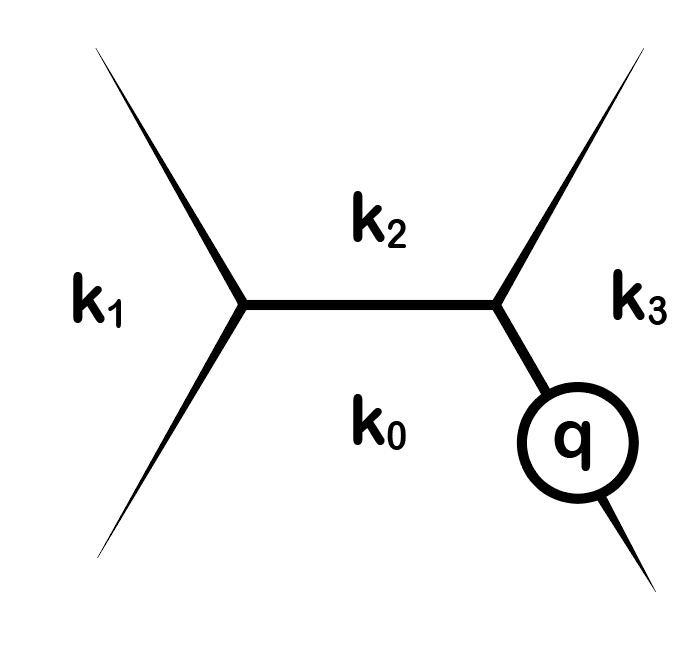}
\end{figure}

Here, the loop momentum is $\pvec=\qvec-\kvec_0$ and external momenta remain to be the same as $\pvec_i=\kvec_i-\kvec_{i-1}$. Then the bubble on leg-$i$ denoted as $\Gamma_{\bigcirc}^{i}$ reads (remember that we have two different channels for each due to color ordering)
\begin{align}
    \Gamma_{\bigcirc}^1&=-\sum_{\omega}\frac{(l_p)^{\Lambda_4-4}\alpha_4^{\Lambda_4-4}}{\prod_{i=1}^4\beta_i^{\lambda_i}\Gamma(\Lambda_4-3)}\frac{\PPb_{23}\PPb_{34}\beta_1^2(\bar{k}_0^2+\bar{k}_0\bar{k}_1+\bar{k}_1^2)}{96\pi^2 s_{12}s_{23}}\,,\\
    \Gamma_{\bigcirc}^2&=-\sum_{\omega}\frac{(l_p)^{\Lambda_4-4}\alpha_4^{\Lambda_4-4}}{\prod_{i=1}^4\beta_i^{\lambda_i}\Gamma(\Lambda_4-3)}\frac{\PPb_{34}\PPb_{41}\beta_2^2(\bar{k}_1^2+\bar{k}_1\bar{k}_2+\bar{k}_2^2)}{96\pi^2 s_{12}s_{23}}\,,\\
    \Gamma_{\bigcirc}^3&=-\sum_{\omega}\frac{(l_p)^{\Lambda_4-4}\alpha_4^{\Lambda_4-4}}{\prod_{i=1}^4\beta_i^{\lambda_i}\Gamma(\Lambda_4-3)}\frac{\PPb_{41}\PPb_{12}\beta_3^2(\bar{k}_2^2+\bar{k}_2\bar{k}_3+\bar{k}_3^2)}{96\pi^2 s_{12}s_{23}}\,,\\
    \Gamma_{\bigcirc}^4&=-\sum_{\omega}\frac{(l_p)^{\Lambda_4-4}\alpha_4^{\Lambda_4-4}}{\prod_{i=1}^4\beta_i^{\lambda_i}\Gamma(\Lambda_4-3)}\frac{\PPb_{12}\PPb_{23}\beta_4^2(\bar{k}_3^2+\bar{k}_3\bar{k}_0+\bar{k}_0^2)}{96\pi^2 s_{12}s_{23}}\,.
\end{align}
Equivalently, we can write them as
\small
\begin{align}
        \Gamma^1_{\bigcirc}&=-\nu_0\mathcal{N}_{\square}\frac{\PPb_{23}\PPb_{34}(\beta_1\beta_3\PPb_{41}\PPb_{12}+\beta_1(\beta_1+\beta_4)\PPb_{12}\PPb_{34}+\beta_1(\beta_1+\beta_2)\PPb_{23}\PPb_{41})(\bar{k}_0^2+\bar{k}_0\bar{k}_1+\bar{k}_1^2)}{96\pi^2 s_{12}s_{23}}\,,\\
        \Gamma_{\bigcirc}^2&=-\nu_0\mathcal{N}_{\square}\frac{(\beta_2\beta_3\PPb_{41}\PPb_{12}+\beta_2(\beta_1+\beta_2)\PPb_{23}\PPb_{41})(\bar{k}_1^2+\bar{k}_1\bar{k}_2+\bar{k}_2^2)}{96\pi^2 s_{12}s_{23}}\,,\\
        \Gamma_{\bigcirc}^3&=-\nu_0\mathcal{N}_{\square}\frac{\PPb_{41}\PPb_{12}\beta_3^2(\bar{k}_2^2+\bar{k}_2\bar{k}_3+\bar{k}_3^2)}{96\pi^2 s_{12}s_{23}}\,,\\
        \Gamma_{\bigcirc}^4&=-\nu_0\mathcal{N}_{\square}\frac{(\beta_3\beta_4\PPb_{41}\PPb_{12}+\beta_4(\beta_1+\beta_4)\PPb_{12}\PPb_{34})(\bar{k}_3^2+\bar{k}_3\bar{k}_0+\bar{k}_0^2)}{96\pi^2 s_{12}s_{23}}\,.
\end{align}
\normalsize
Collecting numerator coefficients and remember that $\pvec_i=\kvec_i-\kvec_{i-1}$. All together they provide 
\begin{equation}
    \Gamma_{\text{bubbles}}=\sum_{i=1}^4\Gamma^i_{\bigcirc}+2\Gamma^{\text{in}}_{\bigcirc}=-\nu_0\,\frac{(l_p)^{\Lambda_4-4}\alpha_4^{\Lambda_4-4}}{96\pi^2\Gamma(\Lambda_4-3)\prod_{i=1}^{\lambda_i}\beta_i^{\lambda_i}}\frac{\PPb_{12}\PPb_{23}\PPb_{34}\PPb_{41}}{s_{12}s_{23}}
\end{equation}
Hence,
\begin{equation}
    \Gamma_4=\parbox{2.5cm}{\includegraphics[scale=0.13]{4ptfull.png}}+2\times \parbox{2.4cm}{\includegraphics[scale=0.2]{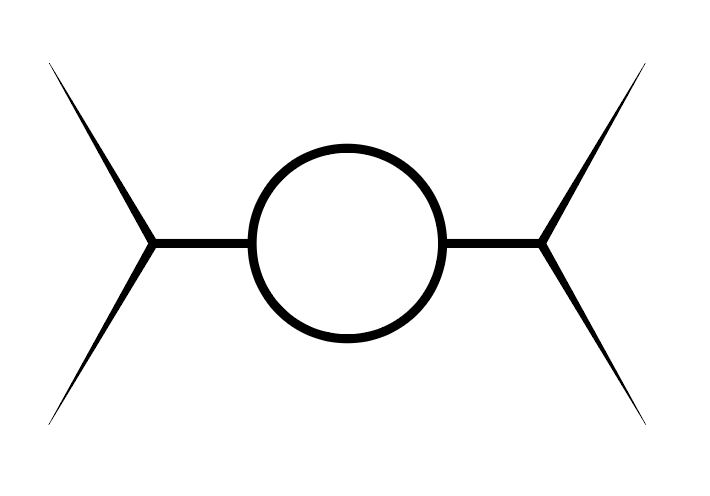}}+8\times \parbox{2cm}{\includegraphics[scale=0.18]{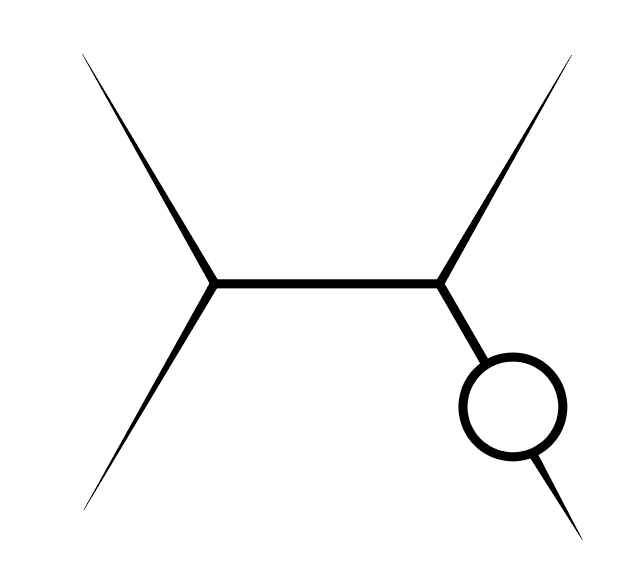}}=0
\end{equation}
Therefore, the 4-point function at one loop does not have any UV-divergences since it can be reduced to UV-convergent integrals we have already analyzed. The complete 4-point amplitude vanishes due to the same $\nu_0$ factor.
\subsection{Sunrise Diagrams and Multiloop Amplitudes}
\label{sec:multiloop}
For multiloop amplitudes in the large $N$ limit, one can start with the sunrise diagrams that have some of the legs off-shell and glue them together.
\begin{figure}[h]
    \centering
    \includegraphics[scale=0.3]{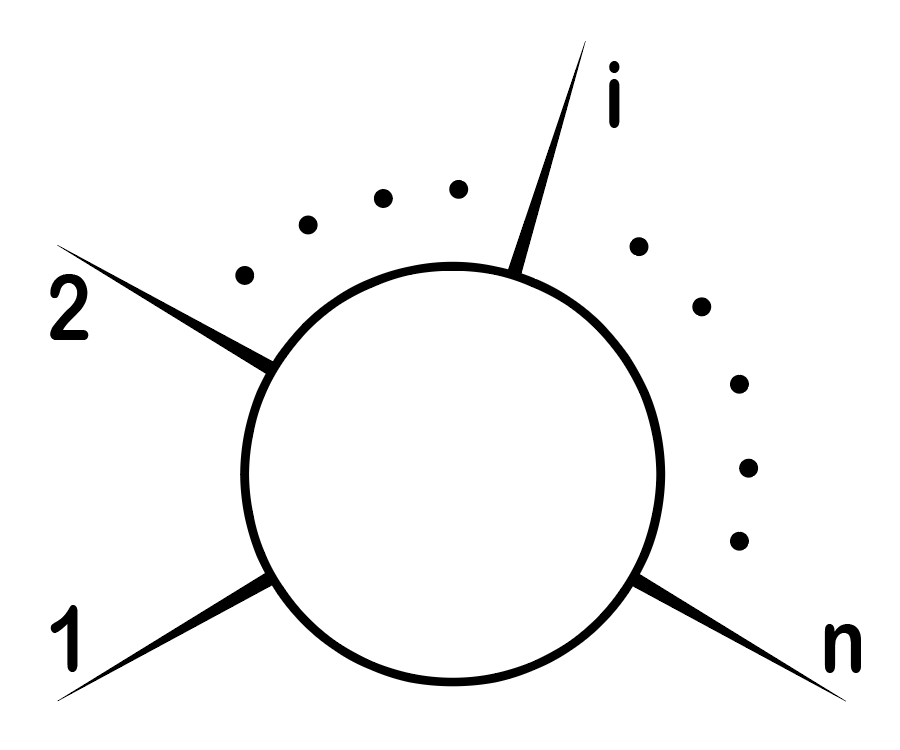}
\end{figure}
The kinematic part of the sunrise diagrams can be simply written as (we omit $\beta_i^{\lambda_i}$ at the moment for simplicity)
\small
\begin{equation}
    \sum_{\{\omega_i\}}\frac{\PPb_{\pvec_1,\pvec,-\pvec-\pvec_1}^{\lambda_1+\omega_1-\omega_n}}{\Gamma(\lambda_1+\omega_1-\omega_n)}\frac{\PPb_{\pvec_2,\pvec+\pvec_1,-\pvec-\pvec_1-\pvec_2}^{\lambda_2-\omega_1+\omega_2}}{\Gamma(\lambda_2-\omega_1+\omega_2)}\cdots \frac{\PPb_{\pvec_n,\pvec-\pvec_n,-\pvec}^{\lambda_n-\omega_{n-1}+\omega_n}}{\Gamma(\lambda_n-\omega_{n-1}+\omega_n)}=\sum_{\omega_n}\frac{\alpha_n^{\Lambda_n-n}\mathcal{K}_n}{\Gamma(\Lambda_n-(n-1))}
\end{equation}
\normalsize
where $i=1,...n$ and
\begin{align}
    \alpha_n=\sum_{i<j=2}^{n-2}\PPb_{ij}+\PPb_{n-1,n}\,,\qquad  \mathcal{K}_n(\PPb)=\prod_{j=1}^{n}\Big(\PPb_{pj}+\sum_{i<j}\PPb_{ij}\Big)\,.
\end{align}
Putting the propagator and coupling constant together, one get the general form of one loop correction with $n$-external legs, some of which can be off-shell
\begin{equation}
    \Gamma_{n}=\nu_0\,\frac{(l_p)^{\Lambda_n-n}\alpha_n^{\Lambda_n-n}}{\Gamma(\Lambda_n-(n-1))\prod_{i=1}^n\beta_i^{\lambda_i}}\int\frac{d^4p}{(2\pi)^4}\frac{\mathcal{K}_n(\PPb)}{\pvec^2(\pvec+\pvec_1)^2 ... (\pvec-\pvec_n)^2}\,.
\end{equation}
The sum over helicities is crucial to make the contribution vanish even though we do not evaluate the integral explicitly. The integral itself has to be UV-convergent due to vanishing of the three-level amplitudes. Consequently, all multiloop amplitudes vanish confirming that $S=1$.
\subsection{One-loop Finiteness of Chiral HSGRA}
In this subsection, we show that chiral HSGRA is one-loop finite. The result is that the complete n-point one-loop S-matrix element consists of three
factors: the all-plus helicity one-loop amplitude in QCD (or self-dual Yang-Mills), which can be anticipated from \cite{Ponomarev:2017nrr};\footnote{As a side remark, the computation in the paper, after erasing the higher spin modes, can give a simple way to compute one-loop amplitudes in self-dual Yang-Mills. } a certain higher spin dressing --- an overall kinematical factor that accounts for the helicities on the external legs; a purely numerical factor of the total number of degrees of freedom:
\begin{align}\tag{$\Diamond$}\label{result}
    \Gamma_{\text{Chiral HSG, 1-loop}}&= \Gamma^{++...+}_{\text{QCD, 1-loop}} \times \left[\,\parbox{3.5cm}{kinematical \\ higher spin dressing}\right] \times \nu_0\,.
\end{align}
The evaluation of one-loop integrals with $2,3,4$-legs reveals the nuts and bolts of how higher spin fields eliminate UV-divergences: the specific structure of higher derivative interactions helps to factor enough momenta out of the integrand to make the integral UV-convergent, which is somewhat reminiscent of $\mathcal{N}=4$ Yang-Mills Theory \cite{Mandelstam:1982cb,Brink:1982wv} where one power of the momentum suffice. The final one-loop scattering amplitude vanishes, due to the total number of effective degrees of freedom $\nu_0=0$ \cite{Beccaria:2015vaa}, which is consistent with the Weinberg and Coleman-Mandula theorems. We note that the tree-level holographic $S$-matrix of Chiral Theory in $\ads_4$ does not vanish and is related \cite{Skvortsov:2018uru} to the correlation functions in Chern-Simons Matter Theories, which supports the dualities they were conjectured to exhibit \cite{Giombi:2011kc, Maldacena:2012sf, Aharony:2012nh,Aharony:2015mjs,Karch:2016sxi,Seiberg:2016gmd}.  

Let us now take a sum of integrands of all one-loop Feynman diagrams with $n$ external on-shell momenta $\pvec_i$, $\pvec_i^2=0$. We denote this sum $F$. The loop momentum is $\lvec$. $F$ is a rational function of momenta $\pvec_i$, $\lvec$. Note, that the vertices do not contain the minus-component of the momenta. Therefore, $p^-_i$, $\ell^-$ appear only in the denominators, as a part of the propagator, $\pvec^2=2p^+p^- +2 p\pb$. Now, $F$, as a function of $\ell^-$, vanishes at infinity and has only simple poles. The poles correspond to some momenta along the loop going on-shell in various diagrams that contribute to $F$. Since the loop momenta is to be integrated over, there is a an ambiguity in the momenta assigned to the lines going around any loop. Indeed, we can simply add any amount $\qvec$ to all momenta of the loop. We would like to choose the momenta around the loop in such a way that the residues of $F$ at the poles in $\ell^-$ give the complete $(n+2)$-point tree-level amplitude:
\begin{align*}
    F=\sum\parbox{3.cm}{\includegraphics[scale=0.27]{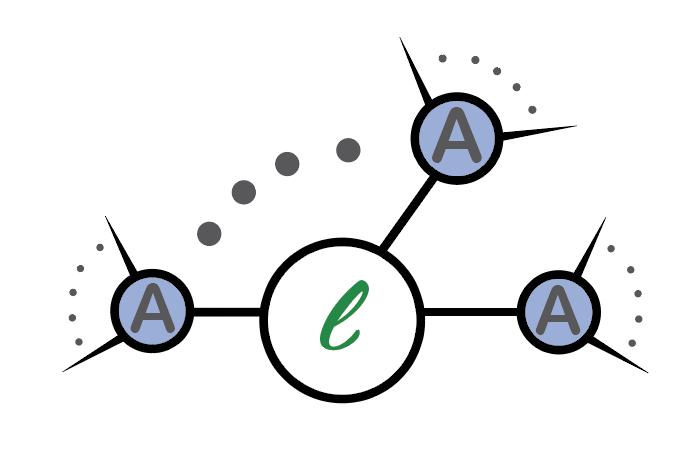}}\xrightarrow[\text{residue}]{\ell^2\rightarrow0 }\sum\parbox{3.cm}{\includegraphics[scale=0.27]{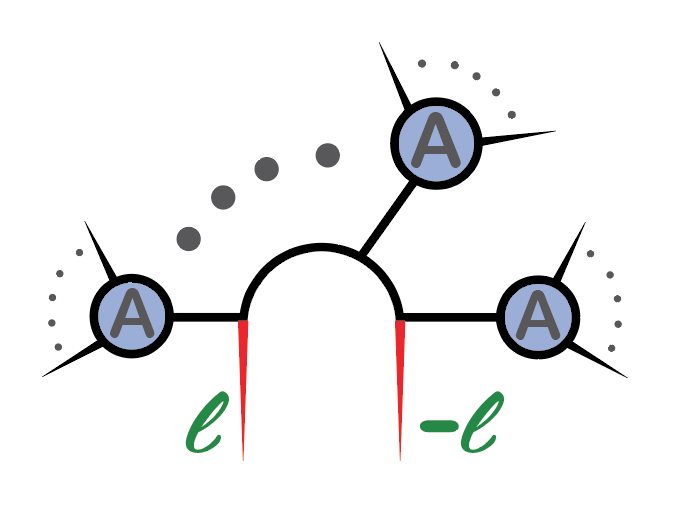}}=A_{\text{tree}}(\pvec_1,...,\lvec,...,-\lvec,...,\pvec_n)\,.
\end{align*}
The relation between the original momenta $\lvec,\pvec_i$ and the dual momenta $\qvec,\kvec_i$ reads
\begin{align}
    \lvec=\qvec-\kvec_0,\qquad \pvec_i=\kvec_i-\kvec_{i-1},\quad \kvec_n\equiv\kvec_0\,.
\end{align}
Note that for an $n$-point amplitude there are $n$ independent $\kvec_i$ instead of $n-1$ independent $\pvec_i$ (due to momentum conservation). Therefore, there should be a translation symmetry in the dual space to compensate for this redundancy in $\kvec_i$. The physical amplitude must be translation invariant in $\kvec_i$. If this is so, then it is possible to solve for all $\kvec_i$ in terms of external momenta $\pvec_i$. At this point we move to the dual space. Each term in $F$ has a loop and now each segment of the loop has $\qvec-\kvec_i$ flowing through it for a certain $i$. The dual space automatically leads to the correct routing of the momenta. Now, we consider $F$ to be a function of $\qvec$, $\kvec_i$ and are interested in the poles with respect to $\qvec^-$. The residue at each pole gives the sum over all tree level diagrams with the same momenta on the external lines. The latter is crucial for getting the complete tree-level amplitude as the residue (rather than just a random sum of tree-level diagrams with different momenta on some of the external lines). 

It turns out that the interactions fine-tuned by the higher spin symmetry make all tree-level amplitudes vanish \cite{Skvortsov:2018jea,Skvortsov:2020wtf}. Therefore, we have a meromorphic function $F$, whose residues vanish. Therefore, $F\equiv 0$. Note that $F$ is just the total one-loop integrand. However, we do not need all terms of $F$ to get the $S$-matrix element. The self-energy corrections and the tadpoles should be excluded. To this end, we represent $F$ as follows
\begin{align}
    F&=F^{1-\text{loop}}_{S} +F^{1-\text{loop}}_{\text{bubbles}}+F^{1-\text{loop}}_{\text{tadpoles}}=0\,,
\end{align}
where $F^{1-\text{loop}}_{S}$ is the complete integrand for the one-loop $S$-matrix element (that includes triangles, boxes and up to $n$-gon diagrams) and $F^{1-\text{loop}}_{\text{bubbles}}$, $F^{1-\text{loop}}_{\text{tadpoles}}$ are self-evident. The tadpoles and the cuts of tadpoles vanish by themselves. Indeed, the tadpole has $V(\boldsymbol{0},\mu;\lvec,\lambda;-\lvec,-\lambda)\equiv0$ as a vertex. It is important that the cubic self-interaction of the scalar field is absent, i.e. $V(\pvec_1,0;\pvec_2,0;-\pvec_1-\pvec_2,0)\equiv0$.

There is a nontrivial, but finite, contribution from the self-energy insertions into various external and internal lines, see below. As a result we have
\begin{align}
    F^{1-\text{loop}}_{S} +F^{1-\text{loop}}_{\text{bubbles}}&=0\,, && F^{1-\text{loop}}_{\text{tadpoles}}=0\,.
\end{align}
Therefore, in order to get the full one-loop $S$-matrix element we need to sum over all bubble's insertions. The summation will be done with the help of the tree-level amplitudes that are available \cite{Skvortsov:2018jea,Skvortsov:2020wtf} and we briefly summarize the results. 

To proceed, let us recall the recursion result for $n$-point treel-level amplitude \small
\begin{equation}\label{eq:npointrecursive}
    A_n(1...n)=\frac{(-)^n\,\alpha_n^{\Lambda_n-(n-2)}\beta_3...\beta_{n-1}\,\pvec_1^2}{2^{n-2}\Gamma(\Lambda_n-(n-3))\prod_{i=1}^{n}\beta_i^{\lambda_i-1}\beta_1\PP_{23}...\PP_{n-1,n}}\,, \quad \alpha_n=\sum_{i<j}^{n-2}\PPb_{ij}+\PPb_{n-1,n}\,,
\end{equation}
\normalsize
where $\Lambda_n=\lambda_1+...+\lambda_n$, and also the self-energy correction in the planar limit of the $U(N)$-gauged Chiral Theory or for the $N=1$ theory \cite{Skvortsov:2018jea,Skvortsov:2020wtf}
\begin{equation}\label{eq:2pt}
\begin{split}
    \parbox{3.8cm}{\includegraphics[trim={0cm 0.3cm 0.0cm 0.6cm},clip,scale=0.22]{Thornselfenergy.png}}&=\nu_0\,N\frac{(l_p)^{\Lambda_2-2}}{\beta_1^{\lambda_1}\beta_2^{\lambda_2}\Gamma[\Lambda_2-1]}\int \frac{d^4q}{(2\pi)^4} \frac{\PPb_{q-k_0,p_1}^2\delta_{\Lambda_2,2}}{(\qvec-\kvec_0)^2(\qvec-\kvec_1)^2}\\
    &=\nu_0 N(\kb_0^2+\kb_0\kb_1+\kb_1^2)\frac{\delta_{\Lambda_2,2}(l_p)^{\Lambda_2-2}}{96\pi^2\beta_1^{\lambda_1-1}\beta_2^{\lambda_2-1}\Gamma[\Lambda_2-1]}\,,
    \end{split}
\end{equation}
where $\nu_0=\sum_{\lambda} 1$. It is important to note that the result is non-vanishing only when $\Lambda_2=\lambda_1+\lambda_2=2$. Below, we set Planck's length $l_p=1$ for simplicity. We note that $\nu_0$ counts the number of degrees of freedom in the theory and has nothing to do with the UV-convergence. Moreover, the amplitude is not translation invariant in the dual space, i.e. it is anomalous. Therefore, it has to removed by a counterterm, which will be important later.

\noindent\textbf{Inserting bubbles into tree-level diagrams.} As for the tree-level amplitudes, the direct summation over all tree-level diagrams with the bubble inserted is hardly feasible. Instead, in order to compute $F^{1-\text{loop}}_{\text{bubbles}}$ we apply another recursive relation, which can be depicted as 
\begin{align}\label{eq:recursivegraph}
   \sum_{i=1}^{[n/2]} \parbox{1.8cm}{\includegraphics[scale=0.2]{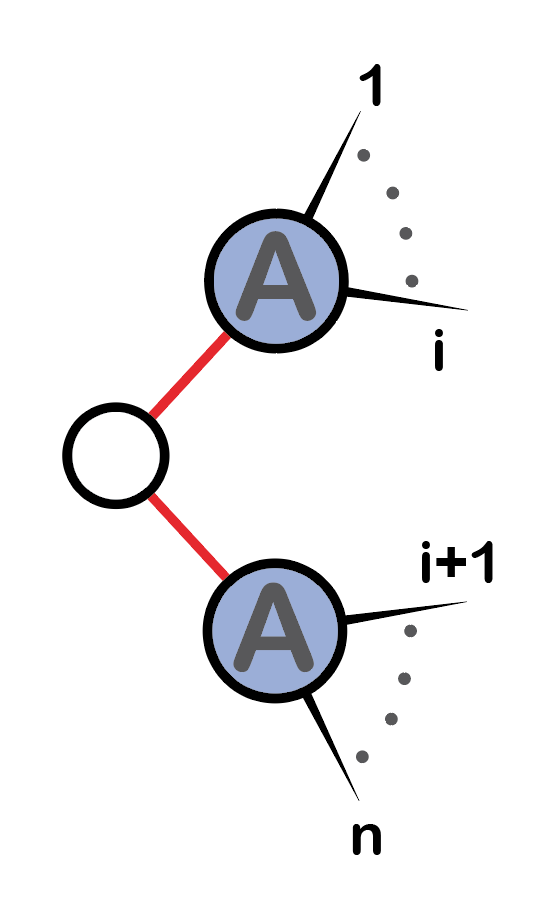}}=\Bigg[\parbox{2.82cm}{\includegraphics[scale=0.27]{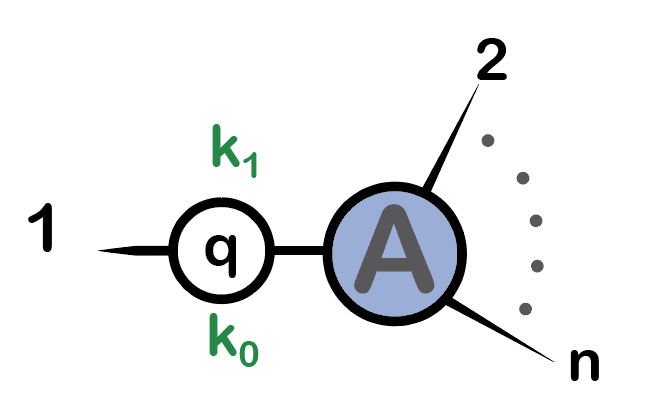}}+\parbox{2.88cm}{\includegraphics[scale=0.27]{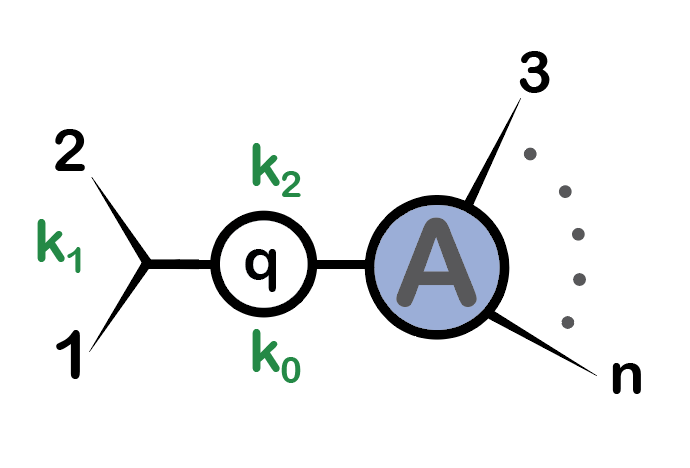}}+\parbox{3.6cm}{\includegraphics[scale=0.265]{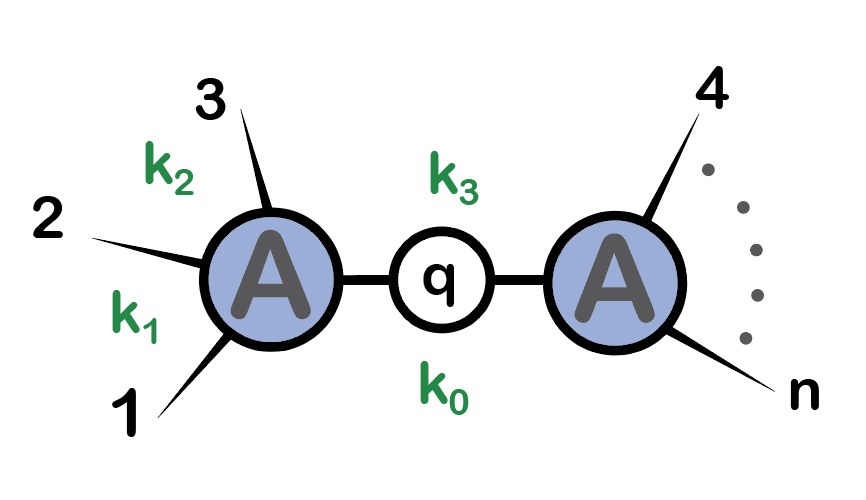}}+...\Bigg]\,.
\end{align}
Here, the blue blobs are the tree-level sub-amplitudes that are being glued to the bubble (the white blob). First, the white blob sits on the leftmost external line. In the second term it is one vertex away from the external lines on the left. In the third term it has passed three external lines on the left and so on. The final $\ldots$ also implies the sum over the cyclic permutations. Inserting the self-energy integral \eqref{eq:2pt} will give a contribution of
\begin{align}\label{eq:qintegral}
    &(k_j^+-k_i^+)^2(\kb_i^2+\kb_i\kb_j+\kb_j^2)\,,&& k_j^+-k_i^+=\sum_{m=i+1}^{j}\beta_m\,,
\end{align}
where $\kvec_{i,j}$ are the regional dual momenta that are adjacent to the inserted bubble. Note that once we insert the bubble into an internal line, the two propagators get cancelled against the $\pvec^2$-factors of the two tree-level diagrams \eqref{eq:npointrecursive} being glued. We also note that the bubble is slightly off-diagonal in the helicity space since it has $\delta_{\lambda_1+\lambda_2,2}$ instead of $\delta_{\lambda_1+\lambda_2,0}$ for the propagators. 

\noindent\textbf{One-loop amplitude.} What remains is to massage the sum over the bubble's insertions and to put the minus sign in front. Let us write \eqref{eq:recursivegraph} in terms of $\PP,\kb$ and $\beta$ components by using \eqref{eq:npointrecursive} and \eqref{eq:qintegral}. The diagrams in \eqref{eq:recursivegraph} correspond to gluing the bubble to the two sub-amplitudes with the total number of external legs equal $n$ and then taking the cyclic permutations. We arrive at
\small
\begin{equation}\label{eq:recursiveexplicit}
\begin{split}
    (2.10)&=\mathcal{N}_n\Bigg[\sum_{i=1}^{[n/2]}\frac{(\kb_0^2+\kb_0\kb_i+\kb_i^2)(\sum_{k=1}^i\beta_k)^2}{\beta_1\beta_i\beta_{i+1}\beta_n}\PP_{i,i+1}\PP_{n1}+\text{cyclic permutations}\Bigg]
    \end{split}
\end{equation}
\normalsize
where 
\begin{align}
    \mathcal{N}_n=\nu_0\,\frac{(-1)^{n}\alpha_n^{\Lambda_n-n}}{2^{n+3}3\pi^2\Gamma[\Lambda_n-(n-1)]\prod_{i=1}^n\beta_i^{\lambda_i-2}\PP_{12}\PP_{23}...\PP_{n1}}\,.
\end{align}
As we have already stressed, all physical quantities must be translation invariant in the dual space. Therefore, \eqref{eq:recursiveexplicit} should not change if we replace $\kvec_i$ by $\kvec_i+\boldsymbol{a}$ for any $\boldsymbol{a}$. One way to see it is to solve for all $\kvec_i$ except for $\kvec_0$ via $\kvec_i=\kvec_0+\sum_{j=1}^i\pvec_j$. In order to see that the resulting expression $f(\kb_0)$ does not depend on $\kb_0$ we can take its derivative $f'(\kb_0)$ to get
\begin{align}
    \mathcal{N}_n\kb_0&\Bigg[\sum_{i=1}^{[n/2]}\frac{(\sum_{k=1}^i\beta_k)^2}{\beta_1\beta_i\beta_{i+1}\beta_n}\PP_{i,i+1}\PP_{n1}+\text{cyclic permutations}\Bigg]\,.
\end{align}
This is nothing but \eqref{eq:recursiveexplicit} with all $(\kb_i^2+\kb_i\kb_j+\kb_j^2)$ factors erased, times $\kb_0$. It is easy to show that this expression is indeed zero with the help of the momentum conservation, see various identities in \cite{Skvortsov:2020wtf}. Once \eqref{eq:recursiveexplicit} is shown to be translation invariant, it can be expressed in terms of external momenta $\pvec_i$ only. This is quite remarkable since the self-energy diagram itself, \eqref{eq:2pt}, is not translation invariant, it is anomalous. 

Due to many kinematical identities involving $\beta_i$ and $\PPb_{ij}$, there is no unique way to write the final result, but the following form is very suggestive
\begin{align}\label{eq:result}
    A_{\text{1-loop}}^{\text{HSG}}=\Big[\sum_{1\leq i_1<i_2<i_3<i_4\leq n}\frac{ \check{\beta}^{n-4}_{i_1,i_2,i_3,i_4}\PP_{i_1i_2}\PPb_{i_2i_3}\PP_{i_3i_4}\PPb_{i_4i_1}}{2^{\tfrac{n}2-2}\PP_{12}\PP_{23}...\PP_{n1}}\Big]\times D^{\text{HSG}} \times\nu_0\,,
\end{align}
where 
\begin{align}
    \check{\beta}^{n-4}_{i_1,i_2,i_3,i_4}&=\frac{\prod_{j=1}^{n}\beta_j}{\beta_{i_1}\beta_{i_2}\beta_{i_3}\beta_{i_4}}\,, & D^{\text{HSG}}&=\frac{(-)^{n}\alpha_n^{\Lambda_n-n}}{2^{\frac{n}{2}+5}3\pi^2\Gamma[\Lambda_n-(n-1)]\prod_{i=1}^n\beta_i^{\lambda_i-1}}\,.
\end{align}
Clearly, the one-loop amplitude in Chiral Higher Spin Gravity consists of (i) a factor that has a lower spin origin as it does not have enough $\PP$ to account for $\lambda_i$; (ii) kinematical higher spin dressing factor $D^{\text{HSG}}$ that accounts for helicities $\lambda_i$ on the external lines, which the first factor cannot accomplish; (iii) the total number of physical degrees of freedom $\nu_0$. 

The first factor is telling. Applying the light-cone vs. spinor-helicity dictionary \eqref{spinor-lightcone}, we discover the all-plus helicity one-loop amplitude in QCD or in self-dual Yang-Mills \cite{Bern:1993sx,Bern:1993qk,Mahlon:1993fe}:
\begin{align}
    A_{\text{SDYM, 1-loop}}=A_{\text{QCD,1-loop}}^{++...+}=\sum_{1\leq i_1<i_2<i_3<i_4\leq n}\frac{\langle i_1i_2\rangle[i_2i_3]\langle i_3i_4\rangle[i_4i_1]}{\langle 12\rangle \langle 23 \rangle ... \langle n1\rangle}\,.
\end{align}
In other words, the one-loop amplitude in Chiral Higher Spin Gravity is found to be
\begin{align}\label{finalres}
    A_{\text{1-loop}}^{\text{HSG}} =A_{\text{QCD,1-loop}}^{++...+}\times D^{\text{HSG}} \times \nu_0\,.
\end{align}
Therefore, we get precisely the structure \eqref{result} that is sketched in the introduction. Moreover, when we set $\lambda_i=1$, \eqref{finalres} reduces to just the SDYM/QCD amplitude times an overall numerical factor, i.e. the higher spin dressing disappears. To conclude, both the Weinberg, Coleman-Mandula theorems and the one-loop determinants instruct us to set $\nu_0=0$ and get $S=1$. This can safely be done since the one-loop amplitude is shown to be UV-finite.  

\section{Conclusions and Discussion}
\label{sec:ads}
Chiral Theory reveals a remarkable cancellation mechanism for UV-divergences and should be an example of a consistent quantum HSGRA. This is the only higher spin model with propagating massless higher spin fields where quantum corrections can be computed. 

We showed in this chapter that the tree-level amplitudes vanish on-shell, which is a result of highly nontrivial cancellations after the summation over Feynman diagrams. This is the requirement of the Weinberg low energy theorem. Another interesting property of chiral theory is that the spin sums are bounded from both above and below assuming the external helicities are fixed. In generic higher spin theories we would expect an infinite sum over all spins already for tree level diagrams.\footnote{This corresponds to gluing chiral and anti-chiral vertices together.} This does not happen for chiral HSGRA and infinite spin sums show up only at the loop level as an overall factor as shown in the body of this chapter. 

The loop diagrams turn out to consist of two factors: the UV-convergent integral and a purely numerical factor $\nu_0=\sum_{\lambda} 1$. The UV-convergence is a very important property that again relies on the presence of higher spin fields. This effect is reminiscent of $\mathcal{N}=4$ Yang-Mills theory \cite{Mandelstam:1982cb,Brink:1982wv}, in which the supersymmetry forces one momentum to eventually factor out and makes the integrals convergent. Higher spin symmetry amplifies this effect. Chiral HSGRA has infinitely many non-renormalizable interactions, which include the two-derivative graviton self-coupling. Higher spin symmetry forces enough momenta to factor out in every loop integral and makes all loop integrals free of UV-divergences. Overall factor $\nu_0$ is to be expected in any theory with infinitely many fields and some value needs to be assigned to the sum. Based on zeta-function regularization, it is natural to set $\nu_0=0$. Such an assignment is consistent both with the Weinberg soft theorem and with a large bulk of results on one-loop determinants in HSGRA/Vector Models duality context. 

The result $S=1$ agrees with our expectation for any HSGRA in flat space. It is, however, no longer true once the cosmological constant is switched on. The holographic S-matrix turns out to be nontrivial \cite{Skvortsov:2018uru}. Therefore, we consider chiral HSGRA in flat space as a useful toy model to check the cancellation of UV-divergences thanks to higher spin symmetry. It is exactly the effect that HSGRAs have long been expected to have.

We also extend chiral HSGRA in such a way that it incorporates Yang-Mills gaugings, see Appendix \ref{app:chap4}. Even though we do not see any immediate relation to string theory, it is quite surprising that higher spin fields can be made matrix-valued fields via the method that is very similar to the Chan-Paton approach. Higher spin symmetry seems to be restrictive enough to make theories with a graviton in the spectrum to be quantum consistent. It was recently shown that one can extend chiral HSGRA to  supersymmetric chiral HSGRAs \cite{Metsaev:2019aig}. However, the mechanism that cancels UV-divergences should be the same with the pure bosonic case we investigated in this chapter.

Chiral HSGRA is the only class of HSGRAs at present with propagating massless higher spin fields and an action. Nevertheless, there is a handful of other higher spin models with an action that are of great interest. There are topological theories, which are free of UV-divergences, in three dimension: purely massless \cite{Blencowe:1988gj,Bergshoeff:1989ns,Campoleoni:2010zq,Henneaux:2010xg} and conformal \cite{Pope:1989vj,Grigoriev:2019xmp}. Another class is $4d$ conformal higher spin gravity \cite{Segal:2002gd,Tseytlin:2002gz,Bastianelli:2012bn}, which is an extension of conformal gravity. There also has been some progress in two dimensions \cite{Alkalaev:2020kut}. There are encouraging results on quantum checks for conformal higher spin gravity \cite{Beccaria:2016syk,Joung:2015eny} that indicate that the conformal higher spin symmetry also makes the S-matrix trivial in flat space. The 2d-models of \cite{Alkalaev:2020kut} involve propagating matter fields with interactions mediated via topological higher spin fields, thereby providing interesting toy models for quantum checks. Lastly, it would be very important to directly verify that $\ads_4$ chiral HSGRA is free of UV-divergences.

  \chapter{Formal HSGRA in \texorpdfstring{$\ads_5$}{ads5}}\label{chapter5}
In this chapter, we construct a \textit{formal} bosonic HSGRA in $\ads_5$ in terms of formally consistent classical equations of motion. Finding the equations of motion was shown to be equivalent to a certain deformation of a given higher-spin algebra \cite{Sharapov:2019vyd,Sharapov:2018kjz}. There are two different realization of the deformed higher-spin algebra: (i) through the universal enveloping algebra of $su(2,2)$; (ii) through oscillator variables. Both of the new realizations admit supersymmetric extensions and the $\mathcal{N}=8$ extension should describe the massless sector of tensionless Type-IIB strings on $AdS_5\times S^5$.

\section{Motivation}
In the previous chapters we have seen how higher-spin symmetry can render HSGRA renormalizable and even finite in both AdS and flat spaces. We also understood that the (holographic) S-matrix is fixed by this \textit{rich} symmetry
\cite{Maldacena:2011jn,Boulanger:2013zza}, which eventually led us to the conjecture that HSGRA is UV-finite. Therefore, if we believe the symmetry arguments, i.e. that the higher spin symmetry alone forbids all relevant counterterms, our task of finding a quantum consistent theory downgrades to a task of constructing a purely classical HSGRA. 
The AdS/CFT correspondence is a crucial reference point for the construction of the bulk theory since the holographic S-matrix should precisely match the free/weakly coupled CFT's correlation functions.

However, free (or weakly coupled) CFT’s do not have a large gap in the dimensions of single-trace operators and hence the existence of the gravitational dual requires justification \cite{Heemskerk:2009pn}. Due to severe nonlocalities required by the higher spin symmetry \cite{Bekaert:2015tva,Sleight:2017pcz,Ponomarev:2017qab}, we can understand that HSGRAs are not conventional field theories. Yet, the existence of CFT dual descriptions should in principle allow one to reconstruct the bulk theory from the CFT correlation functions \cite{Das:2003vw,deMelloKoch:2018ivk}, i.e. to write down certain interaction vertices in $\ads$ that, via Witten diagrams, compute exactly the correlation functions of the required CFT. There are at least two issues here: (i) reconstruction does not give a definition of the bulk theory that would be independent of its CFT dual, thereby trivializing AdS/CFT duality;\footnote{An important question is what are the bulk questions that cannot be immediately answered from the reconstruction vantage point. One such question is about quantum corrections in the bulk. Given that the reconstruction gives a classical action that computes the required CFT correlation functions at tree-level in the bulk, it is still a challenge to prove that the quantum corrections come out right. See, for example, \cite{Ponomarev:2019ltz} for the analysis of the one-loop corrections in holographic HSGRA. } (ii) still, the interactions that are required to get the free (weakly-coupled) CFT's correlation functions are too non-local to treat them as local field theories and there are ambiguities that do not even allow one to compute tree-level amplitudes without further prescriptions \cite{Bekaert:2015tva,Sleight:2017pcz,Ponomarev:2017qab}. The latter calls for a better understanding of the bulk locality in HSGRA. 

It is worth stressing that conformal HSGRA and chiral HSGRA avoid the aforementioned problems. Nevertheless, it is important to understand how to stretch the axioms of local field theory as to be able to define holographic HSGRA's, e.g. those that are dual to free and critical vector models. 

In this chapter, we will construct the bulk theory by studying the deformation of higher-spin algebra --- an extension of conformal algebra $so(d,2)$ in generic dimensions and $su(2,2)$ in $\ads_5/\cft^4$. Our starting point is any free CFT. As a known fact, free CFTs come with higher-spin algebra {\bf hs} --- the symmetry algebra of the free equations of motion \cite{Eastwood:2002su}. Higher-spin algebra is associative and is the quotient of $U(so(d,2))$ (or its supersymmetric extension) by the two sided Joseph ideal $\mathcal{I}$. 

Being identified with a global symmetry on the CFT side, {\bf hs} carries complete information about the spectrum of single-trace operators and their correlators.\footnote{The correlation functions are just the simplest {\bf hs} invariants \cite{Colombo:2012jx,Didenko:2013bj,Didenko:2012tv,Bonezzi:2017vha}.} It has to be gauged in the gravitational dual producing thus inevitable non-localities. If we \textit{sacrifice} locality, then there is a formal way to make the problem of finding vertices for HSGRA well-defined mathematically. It involves writing down formally consistent classical equations of motion which take the form 
\begin{align}\label{eq:SFTformal}
    d\Phi=\Vcal_2(\Phi,\Phi)+\Vcal_3(\Phi,\Phi,\Phi)+...\,, \qquad d^2=0\,,
\end{align}
where $\Phi$ is some field.\footnote{The equations above look similar to those of String Field Theory, see e.g. \cite{Gaberdiel:1997ia,Erler:2013xta}.} The construction of $\Vcal_n$ are heavily based on strong homotopy algebras a.k.a. $A_{\infty}/L_{\infty}$-algebras (see e.g. \cite{Lada:1992wc,Kajiura:2003ax}) and the quantization deformation \cite{Kontsevich:1997vb}. 

The formal HSGRA approach was initiated by Vasiliev, who constructed the first example of such a system \cite{Vasiliev:1990en}. At present there are several examples of formally consistent equations of motion \cite{Prokushkin:1998bq,Vasiliev:2003ev,Bonezzi:2016ttk,Bekaert:2017bpy,Arias:2017bvi,Grigoriev:2018wrx,Sharapov:2019vyd} that deal with different higher spin algebras or provide a different realization of the same system. The general problem of how to construct a formal HSGRA, i.e. the vertices,  starting from any higher spin algebra was solved in \cite{Sharapov:2018kjz,Sharapov:2019vyd}, where it was shown that constructing $\Vcal_n$ is equivalent to deforming a certain extension of {\bf hs} as an associative algebra.\footnote{See also \cite{Sharapov:2018ioy,Sharapov:2018hnl,Sharapov:2017lxr} for various closely related mathematical aspects, in particular,  \cite{Sharapov:2017yde} for a relation to the Kontsevich–Shoikhet–Tsygan formality theorem. }

In this chapter, we will construct a formal HSGRA in five dimensions, which has been an open problem since the late 1990s. The relevant {\bf hs} had been known \cite{Fradkin:1989yd}. Free fields that comprise the spectrum of HSGRA's in $\ads_5$, including the mixed-symmetry ones, were studied in \cite{Metsaev:2002vr,Metsaev:2004ee,Metsaev:2014sfa}. Certain cubic vertices for the $\mathcal{N}=0,1,2$ cases were constructed in \cite{Vasiliev:2001wa,Alkalaev:2002rq,Alkalaev:2010af}. The free equations of type \eqref{eq:SFTformal} were analyzed in \cite{Sezgin:2001zs}. However, when it comes to classical equations the previously known methods do not work. Our solution heavily relies on the work \cite{Sharapov:2019vyd}.

In a few words, \cite{Sharapov:2019vyd} allows to construct all the vertices once we are able to deform an extension of a higher-spin algebra {\bf hs}. There are two different ways to deform it. The first one is to deform relations coming from the Joseph Ideal together with the commutator of the translation generators. This leads to an interesting way to deform (quotients
of) universal enveloping algebras. The second one is to utilize the quasi-conformal
realizations \cite{Gunaydin:2000xr,Fernando:2009fq,Govil:2013uta} that were previously underrated in the higher spin context. The
main feature is that they resolve all of the Joseph relations and give the minimal oscillator
realization of the free field and of the corresponding higher spin algebra. We found a way to deform the quasi-conformal realization so that the deformed Joseph’s relations are satisfied.

The study of five-dimensional HSGRA is also well motivated by the relation to string theory. In particular, $\mathcal{N}=8$ HSGRA is believed to describe a massless subsector of tensionless type-IIB superstring theory on $AdS_5\times S^5$, see e.g. \cite{Sezgin:2002rt,Sezgin:2001yf} for the important development towards this theory. One can start with a purely bosonic model in $\ads_5$ and, then, try to construct its supersymmetric extension. 

If we take AdS/CFT as the \textit{guiding compass} then the free limit of $\mathcal{N}=4$ SYM is anticipated to be dual to the tensionless limit of the Type-IIB
string theory on $AdS_5\times S^5$ \cite{Sundborg:2000wp,HaggiMani:2000ru}. The tensionless limit corresponds to very long strings $l_s\gg R$ ($R$ is AdS radius), see e.g. \cite{Bonelli:2003zu,Tseytlin:2002ny} for discussion.\footnote{There is a worldsheet description of the tensionless strings on $\ads_3$ \cite{Eberhardt:2018ouy}, which comes as a surprise since the limit is somewhat singular and not well-understood in higher dimensions, e.g. for $\ads_5$ that we are discussing.}

The outline of chapter 5 is as follows. In Section 2, we review the algebraic construction of HSGRA via a deformation of the extended higher spin algebra. In particular, we show how to construct $\Vcal_n$. In section 3, we discuss the input that is needed in $\ads_5$ case paying attention to the quasi-conformal realization. In section 4, we deform the algebra by either deforming the extension of {\bf hs} in terms of Joseph relations or the quasi-conformal realization, which eventually leads to the equations of motion. In section 5, we briefly review the non-locality problem in HSGRA and discuss how this construction may bypass it. We summarize the results in section 6 and discuss possible future developments.

\section{Algebraic Construction of HSGRA via Higher Spin Algebra}
In chapter 4, we have completely determined the interaction vertices for chiral HSGRA in four dimensional Minkowski and AdS spacetime. In this section, we would like to return to the question posed in chapter 2 on how to determine vertices for holographic HSGRAs in $\ads_{d+1}$, i.e. for those theories that have a free CFT dual. To begin with, let us recall that the equation that describes higher-spin background in $\ads$ is $d\omega=\omega\star\omega$, where $\star$ is the product in the higher spin algebra {\bf hs}. The appearance of the 0-form $\mathcal{W}^{a(s),b(s)}$ --- the generalized Weyl tensors which are built out of $s$-derivatives of the Fronsdal fields, suggests that we can introduce a \textit{master} 0-form field, call it $C$, to capture every $\mathcal{W}^{a(s),b(s)}$.\footnote{It is easy to notice that for $s=2$, we have a 0-form field $\mathcal{W}^{ab,cd}$, which have the same properties of the usual Weyl-tensor in GR.} Then, the free equations of motion for higher-spin fields are\footnote{See \cite{Vasiliev:1986td} for the very first equations of this  form in the context of the $4d$ HSGRA.}
\begin{subequations}\label{eq:freeHSeqs}
\begin{align}
    d\omega&=\omega \star \omega\,,\\
    dC&=\omega\star C-C\star \pi(\omega)\,.
\end{align}
\end{subequations}
As noted in Chapter 2, all elements of {\bf hs} can be written in terms of $P^a,L^{ab}$, the generators of $so(d,1)\subset so(d,2)$. Moreover, $\pi$ is an automorphism that flips the sign of translation generator $P^a$ and preserves the sign of the Lorentz generator $L^{ab}$, i.e. $\pi f(P^a,L^{ab})=f(-P^a,L^{ab})\,$. In components,
\begin{align}
    \omega(P_a,L_{ab})&=A\,1+e^aP_a+\varpi^{ab}L_{ab}+\cdots\,,\\
    C(P_a,L_{ab})&=\Phi\,1+F^{ab}L_{ab}+\mathcal{W}^{ab,cd}L_{ab}L_{cd}+\cdots\,.
\end{align}
It is easy to see that $\omega$ carries gauge degree of freedom in terms of the spin-1 gauge potential $A$, the vielbein $e^a$, spin connection $\varpi^{ab}$ and their higher spin generalizations. On the other hand, the 0-form $C$ describes physical d.o.f. in terms of the scalar field $\Phi$, the Maxwell field-strength $F^{ab}$, the Weyl tensor $\mathcal{W}^{ab,cd}$ and other higher spin generalizations thereof. 

The system \eqref{eq:freeHSeqs} describes free fields, and this is the starting point for deformation. A useful observation is that the automorphism $\pi$ can be completely absorbed if we consider an \textit{extended} associative algebra $\mathbf{A}^{\Gamma}=\mathbf{hs}\rtimes \Gamma$, that is a smash product algebra of {\bf hs} and its finite group of automorphism $\Gamma$  that contains $\pi$ \cite{Sharapov:2018kjz}. For the bosonic HSGRA, e.g. type-A, $\Gamma=\mathbb{Z}_2=\{1,\kappa\}$ with $\kappa^2=1$. The generator $\kappa\in \mathbb{Z}_2$ acts on $P_a$ and $L_{ab}$ as
\begin{align}
    \kappa P_a\kappa=-P_a\,,\qquad\qquad \kappa L_{ab}\kappa = L_{ab}\,.
\end{align}
Then, any elements of $\mathbf{A}^{\Gamma}$ can be written as $a=a_1\cdot 1+a_2\cdot \kappa$ for $a_1,a_2\in {\bf hs}$ and the $\star$-product in $\mathbf{A}^{\Gamma}$ reads
\begin{align}
    a\star b=(a_1 b_1+a_2\pi(b_2))\cdot 1 + (a_2\pi(b_1)+a_1b_2)\cdot \kappa\,,
\end{align}
where $\kappa a_i \kappa=\pi(a_i),\, a_i\in {\bf hs}$. By making the substitution $\omega=\omega\cdot 1$ and $C=C\cdot \kappa$ we can always go back to the fields taking values in the higher spin algebra. At this point it is useful to work with the extended algebra so as to eliminate $\pi$ out of the formal equations of motion for HSGRA. 

After eliminating $\pi$, it is easy to see based on form-degree counting that the most general non-linear equations read:
\begin{subequations}\label{eq:interactingAssociative}
\begin{align}
    d\omega&=\omega\star \omega+ \Vcal_3(\omega,\omega,C)+\Vcal_4(\omega,\omega,C,C)+\mathcal{O}(C^3)\,,\\
    dC&=\omega\star C-C\star \omega+\Vcal_3(\omega,C,C)+\mathcal{O}(C^3)\,.
\end{align}
\end{subequations}
Here, $C$ is an \textit{expansion parameter}  and the interaction vertices $\mathcal{V}_n$ should satisfy the Frobenius integrability condition, i.e. they should be compatible with $d^2=0$. The relevant framework for us to construct $\Vcal_n$ turns out to be strong homotopy algebras which are also known as $A_{\infty}/L_{\infty}$-algebras \cite{Kajiura:2003ax,Kajiura:2004xu}. 
\subsection{Vertices from \texorpdfstring{$A_{\infty}$}{ainfinity}-algebra}
The derivation of $\mathcal{V}_n$ is based on the assumption that $\mathbf{A}^{\Gamma}$ can be deformed into a one-parameter family of associative algebras $\mathbf{A}_{\nu}$ with a formal deformation parameter $\nu$. We assume the product in $\mathbf{A}_{\nu}$ to be
\begin{align}\label{eq:deformproduct}
    a\ast b=a\star b+\sum_{n}\phi_n(a,b)\nu^{n}\equiv\mu(a,b)\,,\qquad a,b\in {\bf A}_{\nu=0}\,.
\end{align}
where $\phi_n(\bullet,\bullet)$ are some bilinear maps that satisfy consistency conditions coming from associativity, i.e. $a\ast(b\ast c)=(a\ast b)\ast c\,$. Note that the $\star$ in \eqref{eq:deformproduct} is the product in $\mathbf{A}_{\nu=0}$. At first order in $\nu$, we get
\begin{align}
    a\star \phi_1(b,c)-\phi_1(a\star b,c)+\phi_1(a,b\star c)-\phi_1(a,b)\star c=0\,.
\end{align}
This equation is a Hochschild two-cocycle that induces a deformation of $\mathbf{A}^{\Gamma}$. One can then construct $\mathcal{V}_n$ from bi-linear maps $\phi$ via $A_{\infty}$-algebras.\footnote{We dedicate Appendix \ref{app:chap5} to all the related technicalities in this chapter.} In what follows we will see how this is done. 

\paragraph{$A_{\infty}$-algebra interlude.} Consider a graded vector space $V$ and a space $X=\Hom(TV,V)$ of all multilinear maps on $V$. Here, $TV$ is the tensor algebra on $V$. Then, an $A_{\infty}$-algebra on $V$ is realized by a \textit{master} degree-one map $x\in X$ that obeys the Maurer-Cartan equation:
\begin{align}
    \llbracket x,x\rrbracket =0\,,\qquad x=x_1+x_2+...\,, \qquad x_n\in \Hom(T^nV,V)\,.
\end{align}
The above \textit{double-bracket} is the Gerstenhabar bracket defined as \footnote{We follow the convention in \cite{Kajiura:2003ax,Sharapov:2018kjz}.}
\begin{align}
    \llbracket x_m,x_n\rrbracket = x_m\circ x_n-(-1)^{|x_m||x_n|}x_n\circ x_m\,,
\end{align}
which is graded skew-symmetric and obeys the graded Jacobi identity:
\begin{align}
    \llbracket x_m,x_n\rrbracket&=-(-1)^{|x_m||x_n|}\llbracket x_n,x_m\rrbracket\,,\\
    \llbracket\llbracket x_m,x_n\rrbracket,x_l\rrbracket&=\llbracket x_m,\llbracket  x_n , x_l \rrbracket\rrbracket-(-1)^{|x_m||x_n|}\llbracket x_n,\llbracket x_m,x_l\rrbracket\rrbracket\,.
\end{align}
The non-associative $\circ$ product is the Gerstenhaber product:
\begin{equation}
    \begin{split}
    (x_m\circ x_n)&(a_1,...,a_{m+n-1})\\
    &=\sum_{i=0}^{n-1}(-)^{|x_n|\sum_{j=1}^{i}|a_j|}x_m(a_1,...,a_i,x_n(a_{i+1},...,a_{i+m}),...,a_{m+n-1})\,,
\end{split}
\end{equation}
where $a_i\in V$. Here $|x_m|$ and $|a_i|$ denotes the grading of the map $x_m$ and the vector $a_i$. Pictorially, we can treat $a_i$ as leaves (they have green color in the figure below, and $x_i$ as $i$-tree without any internal branches. By \textit{grafting} $x_i$ and $x_j$ together, we are effectively making the root of $x_j$ become one of the branches of $x_i$ and the total number of leaves we have is $i+j-1$. The sum over all possible insertions of $x_n$ into $x_m$ is taken. As an example, let us consider $x_4\circ x_3$
\begin{align}
    \parbox{2.cm}{\includegraphics[scale=0.3]{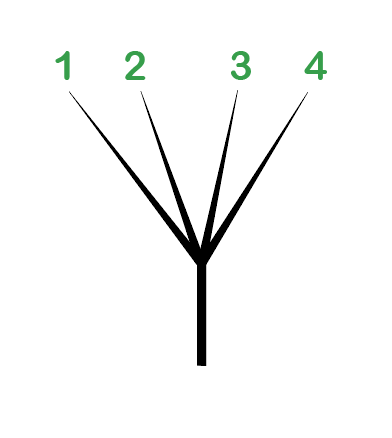}}\circ  \parbox{1.8cm}{\includegraphics[scale=0.3]{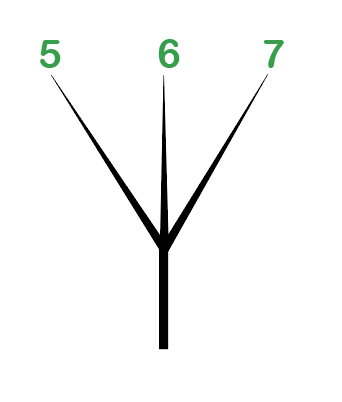}}=\sum \parbox{2cm}{\includegraphics[scale=0.35]{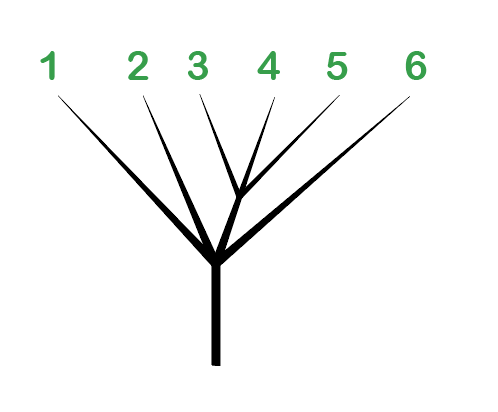}}
\end{align}

It is, then, natural to generalize the Gerstenhaber product to a \textit{braces} operation \cite{Kadeishvili:1998a,Gerstenhaber:1995a} that has the form
\begin{align}
    x_n\{y_1,...,y_k\}(a_1,...)=\sum \pm x_n(a_1,...,a_i,y_1(a_{i+1},...),...,y_i(...),...)\,,
\end{align}
where the total sign factor is a product of the sign factors for each $y_i\in \Hom(T^iV,V)$, which schematically are $|y_k|\sum_j|a_j|$ (the sum is performed for all $a_i$'s that are to the left of $y_k$). Pictorially, this brace operation is the grafting of more $i$-trees together to make a bigger tree with more leaves on top.

The first few relations coming from $A_{\infty}$-algebras are, e.g.
\begin{align}
    &x_1(x_1(a))=0\,,\\
    &x_1( x_2(a_1,a_2))+x_2(x_1(a_1),a_2)+(-1)^{|a_1|}x_2(a_1,x_1(a_2))=0\,.
\end{align}
The first equation simply tells us that $x_1$ is nilpotent, i.e. $x_1$ is a differential. The second equation implies $x_1$ satisfies graded Leibniz rule for a bi-linear product $x_2$. At the next level, we have
\small
\begin{equation}\label{x2x2}
    \begin{split}
        &x_2(x_2(a_1,a_2),a_3)+(-1)^{|a_1|}x_2(a_1,x_2(a_2,a_3))+x_1(x_3(a_1,a_2,a_3))+x_3(x_1(a_1),a_2,a_3)\\
        &+(-1)^{|a_1|}x_3(a_1,x_1(a_2),a_3)+(-1)^{|a_1|+|a_2|}x_3(a_1,a_2,x_1(a_3))=0\,,
    \end{split}
\end{equation}
meaning $x_2$ is associative up to a coboundary that includes $x_3$. In what follows, we will consider \textit{minimal} $A_{\infty}$-algebras, i.e. the $A_{\infty}$-algebras without $x_1$. The reason is that we will match $x_n$ with $\Vcal_n$ --- the vertices of the HSGRA, and the lowest order vertex of the HSGRA is of second order, e.g. $\mathcal{V}_2(\omega,\omega)=\omega\star\omega$. Then, the first non-trivial equation is just \eqref{x2x2} without $x_1$
\begin{align}\label{eq:Stasheff}
    \llbracket x_2,x_2\rrbracket=0\qquad\Leftrightarrow \qquad x_2(x_2(a,b),c)+(-)^{|a|}x_2(a,x_2(b,c))=0\,.
\end{align}
From here, we can perturbatively construct higher order maps $x_n$ via the following system
\begin{align}\label{recursiveAinfity}
    \llbracket x,x\rrbracket=0 \qquad \Leftrightarrow \qquad \delta x_n + \sum_{i+j=n+2}x_i\circ x_j=0\,,\ \ i,j\geq 3\,,
\end{align}
where $\delta=\llbracket x_2,-\rrbracket$ is a differential of degree one in $X$ that is nilpotent. Indeed,
\begin{align}
    \delta^2 f=\llbracket x,\llbracket x,f\rrbracket\rrbracket=\frac{1}{2}\llbracket\llbracket x,x\rrbracket,f\rrbracket=0,\qquad \forall f \in X\,.
\end{align}
It is then easy to construct $x_n$ order by order by solving \eqref{recursiveAinfity} recursively. 

\paragraph{Back to the construction of $\Vcal_n$.} Recall that $\mathbf{A}^{\Gamma}=\mathbf{hs}\rtimes \mathbb{Z}_2$ is an associative algebra (understood as $A_\infty$ algebra it concentrates in degree $-1$). Due to the restrictions imposed by the
grading, there cannot be any interesting $A_{\infty}$-structure on it. We can, however, deform $\mathbf{A}^{\Gamma}$ as an associative algebra. We define the $A_{\infty}$-structure perturbatively and the first step is to extend $\mathbf{A}^{\Gamma}$ by its adjoint bimodule $M$, note that $M$ has degree $0$.
Then, the $A_{\infty}$-structure contains only $x_2$ at its lowest order, where $\llbracket x_2,x_2\rrbracket=0$.\footnote{Here, $x_2$ is defined for  various pairs $\mathbf{A}_{-1}\otimes \mathbf{A}_{-1}$ (the $\star$-product),
$\mathbf{A}_{-1}\otimes\mathbf{A}_{0}$ (the left action of $\mathbf{A}^{\Gamma}$ on $M$), $\mathbf{A}_{0}\otimes \mathbf{A}_{-1}$ (the right action of $\mathbf{A}^{\Gamma}$ on $M$)} Based on \eqref{eq:interactingAssociative}, we can make the following assumption:
\begin{align}\label{eq:initialvalue}
    x_2(a,b)=a\star b,\quad x_2(a,u)=a\star u,\quad x_2(u,a)=-u\star a,\quad x_2(u,v)=0\,,
\end{align}
where $a,b\in \mathbf{A}_{-1}$ and $u,v\in \mathbf{A}_{0}$. Now we try to deform this trivial $A_{\infty}$-structure where the first-order deformations 
can be described in terms of the Hochschild cohomology of $\mathbf{A}^{\Gamma}$. From \eqref{recursiveAinfity}, the first-order deformation should be $x_3(\bullet,\bullet,\bullet)$ with arguments from
$\mathbf{A}_{-1}$ and $\mathbf{A}_{0}$. We have
\begin{align}
    \delta x_3=0\qquad \Longleftrightarrow \qquad \llbracket x_2,x_3\rrbracket=0\,.
\end{align}
If $\mathbf{A}^{\Gamma}$ admits
a deformation to one-parameter family $\mathbf{A}_{\nu}$, then the second Hochschild cohomology group is
nonzero, i.e. $\mathcal{H}^2
(\mathbf{A}^{\Gamma}, \mathbf{A}^{\Gamma}) \neq 0$. Any element $\phi \in  \mathcal{H}^2
(\mathbf{A}^{\Gamma}, \mathbf{A}^{\Gamma})$ can be represented by a cocycle $\phi$. Given these initial data, $\delta x_3=0$ yields \cite{Sharapov:2018kjz}:\footnote{We place the expansion parameter $C$ on the right as a convention.}
\begin{align}
    x_3(a,b,u)=f_3(a,b)\star u, \quad x_3(a,u,v)=f_3(a,u)\star v,\quad x_3(u,a,v)=-f_3(u,a)\star v\,. 
\end{align}
Then, as a consequence of the associativity of the $\ast$-product in $\mathbf{A}_{\nu}$, see \eqref{eq:deformproduct}, we can identify $f_3(a,b)=\phi_1(a,b)$ where $a,b\in \mathbf{A}^{\Gamma}$. We obtain, for instance
\begin{align}
    x_3(a,b,u)=\phi_1(a,b)\star u\,.
\end{align}
The associativity of the $\ast$-product also give us relations between $\phi_n$. The next order, which is $\delta x_4+x_3\circ x_3=0$, is solved by
\begin{align}
    x_4(a,b,u,v)=\phi_2(a,b)\star u\star v+\phi_1(\phi_1(a,b),u)\star v\,.
\end{align}
The last step of the construction is to replace
\begin{align}\label{eq:replacementrule}
    x_n\rightarrow \Vcal_n,\qquad a,b\rightarrow \omega,\qquad u,v\rightarrow C\,.
\end{align}
The above approach, however, possesses difficulty when we try to get higher-order interaction vertices. There is another way to get all $\Vcal_n$ at once.

\paragraph{Generating function of $\Vcal_n$} So far, we only use $\nu$ to obtain relation between $\phi_n$, the cocycle in $ \mathcal{H}^2
(\mathbf{A}^{\Gamma}, \mathbf{A}^{\Gamma})$, to determine $\Vcal_n$ order by order. Let us now consider an auxiliary family $\mathbf{A}_{\nu}(t)$ of $A_\infty$ algebras \cite{Sharapov:2018kjz,Sharapov:2019vyd} where the master degree-one map becomes 
\begin{align}
    \textbf{x}=\textbf{x}_2+t\textbf{x}_3+t^2\textbf{x}_4+...\,,\quad \textbf{x}=\textbf{x}(t,\nu),
\end{align}
and
\begin{align}
    \textbf{x}(0,\nu)(a,b)=\textbf{x}_2\equiv a\ast b=a\star b+\sum_n\phi_n(a,b)\nu^n\,,\qquad a,b\in \mathbf{A}^{\Gamma}\,.
\end{align}
These data can be used to solve the evolution
\begin{align}
    \pl_t \textbf{x}_n=\textbf{x}_i\{\pl_{\nu}\textbf{x}_j,\pl\}\,\ (\text{where}\ \ i+j=n+1)\,, \qquad \llbracket \textbf{x},\pl \rrbracket=0\,,\qquad \llbracket \textbf{x},\textbf{x}\rrbracket=0\,.
\end{align}
Here, $\pl$ is a degree minus one map that maps $M=\mathbf{A}^{\Gamma}$ (that has degree 0) to $\mathbf{A}^{\Gamma}$ and annihilates $\mathbf{A}^{\Gamma}$. We assume that the \textit{flow} in $t$ start from the surface $\llbracket \textbf{x},\textbf{x}\rrbracket=0$. Choosing the initial condition at $t=0$ and $\nu=0$ as \eqref{eq:initialvalue}, we can solve for examples \small
\begin{align}
    &\textbf{x}_3(\omega,\omega,C)=\textbf{x}_2\{\pl_{\nu}\textbf{x}_2,\pl\}\qquad \qquad \Rightarrow\qquad \qquad \quad  x_3=\phi_1(a,b)\star u\,,\\
    &2\textbf{x}_4(a,b,u,v)=\textbf{x}_3\{\pl_{\nu}\textbf{x}_2,\pl\}+\textbf{x}_2\{\pl_{\nu}\textbf{x}_3,\pl\}\ \ \Rightarrow\ \  x_4=\phi_2(a,b)\star u\star v+\phi_1(\phi_1(a,b),u)\star v\,. 
\end{align}
\normalsize
At the last step, we set $\nu=0$ and use the replacement rule \eqref{eq:replacementrule}. We get for instance
\begin{align}
    &\Vcal_3(\omega,\omega,C)=\phi_1(\omega,\omega)\star C\,,\\
    &\Vcal_4(\omega,\omega,C,C)=\phi_2(\omega,\omega)\star C\star C+\phi_1(\phi_1(\omega,\omega),C)\star C\,,\\
    &\Vcal_n(\omega,\omega,C,...,C)=x_n(\omega,\omega,C,...,C)\Big|_{\nu=0}\,.
\end{align}
We note that the vertices $\Vcal_n$ of HSGRA cannot be removed by field redefinitions. Thus, we see that the vertices are completely determined by the associative $\ast$-product \eqref{eq:deformproduct}. For more details, we refer the interested readers to \cite{Sharapov:2018kjz,Sharapov:2019vyd}.
\subsection{Formal Equations of Motion for HSGRA}
The system that describes HSGRAs in $\ads_{d+1}$ is almost identical with \eqref{eq:interactingAssociative} up to a twist $\pi$:
\begin{subequations}\label{eq:interactingHSGRA}
\begin{align}
    d\omega&=\omega\star \omega+ \Vcal_3(\omega,\omega,C)+\Vcal_4(\omega,\omega,C,C)+\mathcal{O}(C^3)\,,\\
    dC&=\omega\star C-C\star\pi( \omega)+\Vcal_3(\omega,C,C)+\mathcal{O}(C^3)\,,
\end{align}
\end{subequations}
where for example
\begin{align}
    \Vcal_3(\omega,\omega,C)=\phi_1(\omega,\omega)\star\pi(C)\,.
\end{align}
Here, $\phi_1$ takes values in the twisted adjoint representation and is a nontrivial solution of 
\begin{align}\label{twisthoch}
    a\star \phi_1(b,c)-\phi_1(a\star b,c)+\phi_1(a,b\star c)-\phi_1(a,b)\star \pi(c)=0\,.
\end{align}
If we remove $\pi$ from the above equation, we get the usual Hochschild two-cocycle that induces a deformation of the associative structure. With the twist $\pi$ in \eqref{twisthoch}, $\phi$ is not a deformation of {\bf hs}. In fact, higher spin algebras are usually rigid and do not have deformations. Note, however, that $\phi$ induces a deformation of the extended algebra $\mathbf{A}^{\Gamma}$ since the twisted representation is a part of it by construction. Therefore, while {\bf hs} is \textit{rigid} and can \textit{not} be deformed, the extended algebra $\mathbf{A}^{\Gamma}$ is \textit{soft} and can be deformed. Since the physical zero-form $C$ takes values in the representation twisted by $\pi$, it is not surprising that the deformation that leads to interaction vertices $\Vcal_n$ has something to do with deforming $\mathbf{A}^{\Gamma}$ along $\pi$. The problem now reduces to the problem of deforming $\mathbf{A}^{\Gamma}$ to all orders so as to find $\Vcal_n$, which we already know how to do.  
\subsection{Consistency Criteria and Physical Implications}
Let us recall some of the information for the undeformed higher spin algebra. The ${\bf hs}$ contains all the higher-spin generators $T^{A(s-1),B(s-1)}$ described by rectangular, two-row, Young diagrams:
\begin{align}
   \textbf{hs}=\bullet\oplus \bigoplus_{s\geq 1}\parbox{70pt}{\bep(\Length{7},20)\put(0,0){\RectT{7}{1}{\TextTop{$s$}}}%
\put(0,10){\RectT{7}{1}{\TextTop{$s$}}}\eep}\,.
\end{align}
This algebra is obtained as a quotient of the universal enveloping algebra $U(so(d,2))$ by the two sided Joseph ideal $\mathcal{I}$ that is generated by
\begin{align}
    \big[C_2-\lambda\big]\oplus \parbox{20pt}{\YoungB}\oplus\parbox{10pt}{\YoungAAAA}\equiv\mathcal{J}\oplus \mathcal{J}^{AB}\oplus \mathcal{J}^{ABCD}\,.
\end{align}
More explicitly,
\begin{align}
    \mathcal{J}^{ABCD}&=T^{[AB}T^{CD]}\,,\\
    \mathcal{J}^{AB}&=T^{A}_{\ C}T^{BC}+T^B_{\ C}T^{AC}-(d-2)\eta^{AB}\,, \\
    \mathcal{J}&=-\frac{1}{2}T_{AB}T^{AB}+\frac{1}{4}(d^2-4)\,.
\end{align}
We can write the above relations in terms of $P_a$ and $L_{ab}$ by identifying $P_a=T^{\blacklozenge}_a$ and $L_{ab}=T_{ab}$, we simply get:\footnote{We set the cosmological constant to 1.  The $so(d,2)$ index $A=\{a,\blacklozenge\}$ with $\blacklozenge$ being the extra direction of $so(d,2)$-vector as compared to the vector of the Lorentz algebra $so(d,1)$.}
\begin{align}
    \mathcal{J}^{abcd}&=L^{[ab}L^{cd]}, \qquad \mathcal{J}^{abc\blacklozenge}=\{L^{[ab},P^{c]}\}\\
    \mathcal{J}^{ab}&=\{L^a_{\ c},L^{bc}\}-\{P^a,P^b\} - (d-2)\eta^{ab}.\\
    \mathcal{J}^{a\blacklozenge}&=\{L^a_{\ c},P^c\}, \qquad \mathcal{J}^{\blacklozenge\blacklozenge}=2P_aP^a+(d-2)\\
    \mathcal{J}&=-\frac{1}{2}L_{ab}L^{ab}+P_aP^a+\frac{d^2-4}{4}
\end{align}
The commutation relations of $so(d,2)$ in terms of $P_a$ and $L_{ab}$ are:
\begin{align}
    [P_a,P_b]=L_{ab},\quad [L_{ab},L_{cd}]=L_{ad}\eta_{bc}+...\,,\quad [L_{ab},P_c]=P_a\eta_{bc}-P_b\eta_{ac}\,.
\end{align}
The remaining task is to show that we can deform the infinite-dimensional $\mathbf{A}^{\Gamma}={\bf hs}\rtimes \Gamma$ consistently. This might be a very complicated problem because there are infinitely many structure constants that will receive correction. However, as the matter of fact, we can deform very few relations by hand and all of the structure constants can be derived easily.

For Type-A HSGRA in generic $\ads_{d+1}$, the simplest relation that needs to be deformed is 
\begin{align}\label{eq:PPtoEinstein}
    [P_a,P_b]=(1+\nu \kappa)L_{ab}\,.
\end{align}
Note that we \textit{want} to keep $[L,L]$ and $[L,P]$ intact to preserve local Lorentz algebra and its action on tensorial objects. $[P,P]$-bracket plays the role of the seed that drives the whole deformation making $\mathbf{A}^{\Gamma}\rightarrow \mathbf{A}_{\nu}$. To see the physical implication of this deformation, let us for the moment turn off all the HS fields (but we will not truncate them) and look at the gravitational sector of the cubic vertex where $C=...+\mathcal{W}^{ab,cd}L_{ab}L_{cd}+...$, then
\begin{align}
    V_3(\omega,\omega,C)=e_c\wedge e_d\,\phi_1(P^c,P^d)\star C\sim e_c\wedge e_d \mathcal{W}^{ab,cd}L_{ab}+...\,.
\end{align}
Here, because $e\wedge e$ is anti-symmetric the appropriate substitution to $\phi_1(P,P)$ is nothing but the first order deformation of the $[P,P]$-bracket. The appearance of the $\mathcal{W}^{ab,cd}$ allows us to get the correct Einstein equations in the frame-like formalism. Indeed, the coefficient of $L^{ab}$ from $d\omega=\Vcal_2+\Vcal_3+...\,$ should lead to 
\begin{align}
    d\varpi^{ab}-\varpi^{a}_{\ c}\wedge \varpi^{cb}-\Lambda e^a\wedge e^bL_{ab}=e_c\wedge e_d \mathcal{W}^{ab,cd}\,,
\end{align}
which is achieved iff the $[P,P]$-bracket reads
\begin{align}
    [P_a,P_b]=(1+\nu \kappa)L_{ab}\qquad \qquad \Leftrightarrow \qquad \qquad \text{Einstein equations}\,.
\end{align}
Therefore, the physical interpretation of the deformed $[P,P]$-bracket is that it leads to the Einstein equations. Together with other higher spin fields, we have obtained consistency bosonic HSGRAs in generic $\ads_{d+1}$ via the above construction. 

\section{Initial Data for \texorpdfstring{$\ads_5$}{ads5space}}
Above we have reviewed how to construct type-A HSGRA starting from free scalar CFT. This construction indeed can work with any free CFT and therefore the only input we need to construct HSGRAs is {\bf hs}. For the case of type-A HSGRA in $\ads_5$, the higher-spin algebra comes from the universal enveloping algebra of $su(2,2)$, whose generators $T_A^{\ B},\, A,B=1,...,4$ obey
\begin{align}\label{eq:su(2,2)algebra}
    [T_A^{\ B},T_C^{\ D}]=\delta_A^{\ D}T_C^{\ B}-\delta_C^{\ B}T_A^{\ D}\,.
\end{align}
Here, the indices $A,B,...$ are the indices of the (anti)-fundamental representation of $su(2,2)$. The ideal is generated by the quadratic relations
\begin{subequations}\label{eq:su(2,2)ideal}
\begin{align}
    C_2&=T_A^{\ B}T_B^{\ A}=-3\,,\\
    T_A^{\ C}T_C^{\ B}&=-2T_A^{\ B}+\frac{1}{4}C_2\delta_A^{\ B}\,,\\
    \{T_{[A}^{\ [B},T_{C]}^{\ D]}\}&=\delta_A^{\ B}\delta_C^{\ D}-\delta_C^{\ B}\delta_A^{\ D}\,.
\end{align}
\end{subequations}
Elements of the higher spin algebra {\bf hs} are polynomials $f(T)$ in $T_A^{\ B}$ modulo the Joseph relations. Therefore, following the definition in subsection \ref{HS/Vec-HSalgebra}, it is easy to see that
\begin{align}\label{eq:elementofsu(2,2)}
    f=\sum_kf^{A_1...A_k}_{B_1...B_k}T_{A_1}^{\ B_1}...\,T_{A_k}^{\ B_k}\,,
\end{align}
where the coefficients are traceless and symmetric in upper and lower indices, i.e., define an irreducible representation of weight $(k,0,k)$\,. 

Although one can take relations \eqref{eq:su(2,2)algebra} and \eqref{eq:su(2,2)ideal} as an initial definition of {\bf hs}. In practice, it is sometimes convenient to resolve (some of) the Joseph relations by passing to an appropriate realization. We will introduce two quartets of oscillator variables $a^A$ and $b^B$
in the fundamental
and anti-fundamental representations of $su(2,2)$ (they generate the Weyl algebra $A_4$):
\begin{align}
    [a^A,a^B]=0\,, \qquad [b_A,b_B]=0\,,\qquad [a_A,b^B]=\delta_A^{\ B}\,.
\end{align}
Then the $su(2,2)$ generators are given by
\begin{align}
    T_A^{\ B}=\frac{1}{2}\{a_A,b^{B}\}-\frac{1}{4}\delta_A^{\ B}N\,,
\end{align}
where the $u(1)$ generator $N=\frac{1}{2}\{a_C,b^C\}$ commutes with $T_A^{\ B}$. We can then define {\bf hs} as a subquotient of the oscillator algebra:
\begin{align}
    f\in {\bf hs}\qquad \Longleftrightarrow \qquad [f,N]_{\star}=0\,,\qquad f\sim f+g\star N\,,
\end{align}
The first relation demands $f(a,b)$ to have an equal number of $a$ and $b$ oscillators. The quotient with respect to $N$ makes the Taylor coefficients effectively traceless, as in \eqref{eq:elementofsu(2,2)}. It is this realization that was used in [62] to study the spectrum and free higher spin equations.\footnote{Note that we need to gauge $u(1)$ generator $N$ in order to have the right algebra. This is because the first two relations of \eqref{eq:su(2,2)ideal} do not satisfy the trace conditions.} 

Another way to resolve all Joseph's relations is to utilize quasi-conformal realization (QCR) [54,55,63]. The idea is to represent {\bf hs} by a minimal possible number of oscillators. We have the following of canonical pairs of oscillators
\begin{align}
    [z,p_z]=[y,p_y]=[x,p_x]=i\,.
\end{align}
In QCR approach, it can be shown that the following two composite operators 
\begin{align}
    Y_L^A=\left\{z,p_z,0,\frac{1}{x}(zp_y-p_zy-\frac{1}{2})\right\},\qquad Y_R^A=\left\{y,p_y,x,p_x\right\}
\end{align}
can be used to define the generators of {\bf hs}. Indeed,
\begin{align}\label{eq:su(2,2)T}
    T_A^{\ B}=-\frac{i}{2}\Big(Y_A^+Y_-^B-\frac{1}{4}\delta_A^{\ B}Y_C^+Y_-^C\Big)
\end{align}
obey the commutation relations \eqref{eq:su(2,2)algebra} as well as \eqref{eq:su(2,2)ideal}.

Recall that the Lorentz subalgebra $so(4,1)\sim sp(4)$ --- the maximally symmetric subalgebra of $so(4,2)$ that remains undeformed. It allows one to split the $su(2,2)$ generators into the Lorentz generators $L_{AB}$ and translations $P_{AB}$:
\begin{align}
    L_{AB}=T_{AB}+T_{BA}\,, \qquad \qquad P_{AB}=T_{AB}-T_{BA}\,.
\end{align}
Here and after, we will raise and lower $sp(4)$-indices with the help of $sp(4)$-invariant tensor  $C_{AB}=-C_{BA}$. Then, the $su(2,2)$ commutation relations \eqref{eq:su(2,2)algebra} read
\begin{align}\label{eq:newsu(2,2)}
    [L_{AB},L_{CD}]&=L_{AD}C_{BC}+L_{BD}C_{AC}+L_{AC}C_{BD}+L_{BC}C_{AD}\,,\\
    [L_{AB},P_{CD}]&=P_{AD}C_{BC}+P_{BD}C_{AC}-P_{AC}C_{BD}-P_{BC}C_{AD}\,,\\
    [P_{AB},P_{CD}]&=L_{AD}C_{BC}-L_{BD}C_{AC}-L_{AC}C_{BD}+L_{BC}C_{AD}\,.
\end{align}
In order to write the equations of motion, we need the automorphism $\pi$ as explained above. Here, $\pi$ acts on $L$ and $P$ generators as $\pi \,f(T)=f(L,-P)$. The spin-two subsector of the master one-form $\omega$ reads
\begin{align}
    \varpi_{\boldsymbol{0}} = \frac{1}{2}\bar{e}^{AB}P_{AB}+\frac{1}{2}\bar{\varpi}^{AB}L_{AB}\,.
\end{align}
Here, $\bar{e}^{AB}=-\bar{e}^{BA}$ is the background f\"unfbein and $\bar{\varpi}^{AB}=\bar{\varpi}^{BA}$ is the background spin-connection. It is worth stressing that $A,B$ are $sp(4)$-indices. 
\section{Deformation of Higher Spin Algebra}
As explained in section 2, the problem of constructing the interaction vertices $\Vcal_n$ is equivalent to finding the deformation of $\mathbf{A}^{\Gamma}={\bf hs}\rtimes \Gamma$. In what follows, we will show how to deform the algebra to obtain bosonic HSGRA in $\ads_5$. 

The usual oscillator realization of \eqref{eq:su(2,2)T} does not allow us to deform the algebra since the $\pi$-map gives us the smash-product algebra $A_4\rtimes \mathbb{Z}_2$. This algebra does not admit any non-trivial deformations \cite{Sharapov:2017lxr,Sharapov:2018a} since its second Hochschild cohomology group vanishes. Therefore, we should deform the {\bf hs} through $U(su(2,2))$ or QCR \cite{Gunaydin:2000xr,Fernando:2009fq,Govil:2013uta}.
\subsection{Deformation through The Universal Enveloping Algebra}
We learnt in section 2 that it is useful to introduce the $\kappa$ operator\footnote{This operator is also known as the Klein operator, see e.g. \cite{Vasiliev:1999ba,Didenko:2014dwa}.} to expand the algebra, and to absorb $\pi$. The universal enveloping algebra is now expanded to $U(su(2,2))\rtimes \mathbb{Z}_2$. The set of Joseph relations \eqref{eq:su(2,2)ideal} split into the triple of finite-dimensional irreducible modules of $su(2,2)$: the first module corresponds to the Casimir operator and is a trivial one; the second module is the 15-dimensional adjoint representation $(1,0,1)$, and the last module is the 20-dimensional representation of $(0,2,0)$. Since the modules are irreducible, we can take the following components as the lowest weight vector
\begin{align}\label{eq:seedofideal}
    I=\frac{1}{2}P_{AB}P^{AB}-m^2\,,\qquad I_{AB}\equiv \{L_{AM},P_{B}^{\ M}\}+\{L_{BM},P_A^{\ M}\}\,
\end{align}
and commute these with $P_{AB}$ to generate other relations. The consistency of the ideal fixes $m^2=-2$ (in which we set the cosmological constant to one). Then, following the discussion in section 2, the only commutation relation that we need to deform from $su(2,2)$ algebra is
\begin{align}
    [P_{AB},P_{CD}]=(1+\nu \kappa)(L_{AD}C_{BC}-L_{BD}C_{AC}-L_{AC}C_{BD}+L_{BC}C_{AD})\,.
\end{align}
This commutation relation acts as a \textit{seed} that drives the whole deformation of the $\mathbf{A}^{\Gamma}$. We also know that it will lead to Einstein's equations. Starting from \eqref{eq:seedofideal}, we act on them with $[P_{AB},\bullet]$ to generate the other components of the deformed Joseph's ideal. We obtain
\begin{equation}\label{eq:deformedJoseph}
\begin{split}
    (0,0,0)&: \quad  2C_2=\frac{1}{2}P_{AB}P^{AB}-\frac{1}{2}L_{AB}L^{AB}=-\frac{1}{2}(6+\nu\kappa)(2+\nu\kappa)\,,\\
    (0,2,0)&: \quad  \begin{cases}
    \frac{1}{2}P_{AB}P^{AB}-m^2=0\,,\\
    \{L_{AM},P_{B}^{\ M}\}-\{L_{BM},P_A^{\ M}\}-2\nu\kappa P_{AB}=0\,,\\
    \{L_{[A}^{\ [B},L_{C]}^{\ D]}\}+\{P_{[A}^{\ [B},P_{C]}^{\ D]}\}+\\
    \qquad +2\nu\kappa(2+\nu\kappa)C_{AC}C^{BD}-(2+\nu\kappa)^2(\delta_A^{\ B}\delta_C^{\ D}-\delta_C^{\ B}\delta_{A}^{\ D})=0\,,
    \end{cases}\\
    (1,0,1)&: \quad \begin{cases}
    \{L_{AM},P_B^{\ M}\}+\{L_{BM},P_A^{\ M}\}=0\,,\\
    \{L_{AM},L_B^{\ M}\}+\frac{1}{2}C_{AB}(2+\nu\kappa)(-4+\nu \kappa)=0\,.
    \end{cases}
    \end{split}
\end{equation}
For the consistency of the deformed Joseph's ideal we find $m^2=-(2+\nu\kappa)(1+\nu\kappa)$. It is easy to see that by setting $\nu=0$, we return to the original Joseph's relations. The above relations \eqref{eq:deformedJoseph} together with $[L,L]$, $[L,P]$ and the deformed $[P,P]$-brackets determine the deformation of $\mathbf{A}^{\Gamma}$. The deformation is smooth in the sense that we can construct the product of $(f\ast g)(L,P,K)$ from any $f,g\in \mathbf{A}^{\Gamma}$ and decompose it into irreducible Lorentz tensors. Therefore, the deformation is well-defined which eventually leads us to the equations of motion \eqref{eq:interactingHSGRA} of HSGRA. \\
${}$\\
To read off the spectrum of the algebra from \eqref{eq:deformedJoseph}, we first notice that there are no singlets except for the unit element itself because $P^2,L^2$ are $\kappa$-dependent numbers. It easy to see that all single contractions, namely $L_{AM}P_B^{\ M}$, $L_{AM}L_B^{\ M}$, and $P_{AM}P_B^{\ M}$, can be transformed into $L_{AB},\ P_{AB}$ and $C_{AB}$. Moreover, relations with un-contracted indices implies that it is equivalent to get $(0,2)$ of $sp(4)$ from either $L_{AB}L_{CD}$ or $P_{AB}P_{CD}$ projections. Therefore, the spectrum of the algebra consists of $sp(4)$-tensors of weight $(2t,m)$
\begin{align}
   \parbox{90pt}{\bep(\Length{8},20)\put(0,0){\RectT{4}{1}{\TextTop{$m$}}}%
\put(0,10){\RectT{8}{1}{\TextTop{$m+2t$}}}\eep}, \qquad m,t=0,1,2,...\,,
\end{align}
which can be thought of as coefficients of the appropriately symmetrized monomials $L^tP^m$\,.
\subsection{Deformation through QCR}
In \cite{Sharapov:2019pdu}, we show that QCR also admit a deformation and it is a minimal one. The automorphism $\pi$ acts as
\begin{align}
    \pi\,f(z,p_z,\bullet)=f(-z,-p_z,\bullet)\,,
\end{align}
meaning $\pi$ flips the sign of $z,p_z$ while leaving other oscillators, denoted as $\bullet$, intact. Therefore
\begin{align}
    \{z,\kappa\}=0,\qquad \{p_z,\kappa\}=0\,.
\end{align}
The desired deformation can be obtained by redefining the momentum $p_z$ as
\begin{align}
    p_z\quad \longrightarrow \quad  \tilde{p}_z=p_z+\frac{i\nu}{2z}\kappa\,.
\end{align}
The deformation of QCR is realized through [75]:\footnote{This deformed oscillators have a long history and were discovered by Wigner, 70 years ago, who asked the  question  whether it is possible to modify  the canonical commutation  relations in such a way that basic commutation relations still remain valid. The answer is yes, and it is precisely the deformed $[z,\tilde{p}_z]$ that allows us to have one-parameter deformation.}
\begin{align}\label{eq:deformedoscillators}
    [z,\tilde{p}_z]=i(1+\nu\kappa)\,,\qquad \{z,\kappa\}=0\,,\qquad \{\tilde{p}_z,\kappa\}=0\,.
\end{align}
Then, the composite operators $Y^A_L$ changes accordingly as
\begin{align}
    Y_L^A=\left\{z,p_z,0,\frac{1}{x}(zp_y-p_zy-\frac{1}{2}-\frac{1}{2}\nu\kappa)\right\}\,,
\end{align}
while $Y_R^A$ stay the same. It can be shown that from the deformed QCR, we will obtain precisely the deformed $su(2,2)$ algebra and the deformed Joseph's relations. This gives an explicit QCR of $\mathbf{A}^{\Gamma}$. Thus, \eqref{eq:deformedJoseph} provide the complete solution of HSGRA in $\ads_5$. 
\subsection{Einstein's Equations}
For completeness, let us show once again that the deformed $[P,P]$-bracket does lead to Einstein's equations and therefore well-motivated. If we look at the spin-two sector from HSGRA equations of motion, we see that\footnote{Note that $C=C\kappa$ in the extended algebra.}
\begin{align*}
    \Vcal_3(\omega,\omega,C)&=\bar{e}_{MC}\wedge \bar{e}_{ND}\phi_1(P^{MC},P^{ND})\star C\kappa\sim \bar{e}_C^{\ M}\wedge \bar{e}_{MD}\,\kappa L^{CD}\mathcal{W}^{ABEF}L_{AB}L_{EF}\kappa\\
    &\sim \bar{e}_C^{\ M}\wedge \bar{e}_{MD}\delta_A^{\ C}\delta_B^{\ D}\mathcal{W}^{ABEF}L_{EF}\sim \bar{e}_C^{\ M}\wedge \bar{e}_{MD}\mathcal{W}^{ABCD}L_{AB}\,,
\end{align*}
where we used the fact that $L_{AB}L^{AB}=8+\mathcal{O}(\nu)$ and therefore $\{L_{AB},L_{CD}\}=C_{AC}C_{BD}+...\,$.The Einstein equations are realized as the coefficients in front of $P_{AB}$ and $L_{AB}$, they are
\begin{align*}
    P_{AB}&:\qquad d\bar{e}^{AB}-\bar{\varpi}^{[A}_{\ C}\wedge \bar{e}^{CB]}=0\,,\\
    L_{AB}&: \qquad d\bar{\varpi}^{AB}-\bar{\varpi}^A_{\ C}\wedge \bar{\varpi}^{CB}-\bar{e}^A_{\ C}\wedge \bar{e}^{CB}=\bar{e}_C^{\ M}\wedge \bar{e}_{MD}\mathcal{W}^{ABCD}\,.
\end{align*}
Note that the potentially dangerous $\bar{\varpi}\bar{\varpi}\mathcal{W}$ and $\bar{\varpi}\bar{e}\mathcal{W}$ terms vanish since the deformation preserves both $[L,L]$ and $[L,P]$ commutators. \\
${}$\\
We will end this section with general discussion using $so(d,2)$ language. The Einstein equations is a part\footnote{We would like to stress that there is not consistent truncation of a HSGRA that eliminates higher spin fields: graviton sources higher spin fields and higher spin fields backreact onto the graviton. } of numerous HSGRAs which is the result of formal deformation of the $[P_a,P_b]$-bracket. The deformation \eqref{eq:PPtoEinstein} is a small part of the Hochschild cocycle $\phi_1$ of {\bf hs} which eventually leads to the $A_{\infty}$-algebra \cite{Sharapov:2018kjz,Sharapov:2019vyd}. For more details, we refer the interested readers to \cite{Sharapov:2019pdu,Sharapov:2019vyd}.
\section{Non-Locality Problem in HSGRAs}
In the body of the thesis we have already mentioned that besides chiral HSGRA, three-dimensional HSGRA and conformal HSGRA, other (holographic) higher-spin theories turn out to be non-local. To understand how to make sense out of these non-localities is one of the main challenges in HSGRA. Let us recall that the formally consistent equations of motion for HSGRAs read as
\begin{subequations}\label{FDAschem}
\begin{align}
    d\omega&=\omega\star \omega+ \Vcal_3(\omega,\omega,C)+\Vcal_4(\omega,\omega,C,C)+\mathcal{O}(C^3)\,,\\
    dC&=\omega\star C-C\star\pi( \omega)+\Vcal_3(\omega,C,C)+\mathcal{O}(C^3)\,.
\end{align}
\end{subequations}
We will translate vertices to the Fronsdal (field theory) language to understand why non-locality appears. First of all, the 1-form $\omega$ contains the Fronsdal field $\Phi_{\ua(s)}$ for $s=1,2,3,...$ and other components that are derivatives of the Fronsdal field up to order-$(s-1)$. Therefore, $\omega$ contains a \textit{finite} number of derivatives of every $\Phi_{\ua(s)}$. On the other hand, the 0-form $C$ starts with the generalized Weyl tensor $W^{\ua(s),\ub(s)}$ --- that are the order-$s$ curl of the Fronsdal field, and there is an infinite tower of fields that are $k$-derivatives of $W^{\ua(s),\ub(s)}$, where $k=0,...,\infty$. The spectrum can simply be presented as the following Young diagrams
\begin{align}
     \omega&:\qquad \parbox{90pt}{\bep(\Length{6},20)\put(0,0){\RectT{3}{1}{\TextTop{$t$}}}%
\put(0,10){\RectT{5}{1}{\TextTop{$s-1$}}}\eep}\sim \nabla^t \Phi_s, \qquad t=0,1,2,...,s-1\,,\\
C&: \qquad \parbox{90pt}{\bep(\Length{8},20)\put(0,0){\RectT{6}{1}{\TextTop{$s$}}}%
\put(0,10){\RectT{8}{1}{\TextTop{$s+k$}}}\eep}\sim \nabla^{s+k} \Phi_s,\qquad k=0,...,\infty\,,
\end{align}
or pictorially as
\begin{figure}[h!]
    \centering
    \includegraphics[scale=0.36]{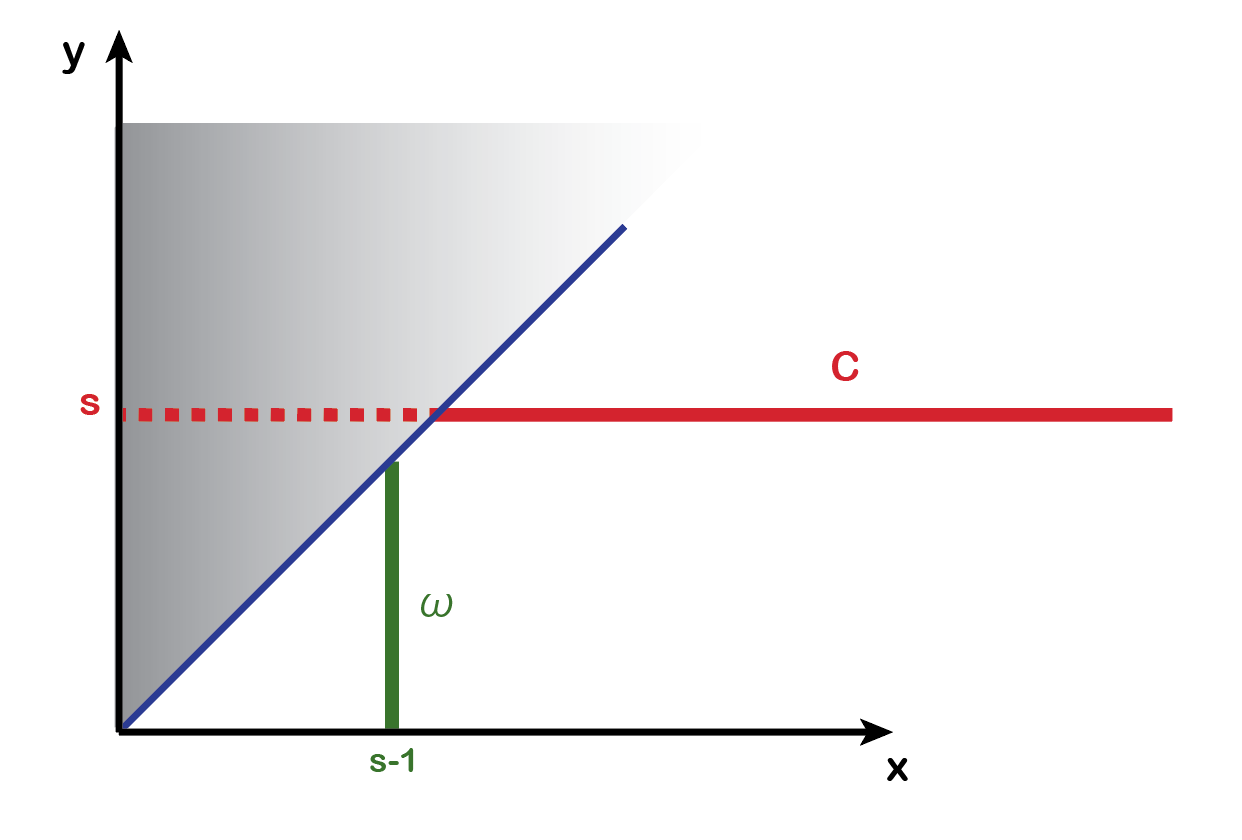}
    \captionsetup{labelformat=empty}
    \caption{The $x$-axis represents the number of boxes of the first row of Young diagrams while the $y$-axis is the number of boxes in the second row. The grey zone is the forbidden region due to Young symmetry.}
\end{figure}

Naively, the system is non-local in the field theory context since there is an unbounded number of derivatives. It is known, see e.g. \cite{Boulanger:2015ova}, that the only dynamical equations contained in \eqref{FDAschem} are the Fronsdal equations with sources that take the following schematic form:
\begin{align}\label{cubicschem}
    (\square-M^2_s)\Phi_{\ua(s)}=\sum_{k,\ell}a_{k,\ell}\nabla_{c(\ell)}\nabla_{\ua(s-k)}\Phi\nabla^{c(\ell)}\nabla_{\ua(k)}\Phi+...\,, \ \ \text{with}\ \ \ell=0,...,\infty\,.
\end{align}
The sources above correspond to $\mathcal{V}_3(\omega,\omega,C)$ and $\mathcal{V}_4(\omega,\omega,C,C)$ vertices in the frame-like formalism. For simplicity, we omit the spin labels on the right hand side (for example, we can think that only the backreaction of the scalar field is taken into account). Moreover, the sum over derivatives can in principle be infinite. This certainly can happen when at least two $C$'s are found in a vertex, e.g. it is so for $\mathcal{V}_4(\omega,\omega,C,C)$, see  \cite{Boulanger:2015ova,Skvortsov:2015lja}.\footnote{See also, \cite{Giombi:2009wh} for the early discussion about non-local behaviour of $\Vcal_3(\omega,C,C)$.} 

Now we need to distinguish between cubic and higher order terms from the action point of view (or bilinear and higher from the equations of motion vantage point). 
HSGRA's (and any other theory in $\ads$) are local at the cubic level: given any three spins there is a finite number of independent cubic vertices each of which contains a finite number of derivatives. Starting for the quartic order there are infinitely many independent quartic structures that can contribute and, more importantly, the number of derivatives is unbounded, i.e. each such quartic interaction contain a finite number of derivatives, but there exist interactions with any given number of derivatives. 

As was shown in \cite{Bekaert:2015tva} (see also \cite{Sleight:2017pcz,Ponomarev:2019ltz}) the quartic vertex in the Type-A HSGRA is non-local. Moreover, it is proportional to the contribution of exchanges to the quartic amplitude. This means that: (i) the complete quartic interaction in the Type-A HSGRA has an unbounded number of derivatives; (ii) the coefficients do not decay fast enough with the number of derivatives. Therefore, there is no difference between the contribution of the contact vertex and of the exchanges to the quartic amplitude. This invalidates the Noether procedure \cite{Sleight:2017pcz,Ponomarev:2019ltz}: one cannot construct the Type-A HSGRA by writing the most general ansatz for interactions and fixing it by the requirement of gauge invariance.  

The situation with \eqref{FDAschem} and \eqref{cubicschem} is more subtle. The vertices $\mathcal{V}$ are fixed by the formal consistency, which is equivalent to a formal gauge invariance. The formal consistency by itself does not constrain the number of derivatives. Therefore, unless one controls the number of derivatives by hand one can easily get formally consistent equations that contain too many derivatives for them to make sense as a field theory. This issue has nothing to do with HSGRA and will be faced even for low spin theories once trying to write them as \eqref{FDAschem}. This is exactly what happens with the original proposal \cite{Vasiliev:1990cm}, as was shown in \cite{Boulanger:2015ova,Skvortsov:2015lja}, based on the earlier observation \cite{Giombi:2009wh} that certain holographic correlation functions, as computed from \eqref{FDAschem}, are infinite and/or inconsistent. While the issue can clearly be resolved at the cubic level, it is an open question if the non-locality can be tamed at higher orders. 

As an example, we can consider $\Phi^3$ theory. Then, $C$ consists only of $C^{a(k)}$ that encode derivatives of $\Phi$. Let's us also go flat space to simplify the algebra
\begin{align}\label{cubicschemA}
    (\square-M^2)\Phi=\sum_{\ell}\left(\frac{2}{M^2}\right)^{\ell}a_{\ell}\pl_{c(\ell)}\Phi\pl^{c(\ell)}\Phi\,.
\end{align}
This what one generically gets unless no locality constraints are imposed on the vertices --- general infinite series corresponding to terms $C_{a(\ell)}C^{a(\ell)}$ in $\mathcal{V}(\omega,C,C)$ for $\omega$ being the background vielbein. Suppose we would like to compute the cubic amplitude $A_3$ in this theory. It is easy to do that in momentum space where $\pl$ turns into $p_{i}$, $p_i^2=-M^2$ and, hence, $p_1\cdot p_2=M^2/2$. Therefore, all derivatives disappear and the cubic amplitude will get a contribution proportional to\footnote{We should have started with an action where the derivatives are effectively symmetrized over the three fields. We will ignore this complication. }
\begin{align}
    A_3 = \sum_{\ell} a_{\ell}\,.
\end{align}
It is easy to anticipate that $A_3$ gets a  contribution from every term on the r.h.s. of \eqref{cubicschemA} since there exists only one independent cubic amplitude in the scalar theory, the one arising simply from the $\Phi^3$ coupling in the action. All the other terms, which involve higher derivatives, are not independent from this one and can be reduced to $\Phi^3$, one by one, via field redefinitions. 

Now we make a worrisome observation: (i) Eq. \eqref{cubicschemA}, and hence \eqref{FDAschem}, are certainly formally consistent for any choice of $a_{\ell}$; (ii) for most $a_{\ell}$ the cubic amplitude is infinite and does not make any sense. This gives a simple example to illustrate the fact that formal consistency does not imply actual consistency. The observation has nothing to do with HSGRA. It is just the fact that field redefinitions can generate infinitely many higher derivative \textit{avatars} of the same basic interaction and all such avatars will contribute to the physical observables. In reality one would like $a_0$ to be the actual coupling and constrain $\mathcal{V}(\omega,C,C)$ in such a way that $a_{\ell>0}=0$. Therefore, one has to keep by hand under control various terms in vertices $\mathcal{V}$ that are related via field redefinitions. This issue is present for all vertices $\mathcal{V}$ that have at least two $C$ fields.

Another way to understand the non-locality problem in HSGRA is to look at the higher spin gauge transformations: higher-spin symmetry mixes not only fields together, it intertwines also derivatives. Indeed, the gauge transformation for $\omega$ reads
\begin{align}
    \delta_{\xi} \omega=d\xi -[\omega,\xi]+\xi\frac{\pl}{\pl \omega}\Vcal(\omega,\omega,C)+...\,,
\end{align}
It is clear that with the help of $C$, which is a generating function of infinite number of derivatives, one easily change the number of derivatives in vertices. Therefore, even though higher-spin symmetry demands the present of all vertices $\Vcal_n$ for consistency, it is unclear how to give them physical interpretation at present. 

There are several observations that help to tame the non-localities and may eventually solve the problem. One can argue that if the observables, e.g. (holographical) S-matrix is well-defined, then non-locality is just an \textit{artifact} of HSGRA. One can directly focus on the constraints imposed by the higher spin symmetry on physical observables. For example, the holographic correlation functions are the simplest invariants of the higher spin algebra \cite{Colombo:2012jx,Didenko:2013bj,Didenko:2012tv,Bonezzi:2017vha}. 

If we are only interested in the solutions coming out of formal HSGRAs, we can then solve explicitly the equations of motion for HSGRA in terms of Lax pair system \cite{Sharapov:2019pdu,Sharapov:2019vyd}
\begin{subequations}\label{eq:Lax}
\begin{align}
    d\boldsymbol{\omega}=\boldsymbol{\omega}\ast \boldsymbol{\omega},\qquad d\boldsymbol{C}=\boldsymbol{\omega}\ast \boldsymbol{C}-\boldsymbol{C}\ast \boldsymbol{\omega}\,,
\end{align}
\end{subequations}
where $\boldsymbol{\omega}$ and $\boldsymbol{C}$ take values in the deform algebra $\mathbf{A}_{\nu}$. The system \eqref{eq:Lax} can be solved in a pure gauge form, namely
\begin{align}\label{eq:puregauge}
    \boldsymbol{\omega}=\boldsymbol{g}^{-1}\ast d\boldsymbol{g}, \qquad \qquad \boldsymbol{C}=\boldsymbol{g}^{-1}\ast \boldsymbol{C}_0\ast \boldsymbol{g}\,.
\end{align}
It is important to stress that even though the solutions \eqref{eq:puregauge} look like ones from a free system they are not. The reason is that the fields $(\boldsymbol{\omega},\boldsymbol{C})\equiv(\boldsymbol{\omega}(\nu,x),\boldsymbol{C}(\nu,x))$ are sources of the following system \cite{Sharapov:2019vyd}:
\begin{align}
    d\underline{\omega}&=\underline{\omega}\ast\underline{\omega}+t\pl_{\nu}\mu(\underline{\omega},\underline{\omega})+\frac{t^2}{2}\pl_{\nu}^2\mu(\underline{\omega},\underline{\omega})\ast \underline{C}\ast\underline{C}+t^2\pl_{\nu}\mu(\pl_{\nu}\mu(\underline{\omega},\underline{\omega}),\underline{C})\ast\underline{C}+...\,,\\
    d\underline{C}&=\underline{\omega}\ast \underline{C}-\underline{C}\ast \underline{\omega}+t\pl_{\nu}\mu(\underline{\omega},\underline{C})-t\pl_{\nu}\mu(\underline{C},\underline{\omega})\ast \underline{C}+...\,,
\end{align}
where $\mu(a,b)=a\star b+\sum \phi_n(a,b)\nu^n$ and
\begin{align}
    \underline{\omega}&=\boldsymbol{\omega}+t\pl_{\nu}\boldsymbol{\omega}\ast \boldsymbol{C}+\frac{t^2}{2}\pl_{\nu}^2\boldsymbol{\omega}\ast \boldsymbol{C}\ast \boldsymbol{C}+t^2\pl_{\nu}\boldsymbol{\omega}\ast \pl_{\nu}\boldsymbol{C}\ast \boldsymbol{C}+t^2\pl_{\nu}\mu(\pl_{\nu}\boldsymbol{\omega},\boldsymbol{C})\ast C+...\,,\\
    \underline{C}&=\boldsymbol{C}+t\pl_{\nu}\boldsymbol{C}\ast \boldsymbol{C}+...\,.
\end{align}
By setting $\nu=0$, we will return to the formal equations of motion for HSGRA. In other words, one can obtain well-defined solutions of the formally consistent equations that are not by themselves well-defined in being too non-local. From here, one can construct observables in terms of traces a.k.a invariants. There are scalar invariants, for example, which given by the on-shell closed 0-forms
\begin{align}
    I_n(\nu)=\Tr\big[\underbrace{\boldsymbol{C}\ast\boldsymbol{C}\ast ...\ast \boldsymbol{C}}_n\big]\,.
\end{align}
Indeed, at $\nu=0$, they reduce to correlation functions of higher spin currents in the dual free CFT \cite{Colombo:2012jx,Didenko:2013bj,Didenko:2012tv,Bonezzi:2017vha}. Therefore, we conclude that while the equations of motion exhibit problematic non-locality that needs to be understood, the solution space seems to be well-defined. We summarize the whole procedure in this chapter as follows:
\begin{figure}[h!]
    \centering
    \includegraphics[scale=0.6]{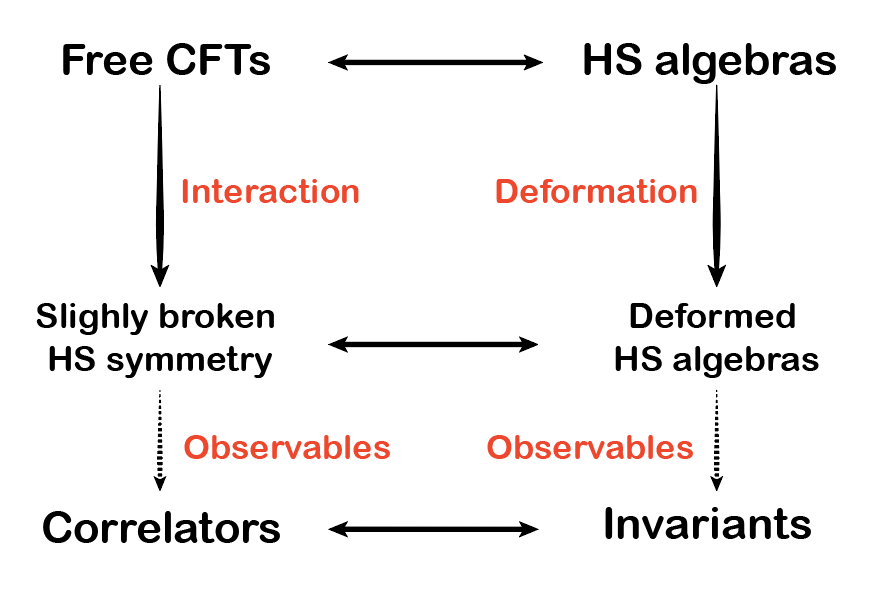}
\end{figure}
\section{Summary of Chapter 5}
In this chapter, we reviewed the construction that leads to formally consistent equations of motion for HSGRAs starting from the higher-spin algebra. The construction captures certain algebraic aspects of the higher spin problems that survives even in the presence of non-localities. It is also hard, if not impossible, to detect these algebraic structures in any perturbative approach like the Noether procedure. We also showed that we only need to deform the $[P,P]$-bracket to drive the whole deformation of the algebra. Moreover this deformed $[P,P]$-bracket leads to Einstein's equation and therefore is well-motivated. 

We constructed the Type-A HSGRA in $\ads_5$ at the level of formally consistent equations. There, we employ two approach to deform the algebra: (i) we deform the universal enveloping algebra via the deformed Joseph's relations and the deformed $[P,P]$-bracket of $su(2,2)$; (ii) utilizing QCR, we found a minimal set of canonical pairs of oscillators that generate the deformation. 

It is worth mentioning, that there are not so many ways to deform enveloping algebras. A well-known approach is to deform the Hopf algebra structure. In this chapter, we presented another way: enveloping algebras evaluated in certain irreducible representations turn out to admit
a deformation as associative algebras once they are extended with a group of automorphism{\color{red}s} $\Gamma$. This is closely related to the Deformation quantization of Poisson manifolds \cite{Kontsevich:1997vb}.
\subsubsection{Outlook}
We can extend the results of this chapter to the case of supersymmetric HSGRAs. Some interesting examples include the $5,7$-dimensional supersymmetric theories \cite{Sezgin:2001yf,Sezgin:2001ij}, and the 6-dimensional exceptional HSGRA based on $F(4)$ superalgebra \cite{Gunaydin:2016amv}. We also expect that the massless sector of tensionless strings should be described by a theory based on the higher spin extension of the gauge symmetry $psu(2,2|4)$. The approaches we used in this chapter should, in principle, admit a straightforward supersymmetric extension.


  \chapter{Summary and Discussion}\label{chapter6}
\section{Summary of Results}
In this thesis, we studied three different approaches to HSGRAs that including the metric-like formalism, light-front formalism and frame-like formalism. Each of them has their own advantages and drawbacks when we tackle a specific problem in HSGRA. However, the main messages and results of this thesis indicate that HSGRAs are UV-finite thanks to higher-spin symmetry. Therefore, HSGRAs can be thought of as toy models for Quantum Gravity. We summarise our results as follows:

\textbf{One loop Tests of HSGRAs/Vector Models Duality}\\
We derive the spectral zeta-functions for various HSGRAs where fields are totally symmetric or mixed-symmetric. Using zeta-regularization, we computed the one-loop vacuum contributions for several HS theories and most of them precicely match the predictions from the CFT duals. The test failed naively for type-B theory in even dimension and calls for better understanding of the duality.

We computed the vacuum one-loop energy in Type-A HSGRA in all (including fractional) dimensions and showed that it gives exactly the generalized sphere free energy of a scalar field. Upon changing the boundary condition, the Type-A theory gives a change in the generalized sphere free energy of the critical $O(N)$ vector model as compared to the free one to the leading order in $1/N$ expansion. 

\textbf{UV-finiteness of Quantum Chiral HSGRA}\\
Chiral HSGRA is a local quantum higher-spin theory that has a simple action in light-cone gauge. Due to the specific form of the coupling constant $C\sim \frac{1}{\Gamma[\lambda_1+\lambda_2+\lambda_3]}$, the interactions conspire as to make the S-matrix trivial at the tree level. We also showed that there are no UV divergences at one-loop order in all diagrams we have analyzed. Therefore, our results showed that chiral HSGRA is a consistent quantum theory in flat space. 

It is important to stress that our results confirm the expectation that higher-spin symmetry is rich enough to forbid all counterterms that can spoil renormalizability of the model. 

\textbf{Formal HSGRA in $\ads_5$}\\
We constructed formally consistent HSGRA in $\ads_5$. We found two solutions. The first is to deform Joseph relations and $su(2,2)$-algebra, which ultimately deforms the whole higher-spin algebra. The second solution is via the quasi-conformal realization (QCR) which is built from the minimal number of canonical pairs of oscillators. We deform some of the commutation relations, and as the result, the deformed (QCR) also gives us the deformed higher-spin algebra which allows us to construct interaction vertices. Our construction should admit a simple supersymmetric extension. The most interesting case is a HSGRA based on universal enveloping algebra of the gauge symmetry $psu(2,2|4)$ that should describe massless sector of tensionless strings in $\ads_5$. 

\section{Discussion}
Inspired by the original work of Fronsdal and Fang \cite{Fronsdal:1978rb,Fang:1978wz} on free higher-spin fields, there has been a lot of development to uplift the free theories to interacting ones. They include the frame-like formalism and its extensions \cite{Vasiliev:1999ba,Vasiliev:2004qz,Bekaert:2005vh}, the Noether procedure \cite{Berends:1984rq,Barnich:1993vg,Joung:2011ww,Fredenhagen:2019hvb}, the holographic reconstruction of higher-spin theories \cite{Das:2003vw,Bekaert:2015tva,Skvortsov:2015pea,Sleight:2016dba,deMelloKoch:2018ivk} and the light-front formalism \cite{Bengtsson:1983pd,Bengtsson:1983pg,Metsaev:1991mt,Metsaev:1991nb,Ponomarev:2016lrm,Metsaev:2018xip,Skvortsov:2018uru}.

There are very few higher spin theories that are local enough as to be treated by the field theory methods: chiral HSGRA, conformal HSGRA and purely massless HSGRA in $3d$. Generic holographic HSGRA's that are dual to free or weakly-coupled CFT's like vector models were shown to be too non-local. This is not an end of the story and calls for a better understanding of locality in HSGRA. It is also clear at present that there is not much difference between the problems of HSGRA in flat space and in (anti)-de Sitter space. Indeed, (i) the main no-go theorems, e.g. Weinberg and Coleman-Mandula theorems, have a direct counterpart in anti-de Sitter space: the (holographic) S-matrix is fixed by the higher spin symmetry to be $S=1$ in flat space and $S=$free CFT in $AdS$ \cite{Maldacena:2011jn,Boulanger:2013zza}; (ii) the obstructions for HSGRA in flat space that arise at the quartic order \cite{Fotopoulos:2010ay,Bekaert:2010hp,Roiban:2017iqg} indicate that the theory becomes badly non-local, which is exactly what has been established recently in $AdS$ \cite{Bekaert:2015tva,Maldacena:2015iua,Sleight:2017pcz,Ponomarev:2017qab}. The only difference between flat space and $AdS$ is that, thanks to the existence of simple CFT duals, we expect to tame holographic HSGRA's one way or another.

At very high energies all particles should effectively become massless. Therefore, HSGRA can be good probes of the quantum gravity problem since many purely quantum issues seem to find their counterparts already at the classical level. For example, if higher spin symmetry is powerful enough as to forbid the relevant counterterms, then constructing a classical HSGRA is equivalent to having a consistent quantum gravity model (in the sense that there is nothing to be analyzed at the quantum level, at least perturbatively). Not to forget that HSGRA's exhibit certain features that make them closer to string theory than to a field theory (e.g. infinite number of states, Chan-Paton factors, ...). Due to these stringy features, one should not expect to look at HSGRAs as conventional field theories. There are, however, some classes of HSGRAs that are very close to field theories. 

The study of HSGRAs should help to better understand some aspects of the Quantum Gravity Problem. HSGRAs are also useful in view of their relation to the weakly coupled CFTs that describe the critical phenomena. It would be interesting to further investigate the tensionless limit of string theory in order to understand the bizarre behaviour of HSGRA's. Indeed, a world-sheet model should resolve the non-locality problem of HSGRAs. 
  \begin{appendix}
\chapter{Appendix for One-loop Tests in Integer Dimensions}\label{app:chap3}
\section{Characters, Dimensions and all that}
\label{app:dimschars}
Below are some useful formulas for the dimensions of various irreducible representations. The general formulae for the dimensions of irreducible representations for the case of $so(2k)$ and $so(2k+1)$ read:
{\allowdisplaybreaks\besubeqs\label{dimweyl}
\begin{align}
    \mathbb{Y}^{so(2k)}(s_1,...,s_k)&: && \prod_{1\leq i<j\leq k} \frac{(s_i-s_j-i+j)(s_i+s_j-i-j+2k)}{(j-i)(2k-i-j)}\,,\\
    \mathbb{Y}^{so(2k+1)}(s_1,...,s_k)&: && \prod_{1\leq i<j\leq k} \frac{(s_i-s_j-i+j)}{(j-i)}  \prod_{1\leq i\leq j\leq k} \frac{(s_i+s_j-i-j+2k+1)}{(2k+1-i-j)}\,,
\end{align}
\esubeqs}\noindent
where the representation is defined by Young diagram $\mathbb{Y}(s_1,...,s_k)$ with the $i$-th row having length $s_i$ or $s_i-\tfrac12$ if all $s_i$ are half-integer. For some of the particular cases of use we find for $so(d)$:
{\allowdisplaybreaks\besubeqs\label{soddims}
\begin{align}
    \mathbb{Y}(s)&: && \frac{(d+2 s-2) \Gamma (d+s-2)}{\Gamma (d-1) \Gamma (s+1)}\,,\\
    \mathbb{Y}_{\tfrac12}(s)&: && \frac{\Gamma (d+s-1) 2^{\left[\frac{d}{2}\right]}}{\Gamma (d-1) \Gamma (s+1)}\,,\\
    \mathbb{Y}(a,b)&: && \frac{(a-b+1) (2 a+d-2) (2 b+d-4) (a+b+d-3) \Gamma (a+d-3) \Gamma (b+d-4)}{\Gamma (a+2) \Gamma (b+1) \Gamma (d-3) \Gamma (d-1)}\,,\\
    \mathbb{Y}_{\tfrac12}(a,b)&: && \frac{(a-b+1) (a+b+d-2) \Gamma (a+d-2) \Gamma (b+d-3) 2^{\left[\frac{d}{2}\right]}}{ (a+1)!  b! \Gamma (d-3) \Gamma (d-1)}\,,\\
    \mathbb{Y}(s,1^p)&: && \frac{(N+2 s-2) \Gamma (N+s-1)}{(p+s) \Gamma (p+1) \Gamma (s) (N-p+s-2) \Gamma (N-p-1)}\,,\\
    \mathbb{Y}(a,b,1^h)&: &&\frac{(a-b+1) (2 a+d-2) (2 b+d-4) (a+b+d-3) \Gamma (a+d-2) \Gamma (b+d-3)}{(a+h+1) a! (b+h) \Gamma (b) \Gamma (d-1) h! (a+d-h-3) (b+d-h-4) \Gamma (d-h-3)}\,,
\end{align}
\esubeqs}\noindent
where we use $\mathbb{Y}_{\tfrac12}(m_1,...)$ to denote spinorial representations. For example, $\mathbb{Y}_{\tfrac12}(m)$ is a symmetric rank-$m$ spin-tensor $T^{a(s);\alpha}$, i.e. it has spin $s=m+\tfrac12$. Similar formula for $sp(N)$ yields:
\begin{align}
    \mathbb{Y}(a,b)&: && \frac{(a-b+1) (a+b+N-1) \Gamma (a+N-1) \Gamma (b+N-2)}{\Gamma (a+2) \Gamma (b+1) \Gamma (N-2) \Gamma (N)}\,,
\end{align}
which allows to compute the dimension of any representation of $so(5)\sim sp(4)$:
\begin{align}
    \mathbb{Y}(a,b)&: &&
\frac{1}{6} (3 + 2 a) (1 + a - b) (2 + a + b) (1 + 2 b)\,,\\
    \mathbb{Y}_{\tfrac12}(s)&: &&\frac{2}{3} (s+1) (s+2) (s+3)\,,
\end{align}
where $a$, $b$ can be half-integers. Analogously, for special linear algebra $sl(d)$:
\begin{align}
    \mathbb{Y}(a,b,c)&: && \frac{(b+c) \Gamma (b) c! (a+b-c-2) \Gamma (a-c-1) \Gamma (a+d) \Gamma (b+d-1) \Gamma (c+d-2)}{(a+2 b-2) \Gamma (d-2) \Gamma (d-1) \Gamma (d) \Gamma (a+b-1)}\,.
\end{align}
The isomorphism $su(4)\sim so(6)$ gives for $so(6)$:
\begin{align}\notag
    \mathbb{Y}(a,b,c)&: &&\frac{(2 a-2)! (a+b+3)! (a-c-1)! (a-c+2)! (a+c-2)! (b-c)! (b-c+1)! (a+b-2 c)}{12 (2 a-3)! (3 a+b-2 (c+1)) (2 a+b-c-2)!}\,.
\end{align}
Note that the dimension \eqref{dimweyl} in the even case $so(2k)$ is the dimension of irreducible representation, while \eqref{soddims} formulas pack (anti)-selfdual representations together, so that \eqref{soddims} sometimes gives twice that of \eqref{dimweyl}.

\paragraph{Characters.} We will discuss only one-particle partition-functions without extra chemical potentials. Character of a generic representation with spin $\mathbb{S}$ is obtained by counting $\pl^k$-descendants assuming there are no relations among them:
\begin{align}
    \chi_{\Delta,\mathbb{S}}&=\mathrm{dim}\, \mathbb{S} \times \frac{q^\Delta}{(1-q)^d}\,.
\end{align}
The following short exact sequence is the simplest representations that correspond to partially-massless HS fields:
\begin{align}
    0\longrightarrow V(\Delta,\mathbb{S}') \longrightarrow V(\Delta-t,\mathbb{S}) \longrightarrow D(\Delta-t,\mathbb{S}) \longrightarrow0\,,
\end{align}
where $V(...)$ denotes generalized Verma module, which can be reducible, and $D$ is the irreducible module. Here, $\Delta=d+s_i-1-i$ and $\mathbb{S}'$ is the spin of the gauge parameter in $\ads_{d+1}$ or, equivalently, the symmetry type of the conservation law for a higher-spin current.\footnote{In the case of massless totally-symmetric fields we have
\begin{align}
    0\longrightarrow V(d+s-2,s-1) \longrightarrow V(d+s-2,s) \longrightarrow D(d+s-2,s) \longrightarrow0\,.
\end{align}}
An additional parameter $t$ is the depth of partially-masslessness \cite{Deser:2001us} and $t=1$ for massless fields. \\
${}$\\
In the case of free scalar, $\Rac$, and free fermion, $\Di$, the sequence is short but different. The singular vectors are  associated with $\square\phi$ and $\slashed{\pl}\psi$:
\begin{align}
    \Rac&: &&     0\longrightarrow V(\tfrac{d+2}{2},0) \longrightarrow V(\tfrac{d-2}{2},0) \longrightarrow D(\tfrac{d-2}{2},0) \longrightarrow0\,,\\
    \Di&: &&    0\longrightarrow V(\tfrac{d+1}{2},\tfrac{1}{2}) \longrightarrow V(\tfrac{d-1}{2},\tfrac{1}{2}) \longrightarrow D(\tfrac{d-1}{2},\tfrac{1}{2}) \longrightarrow0\,.
\end{align}
Below we collect some of the blind characters of $so(d,2)$. The dimensions of irreducible $so(d)$ representations can be found above
\begin{align*}
    \chi(\phi_\Delta)&=(1-q)^{-d} q^{\Delta }\,, && \text{scalar of dimension } \Delta\,,\\
    \chi(Rac)&=\chi(\phi_\Delta)-\chi(\phi_{\Delta+2})\Big|_{\Delta=\frac{d-2}2}=\left(1-q^2\right) (1-q)^{-d} q^{\frac{d}{2}-1}\,,\\
    \chi(O_{\Delta,s})&=\frac{(1-q)^{-d} (d+2 s-2) q^{\Delta } \Gamma (d+s-2)}{\Gamma (d-1) \Gamma (s+1)}\,,&& \text{symmetric tensor operator}\,,\\
    \chi(J_s)&=\chi(O_{\Delta,s})-\chi(O_{\Delta+1,s-1})\Big|_{\Delta=d+s-2}\,, && \text{conserved tensor}\,,\\
    \chi(\psi_\Delta)&=(1-q)^{-d} q^{\Delta }2^{[\tfrac{d}2]}\,,&& \text{fermion of dimension } \Delta\,,\\
    \chi(Di)&=\chi(\psi_\Delta)-\chi(\psi_{\Delta+1})\Big|_{\Delta=\frac{(d-1)}{2}}\,.
\end{align*}
The simplest instance of the Flato-Fronsdal theorem then follows from
\begin{align}
    \chi^2(Rac)=\sum_s \chi(J_s)\,.
\end{align}
Given a character $Z(q=e^{-\beta})$, the (anti)-symmetric parts of the tensor product can be extracted in a standard way:
{\allowdisplaybreaks\begin{align}
    \text{symmetric}&: && \frac12 Z^2(\beta)+\frac12 Z(2\beta)\,,\\
    \text{anti-symmetric}&: && \frac12 Z^2(\beta)-\frac12 Z(2\beta)\,.
\end{align}}\noindent
The character of the weight-$\Delta$ spin-$(s,1^h)$ operator and the associated conserved current are:
\begin{align}
   \chi(O_{s,1^h})&=\frac{(1-q)^{-d} (d+2 s-2) q^{\Delta } \Gamma (d+s-1)}{(h+s) \Gamma (h+1) \Gamma (s) (d-h+s-2) \Gamma (d-h-1)}\,,\\
   \chi(J_{\Delta,s,1^h})&=\chi(O_{\Delta,s,1^h})-\chi(O_{\Delta+1,s-1,1^h})\Big|_{\Delta=d+s-2}\,.
\end{align}
Fermionic spin-tensor conformal quasi-primary operator $O_{\alpha;a(s)}$ obeys $\gamma^m{}\fud{\beta}{\alpha}O_{\beta;ma(s-1)}=0$, which allows to compute its character and the character of the conserved higher-spin super-current:
\begin{align*}
    \chi(O)&=\frac{(1-q)^{-d} q^{\Delta } \Gamma (d+s-1) 2^{\left[\frac{d}{2}\right]}}{\Gamma (d-1) \Gamma (s+1)}\,,\\
    \chi(J_s^F)&=\chi(O_{\Delta,s})-\chi(O_{\Delta+1,s-1})\Big|_{\Delta=d+s-3/2}=\frac{(1-q)^{-d} q^{d+s-\frac{3}{2}} (d-q s+s-2) \Gamma (d+s-2) 2^{\left[\frac{d}{2}\right]}}{\Gamma (d-1) \Gamma (s+1)}\,.
\end{align*}

\paragraph{Tensor Products of Spinors.} To derive the decomposition of $\Di\otimes \Di$ together with its (anti)-symmetric projections we need to know how to take tensor product of two $so(d)$ spinors. For $d$ odd we have Dirac spinors, which we denote $\sD$. For $d$ even there are two Weyl spinors, which we denote $\sW$ and $\sWb$.\footnote{Various other possibilities like  symplectic Majorana-Weyl spinors in some dimensions will be ignored.} There are three distinct cases: $so(2k+1)$, $so(4k)$ and $so(4k+2)$. Consulting math literature we can find out that:
\begin{align}
    so(2k+1)&: &&
    \left\{\begin{aligned}
    (\sD\otimes \sD)_S&=\bigoplus \Yy{1^{k-4i}}\oplus\Yy{1^{k-4i-3}}\\
    (\sD\otimes \sD)_A&=\bigoplus \Yy{1^{k-4i-1}}\oplus\Yy{1^{k-4i-2}}
    \end{aligned}\right.\\
    so(4k)&: &&
    \left\{\begin{aligned}
    (\sW\otimes \sW)_S&=\Yy{1^{2k}}_+\oplus\bigoplus \Yy{1^{2k-4i}}\\
    (\sW\otimes \sW)_A&=\bigoplus \Yy{1^{2k-4i-2}}\\
    (\sW\otimes \sWb)_{\phantom{A}}&=\bigoplus \Yy{1^{2k-2i-1}}\\
    \end{aligned}\right.  \\
    so(4k+2)&: &&
    \left\{\begin{aligned}
    (\sW\otimes \sW)_S&=\Yy{1^{2k+1}}_+\oplus\bigoplus \Yy{1^{2k+1-4i}}\\
    (\sW\otimes \sW)_A&=\bigoplus \Yy{1^{2k-4i-1}}\\
    (\sW\otimes \sWb)_{\phantom{A}}&=\bigoplus \Yy{1^{2k-2i}}\\
    \end{aligned} \right.
\end{align}
where the sums are from $i=0$ to the maximal value it can take in each of the cases.  
Defining in even dimensions $\sD=\sW\oplus\sWb$ we observe:
\begin{align}
   so(2k+1)&: & \sD\otimes\sD&=\bigoplus_{i=0} \Yy{1^{k-i}}\,,\\
   so(2k)&: &   \sD\otimes\sD&=\Yy{1^{k}}_+\oplus\Yy{1^{k}}_-\oplus2\bigoplus_{i=1} \Yy{1^{k-i}}\,.
\end{align}
The decomposition of $\Di\otimes \Di$ of the $O(N)$-singlet is known and is quoted in the main text. We simply present the result for other cases:
\begin{align}
    so(2k+1)&: &&
    \left\{\begin{aligned}
    (\Di\otimes \Di)_A=&\bigoplus \Yy{2n+1,1^{k-4i-1}}\oplus\Yy{2n+1,1^{k-4i-4}}\oplus\\ &\bigoplus \Yy{2n,1^{k-4i-2}}\oplus\Yy{2n,1^{k-4i-3}}
    \end{aligned}\right. \\
    so(4k)&: &&
    \left\{\begin{aligned}
    (\Wi\otimes \Wi)_A=&\Yy{2n+1,1^{2k-1}}_+\oplus\bigoplus \Yy{2n+1,1^{2k-4i-1}}\oplus\\
    &\bigoplus \Yy{2n,1^{2k-4i-3}}\oplus \begin{cases}
    \bullet\,, & k=2m+1\\ \emptyset\,,& k=2m \end{cases}
    \end{aligned}\right.  \\
    so(4k+2)&: &&
    \left\{\begin{aligned}
    (\Wi\otimes \Wi)_A=&\Yy{2n+1,1^{2k}}_+\oplus\bigoplus \Yy{2n+1,1^{2k-4i}}\oplus\\
                &\bigoplus \Yy{2n,1^{2k-4i-2}}
    \end{aligned} \right.
\end{align}
where we indicated the $so(d)$-spin of the singlet quasi-primary operators, the conformal weight being obvious from $\Di\otimes \Di$. The above formulae generalize the Flato-Fronsdal theorem to the $O(N)$-singlet sector of free fermion theory in any dimension. Other versions of the singlet constraint follow from the above results.

\section{Amusing Numbers}
\label{app:MrCasimir}
We collect below various numbers associated to the fields discussed in the main text: Casimir Energy, sphere free energy, Weyl $a$-anomaly coefficients.
\paragraph{Casimir Energy.}
Casimir Energy, $E_c$, is given by a formally divergent sum
\begin{align}
    E_c&=(-)^F \frac12\sum_n d_n\omega_n\,,
\end{align}
for which the standard regularization is to use the $\exp[ -\epsilon\omega_n]$ as a cut-off and then remove all poles in $\epsilon$. All the data can be extracted from the characters. We see that the spin degrees of freedom factor out for massive fields and the Casimir energy is given by
\begin{align*}
   (-)^F E_c(\chi_{\Delta,\mathtt{S}})&= \frac12 \mathrm{dim}\, \mathtt{S} \left.\sum \frac{\Gamma[d+n]}{n!\Gamma[d]} (\Delta+n) e^{-\epsilon(\Delta+n)} \right|_{\text{finite}}=\mathrm{dim}\, \mathtt{S} \left.\frac{ e^{-(\Delta +1) \epsilon } \left(d+\Delta(  e^{\epsilon }-1)\right)}{\left(1-e^{-\epsilon }\right)^{d+1} }\right|_{\text{finite}}\,.
\end{align*}
Casimir Energy for a massive scalar field of weight $\Delta$:
\begin{center}
\begin{tabular}{c|c}
     $d$ & $E_c$ \\ \hline
     $2$ & $\frac{1}{24} (\Delta -1) \left(2 \Delta ^2-4 \Delta +1\right) $ \\
     $3$ & $\frac{1}{480} \left(-10 \Delta ^4+60 \Delta ^3-120 \Delta ^2+90 \Delta -19\right) $ \\
     $4$ & $\frac{1}{1440} (\Delta -2) \left(6 \Delta ^4-48 \Delta ^3+124 \Delta ^2-112 \Delta +27\right) $  \\
     $5$ & $\frac{-84 \Delta ^6+1260 \Delta ^5-7350 \Delta ^4+21000 \Delta ^3-30240 \Delta ^2+19950 \Delta -4315}{120960} $  \\
     $6$ & $\frac{(\Delta -3) \left(12 \Delta ^6-216 \Delta ^5+1494 \Delta ^4-4968 \Delta ^3+8112 \Delta ^2-5904 \Delta +1375\right)}{120960} $
\end{tabular}
\end{center}
allows one to get the Casimir Energy for any massive representation by multiplying it by $\mathrm{dim}\,\mathbb{S}$. 
Formulas for massless representations are obtained as differences of the massive ones according to exact sequences. Some of the formulae below can be found in \cite{Gibbons:2006ij,Ozcan:2006jn}. The Casimir Energies for higher-spin bosonic fields in lower dimensions are:
\begin{center}
\begin{tabular}{c|c}
     $d$ & $E_c$ \\ \hline
     $3$ & $\frac{1}{240} \left(30 s^4-20 s^2+1\right)$ \\
     $4$ & $-\frac{1}{1440} s (s+1) \left(18 s^4+36 s^3+4 s^2-14 s-11\right)$  \\
     $5$ & $\frac{(s+1)^2 \left(84 s^6+504 s^5+994 s^4+616 s^3-308 s^2-504 s-31\right)}{120960}$  \\
     $6$ & $-\frac{(s+1)^2 (s+2)^2 \left(12 s^6+108 s^5+338 s^4+408 s^3+32 s^2-282 s-31\right)}{483840}$  \\
\end{tabular}
\end{center}
Note that $d=3$ and $s=0$ case is special in that the fake ghost contribution does not vanish automatically and the right value is $E_c=\tfrac{1}{480}$. Casimir Energies for higher-spin fermionic fields in lower dimensions are:
\begin{center}
\begin{tabular}{c|c}
     $d$ & $E_c$ \\ \hline
     $3$ & $\frac{1}{240} \left(-30 s^4+20 s^2-1\right)$ \\
     $4$ & $\frac{(2 s+1)^2 \left(18 s^4+36 s^3-8 s^2-26 s+3\right)}{2880}$  \\
     $5$ & $-\frac{(2 s+1) (2 s+3) \left(84 s^6+504 s^5+910 s^4+280 s^3-532 s^2-280 s+11\right)}{241920}$  \\
     $6$ & $\frac{(2 s+1) (2 s+3)^2 (2 s+5) \left(12 s^6+108 s^5+314 s^4+264 s^3-144 s^2-162 s-3\right)}{1935360}$  \\
\end{tabular}
\end{center}
Note that $d=3$ and $s=\tfrac12$ the general formula does not oversubtract the fake descendants and the right value is still $E_c=\tfrac{17}{1920}$.
Casimir Energies for $\Rac$'s and $\Di$'s in lower dimensions $d=2,3,...$ are:\footnote{The fermion is always a Dirac one. $E_c$ for the Weyl fermion is half of the value in the table.}
\begin{align}
    E_c(\Rac)&=\left\{-\frac{1}{12},0,\frac{1}{240},0,-\frac{31}{60480},0,\frac{289}{3628800},0,-\frac{317}{22809600},0,\frac{6803477}{2615348736000}\right\}\,,\\
    E_c(\Di)&=\left\{-\frac{1}{24},0,\frac{17}{960},0,-\frac{367}{48384},0,\frac{27859}{8294400},0,-\frac{1295803}{851558400},0,\frac{5329242827}{7608287232000}\right\}\,.
\end{align}
Casimir Energies for massive $\Delta=d-1$ anti-symmetric tensors $\mathbb{Y}(1^h)$, $h=2,3,...$:\footnote{When self-duality applies it is the Casimir energy of the two fields.}
\begin{center}
\begin{tabular}{c|c|c|c}
     $d$ & $E_c$ \\ \hline
     $4$ & $-\frac{1}{20 h! \Gamma (5-h)} $  \\
     $5$ & $\frac{221}{1008 h! \Gamma (6-h)} $  \\
     $6$ & $-\frac{95}{84 h! \Gamma (7-h)} $
\end{tabular}
\end{center}
The Casimir Energies for massless hooks $\mathbb{Y}(s,1^p)$:
\begin{center}
\begin{tabular}{c|c}
     $d$ & $E_c,p=1$ \\ \hline
     $4$ & $\frac{1}{720} (-s (s+1) (2 s (s+1) (9 s (s+1)-22)+19)-3) $  \\
     $5$ & $\frac{3 s (s+2) (42 (s-1) s (s+2) (s+3) (2 s (s+2)+1)+221)+221}{120960} $  \\
     $6$ & $-\frac{(s+1) (s+2) (s (s+3) (2 s (s+3) (s (s+3) (6 s (s+3)-11)-54)+111)+95)}{120960} $
\end{tabular}
\end{center}

\paragraph{Sphere Free Energy.} Also, we will need the free energy on a sphere for free scalar and fermion, see e.g. \cite{Klebanov:2011gs},
\begin{align}
    F^3_\phi&=\frac{1}{16}(2\log 2-\frac{3\zeta(3)}{\pi^2})\,,
    && F^5_\phi=\frac{-1}{2^8}(2\log 2+\frac{2\zeta(3)}{\pi^2}-\frac{15\zeta(5)}{\pi^4})\,,\\
    F^3_\psi&=\frac{1}{16}(2\log 2+\frac{3\zeta(3)}{\pi^2})\,,
    && F^5_\psi=\frac{-1}{2^8}(6\log 2+\frac{10\zeta(3)}{\pi^2}+\frac{15\zeta(5)}{\pi^4})\,.
\end{align}

\paragraph{Weyl Anomaly.} The general formula for Weyl anomaly $a$ for real conformal scalar \cite{Casini:2010kt} and fermion \cite{Aros:2011iz} gives for $d=4,6,8,...$:\footnote{We changed normalization as compared to \cite{Aros:2011iz}.}
\begin{align}
    a_\phi&=\left\{\frac{1}{90},-\frac{1}{756},\frac{23}{113400},-\frac{263}{7484400},\frac{133787}{20432412000}\right\}\,,\\
    a_\psi&=\left\{\frac{11}{180},-\frac{191}{7560},\frac{2497}{226800},-\frac{14797}{2993760},\frac{92427157}{40864824000}\right\}\,.
\end{align}
\paragraph{Volumes.} The volume of $d$-sphere and the regularized volume of the hyperbolic space, which is Euclidean anti-de Sitter space, are \cite{Diaz:2007an}:
\begin{align}
    \mathrm{vol}\, S^d&=\frac{2 \pi ^{(d+1)/2}}{\Gamma \left(\frac{d+1}{2}\right)}\,, &&
    \mathrm{vol}\, \mathbb{H}^{d+1}=
        \begin{cases}
        \frac{2 (-\pi )^{d/2} }{\Gamma \left(\frac{d}{2}+1\right)}\log R\,, & d=2k\,,\\
        \pi ^{d/2} \Gamma \left(-\frac{d}{2}\right)\,, & d=2k+1\,.
        \end{cases}
\end{align}

\section{Other Classes}
\label{app:strangeHS}
In this section we discuss higher-spin doubletons that result in more general mixed-symmetry fields and higher-order singletons that lead to partially-massless fields and mixed-symmetry fields.

\subsection{Higher-Spin Doubletons}
\label{app:HSgletons}
In any $\ads_{2n+1}$, $n\geq 2$, we have higher-spin doubletons \cite{Gunaydin:1984fk,Gunaydin:1984wc,Metsaev:1995jp,Bekaert:2009fg,Fernando:2015tiu} as conformal fields in $\cft^{2n}$. These are parametrized by (half)-integer spin $J$, with $J=0,\tfrac12$ being the usual \Rac{} and \Di.\footnote{The Young diagram of $so(2n)$ that determines the spin of the field has a form of a rectangular block of length $J$ and height $n$, i.e. the labels are $\mathbb{Y}(J,...,J)$. One can also consider higher-spin representations of more complicated symmetry type, however they may be  non-unitary.} The $J=1$ is free massless spin-one field, i.e. Maxwell. For $J>1$ the HS doubletons are unusual CFT's in not having a local stress-tensor, while they still are unitary representations of the conformal algebra.

In \cite{Beccaria:2014zma,Beccaria:2014xda} it was conjectured that there should exist an AdS HS theory that is dual to $N$ free Maxwell fields, called Type-C in analogy with Type-A, $J=0$, and Type-B, $J=\tfrac12$. It was found that one-loop tests are successfully passed, but already the non-minimal theory requires to modify the bulk coupling as $G^{-1}=2N-2$. Similar conclusions were arrived at in \cite{Beccaria:2014qea} for the $J=1$ doubleton in $\ads_7/\cft^6$ \cite{Gunaydin:1984wc}.

Let us show that all Type-D,E,... theories, i.e. those with $J>1$, do not pass the one-loop test. The Casimir Energy of the spin-$J$ doubleton is easy to find:\footnote{For $J=0$ it gives the Casimir Energy of two real scalars. For lower spins $J=0,\tfrac12,1$ we therefore find $E_c=\tfrac{1}{240}, \tfrac{17}{960},\tfrac{11}{120}$. }
\begin{align}
    E_{c,J}&=\frac{1}{120} (-1)^{2 J} \left(30 J^4-20 J^2+1\right)\,.
\end{align}
The spectrum of Type-X theory can be found by evaluating the tensor product of two spin-$J$ doubletons \cite{Dolan:2005wy,Boulanger:2011se,Beccaria:2014zma}:
\begin{align}
    (J,0)\otimes(J,0)&=\sum_{k=0}^{2J}\Verma{2+2J}{k}{0}\oplus \sum_{k=1}\Verma{2+2J+k}{2J+\tfrac{k}2}{\tfrac{k}2}\,,\\
    (J,0)\otimes(0,J)&=\sum_{k=0}\Verma{2+2J+k}{J+\tfrac{k}2}{J+\tfrac{k}2}\,,
\end{align}
where in the first line we see massive and massless mixed-symmetry tensors and massless symmetric HS fields in the second line. The absence of the stress-tensor reveals itself in that the spectrum of massless HS fields is bounded from below by $2J$. In particular, there is no dynamical graviton for $J>1$.

The Casimir Energies for the three parts of the spectrum: massive, mixed-symmetry massless, and symmetric massless, can be computed with the net result:
\begin{align}
    E_c^J&=-\frac{1}{630} J (2 J-1) (2 J+1) \left(288 J^4-208 J^2-3\right)\,.
\end{align}
We see that the total Casimir energy vanishes for $J=0,\tfrac12$ in accordance with \cite{Giombi:2014yra}. It does not vanish for $J=1$ \cite{Beccaria:2014zma,Beccaria:2014xda}, rather it equals that of the two Maxwell fields, which still can be compensated by shifting the bulk coupling. However, for $J>1$ there does not seem to be any natural way of compensating the excess of the Casimir energy.

The same problem can be understood at the level of characters, which is a simpler approach. The blind character of the spin-$j$ doubleton is, see e.g. \cite{Beccaria:2014zma}:
\begin{align}
    Z_j&=\sum_k (2j+k+1)(k+1)q^{j+1+k}=\frac{(2 j (q-1)-q-1) q^{j+1}}{(q-1)^3}\,.
\end{align}
The singlet partition function is $[Z_j]^2$. It is symmetric in $\beta$, $q=e^\beta$, for $j=0,\tfrac12$. For $j=1$ it is not symmetric but the anti-symmetric part can be expressed as a multiple of $Z_1$, which can be compensated by modifying $G^{-1}=N$ \cite{Beccaria:2014zma}. However, for $j>1$ the anti-symmetric part cannot be compensated this way, but can be expanded in terms of $Z_{i\leq j}$.

Therefore, we see that the duals of HS doubletons $J>1$ should have pathologies as quantum theories. Assuming AdS/CFT holds at classical level, by reconstruction, we can manufacture some interaction vertices in AdS \cite{Costa:2011mg,Metsaev:1993ap,Metsaev:2005ar} such that
\begin{align}
    \langle \JJ_{s_1}...\JJ_{s_k}\rangle&=\text{Holographic Amplitudes}\,, && s_i\geq 2J\,.
\end{align}
The generating function of three-point correlators was constructed in \cite{Zhiboedov:2012bm}. The number that counts independent structures is $n=\min(s_1,s_2,s_3)+1$ and is given by the minimal spin, which is related to the fact that the currents that one can construct from a spin-$J$ doubleton must have $s\geq 2J$, see \cite{Gelfond:2006be} for the explicit form in $4d$. Indeed, only those doubletons can give a contribution to $\langle \JJ_{s_1}\JJ_{s_2}\JJ_{s_3}\rangle$ that have $2J\leq \min (s_1,s_2,s_3)$. This fact certainly causes a puzzle in the sense that $V_3(s_1,s_2,s_3)$ obtained by reconstruction cannot be a part of any consistent unitary HSGRA.\footnote{Although HS doubletons can only exist for even boundary dimension, the number of independent correlators $\langle \JJ_{s_1}\JJ_{s_2}\JJ_{s_3}\rangle$ seems to be indifferent to this fact, as if one could formally define HS doubletons in odd dimensions as well.}

\subsection{Partially-Massless Fields}
\label{app:PMfields}
If we sacrifice unitarity, the list of free CFT's becomes  infinitely richer. The simplest one-parameter family corresponds to higher-order singletons is
\begin{align}
    \Rac_k&: & \square^k \phi&=0\,, &&\Delta=\frac{d}2-k\,.
\end{align}
The spectrum of single-trace operators contains partially-conserved currents \cite{Dolan:2001ih}
\begin{align}
    \JJ_s&=\phi\square^i \pl^s\phi+...\,, && \pl^{k-i}\cdot \JJ_s=0\,.
\end{align}
The spectrum is encoded in the tensor product of two $\Rac_k$ \cite{Bekaert:2013zya}:
\begin{align}
    \Rac_k\otimes \Rac_k &= \sum_{s=0}^{\infty}\sum_{i=1}^{k} D(d+s-2i,s)\,.
\end{align}
The fields that are dual to partially-conserved currents are partially-massless fields \cite{Deser:2001us,Skvortsov:2006at}:
\begin{align}
    \pl^m...\pl^m J_{m(t)a(s-t)}&=0 &&\Longleftrightarrow && \delta \Phi^{\ua(s)}=\nabla^\ua...\nabla^{\ua}\xi^{\ua(s-t)}+...\,,
\end{align}
where $t$ is the depth of partially-masslessness. Massless fields occur at $t=1$. Therefore, the spectrum of a theory that is dual to $\Rac_k$ is a nested tower of (partially)-massless fields with the $\Rac_{k-1}$ tower contained in the $\Rac_k$ one. In particular, usual massless HS fields are present. Note that the depth $t$ is an odd number in $\Rac_k\otimes \Rac_k$. We can call the dual theory of $\Rac_k$ (which has weight-$\Delta$) as Type-$A_k$  \cite{Alkalaev:2014nsa}. In the irreducible module $D(d+s-2i,s)$, the operators with $s<i$ are not conserved tensors and are dual to massive fields, which for $k>2$ also contain massive HS fields. Therefore, duals of $\Rac_k$ provide an example of HS theories that contain HS gauges fields and HS massive fields with a spin bounded from above.

To test AdS/CFT duality we can check the vanishing of Casimir Energy in the non-minimal Type-$A_k$ theory, see also \cite{Basile:2014wua}. It is important to stress that the Casimir Energy of $\Rac_k$ should vanish in odd dimensions. We find in $d=3,4,...$ that:{\footnotesize
\begin{align*}
    E_c&=\{0,-\frac{1}{720} t \left(6 t^4-20 t^2+11\right),0,-\frac{t \left(12 t^6-126 t^4+336 t^2-191\right)}{60480},0,-\frac{t \left(10 t^8-240 t^6+1764 t^4-4320 t^2+2497\right)}{3628800}\}
\end{align*}}\noindent
The Casimir Energy of a depth-$t$ partially-massless spin-$s$ field can be computed in a standard way. For example, in the  $d=3$ case we find ($g=2s+1$):
\begin{align}
    E_c&=\frac{t \left(5 g (g-2 t) \left(3 g^2-6 g t+4 t^2-6\right)-17\right)}{1920}\,.
\end{align}
Consider the simplest case of $\Rac_2$. The spectrum contains that of Type-A and massive fields $\Phi$, $\Phi_\ua$, $\Phi_{\ua\ua}$ plus depth-$3$ partially-massless fields $s=3,4,...$. The sum over the Type-A spectrum was already found to vanish \cite{Basile:2014wua}. At least for odd $d$ we have to ensure that the sum over the rest vanishes as well. Using the standard exponential cut-off $\exp[-\epsilon(s+x)]$ we find that this is the case for $x=(d-5)/2$. Therefore, different parts of the spectrum should be summed with different regulators.

The dual of $\Rac_3$ contains the spectrum of Type-$A$=Type-$A_1$, the fields we have just studied plus massive fields $\Phi_{\ua(k)}$, $k=0,1,2,3,4$ and depth-$5$ partially-massless fields. The sum of the Casimir Energies of this last part gives zero for $x=(d-7)/2$.

Let us turn to the minimal Type-$A_k$ theory. It is useful to recall that the Casimir Energy can also be computed as
\begin{align}
    E_c&=(-)^F\frac12 \zeta(-1)\,, &\zeta(z)&=\frac{1}{\Gamma(z)}\int \beta^{z-1}d\beta\, Z(q=e^{-\beta})\,.
\end{align}
The non-zero contribution to $E_c$ comes from the $\beta^{-1}$ pole, which is absent if $Z(\beta)$ is an even function of $\beta$ \cite{Giombi:2014yra}. This is typically the case for the tensor product of two singletons, but is not for the (anti)-symmetric projections, which results in
\begin{align}
    Z_{\text{sing}}&=\frac12 Z^2(\beta)\pm \frac12 Z(2\beta)\,,
\end{align}
where the first term is an even function of $\beta$ in most cases. The contribution to the Casimir Energy is equal to that of the free field due to the last term. A slight  generalization of  \cite{Basile:2014wua,Bekaert:2013zya} implies that the minimal type-$A_2$ contains fields of even spins only. The excess of the Casimir Energy can be reduced to a linear combination of $\Rac_k$ by expressing the $\beta$-odd part of $(\Rac_k\otimes\Rac_k)_S$:
\begin{align}
    \beta-\text{odd part}\left[(\Rac_k\otimes \Rac_k)_S- \frac12 Z_k(2\beta)\right]&=0\,,
\end{align}
where $Z_k$ is the character of $\Rac_k$:
\begin{align}
    Z_k(q)&=(1-q)^{-d} \left(1-q^{2 k}\right) q^{\frac{1}{2} (d-2 k)}\,.
\end{align}
This identity directly implies that the Casimir energy of the minimal type-$A_k$ theory is equal to that of one $\Rac_k$, $E_c^k$. If instead we sum over spins with $\exp[-\epsilon(s+x)]$ cut-off we will have to use $x=(d-3)/2$ for depth-$1$ fields, $x=(d-5)/2$ for depth-$2$ fields etc. In particular, for type-$A_2$ the sum over its type-$A$ sub-sector gives $E_c$ of $\Rac_1$, while the sum over the depth-$2$ fields gives $E^2_c-E^1_c$ with the total result $E^2_c$, as before.

Also, it can be checked that the tensor product $\Rac_n\otimes \Rac_m$ with $m\neq n$ gives zero contribution to the Casimir Energy. Such products should arise in a theory built of several different higher-order singletons.

With the help of the zeta-function we can also check that $-2^{-1}\zeta'(0)$ matches the $a$-anomaly of $\square^k \phi=0$ free field. The latter can be extracted from the same zeta-function according to $a_{CHS}=-2a_{HS}$ where the conformal field dual to the order-$k$ singleton has weight $(d+2k)/2$. The summation over spins can be done as before given that the depth-$t$ partially-massless field of spin-$s$ has weight $\Delta=d+s-t-1$ and the spin-$(s-t)$ ghost has weight $d+s-1$. Lastly, the contribution of the massive fields that appear in the tensor product of two higher-order singletons need to be separated. For example, let us consider $\ads_5$ and set $k=2$ as above. We find:
\begin{align}
    \zeta'_{A}(0)&=0\,, &\zeta'_{PM}(0)&=\frac{{\log R}}{15}\,, &\zeta'_{\text{massive}}(0)&=-\frac{{\log R}}{15}\,,
\end{align}
so that the total contribution is zero. For the minimal Type-$A_2$ model, i.e. the one above truncated to even spins only, we have:
\begin{align}
    \zeta'_{\min,A}(0)&=-\frac{{\log R}}{45}\,, &\zeta'_{PM, \text{even}}(0)&=\frac{{\log R}}{3}\,, &\zeta'_{\text{massive, even}}(0)&=\frac{14\, {\log R}}{45}\,,
\end{align}
the total contribution being $-2^{-1}\zeta'(0)=-\tfrac{1}{45} (14\, {\log R})$, which is exactly the value of the zeta-function
\begin{align}
     \frac{1}{180} (\Delta -2)^3 {\log R} (s+1)^2 \left(5 (s+1)^2-3 (\Delta -2)^2\right)
\end{align}
at $s=0$ and $\Delta=(d+4)/2$. Using the explicit form of $\zeta'(0)$ for $d=2k$ it is easy to extract the $a$-anomaly of higher-order singletons.

Therefore, despite non-unitarity, higher-order singletons that lead to partially-massless fields seem to be consistent at one-loop.

\section{On the Computations in Even Dimensions}
\label{app:generalzetaHS}
\setcounter{equation}{0}
Let us briefly summary the steps for computing $\zeta$ and $\zeta'$ in even dimensions. Recall the full zeta-function that is given in the form
\begin{align}
    \zeta(z)&=\int_0^\infty d\lambda\, \frac{\tilde{\mu}(\lambda)}{[\lambda^2+\nu^2]^z} h(\lambda)\,, && \tilde{\mu}(\lambda)=\sum_k \mu_k \lambda^k\,,
\end{align}
where $\nu=\Delta-d/2$ and $h(\lambda)$ is either $\tanh \pi \lambda$ or $\coth{\pi \lambda}$ as in (\ref{schemeofzeta}). The computation of $\zeta(0)$ can be done by using
\begin{align}
    \tanh x&=1+\frac{-2}{1+e^x}\,, &  \coth x&=1+\frac{2}{-1+e^x}\,,
\end{align}
which leads to
\begin{align}
    \zeta(z)&=\int_0^\infty d\lambda\, \frac{\tilde{\mu}(\lambda)}{[\lambda^2+\nu^2]^z}\mp2\int_0^\infty d\lambda\, \frac{\tilde{\mu}(\lambda)}{[\lambda^2+\nu^2]^z(e^{2\pi \lambda}\pm1)}=I+II\,.
\end{align}
The first integral can be done for large enough $z$ and then continued to $z=0$. The second one is perfectly convergent and we can set $z=0$ and use
\begin{align}
  \int  \frac{-2\lambda^k}{e^{2\pi \lambda}+1}&= -4^{-k} \left(2^k-1\right) \pi ^{-k-1} \zeta (k+1) \Gamma (k+1) \label{D.4}\,,\\
  \int  \frac{2\lambda^k}{e^{2\pi \lambda}-1}&=2^{-k} \pi ^{-k-1} \text{Li}_{k+1}(1) \Gamma (k+1) \label{D.5}\,.
\end{align}
To compute $\zeta'(0)$ we first differentiate $\zeta(z)$ with respect to $z$. This can be directly done for the first part $I$, with two contributions produced:
\begin{align}
  \frac{\pl}{\pl z} I\Big|_{z=0}&= p_1(\nu)+ \log \nu \times p_2(\nu)\,,
\end{align}
where $p_{1,2}$ are polynomials. In the second part $II$ we find no problem with convergence, but a quite complicated integral
\begin{align}
  \frac{\pl}{\pl z} II\Big|_{z=0}&=\pm 2\int_0^\infty d\lambda\, \frac{\tilde{\mu}(\lambda)\log[\lambda^2+\nu^2]}{(e^{2\pi \lambda}\pm1)}\,.
\end{align}
Using $\log[\lambda^2+\nu^2]=\log \lambda^2+ \int_0^{\nu} dx\, 2x(x^2+\lambda^2)^{-1}$ we can split it into two parts:
\begin{align}
  II.1&=\pm2\int_0^\infty d\lambda\, \frac{\tilde{\mu}(\lambda)\log[\lambda^2]}{(e^{2\pi \lambda}\pm1)}=\pm 2\sum_k \mu_k c^{\pm}_k\,,\\
    II.2&=\pm 2\int_0^\infty d\lambda\, \frac{\tilde{\mu}(\lambda)}{(e^{2\pi \lambda}\pm1)}\int_0^{\nu} dx\, \frac{2x}{(x^2+\lambda^2)}\,.
\end{align}
Now we introduce two types of auxiliary integrals
\begin{align}
  c^{\pm}_n&=\int_0^\infty d\lambda\, \frac{\lambda^n\log[u^2]}{(e^{2\pi u}\pm1)}\,, &
  J^{\pm}_n&=\int_0^\infty d\lambda\, \frac{\lambda^n}{(x^2+\lambda^2)(e^{2\pi \lambda}\pm1)}\,.
\end{align}
The first one we will not attempt to evaluate since all $c_n$ will cancel in the final expressions. The second one can be done iteratively by first finding
\begin{equation}
    J_1^{\pm}=\int_0^{\infty}d\lambda\frac{\lambda}{(x^2+\lambda^2)(e^{2 \pi \lambda}\pm 1)}\,,
\end{equation}
where in [3.415, Table of integral],
\begin{equation}
    J_1^-=\int_0^{\infty}d\lambda\frac{\lambda }{(\lambda^2+x^2)(e^{2\pi \lambda}-1)}=\frac{1}{2}\left(\log(x)-\frac{1}{2x}-\psi(x)\right)\,.
\end{equation}
Together with a useful formula in \cite{Camporesi:1991nw}, $J_n^+(2\pi)=J_n^-(2\pi)-2J_n^-(4\pi)$, one can get
\begin{equation}
    J_1^+ = \frac{1}{2}\psi(x+1/2)-\frac{1}{2}\log x\,.
\end{equation}
Consider the following equation
\begin{equation}
    \int_0^{\infty}d\lambda \frac{\lambda^n}{e^{2 \pi \lambda}\pm 1}\log(a \lambda^2 + x^2) = \log a \int_0^{\infty}d\lambda\frac{\lambda^n}{e^{2\pi \lambda}\pm 1}+ \int_0^{\infty}d\lambda \frac{\lambda^n}{e^{2\pi \lambda}\pm 1}\log(\lambda^2+x^2/a)\,.
\end{equation}
Taking the derivative at $a=1$ on both sides, we obtain
\begin{equation}
    J_{n+2}^{\pm}=\int_0^{\infty}d\lambda\frac{\lambda^n}{e^{2\pi \lambda}\pm 1}-x^2 J_n^{\pm}\,.
\end{equation}
Therefore, $J_n^{\pm}$ will contain two types of contributions:
\begin{align}
  J_n^+&= q_n^+(x) \psi(x+1/2)+[\tilde{p}^+_2(x)\log x+\tilde{p}_3^+(x)]\,,\\
  J_n^-&=q_n^-(x)\psi(x)+[\tilde{p}^-_2(x)\log x+\tilde{p}_3^-(x)]\,.
\end{align}
The second terms in each equation can be easily integrated over $x$:
\begin{align}
\pm 2\int_0^{\nu} dx\,2x [\tilde{p}^{\pm}_2(x)\log x+\tilde{p}^{\pm}_3(x)]=p_3(\nu)-p_2(\nu)\log \nu\,.
\end{align}
Importantly, all $\log \nu$ now cancel because $p_2(\nu)$ is the same as the one at $\partial_z I\big|_{z=0}$. The purely polynomial leftovers $p_1$ and $p_3$ from $J_n^{\pm}$ and $\partial_z I\big|_{z=0}$ can be added up. We also need to add $II.1$ to them. Then $\nu$ is replaced with $\Delta-d/2$ and we can sum over all spins as usual. This contribution we call $P=\sum P_{\nu,s}-P_{\nu+1,s-1}$. Importantly, all coefficients $c_n$ will be gone and we do not need to deal with their real form, both for Type-A and Type-B.

Now we are left with the contribution that we call $Q=\sum Q_{\nu,s}-Q_{\nu+1,s-1}$, which consists of either $\psi(x+1/2)$ or $\psi(x)$ times a polynomial in $x$, where
\begin{align}
    Q_{\nu,s}&=\ \ \ 4 \ \ \sum_{s,k}\int_0^{\Delta -d/2} dx\,\mu_k q_k(x)\psi(x+1/2)\,,\quad (\text{for bosons})\,,\\
    Q_{\nu,m}&=-4 \sum_{s=m+\frac{1}{2},k}\int_0^{\Delta -d/2} dx\,\mu_k q_k(x)\psi(x)\,,\qquad \quad (\text{for fermions})\,.
\end{align}
It can be simplified by using the integral representation for $\psi(x)$:
\begin{align}
\psi(x)&=\int_0^{\infty} dt\,\left[\frac{e^{-t}}{t}-\frac{e^{-t x}}{1-e^{-t}}\right] \label{D.21}\,.
\end{align}
Next, the integral over $x$ can be done and the sum over the spectrum is taken. As a result we are left with
\begin{align}
Q&=\sum f^{n,m}_{a,b,c}\int dt\, \frac{e^{bt} t^a}{(1-e^{-t})^{n+1}(1+e^{-t})^{m+1}}\,.
\end{align}
The summands can be expressed as derivatives at $z=1$ and $z=-1$ of Hurwitz-Lerch function \cite{Giombi:2013fka,Giombi:2014iua}
\begin{equation}
    \Phi(z,s,\nu)=\frac{1}{\Gamma(s)}\int_0^{\infty}dt \frac{t^{s-1}e^{\nu t}}{1-ze^{-t}}\,,
\end{equation}
which in return, can be analytically continued into Hurwitz zeta function $\zeta(s,\nu)$. It is worth noting that only in the minimal higher-spin theories there will be $(1+e^{-t})^m$ in the denominator. Using this zeta regularization scheme, we will display the results of for HS theories in different even dimensions, which are subdivided into four categories in the following appendices: Type-A (non-minimal and minimal), HS fermions, Hook fields and the result for Hooks and Type-A can be added up to get Type-B theories (non-minimal and minimal). The case of $\ads_6$ is presented in more detail while for other dimensions we only show the main intermediate steps.
\subsection{Zeta Function in \texorpdfstring{$\ads_6$}{AdS(6)}}
\setcounter{equation}{0}
Following the previous steps, let us show explicitly how to calculate the zeta function in $AdS_6$ for Type-A, fermionic HS theory, hook fields and Type-B.
\subsubsection{Type-A}
\paragraph{Zeta.} Starting with Vasiliev type A theory, we recall the zeta-function in the main text
\begin{equation}
     \tilde{\mu}(u)=-\frac{u  \left(u ^2+\frac{1}{4}\right) (s+1) (s+2) (2 s+3) \tanh (\pi  u ) \left(u ^2+\left(s+\frac{3}{2}\right)^2\right)}{720 }\,.
\end{equation}
With $\tanh x= 1 - \frac{2}{e^{2\pi x}+1}$, we can write the spectral zeta function as
\begin{equation}
\begin{split}
\zeta^{\mathbb{H}}(z) &= -\frac{1}{720} (s+1)(2s+3)(s+2) \Bigg[\lim_{z\rightarrow 0} \int_0^{\infty} du \frac{u(u^2+1/4)\left(u^2+(s+3/2)^2\right)}{(u^2+\nu^2)^z}  \\
&- 2\int_0^{\infty} du \frac{u(u^2+1/4)\left(u^2+(s+3/2)^2 \right)}{(1+e^{2\pi u})} \Bigg]\,.
\end{split}
\end{equation}
Using \eqref{D.4}, one can obtain easily the zeta function for the Type-A HS theory \cite{Giombi:2014iua}
\begin{equation} \label{E.3}
\begin{split}
\zeta_{(\Delta,s)}(0) &=- \frac{(s+1)\left(2s+3\right)(s+2)}{29030400}\Big[-1835-714s(s+3) \\&-420 \nu^2(27-60\nu^2+16\nu^4+s(36-72\nu^2)+s^2(12-24\nu^2)) \Big]\,.
\end{split}
\end{equation}
The total contribution from HS fields and ghosts is
{\allowdisplaybreaks
\begin{align}
    \zeta^{A}(0) &= \sum_{s=0}^{\infty}\zeta_{(\Delta,s)}(0) - \zeta_{(\Delta+1,s-1)}(0)\notag\\  \label{E.4} &=\zeta_{(3,0)}+\sum_{s=1}^{\infty} \zeta_{(\Delta,s)} - \zeta_{(\Delta+1,s-1)}\\
    &=\frac{1}{1512} -\sum_{s=1}^{\infty} \frac{(1+s)^2(-20+28s+378s^2+868s^3+847s^4+378s^5+63s^6)}{30240}\,,\notag
    \end{align}}\noindent
where $\Delta=s+3$ and $\nu=s+\frac{1}{2}$. We use the exponential cut-off $\text{exp}[-\epsilon(s+\frac{d-3}{2})]$ to take the summation with $d=5$. A straightforward calculation shows that $\zeta^A=0\,$. The vanishing of zeta function is also true for the minimal Type-A theory, where $s=0,2,...$.
\begin{eqnarray}
    \zeta^{A}_{\min}=\zeta^A_{\min}=\zeta_{(3,0)}+\sum_{s=2,4,...}^{\infty}  \zeta_{(\Delta,s)} - \zeta_{(\Delta+1,s-1)} = 0\,.
\end{eqnarray}

\paragraph{Zeta-prime.} After making sure that the  conformal anomaly does not contribute to the free energy, we now can take the $z$-derivative of $\zeta$ at $z=0$ to calculate $\zeta'(0)$. One can easily obtain
\begin{equation}\notag
\begin{split}
    \zeta'(0)&=-\frac{(s+1)(s+2)(2s+3)}{720}\Bigg[\frac{1}{288}\nu^2\Big(-81+270\nu^2-88\nu^4+108s(-1+3\nu^2)+36s^2(-1+3\nu^2)\\
    &+3\big(27-60\nu^2+16\nu^4+s(36-72\nu^2)+s^2(12-24\nu^2)\big)\log(\nu^2)\Big)\\
    &+2 \int_0^{\infty}du \frac{u(u^2+\frac{1}{4})(u^2+(s+\frac{3}{2})^2)\log(u^2)}{e^{2\pi u}+1}
    +4\int_0^{\infty} du \int_{0}^{\nu}dx x\frac{u(u^2+\frac{1}{4})(u^2+(s+\frac{3}{2})^2)}{(e^{2\pi u}+1)(u^2+x^2)} \Bigg]\,.
    \end{split}
\end{equation}
Following Appendix \ref{app:generalzetaHS}, the first integral is therefore
\begin{equation}
    II.1=-\frac{(s+1)(s+2)(2s+3)}{360}\left[c_5^+ +c_3^+ \left(\frac{1}{4}+\left(s+\frac{3}{2}\right)^2\right)+\frac{c_1^+}{4}\left(s+\frac{3}{2}\right)^2\right]\,.
\end{equation}
The second integral is just
{\allowdisplaybreaks
\begin{align*}
    II.2&=-\frac{(s+1)(s+2)(2s+3)}{180}\int_0^v dx x \left(J_5^+ +\left(\frac{1}{4}+\left(s+\frac{3}{2}\right)^2\right)J_3^+ +\frac{1}{4}\left(s+\frac{3}{2}\right)^2 J_1^+ \right) \\
    &=-\frac{(s+1)(s+2)(2s+3)}{720}\Bigg[\frac{1}{2880} \nu^2 \Big(3 (377 + 160 s (3 + s)) - 120 (8 + 3 s (3 + s)) \nu^2 +
    160 \nu^4\\
    &+
    60 (-3 (3 + 2 s)^2 + 12 (5 + 2 s (3 + s)) \nu^2 - 16 \nu^4) \log(\nu)\Big)\\ &-\frac{1}{8}\int_0^{\nu} x (9 + 12 s + 4 s^2 - 4 x^2) (-1 + 4 x^2) \psi(1/2 + x)\Bigg]\,.
    \end{align*}}\noindent
It is easy to see that the $\log$ constribution in (E.7) and (E.9) cancel each other. In the end, we are left with
\begin{equation}
    \zeta'^A(0)= P_{\nu,s}+Q_{\nu,s}\,,
\end{equation}
where,
\begin{equation}
\begin{split}
    P_{\nu,s}&=-\frac{(s + 1) (s + 2) (2 s + 3)}{720} \Bigg[\frac{
   \nu^2 (107+580\nu^2-240\nu^4+120s(1+6\nu^2)+40s^2(1+6\nu^2))}{960}\\
      &+ \frac{ c_1^+}{2} \big(s + \frac{3}{2}\big)^2 +
   2 c_3^+ \big(\big(s + \frac{3}{2}\big)^2 + 1/4\big) + 2 c_5^+\Bigg]\,,
   \end{split}
   \end{equation}
   \begin{equation}
       Q_{\nu,s}=\frac{(s+1)(s+2)(2s+3)}{5760}\int_0^{\nu} x (9 + 12 s + 4 s^2 - 4 x^2) (-1 + 4 x^2) \psi(1/2 + x)\,.
   \end{equation}
Using the cut-off method, the evaluation of $P=\sum_s P_{\nu,s} - P_{\nu+1,s-1}$ in the case of all spins and in the case of even spins only leads to the same result of zero, i.e the contribution of $P_{\nu,s}$ to $\zeta'(0)$ vanishes for both cases. The evaluation of $Q_{\Delta,s}$ is a little bit harder if one wishes to obtain an analytical result. We write the di-gamma function in its integral representation \eqref{D.21} and obtain
\begin{equation}
     Q = \sum_{s=0}^{\infty}Q_{\nu,s}-Q_{\nu+1,s-1} =0\,.
\end{equation}
Hence,
\begin{equation}
    \sum_{s=1}^{\infty}Q_{\nu,s}-Q_{\nu+1,s-1}=-Q_{\frac{1}{2},0}\,,
\end{equation}
where,
\begin{equation}
    Q_{\frac{1}{2},0}=-\frac{1}{120}\left(\frac{1181}{11520} - \frac{211 \log(2)}{4032} - \frac{23 \log A}{16} + \frac{5 \zeta(3)}{
 4 \pi^2} + \frac{15 \zeta(5)}{4 \pi^4} -
 \frac{63}{16} \zeta'(-5) + \frac{35}{8} \zeta'(-3)\right)
\end{equation}
here, $A=e^{\frac{1}{12}-\zeta'(-1)}$ is the Glaisher-Kinkelin constant. Above, we used the exponential cut-off $\text{exp}[-\epsilon \nu]$ to evaluate the sum over all spins.
For minimal Type-A theory, a straightforward calculation shows that the $\zeta'(0)_{\min}$ is just
\begin{equation}
   \zeta'(0)_{min}= Q_{\frac{1}{2},0}+\sum_{s=2,4...} Q_{s+\frac{1}{2},s}-Q_{s+\frac{3}{2},s-1} = \frac{1}{2^7}\left(2\log 2 + \frac{2\zeta(3)}{\pi^2}-\frac{15\zeta(5)}{\pi^4}\right)=-2F^\phi_5\,,
\end{equation}
where,
\begin{equation}
\begin{split}
    \sum_{s=2,4...}Q_{s+\frac{1}{2},s}-Q_{s+\frac{3}{2},s-1} &=-\frac{1}{180}\Bigg[-\frac{1181}{7680} - \frac{7349 \log(2)}{2688} + \frac{69 \log A}{32} - \frac{
 75 \zeta(3)}{16 \pi^2} \\
 &+ \frac{495 \zeta(5)}{32 \pi^4} +
 \frac{189}{32}\zeta'(-5) -\frac{105}{16} \zeta'(-3)\Bigg]\,.
 \end{split}
\end{equation}

\subsubsection{Fermionic HS fields}

\paragraph{Zeta.} Above, we showed explicitly how to evaluate the zeta-function for the Type-A case. For fermionic HS fields, the computation is similar with the change of variable $s=m+1/2$. We recall the spectral function for fermions from the main text
\begin{eqnarray}
   \tilde{\mu}(u)= -\frac{u  \left(u ^2+1\right) \left(s+\frac{1}{2}\right) \left(s+\frac{3}{2}\right) \left(s+\frac{5}{2}\right) \coth (\pi  u ) \left( u^2+\left(s+\frac{3}{2}\right)^2\right)}{180 }\,.
\end{eqnarray}
We write  $s= m+1/2$, so that we can take the sum from $m=0$ to $\infty$. The degeneracy becomes
\begin{equation}
    g(m)\sim (m+1)(m+2)(m+3)\,.
\end{equation}
As we shall see the overall normalization factor does not affect the final result for fermions. Using \eqref{D.5}, we get
{\footnotesize    \begin{align*}
    \zeta_{\frac{1}{2}} \sim \sum_{m=0}^{\infty}\frac{1}{168} (-542 - 99 m + 8094 m^2 + 22806 m^3 + 28497 m^4 + 19404 m^5 +
   7448 m^6 + 1512 m^7 + 126 m^8) = 0\,.
\end{align*}}\noindent
\paragraph{Zeta-prime.} To find $\zeta'_{\frac{1}{2}}$, the integral that one needs to evaluate is
\begin{equation}
\begin{split}
        &\quad \partial_{z}\Big|_{z=0}g(m) \int_0^{\infty} \frac{u(u^2+1)(u^2+(m+2)^2)}{(\nu^2 + u^2)^z}\left(1+\frac{2}{e^{2 \pi u} -1}\right)\\
        &\sim \partial_{z}\Big|_{z=0} \left(\int_0^{\infty} \frac{u(u^2+1)(u^2+(m+2)^2)}{(\nu^2+u^2)^z} + \int_0^{\infty} \frac{2u(u^2+1)(u^2+(m+2)^2)}{(e^{2\pi u}-1)(\nu^2+u^2)^z}\right)\,.
        \end{split}
\end{equation}
We ignore $g(m)$ at the moment for simplicity. The first integral equals with
\begin{equation}
\begin{split}
    I&=\frac{1}{72} \nu^2 \Bigg[-144 + 135 \nu^2 - 22 \nu^4 + 36 m (-4 + 3 \nu^2) +
   9 m^2 (-4 + 3 \nu^2) \\
   &- 6 (-24 + 15 \nu^2 - 2 \nu^4 + 12 m (-2 + \nu^2) + 3 m^2 (-2 + \nu^2)) \log
     \nu^2\Bigg]\,.
    \end{split}
\end{equation}
The second integral is just $II=II.1+II.2$, where
\begin{equation}
 II.1= 2\left(2c_1^-(m+2)^2+2c_3^-((m+2)^2+1)+2c_5^-\right)\,,
\end{equation}
\begin{equation}
\begin{split}
    II.2 &= -4\int_0^{\nu}xdx \int_0^{\infty} du \frac{ u (1 + u^2) ((2 + m)^2 + u^2)}{( -1 + e^{2 \pi u}) (u^2 + x^2)}\\
    &=-4 \int_0^{\nu} xdx \left[(2+m)^2 J_1^-  + ((2+m)^2+1)J_3^- + J_5^-\right]\,.
    \end{split}
\end{equation}
Repeating the same algorithm as in the case of bosonic theory, we get
\begin{equation}
\begin{split}
   P_{\nu,m}&= -g(m) \Bigg[-\frac{1}{120}\nu(-480+51\nu+200\nu^2-155\nu^3-24\nu^4+30\nu^5-40m(12-\nu-4\nu^2+3\nu^3)\,,\\
   &-10m^2(12-\nu-4\nu^2+3\nu^3))-2c_1^-(m+2)^2-2c_3^-((m+2)^2+1)-2c_5^-\Bigg]\,,
   \end{split}
\end{equation}
\begin{equation}
   Q_{\nu,m}= -2g(m) \int_0^{\nu} dxx(x^2-1)(x^2-(m+2)^2)\psi(x)\,,
\end{equation}
where, we have returned the degeneracy into the calculation.
\begin{equation}
     P= \sum_{m=0}^{\infty} \left(e^{-\epsilon(m+1)}P_{m+1,m}-e^{-\epsilon(m+2)}P_{m+2,m-1}\right)=-\frac{1787}{3402000}\,,
\end{equation}
and $Q$ is just
\begin{equation}
\begin{split}
     Q &=  \sum_{m=0}^{\infty}\left( e^{-\epsilon(m+1)}Q_{m+1,m}-e^{-\epsilon(m+2)}Q_{m+2,m-1}\right)\\
    &=-\frac{1}{180}\int_0^{\infty} dt \Bigg[\frac{1440(e^{3t}-7e^{4t}-12e^{5t}-7e^{6t}-e^{7t})}{(-1+e^{t})^9t}-\frac{72e^{2t}(3+47e^t+47e^{2t}+3e^{3t})}{3(1+e^t)^6t^2}\\
    &\qquad \qquad +\frac{120e^{2t}(1+24e^t+33e^{2t})}{(-1+e^t)^9t^3}+\frac{360e^{2t}(1+19e^t+19e^{2t}+e^{3t})}{(-1+e^t)^6t^4}\\
    &\qquad \qquad +\frac{1440e^{2t}(1+4e^t+e^{2t})}{(-1+e^t)^5t^5}
    +\frac{1440e^{2t}(1+e^t)}{(-1+e^t)^4t^6}\Bigg]\\
    &=\frac{1787}{3402000}\,.
\end{split}
\end{equation}
Hence, $\zeta'(0)_{\frac{1}{2}}=0$, which guarantees that the consistency of SUSY HS theories relies on the bosonic part thereof.

\subsubsection{Height-one Hook HS fields}
\paragraph{Zeta.} To get to the Type-B theory we need to calculate the contribution of hook fields in $\ads_6$. The zeta-function is
\begin{equation}
   \tilde{\mu}(u)=-\frac{u  \left(u ^2+\frac{9}{4}\right) s (s+3) (2 s+3) \tanh (\pi u ) \left(u ^2+\left(s+\frac{3}{2}\right)^2\right)}{ 240}\,.
\end{equation}
Since $\Delta=s+3$ with $s=1,2,...$ and $\nu=s+1/2$, we can repeat the same calculation as for bosonic HS fields. The zeta function is therefore
\begin{equation}
    \zeta^{Hook} =-\frac{1}{240}\sum_{s=1}^{\infty} \frac{74}{63} - \frac{58 s}{7} - \frac{1109 s^2}{21} - 94 s^3 - \frac{337 s^4}{6} + 14 s^5 + \frac{
 91 s^6}{3} + 12 s^7 + \frac{3 s^8}{2}= \frac{1}{180}\,.
\end{equation}
While the result of zeta-function for even spin case is
 \begin{equation}
    \zeta^{Hook}_{min} = -\frac{1}{240}\sum_{s=2,4,...}^{\infty} \frac{74}{63} - \frac{58 s}{7} - \frac{1109 s^2}{21} - 94 s^3 - \frac{337 s^4}{6} + 14 s^5 + \frac{
 91 s^6}{3} + 12 s^7 + \frac{3 s^8}{2}= \frac{37}{7560}\,.
\end{equation}
It is easy to see that the zeta function for hook fields is not zero, which is not a problem since they make only a part of the Type-B spectrum.

\paragraph{Zeta-prime.} The $\zeta'=P_{\nu,s}+Q_{\nu,s}$ can be obtained by using the same treatment for bosonic theory, where we find that
\begin{equation}
\begin{split}
    P_{\nu,s}&= -\frac{s (3 + s) (3 + 2 s)}{240} \Bigg[2 c_5^+ + \frac{9}{8} c_1^+ (3 + 2 s)^2 +
   c_3^+ (9 + 6 s + 2 s^2)\\
   &+
   \frac{\nu^2}{960} \Big(187 + 1060 \nu^2 - 240 \nu^4 + 120 s (1 + 6 \nu^2) +
      40 s^2 (1 + 6 \nu^2)\Big)\Bigg]\,,
      \end{split}
\end{equation}
and
\begin{equation}
    Q_{\nu,s}= -\frac{s(s+3)(2s+3)}{1920}\int_0^{\nu} dx\ x(-9+4x^2)(-9-12s-4s^2+4x^2)\psi(x+1/2)\,.
\end{equation}
Summing over all spins, the result of $P$ is
\begin{equation}
    P^{Hook}= \sum_{s=1}^{\infty} P_{s+1/2,s}-P_{s+3/2,s-1} = \frac{1}{300}\,,
\end{equation}
while for the minimal case of Type-B, one needs to have
\begin{equation}
    P^{Hook}_{min}=\sum_{s=2,4,...}^{\infty}P_{s+\frac{1}{2},s}-P_{s+\frac{3}{2},s-1}=\frac{197}{51200} + \frac{3 c_1^+}{320} + \frac{c_3^+}{24} + \frac{c_5^+}{60}\,.
\end{equation}
Next, we evaluate the $Q^{Hook}$ for the non-minimal and minimal Type-B. We find for all spins:
\begin{equation}
    \begin{split}
         Q^{Hook}&= -\frac{623}{21600} + \frac{\log A}{6} + \frac{1}{6} \zeta'(-4) -
 \frac{1}{3} \zeta'(-3) + \frac{1}{3} \zeta'(-2)\\
    &=-\frac{623}{21600} +\frac{\log A}{6} + \frac{\zeta(5)}{8\pi^4}-\frac{\zeta(3)}{12 \pi^2}-\frac{1}{3}\zeta'(-3)\,,
    \end{split}
    \end{equation}
and for even spins only:
    \begin{equation}
         Q^{Hook}_{min}= -\frac{1433}{51200} + \frac{52709 \log(2)}{483840} + \frac{99 \log A}{640} +
\frac{\zeta(3)}{64 \pi^2} - \frac{93 \zeta(5)}{128 \pi^4} -
 \frac{21}{640} \zeta'(-5) - \frac{19}{64} \zeta'(-3)\,,
    \end{equation}
where we utilized,
\begin{equation}
\zeta'(-2n)=\frac{(-1)^n\zeta(2n+1)(2n)!}{2^{2n+1}\pi^{2n}}\,.
\end{equation}
Having these results at hand, we are now able to compute the $\zeta'_B$ for the non-minimal and minimal Type-B theories.

\subsection{Non-minimal Type-B }
In order to calculate the zeta function for Type-B, we need to collect all the information from Type-A, scalar field with $\Delta=4$ and the above hook fields. From (E.4), one can easily obtain the $\zeta^A_{s >0}$ for non-minimal which is $-\frac{1}{1512}$. For the scalar with $\Delta=4$, we simply get from \eqref{E.3} that
\begin{equation}
    \zeta_{4,0} = -\frac{37}{7560}\,.
\end{equation}
The spectrum of non-minimal Type-B involves the spectrum of Type-A theory with $s\geq1$, a scalar with $\Delta=4$ and the hook fields with $s\geq1$.
\begin{equation}
    \zeta^B=\zeta^A+\zeta_{4,0}+\zeta^{Hook}=-\frac{1}{1512}-\frac{37}{7560}+\frac{1}{180}=0\,.
\end{equation}
Below, we will list all the components in terms of their $P$ and $Q$ to calculate the $\zeta'^B$
\begin{equation}
\begin{tabular}{c|c}
     $Type$ & $P$  \\ \hline
     $P^A$ & $\frac{79}{153600}+\frac{3c_1^+}{320}+\frac{c_3^+}{24}+\frac{c_5^+}{60}$ \\
     $P^A_{\frac{3}{2},0}$ & $-\frac{197}{51200} - \frac{3 c_1^+}{320} -\frac{c_3^+}{24} - \frac{c_5^+}{60}$ \\
     $P^{Hook}$ & $\frac{1}{300}$\\
\end{tabular}
\end{equation}
It is easy to recognize that $P^B=P^A+P^A_{\frac{3}{2},0}+P^{Hook}=0$, i.e there is no contribution from $P$ in the Type-B theory. The relevant $Q$-terms are
\begin{equation}
\begin{tabular}{c|c}
     $Type$ & $Q$  \\ \hline
     $Q^A$ & $\frac{1}{120}\left(\frac{1181}{11520} - \frac{211 \log(2)}{4032} - \frac{23 \log A}{16} + \frac{5 \zeta(3)}{
 4 \pi^2} + \frac{15 \zeta(5)}{4 \pi^4} -
 \frac{63}{16} \zeta'(-5) + \frac{35}{8} \zeta'(-3)\right)$ \\
     $Q^A_{\frac{3}{2},0}$ & $\frac{1433}{51200} + \frac{211 \log(2)}{483840} - \frac{99 \log A}{640} + \frac{
 3 \zeta(3)}{32\pi^2} - \frac{3 \zeta(5)}{32 \pi^4}+
 \frac{21}{640} \zeta'(-5)+\frac{19}{64} \zeta'(-3)$ \\
     $Q^{Hook}$ & $-\frac{623}{21600} + \frac{\log A}{6} + \frac{\zeta(5)}{8\pi^4}-\frac{\zeta(3)}{12 \pi^2}-\frac{1}{3}\zeta'(-3)$
\end{tabular}
\end{equation}
Bringing everything together, we obtain
\begin{equation}
    \zeta'_B=\zeta'_{A,s\geq 1} + \zeta'_{Hook, s\geq 1} + \zeta'_{4,0}
    =\frac{\zeta(3)}{48 \pi^2}+\frac{\zeta(5)}{16\pi^4}\,.
\end{equation}
As explaining in the main text, this number is not random.

\subsection{Minimal Type-B}
From \eqref{E.4}, the zeta-function of Type-A with odd spins only is 0. One can read off the minimal Type-B $\zeta^B_{min}$ by considering the symmetric traceless fields with odd spins only, the hook fields with even spin and a scalar with $\Delta=4$.
\begin{equation}
    \zeta^B_{min}=\zeta^A_{odd}+\zeta_{4,0}+\zeta^{Hook}_{even}=0-\frac{37}{7560}+\frac{37}{7560}= 0\,.
\end{equation}
Therefore, the zeta function for Type-B is vanishing in both non-minimal and minimal cases. Next, we list the result for the minimal Type-B in terms of $P$ and $Q$
\begin{equation}
\begin{tabular}{c|c}
     Type & $P$  \\ \hline
     $P^A$ & $0$ \\
     $P^A_{\frac{3}{2},0}$ & $-\frac{197}{51200} - \frac{3 c_1^+}{320} - \frac{c_3^+}{24} - \frac{c_5^+}{60}$ \\
     $P^{Hook}$ & $\frac{197}{51200} + \frac{3 c_1^+}{320} + \frac{c_3^+}{24} + \frac{c_5^+}{60}$
\end{tabular}
\end{equation}
\begin{equation}
\begin{tabular}{c|c}
     Type & $Q$  \\ \hline
     $Q^A$ & $-\frac{\log(2)}{64} - \frac{\zeta(3)}{64 \pi^2} + \frac{15\zeta(5)}{128 \pi^4}$ \\
     $Q^A_{\frac{3}{2},0}$ & $\frac{1433}{51200} + \frac{211 \log(2)}{483840} - \frac{99 \log A}{640} + \frac{
 3 \zeta(3)}{32\pi^2} - \frac{3 \zeta(5)}{32 \pi^4} +
 \frac{21}{640} \zeta'(-5) +\frac{19}{64} \zeta'(-3)$ \\
     $Q^{Hook}$ & $ -\frac{1433}{51200} + \frac{52709 \log(2)}{483840} + \frac{99 \log A}{640} +
\frac{\zeta(3)}{64 \pi^2} - \frac{93 \zeta(5)}{128 \pi^4} -
 \frac{21}{640} \zeta'(-5) - \frac{19}{64} \zeta'(-3)$
\end{tabular}
\end{equation}
The $\zeta'^B_{min}$ for the minimal Type-B theory is just that:
\begin{equation}
    \zeta'^B_{min} =\zeta'_{A,odd} + \zeta'_{Hook, even} + \zeta'_{4,0}
    =\frac{3}{32}\log 2 +\frac{3\zeta(3)}{32\pi^2}-\frac{45\zeta(5)}{64\pi^4}\,.
\end{equation}
In the following appendices, we list the result of zeta function of Type-A, fermions, hook fields and Type-B in various dimensions, which can be used for later work.

\section{Summary of the Results in Other Even Dimensions}

\subsection{Type-A}
We first evaluate the zeta function in term of spin-$s$. Following the algorithm in the Appendix D, the results are listed below
\begin{equation}
\begin{tabular}{c|c}
     $d$ & $\zeta_{\Delta,s}-\zeta_{\Delta+1,s-1}$  \\ \hline
     $3$ & $\frac{1}{180} (-2 + 15 s^2 - 75 s^4)$ \\
     $5$ & $\frac{(1+s)^2(-20+28s+378s^2+868s^3+847s^4+378s^5+63s^6)}{30240}$\\
     $7$ & $\frac{(2+s)^2(-3048+1024s+55568s^2+162632s^2+228337s^4+188892s^5+98397s^6+32688s^7+6723s^8+780s^9+39s^{10})}{21772800}$ 
\end{tabular}
\end{equation}
The sum over spins will make $\zeta(0)$ vanish in both non-minimal and minimal cases.\footnote{We used the cut-off exponential $\exp[-\epsilon(s+\frac{d-3}{2})]$. The case with $d=3$ is special since one should start the sum from $s \geq 1$ and then add the scalar to have vanishing zeta function.} Next, we compute $P_{\nu,s}$ and $Q_{\nu,s}$ \\ \\
\textbf{Table for $P_{\nu,s}$:\footnote{From here, it is very easy to evaluate $P=\sum_s P_{\nu,s}-P_{\nu+1,s-1}$ by the exponential cut-off.}} \\
{\small$
\begin{aligned}
     d=3: \frac{(2s+1)(12 c_1^+ + 48 c_3^+ + 48 c_1^+ s + 48 c_1^+ s^2 + \nu^2 + 6 \nu^4)}{144}
\end{aligned}$}\\ \\
{\small $\begin{aligned}
     d=5: -\frac{(s + 1) (s + 2) (2 s + 3)}{691200} &\Big[1080 c_1^+ + 4800 c_3^+ + 1920 c_5^+ + 1440 c_1^+ s + 5760 c_3^+ s + 480 c_1^+ s^2 +
  1920 c_3^+ s^2 \\
  &+ 107 \nu^2 + 120 s \nu^2 + 40 s^2 \nu^2 + 580 \nu^4 +
  720 s \nu^4 + 240 s^2 \nu^4 - 240 \nu^6\Big]
\end{aligned}$}\\ \\
{\small$\begin{aligned} d=7:&\frac{(1 + s) (2 + s) (3 + s) (4 + s) (5 + 2 s) }{48771072000}\Big[567000 c_1^+ + 2610720 c_3^+ + 1411200 c_5^+ + 161280 c_7^+ + 453600 c_1^+ s \\
&+ 2016000 c_3^+ s + 806400 c_5^+ s + 90720 c_1^+ s^2 + 403200 c_3^+ s^2 +
 161280 c_5^+ s^2 + 343345 \nu^2 + 271740 s \nu^2 \\&+ 54348 s^2 \nu^2 -
 667674 \nu^4 - 512400 s \nu^4 - 102480 s^2 \nu^4 + 255920 \nu^6 +
 145600 s \nu^6 + 29120 s^2 \nu^6 - 23520 \nu^8\Big]
    \end{aligned}$}\\ \\
\textbf{Table of $Q_{\nu,s}$:}
\begin{equation}
\begin{tabular}{c|c}
     $d$ & $Q_{\nu,s}$  \\ \hline
     $3$ &  $\frac{1}{3}(2s+1)\int_0^{\nu}dx \left[(s+\frac{1}{2})^2x-x^3\right]\psi(x+\frac{1}{2})$\\
     $5$ & $\frac{(s+1)(s+2)(2s+3)}{5760}\int_0^{\nu} x (9 + 12 s + 4 s^2 - 4 x^2) (-1 + 4 x^2) \psi(1/2 + x)$\\
     $7$ & $\frac{(s+1)(s+2)(s+3)(s+4)(2s+5)}{604800}\int_0^{\nu}dx \frac{x}{32} (25 + 20 s + 4 s^2 - 4 x^2) (9 - 40 x^2 + 16 x^4) \psi(
   x+\frac{1}{2})$
\end{tabular}
\end{equation}
\paragraph{Non-minimal Type-A.}
The result for $P$ in both non-minimal and minimal theory are zero, i.e $P$ vanishes. Hence, one only needs to deal with $Q=\sum_s Q_{\nu,s}-Q_{\nu+1,s-1}$. The sum is evaluated with $\text{exp}[-\epsilon \nu]$ for $Q_{\nu,s}$ and with $\text{exp}[-\epsilon (\nu+1)]$ for $Q_{\nu+1,s-1}$. Analytical computation in the non-minimal Type-A shows that $Q$ also vanishes.

\paragraph{Minimal Type-A.}
In minimal theory, the story is a little bit different. Using the method of analytical continuation of Appendix D, we get
\begin{equation}
\begin{tabular}{c|c}
     $d$ & $Q$  \\ \hline
     $3$ &  $-\frac{1}{2^3}\left(2\log 2 - \frac{3\zeta(3)}{\pi^2}\right)$\\
     $5$ & $\frac{1}{2^7}\left(2\log 2 + \frac{2\zeta(3)}{\pi^2}-\frac{15\zeta(5)}{\pi^4}\right)$\\
     $7$ & $-\frac{1}{2^{11}}\left(4 \log 2 + \frac{82\zeta(3)}{15\pi^2}   -\frac{10\zeta(5)}{\pi^4} - \frac{63\zeta(7)}{\pi^6}\right)$
\end{tabular}
\end{equation}
These results can also be found in \cite{Giombi:2014iua,Klebanov:2011gs}.

\subsection{HS Fermions}
Above, we showed that $\zeta_{\frac{1}{2}}$ and $\zeta'_{\frac{1}{2}}$ is zero for $AdS_6$. In this Appendix, let us rewrite the result in $d=3,5$ and then make a general statement about higher dimensional cases. First of all, one needs to make the change of variable $s=m+\frac{1}{2}$. The zeta-functions with the ghost subtracted are
\begin{equation}
\begin{tabular}{c|c}
     $d$ & $\zeta_{\Delta,s}-\zeta_{\Delta+1,s-1}$  \\ \hline
     $3$ & $\frac{-47 - 360 m - 1560 m^2 - 2400 m^3 - 1200 m^4}{2880}$ \\
     $5$ & $\frac{542 + 99 m - 8094 m^2 - 22806 m^3 - 28497 m^4 - 19404 m^5 -
   7448 m^6 - 1512 m^7 - 126 m^8}{30240}$
\end{tabular}
\end{equation}
Summing over all spin starting from $m=0$ with the cut-off $\exp[-\epsilon(m+\frac{d-2}{2})]$, we see that the total zeta-functions in $d=3,5$ vanished. As a simple check, one can confirm that for higher dimensions this statement is also true.\\
Next, to calculate the $\zeta'$-function, we again split it into $P_{\nu,m}$ and $Q_{\nu,m}$.
\paragraph{Table for $P_{\nu,m}$:}
{\footnotesize
\begin{align}
d=3&:&&\begin{aligned}
      -\frac{(1 + m) (24 c_1^- + 24 c_3^- + 48 c_1^- m + 24 c_1^- m^2 - 12 \nu -
   24 m \nu - 12 m^2 \nu + \nu^2 + 4 \nu^3 - 3 \nu^4)}{36}
\end{aligned}\,,\\
d=5&:&&\begin{aligned}
     &-\frac{(1 + m) (2 + m) (3 + m)}{21600} \Big[960 c_1^- + 1200 c_3^- + 240 c_5^- +
    960 c_1^- m + 960 c_3^- m + 240 c_1^- m^2 + 240 c_3^- m^2 \\&- 480 \nu - 480 m \nu -
    120 m^2 \nu + 51 \nu^2 + 40 m \nu^2 + 10 m^2 \nu^2 + 200 \nu^3 +
    160 m \nu^3 + 40 m^2 \nu^3 - 155 \nu^4 \\&- 120 m \nu^4 - 30 m^2 \nu^4 -
    24 \nu^5 + 30 \nu^6\Big]\,.
\end{aligned}\end{align}}\noindent
Summing over all spins leads to
\begin{center}
\begin{tabular}{c|c}
     $d$ & $P$  \\ \hline
     $3$ & $-\frac{11}{270}$ \\
     $5$ & $\frac{1787}{3402000}$\\
\end{tabular}
\end{center}
One can see that for fermions $P$ is non-zero which is different from Type-A theories. For $Q_{\nu,m}$ we get
\begin{center}
\begin{tabular}{c|c|c}
     $d$ & $Q_{\nu,m}$&$Q$  \\ \hline
     $3$ &$-\frac{2(m+1)}{3}\int_0^{\nu} dx (x^3-(m+1)^2x)$ &$\frac{11}{270}$\\
     $5$ &$\frac{(m+1)(m+2)(m+3)}{90}\int_0^{\nu}dx(x^3-x)(x^2-(m+2)^2)\psi(x)$ & $-\frac{1787}{3402000}$\\
\end{tabular}
\end{center}
It is easy to see that $P$ and $Q$ always cancel each other. A further check confirms that $\zeta'(0)$ is zero in higher  dimensions.
\subsection{Hook fields}
The hook fields only appear in dimensions higher than four. For the computation of the spectral density function $\mu(u)$ of hooks with different $p$, the reader can refer to Section 3.2.2.
\subsubsection{Zeta}
In $d=5$, we only have $p=1$, while in $d=7$, $p$ can be one or two.\footnote{Due to the length of the final results, we only list the zeta function for $d=5,7$ here.}
{\footnotesize\begin{align*}
d=5\,,p=1&:&&\begin{aligned}
      \frac{148 - 1044 s - 6654 s^2 - 11844 s^3 - 7077 s^4 + 1764 s^5 +
 3822 s^6 + 1512 s^7 + 189 s^8}{30240}
\end{aligned}\,, \\
d=7\,,p=1&:&&\begin{aligned}
      &-\frac{(2 + s)}{5573836800} \Big[-81336637326 - 260554380359 s -
    287920256390 s^2 - 124396596105 s^3 \\
    &+ 7147903040 s^4 +
    30702694976 s^5 + 14557085760 s^6 + 3622437600 s^7 +
    540003840 s^8 \\&+ 48318720 s^9 + 2388480 s^{10} + 49920 s^{11}\Big]
\end{aligned}\,,\\
d=7\,,p=2&:&&\begin{aligned}
      &-\frac{
 s (4 + s)}{2786918400} \Big[-79449809509 - 151977792308 s - 101475411753 s^2 -
    17276191808 s^3 \\&+ 13378662464 s^4 + 9277153920 s^5 +
    2721896160 s^6 + 451660800 s^7 + 43687680 s^8 \\&+ 2288640 s^9 +
    49920 s^{10}\Big]\,.
\end{aligned}\end{align*}}\noindent
We will list the result of $\zeta$-function in both the non-minimal and minimal theory for hook fields below since it is important for our computation of Type-B theory\footnote{The hook fields of minimal theory in $d=5$ come with even spins while the hook fields with $p=1$ in $d=7$ come with odd spins and $p=2$ come with even spins.}
\begin{equation}
\begin{tabular}{c|c|c}
     $d$ & $p$ &$(\zeta, \zeta_{min})$  \\ \hline
     $5$ &$1$ &$\left(\frac{1}{180},-\frac{37}{7560}\right)$ \\
     $7$ & $1$ &$\left(\frac{1}{280}, -\frac{23}{226800} \right)$\\
     & $2$ &$\left(\frac{1}{1512},\frac{23}{226800} \right)$
\end{tabular}
\end{equation}
It is interesting that the zeta function for hook fields alone is not zero as in bosonic and fermionic theory. However, when one considers the whole spectrum of Type-B theory, the zeta function will again vanish.
\subsubsection{Zeta-prime}
Below are the tables for $P_{\nu,s}$ and $Q_{\nu,s}$ of hook fields.
\paragraph{Table for $P_{\nu,s}$:}
{\footnotesize\begin{align*}
d=5\,,p=1&:&&\begin{aligned}
      &-\frac{
 s (3 + s) (3 + 2 s)}{230400} \Big[9720 c_1^+ + 8640 c_3^+ + 1920 c_5^+ + 12960 c_1^+ s +
    5760 c_3^+ s + 4320 c_1^+ s^2 \\&+ 1920 c_3^+ s^2 + 187 \nu^2 + 120 s \nu^2 +
    40 s^2 \nu^2 + 1060 \nu^4 + 720 s \nu^4 + 240 s^2 \nu^4 - 240 \nu^6\Big]
\end{aligned}\,,\\
d=7\,,p=1&:&&\begin{aligned}
      &\ \ \frac{s (2 + s) (3 + s) (5 + s) (5 + 2 s)}{9754214400} \Big[1575000 c_1^+ +
   6804000 c_3^+ + 2056320 c5 + 161280 c_7^+ + 1260000 c_1^+ s \\&+
   5241600 c_3^+ s + 806400 c_5^+ s + 252000 c_1^+ s^2 + 1048320 c_3^+ s^2 +
   161280 c_5^+ s^2 + 149557 \nu^2 + 112140 s \nu^2 \\&+ 22428 s^2 \nu^2 +
   828786 \nu^4 + 646800 s \nu^4 + 129360 s^2 \nu^4 - 255920 \nu^6 -
   100800 s \nu^6 - 20160 s^2 \nu^6 + 18480 \nu^8\Big] \end{aligned}\,,\\
d=7\,,p=2&:&&\begin{aligned}
      &\ \ \frac{s (1 + s) (4 + s) (5 + s) (5 + 2 s)}{4877107200} \Big[14175000 c_1^+ +
   10836000 c_3^+ + 2378880 c_5^+ + 161280 c_7^+ + 11340000 c_1^+ s \\&+
   6854400 c_3^+ s + 806400 c_5^+ s + 2268000 c_1^+ s^2 + 1370880 c_3^+ s^2 +
   161280 c_5^+ s^2 + 234733 \nu^2 + 145740 s \nu^2 \\&+ 29148 s^2 \nu^2 +
   1329426 \nu^4 + 848400 s \nu^4 + 169680 s^2 \nu^4 - 296240 \nu^6 -
   100800 s \nu^6 - 20160 s^2 \nu^6 + 18480 \nu^8\Big]\,.
\end{aligned}\end{align*}}\noindent
Summing over spins leads to
\begin{equation}
\begin{tabular}{c|c|c}
     $d$ & $p$ &$(P,P_{min})$  \\ \hline
     $5$ &$1$ &$\left(\frac{1}{300},\frac{197}{51200} + \frac{3 c_1^+}{320} + \frac{c_3^+}{24} + \frac{c_5^+}{60}\right)$ \\
     $7$ & $1$ &$\left(\frac{1361}{264600}, \frac{508061}{6502809600} + \frac{5 c_1^+}{3584} + \frac{37 c_3^+}{5760} + \frac{c_5^+}{288} + \frac{c_7^+}{2520}\right)$\\
     & $2$ &$\left(\frac{61}{158760},-\frac{508061}{6502809600} - \frac{5 c_1^+}{3584} - \frac{37 c_3^+}{5760} - \frac{c_5^+}{288} - \frac{c_7^+}{2520} \right)$
\end{tabular}
\end{equation}
\paragraph{Table for $Q_{\nu,s}$:}
{\footnotesize \begin{align*}
d=5\,,p=1&:&&\begin{aligned}
      &-\frac{s(s+3)(2s+3)}{1920}\int_0^{\nu} dx\ x(-9+4x^2)(-9-12s-4s^2+4x^2)\psi(x+1/2)
\end{aligned}\,, \\
d=7\,,p=1&:&&\begin{aligned}
      &\ \ \frac{s(s+2)(s+3)(s+5)(2s+5)}{120960}\int_0^{\nu} dx \frac{x}{32}(25+20s+4s^2-4x^2)(25-104x^2+16x^4)\psi(x+\frac{1}{2}) \end{aligned}\,,\\
d=7\,,p=2&:&&\begin{aligned}
      &\ \ \frac{s (s + 1) (s + 4) (s + 5) (2 s +
   5)}{60480}\int_0^{\nu}dx\frac{x}{32} (25 + 20 s + 4 s^2 - 4 x^2) (225 - 136 x^2 + 16 x^4)\psi(x+\frac{1}{2})\,.
\end{aligned}\end{align*}}
\paragraph{Non-minimal Type-B.}
Following the method in appendix D, we list the results of $Q$ in $d=5,7$.
\begin{equation}
\begin{tabular}{c|c|c}
     $d$ & $p$ &$Q$  \\ \hline
     $5$ & $1$& $-\frac{623}{21600} +\frac{\log A}{6} + \frac{\zeta(5)}{8\pi^4}-\frac{\zeta(3)}{12 \pi^2}-\frac{\zeta'(-3)}{3}$\\
     $7$ &$1$ &$-\frac{26777}{1058400} + \frac{7 \log A}{60} -\frac{113\zeta(3)}{1440\pi^2}+\frac{13\zeta(5)}{96\pi^4} -\frac{\zeta(7)}{32\pi^6}-\frac{\zeta'(-3)}{3}-\frac{\zeta'(-5)}{20} $ \\
     & $2$ & $-\frac{991}{317520} + \frac{\log A}{60}-\frac{7\zeta(3)}{1440\pi^2}-\frac{\zeta(5)}{96\pi^4}+\frac{\zeta(7)}{32\pi^6}+\frac{\zeta'(-5)}{60}  $
\end{tabular}
\end{equation}

\paragraph{Minimal Type-B.} In the minimal theory, the computations are much longer since there are more derivatives involved when one calculates the Hurwitz-Lersch functions.
\begin{equation}\notag
\begin{tabular}{c|c|c}
     $d$ & $p$ &$Q$  \\ \hline
     $5$ & $1$&  {\footnotesize${-\frac{1433}{51200} + \frac{52709 \log(2)}{483840} + \frac{99 \log A}{640} +
\frac{\zeta(3)}{64 \pi^2} - \frac{93 \zeta(5)}{128 \pi^4} -
 \frac{21\zeta'(-5)}{640}  - \frac{19\zeta'(-3)}{64} }$}\\
     $7$ &$1$ &{\footnotesize$\frac{2545 \log (A)}{21504}+\frac{535 \zeta '(-4)}{2304}+\frac{4787 \zeta '(-2)}{11520}-\frac{139 \zeta '(-5)}{3072}-\frac{1037 \zeta '(-3)}{3072}-\frac{487 \zeta '(-6)}{11520}-\frac{17 \zeta '(-7)}{21504}-\frac{6610955}{260112384}-\frac{4067243 \log (2)}{232243200} $} \\
     & $2$ & {\footnotesize$\frac{181 \log (A)}{107520}+\frac{73 \zeta '(-5)}{15360}+\frac{113 \zeta '(-4)}{1152}+\frac{389 \zeta '(-2)}{1152}-\frac{13 \zeta '(-3)}{3072}-\frac{17 \zeta '(-7)}{21504}+\frac{1205 \zeta (7)}{1024 \pi ^6}-\frac{755987}{6502809600}-\frac{13592843 \log (2)}{232243200} $}
\end{tabular}
\end{equation}

\subsection{Type-B}
We can now combine the results above to get the results for Type-B models. The spectrum of such models is given in Section \ref{sec:mixedsymmetrytestsE}.
\subsubsection{Non-minimal}
\paragraph{Scalar Field.}
The scalar in Type-B has $\Delta_B^{\phi}=\Delta_A^{\phi}+1$, where $\Delta_A^{\phi}$ is the conformal weight of the scalar in Type-A theory. One can use this to compute $\zeta,P,Q$ using all the formulas in Type-A:
\begin{equation}\notag
\begin{tabular}{c|c}
     $d$ & $\zeta_{\Delta_B,0}$  \\ \hline
     $5$ &$ -\frac{37}{7560}$\\
     $7$ & $-\frac{119}{32400}$
\end{tabular}
\qquad \qquad \qquad
\begin{tabular}{c|c}
     $d$ & $P^{\phi}$  \\ \hline
     $5$ & $-\frac{197}{51200} - \frac{3 c_1^+}{320} -\frac{c_2^+}{24} - \frac{c_5^+}{60}$\\
     $7$ &  $-\frac{1317595}{260112384}+\frac{5c_1^+}{3584}+\frac{37c_3^+}{5760}+\frac{c_5^+}{288}+\frac{c_7^+}{2520}$
\end{tabular}
\end{equation}
\begin{equation}\notag
\begin{tabular}{c|c}
     $d$ & $Q^{\phi}$  \\ \hline
     $5$ & $\frac{1433}{51200} + \frac{211 \log(2)}{483840} - \frac{99 \log A}{640} + \frac{
 3 \zeta(3)}{32\pi^2} - \frac{3 \zeta(5)}{32 \pi^4}+
 \frac{21\zeta'(-5)}{640} +\frac{19 \zeta'(-3)}{64}$\\
     $7$ &  $\frac{6610955}{260112384}-\frac{15157\log(2)}{232243200}-\frac{2545\log A}{21504}+\frac{23\zeta(3)}{288\pi^2}-\frac{25\zeta(5)}{192\pi^4}+\frac{5\zeta(7)}{128\pi^6}+\frac{1037\zeta'(-3)}{3072}+\frac{139\zeta'(-5)}{3072}+\frac{17\zeta'(-7)}{21504}$
\end{tabular}
\end{equation}

\paragraph{Summary.}
    In non-minimal Type-B theory, we have one scalar with $\Delta_B=\Delta_A+1$, Type-A with $s\geq 1$, and the hook fields with $s\geq 1$. The total contribution to the zeta-function gives zero
\begin{center}
\begin{tabular}{c|c|c}
     $d$ & $\zeta_A + \zeta_{Hook} + \zeta^{\phi}_{\Delta,s}$ & $\zeta_B$ \\ \hline
     $5$ & $-\frac{1}{1512}+\frac{1}{180}-\frac{37}{7560}$ & $0$\\
     $7$ & $ -\frac{127}{226800}+\frac{1}{280}+ \frac{1}{1512}-\frac{119}{32400}$ & $0$ \\
\end{tabular}
\end{center}
For higher dimensions, this is also true and we can confirm that the zeta-function for non-minimal Type-B is always zero by combining all the component fields. Next, we need $\zeta'_B=\zeta'_{\Delta_B,0}+\zeta'_{A,s\geq1}+\zeta'_{Hook}$:
\begin{equation}
\begin{tabular}{c|c}
     $d$ & $\zeta'_B$  \\ \hline
     $5$ & $\frac{\zeta(3)}{48 \pi^2}+\frac{\zeta(5)}{16\pi^4}$\\
     $7$ &  $\frac{\zeta(3)}{360 \pi^2} + \frac{\zeta(5)}{96 \pi^4} +
 \frac{\zeta(7)}{64 \pi^6}$
\end{tabular}
\end{equation}
In the main text, our results were generated up to $AdS_{12}$ or $d=11$, but we checked up to $AdS_{18}$ that they agree with the change of $F$-energy.

\subsubsection{Minimal}
We need to combine the scalar field from the previous sub-section with the results for odd/even spins that can be found above. The final results can be found in the main text.

\chapter{Appendix for One-loop Tests in Fractional Dimensions}\label{app:chap35}
\section{From Intermediate to Final Form}
\label{app:Proof}
As a result of the $\ads$ computation we arrived at the intermediate form (\ref{eq:intermediatestep}), which can easily be seen to arise in the computation of the determinant on the CFT side. Let us now show how to reach the (generalized) sphere free energy $F_{\phi}$ in its final form. In order to compute the $\beta$-integral we use 
\begin{equation}
    \frac{1}{\beta}=\frac{1}{2}\left(\frac{1}{1-e^{-\beta}}\int_0^1 du e^{-u\beta} - \frac{1}{1-e^{\beta}}\int_0^1 du e^{u \beta} \right)\,.
\end{equation}
This allows for an analytic evaluation of the $\beta$ integral. One obtains
\begin{equation}\label{eq:doublepair}
    F_{\text{min.}}^{\phi} = \frac{\Gamma(-d)}{4} \int_0^1 du (d + 4 (-1 + u) u)\left(\frac{\Gamma\left(-1+\frac{d}{2}+u\right)}{\Gamma\left(1-\frac{d}{2}+u\right)}+\frac{\Gamma\left(\frac{d}{2}-u\right)}{\Gamma\left(2-\frac{d}{2}-u\right)} \right)
\end{equation}
After some straightforward algebra (\ref{eq:doublepair}) can be shown to split in two parts, the first one we can bring to the form of (for $\Delta=\frac{d}{2}-1$) \cite{Diaz:2007an,Giombi:2014xxa}:
\begin{equation}
\begin{split}
    F_{\Delta} &= \Gamma(-d) \int_0^{\Delta-\frac{d}{2}} du u \left[\frac{\Gamma(\frac{d}{2}-u)}{\Gamma(1-u-\frac{d}{2})}-\frac{\Gamma(\frac{d}{2}+u)}{\Gamma(1+u-\frac{d}{2})}\right] \\
    &= - \frac{1}{\sin(\frac{\pi d}{2})\Gamma(d+1)}\int_0^{\Delta-\frac{d}{2}} du u \sin(\pi u ) \Gamma\left(\frac{d}{2}+u\right)\Gamma\left(\frac{d}{2}-u\right)\,,
    \end{split}
\end{equation}
where the result for the free scalar field corresponds to $\Delta=\frac{d}{2}-1$. The second part has the form 
\begin{equation}
    \mho  = \frac{1}{4\Gamma(d)\sin\left(\frac{\pi d}{2}\right)}\int_0^1 (1-2u) \sin(\pi u ) \Gamma\left(-1+\frac{d}{2}+u\right)\Gamma\left(\frac{d}{2}-u\right) \,.
\end{equation}
However, this extra term vanishes due to the anti-symmetry of the integrand around $u=1/2$. This shows that
\begin{equation}
    F_{\text{min.}}^{\phi} = \frac{1}{2}\int_0^{\infty} \frac{d\beta}{\beta}\frac{e^{-\beta(2+d)/2} (1+e^{\beta})^2}{(1-e^{-\beta})^d} = \frac{-1}{\Gamma(d+1)\sin\left(\frac{\pi d}{2}\right)} \int_0^1 du\, u \sin(\pi u )\Gamma\left(\frac{d}{2}-u\right)\Gamma\left(\frac{d}{2}+u\right)\notag
\end{equation}
\section{Modified Zeta Function}
\label{app:modified}
In this Appendix we elaborate on the properties of the modified zeta-function we introduced in Section \ref{sec:TypeA}. It follows from the definition that the value of $\zeta_{\Delta,s}(0)$ is unaffected, which is illustrated in \ref{app:zeta}. The value of $\zeta'_{\Delta,s}(0)$ differs in general from its true value. Fortunately, $\zeta'(0)$ is still the same for  for the spectrum of (non)-minimal Type-A, which is studied in \ref{app:deficit}. It is also shown there that there is no deficit for the difference between the scalars with $\Delta=d-2$ and $\Delta=2$ boundary conditions.

\subsection{Zeta}
\label{app:zeta}
From (\ref{eq:modzetaA}), one can easily obtain the full zeta in various odd dimensions with the help of analytical continuation to the Lerch transcendent and then set $z\rightarrow0$. For example,
{\footnotesize
\begin{align*}
    d=3&: &&\tilde{\zeta}_{\nu,s}=\frac{(2s+1)(-17-40s-40s^2+240\nu^4-120(\nu+2s \nu)^2)}{5760}\\
    d=5&: &&\tilde{\zeta}_{\nu,s}=-\frac{(1+s)(2+s)(3+2s)(-1835-2142s-714s^2-1260(3+2s)^2\nu^2+5040(5+2s(s+3))\nu^4-6720\nu^6)}{29030400}
\end{align*}}
It is easy to see that these polynomials in $\nu$ and $s$ are exactly the zeta function for Type-A in \cite{Giombi:2014iua}, see also \cite{Gunaydin:2016amv}. Therefore, there is no deficit at $z^0$ order, i.e
\begin{equation}
   \tilde{\zeta}_{\nu,s}(0) -  \zeta_{\nu,s}(0)=0 + \mathcal{O}(z)\,.
\end{equation}
This explains how we can get all the correct $\tilde{\zeta}_{d,s}(0)$ for individual spins in general odd dimensions.  There are many results on zeta-function at $d=3$, see e.g. \cite{Camporesi:1993mz,Giombi:2013fka,Giombi:2014iua}. Let us illustrate that the modified zeta-function is solid enough to obtain these results. The spin factor in $d=3$ is 
\begin{equation}
    g^A_3(s)= 2s+1
\end{equation}
Together with $\nu=s-\frac{1}{2}$, \eqref{eq:modzetaA} becomes
\begin{equation*}
    \tilde{\zeta}^A_3(z) = -\frac{(2s+1)}{3!\Gamma(2z)} \int_0^{\infty} d\beta \sum_{l=0}^{\infty} e^{-\beta(s-\frac{1}{2})}\beta^{2z-1} e^{-\beta\left(\frac{1}{2}+l\right)}\left(\frac{1}{2}+l\right)
    \left(\left(\frac{1}{2}+s\right)^2-\left(\frac{1}{2}+l\right)^2\right)\,.
\end{equation*}
Now we can sum over $l$ and obtain
\begin{equation*}
    \tilde{\zeta}^A_{3,s}(z)=\frac{1}{12\Gamma(2z)}\int_0^{\infty} d\beta \frac{\beta^{2z-1} e^{-\beta(s-1)} (1 + e^{\beta}) (1 + 2 s) (s (1 + s) + e^{2 \beta} s (1 + s) - 
    2 e^{\beta} (3 + s + s^2))}{(-1 + e^{\beta})^4}
\end{equation*}
In order to get to the actual numbers one needs to plug $s=0,1,2,3,...$ then use the trick of analytical continuation via the Hurwitz-Lerch zeta function \cite{Giombi:2013fka,Giombi:2014iua}. For example,
\begin{equation}
    \tilde{\zeta}^A_{3,s}(0)=\left\{-\frac{1}{180},-\frac{11}{60},-\frac{181}{36},-\frac{6097}{180},...\right\}
\end{equation}
Note that, after the continuation to the Hurwitz-Lerch transcendent, there will be another $\Gamma(2z)$ function in the nominator. This will cancel $1/\Gamma(2z)$ factor in the modified zeta function. Therefore, the modified zeta-function reproduces the correct result, which is expected.

\subsection{Deficit}
\label{app:deficit}
As we already explained in Section \ref{sec:TypeA}, we changed the regularization prescription. As a result the values of $\zeta'_{\Delta,s}(0)$ might be different from the correct ones for individual fields. It was noted in \cite{Bae:2016rgm} that the deficit vanishes for certain representations (with even character). In particular, the deficit is absent for (non)-minimal Type-A theory. The purpose of this Section is to quantify the deficit for a number of cases.

For example, let us take the scalar field in $d=3$. The zeta-prime can be derived by calculating $\zeta(z)$ at $z$ order:
\begin{equation}\label{eq:deficit1}
    \small \zeta_{3,0}(z)=\frac{\zeta(-3+2z)}{6}+\frac{\zeta(-2+2z)}{4}+\frac{\zeta(-1+2z)}{12} = -\frac{1}{180}+\left(\frac{1}{72}-\frac{\log A}{6}+\frac{\zeta'(-3)}{3}+\frac{\zeta'(-2)}{2}\right)z + \mathcal{O}(z^2) 
\end{equation}
One can already notice that there is a deficit between the value of $\tilde{\zeta}'^A_{3,0}(0)$ that is evaluated by the standard zeta function and (\ref{eq:deficit1}). This was also discussed in Appendix (B.1) of \cite{Bae:2016rgm}, when the authors use characters to evaluate $\tilde{\zeta}'(0)$ for different fields. Let us have a look at the deficit in $d=3$ and $d=5$ as to observe the general pattern.
\subsubsection{d=3}
The result before sending $z$ to $0$ for the modified zeta function is
\begin{equation}
\begin{split}
   \tilde{\zeta}^3_{\nu,s}(z)= &\frac{(2s+1)}{24}\Big[\nu\left((1+2s)^2-4\nu^2\right)\zeta(2z,\nu+\frac{1}{2})+4\zeta(-3+2z,\nu+\frac{1}{2})\\
    &-12\nu \zeta(-2+2z,\nu+\frac{1}{2})+(-1-4s(1+s)+12\nu^2)\zeta(-1+2z,\nu+\frac{1}{2})\Big]\,.
    \end{split}
\end{equation}
In order to compute the zeta-prime, one just needs to take the $z$ derivative and set $z=0$:
\begin{equation}
    \begin{split}
   \tilde{\zeta}'^3_{\nu,s}(0)= &\frac{(2s+1)}{12}\Big[\nu\left((1+2s)^2-4\nu^2\right)\zeta'(0,\nu+\frac{1}{2})+4\zeta'(-3,\nu+\frac{1}{2})\\
    &-12\nu \zeta'(-2,\nu+\frac{1}{2})+(-1-4s(1+s)+12\nu^2)\zeta'(-1,\nu+\frac{1}{2})\Big]\,.
    \end{split}
\end{equation}
We then follow the procedure in \cite{Bae:2016rgm} to find the deficit. First, we set $\nu=0$ and obtain
\begin{equation}
    \tilde{\zeta}'^3_{0,s}(0)=\frac{(2s+1)}{12}\left(4\zeta'(-3,\frac{1}{2})-(2s+1)^2\zeta'(-1,\frac{1}{2})\right)\,.
\end{equation}
Recall that for the standard zeta-prime in $d=3$, see \cite{Giombi:2013fka,Giombi:2014iua}, we have
\begin{equation}
    \zeta'^3_{0,s}(0)=\frac{2s+1}{3}\left(c_3 + \left(s+\frac{1}{2}\right)^2c_1\right)\,.
\end{equation}
We note that 
\begin{equation}\label{eq:identity1}
    \zeta'(-n,\frac{1}{2})=(-)^{\frac{n+1}{2}}c_n, \qquad \text{where} \quad c_n=\int_0^{\infty}du \frac{2u^n \log u}{e^{2\pi u}+1}\,.
\end{equation}
Therefore, $\zeta'^3_{0,s}(0)$ and $\widetilde{\zeta}'^3_{0,s}(0)$ do match. Then, we consider the $\nu$ derivatives for each of the zetas:
\begin{equation}
\begin{split}
    \partial_{\nu}\tilde{\zeta}'^3_{\nu,s}(0)=&\frac{(2s+1)}{12}\Bigg(\left((2s+1)^2-12\nu^2\right)\zeta'(0,\nu+\frac{1}{2})-12\zeta'(-2,\nu+\frac{1}{2})+24\nu \zeta'(-1,\nu+\frac{1}{2})\\
    &+\nu((2s+1)^2-4\nu^2)\partial_{\nu}\zeta'(0,\nu+\frac{1}{2})+4\partial_{\nu}\zeta'(-3,\nu+\frac{1}{2})-12\nu\partial_{\nu}\zeta'(-2,\nu+\frac{1}{2})\\
    &+(-1-4s(s+1)+12\nu^2)\partial_{\nu}\zeta'(-1,\nu+\frac{1}{2})+\nu((2s+1)^2-4\nu^2)\partial_{\nu}\zeta'(0,\nu+\frac{1}{2})\Bigg)
    \end{split}
\end{equation}
\begin{equation}\label{zetadev3}
    \partial_{\nu}\zeta'^3_{\nu,s}(0)=\frac{(2s+1)}{3}\left(\frac{\nu^3}{2}+\frac{\nu}{24}+\nu\left(\left(s+\frac{1}{2}\right)^2-\nu^2\right) \psi(\nu+\frac{1}{2})\right)
\end{equation}
Next, using the identities for Hurwitz zeta function
\begin{equation}\label{eq:identity2}
    \partial_{\nu}\zeta(s,\nu)=-s\zeta(s+1,\nu), \qquad \partial_{\nu}\zeta'(0,\nu)=\psi(\nu)
\end{equation}
we can reduce the $\nu$ derivative of the modified zeta-prime to
\begin{equation}\label{modzetadev3}
     \partial_{\nu}\tilde{\zeta}'^3_{\nu,s}(0)=\frac{(2s+1)\nu((2s+1)^2-4\nu^2)\psi(\nu+\frac{1}{2})}{12}
\end{equation}
Subtracting (\ref{modzetadev3}) and (\ref{zetadev3}) together, then integrating over $\nu$, one obtains the deficit for individual fields at order $z$:
\begin{equation}
    \delta \zeta'_{\nu,s}(0)=\tilde{\zeta}'_{\nu,s}-\zeta'_{\nu,s}=-\frac{(2s+1)(\nu^2+6\nu^4)}{144}
\end{equation}
Since the deficit is an even function of $\nu$, we can compute the difference between the scalars with $\Delta=d-2,2$ boundary conditions using the modified zeta function thanks to  $\delta\zeta'_{d-2,0}-\delta\zeta'_{2,0}=0$.
Using the cut-off $e^{-\epsilon(s+\frac{d-3}{2})}$, one can sum over either all spins or even spins and observe that the deficit does vanish:
\begin{equation}
    \sum_{s} \delta \zeta'_{\nu,s}(0)=0\,.
\end{equation}
Therefore, the deficit is absent both for the non-minimal and minimal Type-A theories at order $z$, which is what we need for $\zeta'_{\text{HS}}(0)$. 
\subsubsection{d=5}
In higher dimensions, there is another useful identity that we illustrate on the example of $d=5$. Following the procedure outlined above, we obtain
{\allowdisplaybreaks{\begin{equation*}
\begin{split}
    \small\tilde{\zeta}'^5_{\nu,s}(0)=&\frac{(1 + s) (2 + s) (3 + 2 s)}{5760}\Bigg[-16\zeta'(-5,\nu+\frac{3}{2})+8 \nu (-3 (5 + 2 s (3 + s)) + 20 \nu^2)\zeta'(-2,\nu+\frac{3}{2})\\
    &+80\nu \zeta'(-4,\nu+\frac{3}{2})+(-(3 + 2 s)^2 + 24 (5 + 2 s (3 + s)) \nu^2 - 80 \nu^4)\zeta'(-1,\nu+\frac{3}{2})\\
    &+\nu (-1 + 4 \nu^2) (-9 - 4 s (3 + s) + 4 \nu^2)\zeta'(0,\nu+\frac{3}{2})+8 (5 + 2 s (3 + s) - 20 \nu^2)\zeta'(-3,\nu+\frac{3}{2})\Bigg]\,.
    \end{split}
\end{equation*}}}%
Setting $\nu=0$ we arrive at 
\begin{equation*}
\begin{split}
    \tilde{\zeta}'^5_{0,s}(0)=&\frac{(1 + s) (2 + s) (3 + 2 s)}{5760}\Bigg[-16\zeta'(-5,\frac{3}{2})+8 (5 + 2 s (3 + s))\zeta'(-3,\frac{3}{2})-(3 + 2 s)^2\zeta'(-1,\frac{3}{2})\Bigg]\,.
    \end{split}
\end{equation*}
We massage the formula above as to be able to compare $\zeta'(-k,\tfrac12)$ with $c_n$, which can be done with the help of
\begin{equation}\label{eq:identity3}
    \zeta(s,\nu)=\zeta(s,\nu+m)+\sum_{n=0}^{m-1}\frac{1}{(n+\nu)^s}
\end{equation}
We arrive at
\begin{equation}\label{modzetaprime5}
\begin{split}
    \tilde{\zeta}'^5_{0,s}(0)=&-\frac{(1 + s) (2 + s) (3 + 2 s)}{5760}\Bigg[-16\zeta'(-5,\frac{1}{2})+8 (5 + 2 s (3 + s))\zeta'(-3,\frac{1}{2})-(3 + 2 s)^2\zeta'(-1,\frac{1}{2})\Bigg]\,,
    \end{split}
\end{equation}
which can be compared with the standard zeta-prime: \begin{equation}\label{zetaprime5}
    \zeta'^5_{0,s}(0)=-\frac{(1 + s) (2 + s) (3 + 2 s)}{360}\left(c_5+c_3\left(\frac{1}{4}+\left(s+\frac{3}{2}\right)^2\right)+\frac{c_1}{4}\left(s+\frac{3}{2}\right)^2\right)\,.
\end{equation}
Using the identity (\ref{eq:identity1}), it is easy to realize that  (\ref{modzetaprime5}) and (\ref{zetaprime5}) are the same. Next, one can proceed as in the previous Section and get
\begin{equation}
    \delta \zeta'^5_{\nu,s}=-\frac{(s+1)(s+2)(2s+3)\nu^2(107+580\nu^2-240\nu^4+120s(1+6\nu^2)+40s^2(1+6\nu^2))}{691200}\,.
\end{equation}
The sum over all (even) spins can be found to vanish, which guarantees that the deficit does not contribute to the zeta-prime of the (non)-minimal Type-A. Also the deficit is an even function of $\nu$ and therefore the difference due to $\Delta=d-2,2$ boundary conditions for the scalar field is also free of any deficit.

Let us note that the deficit has already appeared in implicit form in the literature. It is the leftover of $P_{\nu,s}$ in \cite{Gunaydin:2016amv} without the part including $c_n^+$, see also \cite{Giombi:2014iua} where the same structures are present but in different notation.

\chapter{On Chiral HSGRA}\label{app:chap4}

\section{Kinematics}
\label{app:kinematics}
\setcounter{equation}{0}
The four-dimensional Poincare algebra in light-cone gauge introduced in chapter \ref{chapter4} implies that momenta should only enter the game as the following combinations
\begin{align}
   \PP_{km}&=p_k\beta_m-p_m\beta_k\,, & \PPb_{km}&=\pb_k\beta_m-\pb_m\beta_k \,.
\end{align}
Also, it can be shown that only $N-2$ out of $N(N-1)/2$ $\PP_{ij}$ are independent, likewise for $\PPb$. In particular, for the three-point case there is just one independent transverse momenta (and its conjugate). In particular, all $\PP_{ij}$ are anti-symmetric under permutations:
\begin{align}
    \PP^a_{12}&=...=\PP^a=\frac13\left[ (\beta_1-\beta_2)p_3+(\beta_2-\beta_3)p_1+(\beta_3-\beta_1)p_2\right]\,,\\
    \sigma_{123}\PP&=\PP\,, \qquad\qquad \sigma_{12}\PP=\sigma_{23}\PP=\sigma_{13}\PP=-\PP\,.
\end{align}
where conservation of the total momenta has been used. 
Also, for three points we have
 \begin{align}
      -\sum_i \frac{p_i \pb_i}{\beta_i}&=\frac{\PP \PPb}{\beta_1\beta_2\beta_3}= \frac{\PP \cdot \PP}{2\beta_1\beta_2\beta_3}\,.
\end{align}
We have a number of useful identities (we use $d$-dimensional notation sometimes, $a=z,\bar{z}$ in $4d$). Bianchi-like identities:
\begin{align}\label{eq:Bianchilike}
    \sum_i \PP^a_i&=0&
    \beta_{[i} \PP^a_{jk]}&\equiv0&
    \PP^a_{i[j}\PP^a_{kl]}&\equiv0
\end{align}
Other kinematic identities include
\begin{align}
    \sum_j \frac{\PP_{ij} \PPb_{jk}}{\beta_j}&= -\frac12\beta_i\beta_k \sum_j \frac{\pvec_j^2}{\beta_j}\\
    \sum_j \frac{\PP_{ij} \PPb_{jk}}{\beta_j}&= -\beta_i\beta_k \sum_j {p_j \pb_j}\\
    \PP_{ij}\PPb_{ij} &=-\frac12\beta_i\beta_j (\pvec_i+\pvec_j)^2\qquad \text{for}\quad\pvec_i^2,\pvec_j^2=0 \
\end{align}
and one of the most important for dealing with one off-shell leg is ($s_{ik}=(\pvec_i+\pvec_k)^2$):
\begin{equation}\label{eq:magicidentity}
    \PPb_{ik}\PP_{ik}=-\frac{\beta_i\beta_k}{2}s_{ik}+\frac{1}{2}\beta_i(\beta_k+\beta_i)\pvec_k^2, \qquad \pvec_i^2=0,\quad \pvec_k^2\neq 0
\end{equation}

\section{Color Effects on Dynamical Constraints}
\label{sec:color}
As pointed out in chapter \ref{chapter4} and also \cite{Metsaev:1991mt,Metsaev:1991nb,Ponomarev:2016lrm}, the dynamical constraint that allows for closure of Poincare algebra at cubic vertices is
\begin{equation}\label{eq:Dconstraint}
    [H_3(\PPb),J_3]=0.
\end{equation}
Here, we simply give the Hamiltonian $H_3$
\small
\begin{equation}
    H_3=\sum_{\lambda_i}\int \prod_{i=1}^3d^3p_i \delta^3\left(\sum_{i}p_i\right)h_3(p_i,\partial_{p_i})\Phi^{\lambda_1}_{p_1}\Phi^{\lambda_2}_{p_2}\Phi^{\lambda_3}_{p_3}, \quad h_3=C^{\lambda_1,\lambda_2,\lambda_3}\frac{\PPb^{\lambda_1+\lambda_2+\lambda_3}}{\beta_1^{\lambda_1}\beta_2^{\lambda_2}\beta_3^{\lambda_3}}.
\end{equation}
\normalsize
The dynamical boost generator $J_3$ reads
\small
\begin{equation}
    J_3=\sum_{\lambda_i}\int\prod_{i=1}^3 d^3p_i\delta^3\left(\sum_{i}p_i\right)
    \Bigg[j_3(p_i)-\frac{h_3(p_i)}{3}\Big(\sum_k\frac{\partial}{\partial p_k}\Big)\Bigg]\prod_{i=1}^3\Phi^{\lambda_i}_{p_i},
\end{equation}
\normalsize
where
\small
\begin{equation}
    j_3=\frac{2}{3} C^{\lambda_1,\lambda_2,\lambda_3} \frac{\PPb^{\lambda_1+\lambda_2+\lambda_3-1}}{\beta_1^{\lambda_1}\beta_2^{\lambda_2}\beta_3^{\lambda_3}}\chi^{\lambda_1,\lambda_2,\lambda_3}\quad \text{and} \quad \chi=(\lambda_1-\lambda_2)\beta_3+(\lambda_2-\lambda_3)\beta_1+(\lambda_3-\lambda_1)\beta_2.
\end{equation}
\normalsize
While the authors in \cite{Ponomarev:2016lrm} work with colorless minimal chiral HiSGRA\footnote{There are only even spins in the spectrum.}, most of the technical details therein can be directly generalized to colorful non-minimal cases. For a generic case, the constraint (\ref{eq:Dconstraint}) reads
\small
\begin{equation}\label{eq:constraintdensity}
    \begin{split}
        [H_3,J_3]&=\sum_{\lambda_i,\mu_j}\int \D p_i \D q_j  \delta^3\left(\sum_j q_j\right)\Bigg[j_3(q_j)-\frac{h_3(q_j)}{3}\Big(\sum_k \frac{\partial}{\partial q_k}\Big)\Bigg]\\
        &\times \delta^3\left(\sum_i p_i\right)h_3(p_i)\Big[\prod_{i=1}^3\Phi^{\lambda_i}_{p_i}, \prod_{j=1}^3\Phi^{\mu_j}_{q_j}\Big].
    \end{split}
\end{equation}
\normalsize
There are two type of contributions in (\ref{eq:constraintdensity}). The first contribution comes without derivatives, namely 
\small
\begin{equation}\label{eq:M1constraint}
    M_1=\sum_{\lambda_i,\mu_j}\int \D p_i \D q_i \delta^3\left(\sum_j q_j\right)\delta^3\left(\sum_i p_i\right) j_3(q_j)h_3(p_i)\Big[\prod_{i=1}^3\Phi^{\lambda_i}_{p_i}, \prod_{j=1}^3\Phi^{\mu_j}_{q_j}\Big]
\end{equation}
\normalsize
and the second contribution comes with derivatives
\small
\begin{equation}
    M_2=-\frac{1}{3}\sum_{\lambda_i,\mu_j}\int \D p_i \D q_i \Bigg[\delta^3\left(\sum_j q_j\right)h_3(q_j)\Big(\sum_k\frac{\partial}{\partial q_k}\Big) \Bigg]\delta^3\left(\sum_i p_i\right)h_3(p_i)\Big[\prod_{i=1}^3\Phi^{\lambda_i}_{p_i}, \prod_{j=1}^3\Phi^{\mu_j}_{q_j}\Big],
\end{equation}
\normalsize
where $\D p_i=\prod_{i=1}^3 d^3p_i$. Note that the derivatives $\partial_{q_k}$ in $M_2$ also act on $\Phi^{\mu_j}_{q_j}$. To make the fields only interact with themselves through Poisson brackets, we will integrate by part the operator $\left(\sum_k \partial_{q_k}\right)$. So, $M_2$ becomes
\small
\begin{equation}\label{eq:M2constraint}
    M_2=\frac{1}{3}\sum_{\lambda_i,\mu_j}\int \D p_i \D q_i \Big(\sum_k\frac{\partial}{\partial q_k}\Big)\Bigg[\delta^3\left(\sum_j q_j\right)h_3(q_j) \Bigg]\delta^3\left(\sum_i p_i\right)h_3(p_i)\Big[\prod_{i=1}^3\Phi^{\lambda_i}_{p_i}, \prod_{j=1}^3\Phi^{\mu_j}_{q_j}\Big].
\end{equation}
\normalsize
Now, since both $h_3$ and $j_3$ are cyclic invariant, the associated fields can be reorganized with the same ordering. Hence, the Poisson bracket in (\ref{eq:constraintdensity}) can be written as
\begin{equation}\label{eq:fieldsPoisson}
    \Big[\prod_{i=1}^3\Phi^{\lambda_i}_{p_i}, \prod_{j=1}^3\Phi^{\mu_j}_{q_j}\Big]=\prod_{i,j=1}^2\Phi^{\lambda_i}_{p_i}\Phi^{\mu_j}_{q_j}\Big[\Phi^{\lambda_3}_{p_3},\Phi^{\mu_3}_{q_3}\Big].
\end{equation}
where we choose the "contract" the last fields $\Phi^{\lambda_3}_{p_3}$ and $\Phi^{\mu_3}_{q_3}$ in $H_3$ and $J_3$, respectively. Below, we present solutions of (\ref{eq:Dconstraint}) in various colorful cases. 
\subsubsection{\texorpdfstring{$U(N)$}{U(N)} gauging}
\label{sec:u(N)color}
We now assume that the fields take values in some algebra and the generators of these algebra are labelled as $T_a$. We first look at the case where fields take $U(N)$-valued
\begin{align}
    \Phi^{\lambda}(\pvec)\equiv \Phi_a^{\lambda}(\pvec)T^a \equiv (\Phi^{\lambda}_{\pvec})^A_{\ B},
\end{align}
so that the trace in \eqref{eq:flatchiralaction} is over $U(N)$ indices. The Poisson bracket in this case can be defined as
\begin{equation}\label{eq:PoissonU(N)}
    [(\Phi^{\lambda}_{p})^A_{\ B},(\Phi^{\mu}_{q})^C_{\ D}]=\frac{\delta^{\lambda,-\mu}\delta^3(p+q)}{2q^+}\times [\theta_{\lambda}\delta^C_{ \ B}\delta^A_{\ D}]
\end{equation}
where $\theta_{\lambda}$ is some phase factor that can be used to rescaling fields in order to obtain (\ref{eq:magicalcoupling}). Explicitly solving the commutator in (\ref{eq:Dconstraint}) by using (\ref{eq:M1constraint}), (\ref{eq:M2constraint}) and (\ref{eq:fieldsPoisson}) gives
\small
\begin{equation}\label{eq:dynamicU(N)}
\begin{split}
   0= &\sum_{\omega}\text{Sym}\,(-)^{\omega}\theta_{\omega}\Tr(\Phi_1\Phi_2\Phi_3\Phi_4)\\
    &\times\Big[\frac{(\lambda_1+\omega-\lambda_2)\beta_1-(\lambda_2+\omega-\lambda_1)\beta_2}{\beta_1+\beta_2}C^{\lambda_1,\lambda_2,\omega}C^{\lambda_3,\lambda_4,-\omega}\PPb_{12}^{\lambda_1+\lambda_2+\omega-1}\PPb_{34}^{\lambda_3+\lambda_4-\omega}\Big]
    \end{split}
\end{equation}
\normalsize
Next, we let $\theta_{\omega}=e^{ix\omega}$ to be an arbitrary phase factor and determine the value of $x$ so that the coupling constant (\ref{eq:magicalcoupling}) is the solution of (\ref{eq:dynamicU(N)}). Note that the symmetried sum in (\ref{eq:dynamicU(N)}) appears from the contraction between fields \cite{Ponomarev:2016lrm} that preserve the all possible color-orderings. If we denote $\Tr(\Phi_i\Phi_j\Phi_k\Phi_l)E(i,j,k,l)$ as $[i,j,k,l]$  where $E$ are the kinematic parts, then we have in total six \textit{partial color-ordered contribution} (or $partial$-$contribution$ for short) of the constraint (\ref{eq:Dconstraint}) in terms of $[i,j,k,l]$:
\small
\begin{equation}\label{eq:colormodulo}
    0=[1,2,3,4]+[1,3,4,2]+[1,4,2,3]+[1,3,2,4]+[1,2,4,3]+[1,4,3,2]
\end{equation}
\normalsize
Each of the terms in (\ref{eq:colormodulo}) need to vanish in order to make (\ref{eq:dynamicU(N)}) satisfied since there is no way to make different partial contributions canceling each others. We can take $[1,2,3,4]$ as an example. It is a combination of the following permutations that preserve the color-ordering of $\Tr(\Phi_1\Phi_2\Phi_3\Phi_4)$ 
\begin{equation}
    [1,2,3,4]=\{1,2,3,4\}+\{2,3,4,1\}+\{3,4,1,2\}+\{4,1,2,3\}.
\end{equation}
where the curly brackets $\{i,j,k,l\}$ notation is for permutations with $i,j,k,l$ are indices of left-over external sources. First of all, the combination when we consider the permutation $\{1,2,3,4\}\rightarrow \{3,4,1,2\}$ with $\omega\rightarrow-\omega$, two of them combine to be
\begin{equation}\label{eq:dynamic1234}
\begin{split}
    &\sum_{\omega}e^{i\pi\omega}\frac{\PPb_{12}^{\lambda_1+\lambda_2+\omega-1}}{\Gamma(\lambda_1+\lambda_2+\omega)}\frac{\PPb_{34}^{\lambda_3+\lambda_4-\omega-1}}{\Gamma(\lambda_3+\lambda_4-\omega)}\Tr(\Phi_1\Phi_2\Phi_3\Phi_4) \\
    &\times\Bigg[e^{ix\omega}(\lambda_1-\lambda_2)\PPb_{34}+e^{-ix\omega}(\lambda_3-\lambda_4)\PPb_{12}+\omega\Big(e^{ix\omega}\frac{\beta_1-\beta_2}{\beta_1+\beta_2}\PPb_{34}+e^{-ix\omega}\frac{\beta_3-\beta_4}{\beta_1+\beta_2}\PPb_{12}\Big)\Bigg]
    \end{split}
\end{equation}


Secondly, for the combination of $\{2,3,4,1\}\xrightarrow{\omega\rightarrow-\omega}\{4,1,2,3\}$, we get
\begin{equation}\label{eq:dynamic2341}
\begin{split}
     &\sum_{\omega}e^{i\pi\omega}\frac{\PPb_{23}^{\lambda_2+\lambda_3+\omega-1}}{\Gamma(\lambda_2+\lambda_3+\omega)}\frac{\PPb_{41}^{\lambda_4+\lambda_1-\omega-1}}{\Gamma(\lambda_4+\lambda_1-\omega)}\Tr(\Phi_2\Phi_3\Phi_4\Phi_1)\\
     &\times\Bigg[e^{ix\omega}(\lambda_2-\lambda_3)\PPb_{41}+e^{-ix\omega}(\lambda_4-\lambda_1)\PPb_{23}+\omega\Big(e^{ix\omega}\frac{\beta_2-\beta_3}{\beta_2+\beta_3}\PPb_{41}+e^{-ix\omega}\frac{\beta_4-\beta_1}{\beta_2+\beta_3}\PPb_{23}\Big)\Bigg]
     \end{split}
\end{equation}

Taking the sum over $\omega$ in (\ref{eq:dynamic1234}) for example, we introduce $\Lambda_4=\lambda_1+\lambda_2+\lambda_3+\lambda_4$ and obtain
\tiny
\begin{equation}\label{eq:U(N)1234constraint}
    \begin{split}
    (2.21)=&\frac{e^{-i\pi(\lambda_1+\lambda_2-1)}}{\Gamma(\Lambda_4-1)}\Bigg[e^{-ix(\lambda_1+\lambda_2-1)}(\PPb_{34}-e^{ix}\PPb_{12})^{\Lambda_4-2}(\lambda_1-\lambda_2)\PPb_{34}+e^{ix(\lambda_1+\lambda_2-1)}(\PPb_{34}-e^{-ix}\PPb_{12})^{\Lambda_4-2}(\lambda_3-\lambda_4)\PPb_{12}\\
    &-e^{-ix(\lambda_1+\lambda_2-1)}(\PPb_{34}-e^{ix}\PPb_{12})^{\Lambda_4-3}\frac{\beta_1-\beta_2}{\beta_1+\beta_2}\PPb_{34}\Big[\PPb_{34}(\lambda_1+\lambda_2-1)+e^{ix}\PPb_{12}(\lambda_3+\lambda_4-1)\Big]\\
    &-e^{ix(\lambda_1+\lambda_2-1)}(\PPb_{34}-e^{-ix}\PPb_{12})^{\Lambda_4-3}\frac{\beta_3-\beta_4}{\beta_1+\beta_2}\PPb_{12}\Big[\PPb_{34}(\lambda_1+\lambda_2-1)+e^{-ix}\PPb_{12}(\lambda_3+\lambda_4-1)\Big]\Bigg]\Tr(h_1h_2h_3h_4)
    \end{split}
\end{equation}
\normalsize
The sum over $\omega$ in (\ref{eq:dynamic2341}) gives somewhat similar result with (\ref{eq:U(N)1234constraint}) by relabelling  $(1,2)\rightarrow(2,3)$ and $(3,4)\rightarrow(4,1)$. Now, as we noted, $[1,2,3,4]$ should vanish by itself. This is only possible if $x=\pi$ or $\theta_{\omega}=(-)^{\omega}$. In this case, the computation above get simplified and it reads
\small
\begin{equation}\label{eq:DU(N)1234constraint}
\begin{split}
    [1,2,3,4]=&\Tr(\Phi_1\Phi_2\Phi_3\Phi_4)(\PPb_{12}-\PPb_{23}+\PPb_{34}-\PPb_{41})\times \frac{(\PPb_{12}+\PPb_{34})^{\Lambda_4-3}}{\Gamma(\Lambda_4-1)}\\
    &\times\Big[\lambda_1(\PPb_{23}+\PPb_{34})-\lambda_2(\PPb_{34}+\PPb_{41})+\lambda_3(\PPb_{41}+\PPb_{12})-\lambda_4(\PPb_{12}+\PPb_{23})\Big]\\
    =&0
    \end{split}
\end{equation}
\normalsize
In order to obtain the above result we used  momentum conservation and noticed that $\PPb_{12}+\PPb_{34}=\PPb_{23}+\PPb_{41}$. Without having the common factor $(\PPb_{12}+\PPb_{34})^{\Lambda_4-4}=(\PPb_{23}+\PPb_{41})^{\Lambda_4-4}$, one can not make another choice for $\theta_{\omega}$ to have (\ref{eq:magicalcoupling}) as the solution of $[1,2,3,4]=0$. For other partial contribution in (\ref{eq:colormodulo}), we also see that they are vanishing if $\theta_{\omega}=(-)^{\omega}$. Hence, $\theta_{\omega}=(-)^{\omega}$ is the unique solution of (\ref{eq:Dconstraint}) for $U(N)$ color chiral HSGRA that has (\ref{eq:magicalcoupling}) as the coupling constants. 
\subsubsection{\texorpdfstring{$SO(N) \ \text{and} \ USp(N)$}{SO(N) and USp(N)} gauging}
\label{sec:o(N)/usp(N)color}
In the case where fields have $O(N)$ color, the trace is understood as
\begin{equation}
    \Tr(\Phi_{\pvec_1}^{\lambda_1} ... \Phi^{\lambda_n}_{\pvec_n})=\Phi^1_{AB_1}\Phi^2_{B_1B_2}...\Phi^n_{B_nA}, \quad \Phi^i\equiv \Phi^{\lambda_i}_{\pvec_i}.
\end{equation}
For the $O(N)$ case where the Poisson brackets is defined as
\begin{equation}\label{eq:PoissonO(N)}
    [(\Phi^{\lambda}_{p})_{AB},(\Phi^{\mu}_{q})_{CD}]=\frac{\delta^{\lambda,-\mu}\delta^3(p+q)}{2q^+}\times [\delta_{AC}\delta_{BD}+\theta_{\lambda}\delta_{AD}\delta_{BC}].
\end{equation}
where $\delta_{AB}$ are invariant symmetric tensor. Note that $\phi_{\lambda}$ is a phase factor that enter the Poisson bracket. Next, we solve the constraints (\ref{eq:Dconstraint}) and get
\begin{equation}\label{eq:O(N)Dynamic}
    \begin{split}
        0=&\sum_{\omega}\text{Sym}(-)^{\omega}\Big[\theta_{\lambda_3}\theta_{\lambda_4}\Tr(\Phi_1\Phi_2\Phi_4\Phi_3)+\theta_{\omega}\Tr(\Phi_1\Phi_2\Phi_3\Phi_4)\Big]\\
        &\times \Big[\frac{(\lambda_1+\omega-\lambda_2)\beta_1-(\lambda_2+\omega-\lambda_1)\beta_2}{\beta_1+\beta_2}C^{\lambda_1,\lambda_2,\omega}C^{\lambda_3,\lambda_4,-\omega}\PPb_{12}^{\lambda_1+\lambda_2+\omega-1}\PPb_{34}^{\lambda_3+\lambda_4-\omega}\Big]
    \end{split}
\end{equation}
Now, we repeat the same treatment with the above analysis for $U(N)$-case to determine the phase factor $\theta_{\lambda_i}=e^{ix\lambda_i}$. However, unlike the $U(N)$-case, $SO(N)$-case contains an extra trace that comes from the \textit{M\"obius twist} in the Poisson brackets \eqref{eq:PoissonO(N)}. As a consequence, there will be mixing between $[i,j,k,l]$ partial contributions. First, let us look at $\{1,2,3,4\}\xrightarrow{\omega\rightarrow -\omega}\{3,4,1,2\}$ in $[1,2,3,4]$
\small
\begin{equation}\label{eq:O(N)1234}
\begin{split}
    &\sum_{\omega}\frac{e^{i\pi\omega}\,\PPb_{12}^{\lambda_1+\lambda_2+\omega-1}\PPb_{34}^{\lambda_3+\lambda_4-\omega-1}}{\Gamma(\lambda_1+\lambda_2+\omega)\Gamma(\lambda_3+\lambda_4-\omega)}\\
    \times&\Bigg[\Tr(1234)\Big[e^{ix\omega}(\lambda_1-\lambda_2)\PPb_{34}+e^{-ix\omega}(\lambda_3-\lambda_4)\PPb_{12}+\omega\frac{e^{ix\omega}(\beta_1-\beta_2)\PPb_{34}+e^{-ix\omega}(\beta_3-\beta_4)\PPb_{12}}{\beta_1+\beta_2}\Big]\\
    &+\Tr(1243)e^{ix(\lambda_3+\lambda_4)}\Big[(\lambda_1-\lambda_2)\PPb_{34}+(\lambda_3-\lambda_4)\PPb_{12}+\omega\frac{(\beta_1-\beta_2)\PPb_{34}+(\beta_3-\beta_4)\PPb_{12}}{\beta_1+\beta_2}\Big]\Bigg]
    \end{split}
\end{equation}
\normalsize
where we denote $\tr(ijkl)\equiv\Tr(\Phi_i\Phi_j\Phi_l\Phi_k)$ for simplicity. Similarly, the permutation $\{2,3,4,1\}\xrightarrow{\omega\rightarrow -\omega}\{4,1,2,3\}$ in $[1,2,3,4]$ reads
 \small
\begin{equation}\label{eq:O(N)2341}
\begin{split}
    &\sum_{\omega}\frac{e^{i\pi\omega}\,\PPb_{23}^{\lambda_2+\lambda_3+\omega-1}\PPb_{41}^{\lambda_4+\lambda_1-\omega-1}}{\Gamma(\lambda_2+\lambda_3+\omega)\Gamma(\lambda_4+\lambda_1-\omega)}\\
    \times&\Bigg[\Tr(2341)\Big[e^{ix\omega}(\lambda_2-\lambda_3)\PPb_{41}+e^{-ix\omega}(\lambda_4-\lambda_1)\PPb_{12}+\omega\frac{e^{ix\omega}(\beta_2-\beta_3)\PPb_{41}+e^{-ix\omega}(\beta_4-\beta_1)\PPb_{23}}{\beta_2+\beta_3}\Big]\\
    &+\Tr(2314)e^{ix(\lambda_1+\lambda_4)}\Big[(\lambda_2-\lambda_3)\PPb_{41}+(\lambda_4-\lambda_1)\PPb_{23}+\omega\frac{(\beta_2-\beta_3)\PPb_{41}+(\beta_4-\beta_1)\PPb_{23}}{\beta_2+\beta_3}\Big]\Bigg]
    \end{split}
\end{equation}
\normalsize
One can notice that there are additional contributions (compared to the $U(N)$-case) in the equation \eqref{eq:O(N)2341} that combine two traces inside $[1,2,3,4]$: namely $\Tr(2314)$ and an \textit{exotic} one $\Tr(1243)$. Hence, $[1,2,3,4]$ can not vanish by itself and we need to borrow some contributions from others $[i,j,k,l]$ in order to satisfy (\ref{eq:O(N)Dynamic}). Take a look at the permutation $\{1,3,2,4\}\xrightarrow{\omega\rightarrow -\omega}\{2,4,1,3\}$ in $[1,3,2,4]$
 \small
\begin{equation}\label{eq:O(N)1324}
\begin{split}
    &\sum_{\omega}\frac{e^{i\pi\omega}\,\PPb_{13}^{\lambda_1+\lambda_3+\omega-1}\PPb_{24}^{\lambda_2+\lambda_4-\omega-1}}{\Gamma(\lambda_1+\lambda_3+\omega)\Gamma(\lambda_2+\lambda_4-\omega)}\\
    \times&\Bigg[\tr(1324)\Big[e^{ix\omega}(\lambda_1-\lambda_3)\PPb_{24}+e^{-ix\omega}(\lambda_2-\lambda_4)\PPb_{13}+\omega\frac{e^{ix\omega}(\beta_1-\beta_3)\PPb_{24}+e^{-ix\omega}(\beta_2-\beta_4)\PPb_{13}}{\beta_1+\beta_3}\Big]\\
    &+\tr(3124)e^{ix(\lambda_1+\lambda_3)}\Big[(\lambda_1-\lambda_3)\PPb_{24}+(\lambda_2-\lambda_4)\PPb_{13}+\omega\frac{(\beta_1-\beta_3)\PPb_{24}+(\beta_2-\beta_4)\PPb_{13}}{\beta_1+\beta_3}\Big]\Bigg]
    \end{split}
\end{equation}
\normalsize
Then, we have in total 6 different color-ordered terms. For the combination of permutation $\{1,2,3,4\}\xrightarrow{\omega\rightarrow -\omega}\{3,4,1,2\}$ and $\{2,3,4,1\}\xrightarrow{\omega\rightarrow -\omega}\{4,1,2,3\}$ to get (\ref{eq:DU(N)1234constraint}) for the color-ordering $\Tr(1234)$ we need to set $x=\pi$ or $\theta_{\lambda_i}=(-)^{\lambda_i}$. Next, it is easy to see that the contribution coming from $\Tr(1243)(-)^{\omega}\theta_{\lambda_4}\theta_{\lambda_3}$ and $\Tr(1324)$ also cancel each others with this choice of the phase factors in \eqref{eq:PoissonO(N)}. Similar argument works for $\Tr(2314)(-)^{\omega}\theta_{\lambda_4}\theta_{\lambda_1}$ and $\Tr(3124)(-)^{\omega}\theta_{\lambda_3}\theta_{\lambda_1}$. Hence, even though $[i,j,k,l]$ can not vanish by themselves in the case of $SO(N)$-gauging, the total contribution vanish by combining all the partial contributions together. This indicates that $\theta_{\omega}=(-)^{\omega}$ is the right choice for the phase factors of Poisson bracket (\ref{eq:PoissonO(N)}). \\
${}$\\
Finally, in the case of $USp(N)$, the Poisson bracket reads
\begin{equation}\label{eq:PoissonUsp(N)}
    [(\Phi^{\lambda}_{p})_{AB},(\Phi^{\mu}_{q})_{CD}]=\frac{\delta^{\lambda,-\mu}\delta^3(p+q)}{2q^+}\times [C_{AC}C_{BD}+\theta_{\lambda}C_{AD}C_{BC}].
\end{equation}
where $C_{AB}$ are anti-symmetric matrices invariant tensor
\begin{equation}
    C_{AB}=-C_{BA}, \quad C_{AB}C^{CB}=\delta^C_A
\end{equation}
We can use the $C$-tensors to raise and lower indices as $V^A=C^{AB}V_B$, $V^B C_{BA}=V_A$. Finally, the trace for $USp(N)$ case can be understood as
\begin{equation}
    \tr(\Phi \Phi...)=\Phi\fdu{A}{B}\Phi\fdu{B}{C}...
\end{equation}

One then solves (\ref{eq:Dconstraint}) and get
\begin{equation}\label{eq:USp(N)constraint}
\begin{split}
    0&=\sum_{\omega}\text{Sym}(-)^{\omega+1}\Big[\theta_{\lambda_3}\theta_{\lambda_4}\tr(\Phi_1\Phi_2\Phi_4\Phi_3)+\theta_{\omega}\tr(\Phi_1\Phi_2\Phi_3\Phi_4)\Big]\\
        &\times \Big[\frac{(\lambda_1+\omega-\lambda_2)\beta_1-(\lambda_2+\omega-\lambda_1)\beta_2}{\beta_1+\beta_2}C^{\lambda_1,\lambda_2,\omega}C^{\lambda_3,\lambda_4,-\omega}\PPb_{12}^{\lambda_1+\lambda_2+\omega-1}\PPb_{34}^{\lambda_3+\lambda_4-\omega}\Big]
    \end{split}
\end{equation}
Repeating the same treatment as in the $SO(N)$-case with the requirement that (\ref{eq:magicalcoupling}) is the solution of \eqref{eq:USp(N)constraint}, one obtains $\theta_{\omega}=(-)^{\omega+1}$. To summarize the $SO(N)/USp(N)$-valued fields have the following properties under interchanging $SO(N)/USp(N)$ indices,
\begin{align}
    O(N)&:\quad (\Phi^{\lambda}_{\pvec})_{AB}=(-)^{\lambda}(\Phi^{\lambda}_{\pvec})_{BA}\\
    USp(N)&:\quad (\Phi^{\lambda}_{\pvec})_{AB}=(-)^{\lambda+1}(\Phi^{\lambda}_{\pvec})_{BA}&
\end{align}
Here, fields with odd-spin in $O(N)/USp(N)$ case have odd/even parity, while fields with even-spin have even/odd parity. Fields with odd spins always take values in the adjoint representation.\\
${}$\\
It is important to stress that, the constraint (\ref{eq:Dconstraint}) with the coupling constants (\ref{eq:magicalcoupling}) can only be satisfied with the above choices of $\theta_{\omega}$ for $U(N)$ and $O(N)/USp(N)$-colored chiral HSGRAs.  Interestingly enough the allowed gauge groups as well as the allowed representations coincide with the allowed Chan-Paton symmetry groups and the representations in open string theory \cite{Marcus:1982fr}.

\section{Worldsheet-Friendly Regularization}
\label{app:Thornregulator}
\setcounter{equation}{0}
In practice we face integrals of the following type:
\begin{align}
    \int \frac{d^4 q}{(2\pi)^4} F(\beta, q^a) \frac{1}{\prod_i (\qvec-\kvec_i)^2}
\end{align}
where the polynomial prefactor $F$ depends on external momenta (not shown here), and the loop momentum $\qvec$. Importantly, the $q^-$-component does not enter the vertex. The regularization proposed in \cite{Thorn:2004ie} is to introduce the Gaussian cutoff in the transverse part of the loop momentum $q$, i.e. $q_\perp\equiv (q,\bar{q})$:
\begin{align}
    I=\int \frac{d^4 q}{(2\pi)^4} F(\beta, q_{\perp}) \frac{1}{\prod_i (\qvec-\kvec_i)^2} e^{-\xi q_\perp^2}
\end{align}
The integral can be performed by first using the Schwinger trick with parameters $T_i$, then doing the $q_\perp$ Gaussian integral. Integration over $q^-$ gives a delta function:
\begin{equation}
    \begin{split}
 I= &\frac{2\pi^2}{2(2\pi)^4} \int d\beta \,F(\beta, \sum_i T_ik_i/(T+\xi)) \delta(\sum T_i \beta -\sum T_i \beta_i^+) \\
    &\times\exp{\Big[ 2\sum T_i(\beta-\beta_i^+) k_i^- -\sum T_i k_{i\perp}^2 +\frac{1}{\sum T_i+\xi} (\sum T_i k_{i\perp})^2\Big]}
    \end{split}
\end{equation}
If there are no IR divergences, we can safely solve for $\beta$. It is also convenient to change variables as $T_i=Tx_i$, $\sum x_i=1$, which gives Jacobian $T^{n-1}$. This way we get
\begin{align}
    I= \pi \int \frac{dT\, \prod dx_i }{T\,(2\pi)^4}& F(\beta=\sum x_i \beta_i^+, q_{\perp}=\tfrac{T}{T+\xi}\sum x_ik_{i\perp}) \delta(\sum x_i-1) \frac{\pi}{T+\xi}\\
    &\times\exp{\Big[ -T \sum_{i\leq j} x_i x_j (\kvec_i-\kvec_j)^2 -\frac{T \xi}{T+\xi} (\sum x_i k_{i\perp})^2\Big]}
\end{align}
In a lucky case when the integral is not divergent at all, we simply find
\small
\begin{equation}
    \begin{split}
         I= &\frac{1}{(4\pi)^2} \int dT\,T^{n-3} \prod dx_i\times\\ &\times F(\beta=\sum x_i \beta_i, q^a=\sum x_ik_i^a) \delta(\sum x_i-1) 
    \exp{\Big[ -\frac{T}{2} \sum_{i\leq j} x_i x_j (\kvec_i-\kvec_j)^2 \Big]}
    \end{split}
\end{equation}
\normalsize

Next, to understand how to work with dual momenta is also simple. We choose the direction of the dual loop momentum $\kvec_i$ to be clock wise and consider the self-energy diagram as an example.
\begin{figure}[h]
    \centering
    \includegraphics[scale=0.3]{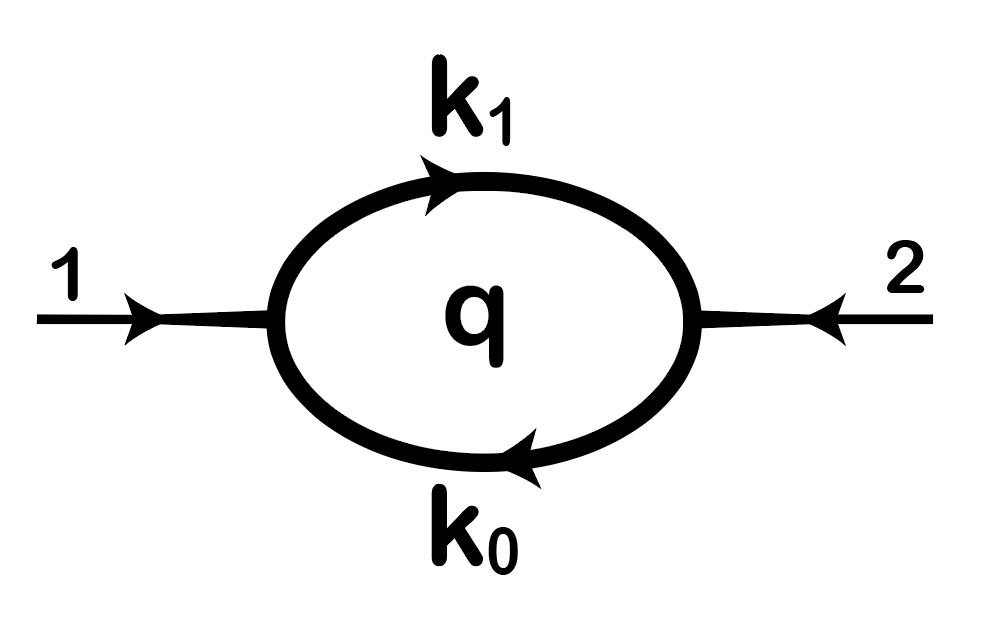}
\end{figure}\\
The dual momentum is related with original momentum as follows. Take the external leg-1, one can define $\pvec_1=\kvec_1-\kvec_0$. We can continue with this pattern for other external momenta as $\pvec_i=\kvec_i-\kvec_{i-1}$ at each vertices. The loop momentum is defined as the different of $\qvec$ with its nearest dual regional momentum $\kvec_i$, where $\qvec$ is the dual momentum that is bounded by a loop. In our example, $\pvec=\qvec-\kvec_0$. With these rules of labeling dual momenta, one can easily compute the quantum correction at one-loop with arbitrary legs like in section \ref{sec:leggedloop}.



\subsection{Anti de-Sitter space}
We can lift the above analysis of chiral HSGRA in flat space to $\ads_4$ as well \cite{Metsaev:2018xip,Skvortsov:2018uru}. The Poincare algebra $iso(3,1)$ now becomes the conformal algebra $so(3,2)$ which contains two new generators that are: dilatation $D$, and conformal boost generator $K$. The conformal algebra reads
\begin{subequations}
\begin{align}
     [L^{AB},L^{CD}]&=L^{AB}\eta^{BC}-L^{BD}\eta^{AC}-L^{AC}\eta^{BD}+L^{BC}\eta^{AD}\,,\\
     [L^{AB},P^C]&=P^A\eta^{BC}-P^B\eta^{AC}\,,  \qquad \quad  [L^{AB},K^C]=K^A\eta^{BC}-K^{B}\eta^{AC}\,,\\
     [D,P^A]&=-P^A\,, \qquad  \qquad \qquad \qquad  \quad \  [D,K^A]=K^A\,,\\
     [P^A,K^B]&=-L^{AB}+\eta^{AB}D\,.
\end{align}
\end{subequations}
The metric of the Poincare patch in light-cone gauge reads
\begin{align}
    ds^2=\frac{1}{z^2}\big[2dx^+dx^-+dx_1^2+dz^2\big]\,.
\end{align}
Once again, we can work directly with in momentum space paying attention to the fact that the $z$ coordinate does not admit Fourier-transformation:
\begin{align}
    \Phi(p|z)=\frac{1}{(2\pi)^{3/2}}\int dx^-dx^1dx^+\,e^{-ipx}\Phi(x^+,x^-,x^1|z)\,.
\end{align}
The two scalar fields that describe a massless spin-$s$ gauge field obey the conjugation rules as
\begin{align}
    \Phi^{\lambda}_{p,z}\equiv \Phi^{\lambda}(p|z),\qquad (\Phi^{\lambda}_{p,z})^{\dagger}=\Phi^{-\lambda}_{-p,z}\,.
\end{align}
The action up to cubic level was found in \cite{Metsaev:2018xip,Skvortsov:2018uru}:
\begin{align}
    S=\frac{1}{2}\intergrallambda \ \,\Tr\Big[\Phi^{\dagger}_{\lambda}(\pl_z^2-p^2)\Phi_{\lambda}\Big]+\intergrallambda\ \D \Gamma_3\, \Tr\Big[\Big(\prod_a   \Phi_a^{\dagger}\Big)\Big]\, V_3(p_a|z_a,\pl_{z_a})\, \prod_a\,\delta(z-z_a)\,.
\end{align}
where 
\begin{align}
    \intergrallambda\ \,\equiv \sum_{\lambda}\int\,,\qquad \D \Gamma_3\equiv \delta^3(\sum   p_a)\prod_a d^3p_a\,dz_a\,dz\,.
\end{align}
The kernel $V_3$ takes the following form 
\begin{align}
    V_3=\begin{cases}
    C_0z^{-1}, \qquad \qquad \qquad \qquad \qquad \qquad \ \ \lambda_i=0\,,\\
    C_L^{\lambda_1,\lambda_2,\lambda_3}U_LV_L^{0},\qquad V_L^{0}=\frac{(z\PP^L)^{-\Lambda_3}}{z\,\beta_1^{-\lambda_1}\beta_2^{-\lambda_2}\beta_3^{-\lambda_3}},\quad \Lambda_3<0\,,\\
    C_R^{\lambda_1,\lambda_2,\lambda_3}U_RV_R^0,\qquad V_R^0=\frac{(z\PP^R)^{\Lambda_3}}{z\,\beta_1^{\lambda_1}\beta_2^{\lambda_2}\beta_3^{\lambda_3}}\,,\qquad \ \Lambda_3>0\,.
    \end{cases}
\end{align}
The holomorphic and anti-holomorphic momenta of flat space get lifted to
\begin{align}
    \PP^L=\frac{1}{\sqrt{2}}(\PP+\PP_z),\qquad \qquad \PP^R=\frac{1}{\sqrt{2}}(\PP-\PP_z)\,,
\end{align}
where
\begin{align}
    \PP=\frac{1}{3}\sum_a\check{\beta}_a\,p_a^1\,,\qquad  \PP_z=\frac{1}{3}\sum_a\check{\beta}_a\, \pl_{z_a}\,,\qquad  \check{p}_a=p_{a+1}-p_{a-1}\ \ (a\mod \, 3)\,.
\end{align}
The main differences with flat space are the following:
\begin{enumerate}
    \item The space we are integrating over looks like a \textit{half} four dimensional Minkowski space, meaning $z\geq 0$. These $z$-factor accounts for the Planck length in flat space with exactly the same power $z^{\Lambda_3-1}\leftrightarrow (l_p)^{\Lambda_3-1}$. Therefore, our coupling constant $C_{\lambda_1,\lambda_2,\lambda_3}$ in $\ads_4$ is \textit{dimensionless}.
    \item There are $U$-maps (unitary-like operators) which address the \textit{tails} (sub-leading terms with lower derivatives) of cubic interactions. They are \cite{Metsaev:2018xip}
    \begin{subequations}
    \begin{align}
        U_{L,R}&=\mathcal{T}\,\exp\Big[\int_0^1d\tau \,u_{\tau}^{L,R}\Big]\,,\\
        u_t^{L}&=\sqrt{2}\MM Y_{L}-\frac{t}{3}\Big[\beta\Mcal+\frac{\delbeta \Lambda_3}{12}\Big]Y_L^2-\frac{t\delbeta}{12}Y_L^2N_{\PP^L}-\frac{\sqrt{2}t^2\check{\beta}}{108}Y_L^3N_{\PP^L}\,,\\
        u_t^{R}&=\sqrt{2}\MM Y_{R}+\frac{t}{3}\Big[\beta\Mcal+\frac{\delbeta \Lambda_3}{12}\Big]Y_R^2-\frac{t\delbeta}{12}Y_R^2N_{\PP^L}+\frac{\sqrt{2}t^2\check{\beta}}{108}Y_R^3N_{\PP^L}\,,\\
        Y_L&=\frac{1}{N_z+2}\pl_z\pl_{\PP^L},\qquad \qquad \qquad Y_R=\frac{1}{N_z+2}\pl_z\pl_{\PP^R}\,,
    \end{align}
    \end{subequations}
    where
    \begin{align}
        \beta&=\beta_1\beta_2\beta_3,\qquad \delbeta=\beta_1^2+\beta_2^2+\beta_3^2,\qquad \check{\beta}=\check{\beta}_1\check{\beta}_2\check{\beta}_3,\\
        \MM&=\frac{1}{3}\sum_a\check{\beta}_a\lambda_a,\quad \Mcal= \sum_a\frac{\lambda_a}{\beta_a}, \quad \NN_{\PP^{L,R}}=\PP^{L,R}\pl_{\PP^{L,R}}, \quad N_z=z\pl_z\,.
    \end{align}
    Aesthetically speaking, the vertices do not look pleasing at first, they are, however, describe the complete cubic interactions in $\ads_4$ of higher spin fields in light-cone gauge. In the covariant formulation the gauge invariance requires these sub-leading terms as well.
\end{enumerate}
To fix the coupling constants $C^{L,R}_{\lambda_1,\lambda_2,\lambda_3}$, one can repeat the same computation as in flat space paying attention to integration by parts. The leading terms (highest power in $z$) yield exactly the same equations to solve for $C_{\lambda_1,\lambda_2,\lambda_3}$ (see chapter \ref{chapter4}), we get
\begin{align}
    C^L_{\lambda_1,\lambda_2,\lambda_3}=\frac{g}{\Gamma[\Lambda_3]}, \qquad C^R_{-\lambda_1,-\lambda_2,-\lambda_3}=\frac{g}{\Gamma[-\Lambda_3]}\,,
\end{align}
where $g$ is a dimensionless coupling constant which, in principle, can be set to one. For more details, see \cite{Metsaev:2018xip,Skvortsov:2018uru}.
\subsubsection{Correlation Functions}
As shown in \cite{Skvortsov:2018uru}, the bulk-to-boundary propagators of the scalar fields with conformal weight $\Delta=1,2$ in $\ads_4$ read
\begin{align}
    K_{\Delta=1}(p|z)=-\frac{1}{|p|}e^{-z|p|}, \qquad \qquad K_{\Delta=2}(p|z)=e^{-z|p|}\,.
\end{align}
Then, one gets the two-point function by sending $z\rightarrow 0$
\begin{align}
    \langle \JJ_{p}^{\lambda}\JJ_q^{\mu}\rangle = \delta^3(p+q)\frac{\delta^{\lambda+\mu,0}}{|p|}\,.
\end{align}
It is not hard to work out the three-point function 
\begin{align}
    \langle \JJ_{\lambda_1}\JJ_{\lambda_2}\JJ_{\lambda_3}\rangle \equiv \int_{\ads_4}V_3^L\prod_a\delta(z-z_a)=\frac{\delta^3(\sum_a p_a)}{\prod_a|p_a|\beta_a^{\lambda_a}}\,\widetilde{U}^L\,\frac{\Pcal^{\Lambda_3}}{|\Pu|^{\Lambda_3}}\,.
\end{align}
where
\small
\begin{align}
    \Pcal=\sum_a\frac{\check{\beta}_a(|p_a|+p_a^1)}{3\sqrt{2}}, \qquad \qquad  |\Pu|=\sum_a|p_a|\,.
\end{align}
Note that the new $\widetilde{U}^L$-map reads
\begin{align}
    &\widetilde{U}^L=\mathcal{T}\,\exp\Big[\int_0^{1}dt\, \tilde{u}_{t}^L\Big]\,.\\
    &\tilde{u}_{t}^L=\sqrt{2}\MM(N_{\Pcal})_1\boldsymbol{\partial}_{\Pcal}-\frac{t}{3}\Big[\beta\Mcal+\frac{\delbeta\Lambda_3}{12}\Big](N_{\Pcal})_2\boldsymbol{\partial}_{\Pcal}^2-\frac{t\delbeta(N_{\Pcal})^3}{12}\boldsymbol{\partial}^2_{\Pcal}-\frac{\sqrt{2}t^2\check{\beta}}{108}(N_{\Pcal})_4\boldsymbol{\partial}^3_{\Pcal}\,.
\end{align}
\normalsize
We also employ the Pochhammer symbol notation that $(N_{\Pcal})_a=N_{\Pcal}(N_{\Pcal}+1)...(N_{\Pcal}+a-1)$ with $N_{\Pcal}=\Pcal\pl_{\Pcal}$, and lastly $\boldsymbol{\partial}_{\Pcal}=|\Pu|\partial_{\Pcal}\,$. These results look remarkably simple when we compare them to the answers in covariant gauge \cite{Maldacena:2011nz,Bzowski:2013sza}. Notice that there is no appearance of the coupling constant $C$ after we carrying out the integration. 
\subsubsection{Flat Limit}
If we look at the three-point function at its most singular pole when $|\Pu|\rightarrow 0$, then, according to the prescription in \cite{Raju:2012zr}, the flat limit reads 
\begin{align}
     (\prod_a|p_a|)|\Pu|^{\Lambda_3}\times\langle \JJ_{\lambda_1}\JJ_{\lambda_2}\JJ_{\lambda_3}\rangle\Big|_{|\Pu|\rightarrow 0}\sim \frac{\Pcal^{\Lambda_3}}{\prod_a\beta_a^{\lambda_a}}\,.
\end{align}
The rhs. of (3.72) can be \textit{interpreted} as three-point scattering amplitude in flat space. The basic argument for this limit is that $1/|\Pu|^{\Lambda_3}$ will play the role of the \textit{fourth} delta functions.




\chapter{Appendix for Formal Construction of HSGRA}\label{app:chap5}
\section{Important Concepts}

\paragraph{Associative Algebra} is a vector space with bilinear map $\star:\mathbf{A}\times \mathbf{A}\rightarrow \mathbf{A}$ that satisfies associativity 
\begin{align}
    x\star(y\star z)=(x\star y)\star z, \qquad \forall x,y,z\in \mathbf{A}.
\end{align}
We shall assume that $\mathbf{A}$ is also unital
\begin{align}
    \exists e\in \mathbf{A}: \quad e\star a=a\star e=a.
\end{align}

\paragraph{Two-sided ideal} is a sub-algebra of $\mathbf{A}$ denoted as $\mathcal{I}$ such that
\begin{align}
    \mathcal{I}\star \mathbf{A}\subset \mathcal{I}\qquad \text{and}\qquad \mathbf{A}\star \mathcal{I}\subset \mathcal{I}\,.
\end{align}
In other words, $\mathcal{I}$ absorbs multiplication from the left and from the right by elements of the associative algebra $\mathbf{A}$. 

\paragraph{Quotient algebra} is defined by equivalent classes $[a]$ with equivalence relation
\begin{align}
    a\sim a+\mathcal{I}\,.
\end{align}
We denote the quotient algebra as $\mathbf{A}/\mathcal{I}$.

\paragraph{Lie algebra} is a vector space $\mathbf{g}$ equipped with a Lie bracket $[\cdot,\cdot]: \mathbf{g}\times \mathbf{g}\rightarrow \mathbf{g}$ such that
\begin{align}
    [x+y,z]&=[x,z]+[y,z], \qquad &[z,x+y]&=[z,x]+[z,y],\\
    [x,x]&=0\,,\qquad &[x,y]&=-[y,x]\,
\end{align}
for all $x,y,z\in \mathbf{g}$. Moreover, the Lie bracket satisfies the Jacobi identity 
\begin{align}
    [a,[b,c]]+[b,[c,a]]+[c,[a,b]]=0\,.
\end{align}
An associative algebra $\mathbf{A}$ can be turned into a Lie algebra $\mathbf{g}$ by equipping $\mathbf{A}$ with the Lie bracket via the commutator $[a,b]=a\star b-b\star a$ where $a,b\in \mathbf{A}$.

\paragraph{Universal enveloping algebra} of a Lie algebra $\mathbf{g}$, denoted as $U(\mathbf{g})$, is an associative algebra.  Define $\mathcal{I}(\mathbf{g})$ to be the two-sided ideal of the tensor algebra $T(\mathbf{g})$ generated by all elements of the form $xy-yx-[x,y]$ where $x,y\in \mathbf{g}$. The universal enveloping algbera is defined as
\begin{align}
    U(\mathbf{g})=T(\mathbf{g})/\mathcal{I}(\mathbf{g})\,.
\end{align}

\paragraph{Tensor algebra} is the algebra of tensors on a vector space $V$ and denoted as $T(V)$. For $k\geq 0$, we can define 
\begin{align}
    T^kV=V^{\otimes k}=\underbrace{V\otimes V\otimes \cdots \otimes V}_{k\ \text{times}}\,.
\end{align}
Then, $T(V)$ is just a direct sum of $T^kV$
\begin{align}
    T(V)\equiv TV=\bigoplus_{k=0}^{\infty}T^kV=\bullet\oplus V\oplus(V\otimes V)\oplus(V\otimes V\otimes V)\oplus ...\,.
\end{align}
In general, the vector space $V$ can also carry some grading. 

\paragraph{Module homomorphism} is a space of all maps between modules $A,B$ that preserves the module structures
\begin{align}
    f(a+b)&=f(a)+f(b)\,, \\
    f(sa)&=sf(a)\ \ (\text{left-module}), \qquad f(as)=f(a)s\ \ (\text{right-module})\,.
\end{align}
We denote the module homomorphism as $\Hom(A,B)$. In general, $A,B$ can also be algebras. 
\section{Q-Manifolds and Strong Homotopy Algberas}
\paragraph{Q-Manifold} is a supermanifold $\mathcal{M}$ equipped with a differential $Q$ that is nilpotent \cite{Alexandrov:1995kv}, i.e. $Q^2=0$. Consider some local coordinates $x^a$ on $\mathcal{M}$, then
\begin{align}
    Q=Q^a\frac{\pl}{\pl x^a} \qquad \Rightarrow \qquad Q^2=Q^b\frac{\pl Q^a}{\pl x^b}=0\,.
\end{align}
where
\begin{align}
   Q^2&=\frac{1}{2}\llbracket Q,Q\rrbracket\,,\qquad  Q^a(x)=\sum_n \sum_{b_1,...,b_n} f^{a|\boldsymbol{n}}_{b_1...b_n}x^{b_1}...\,x^{b_n}\,
\end{align}
The coefficients $f^{a|\boldsymbol{n}}_{b_1...b_n}$ obey
\begin{align}\label{eq:commutativity}
    f^{a|\boldsymbol{n}}_{b_1...b_ib_{i+1}...b_n}=(-1)^{|b_i||b_{i+1}|}f^{a|\boldsymbol{n}}_{b_1...b_{i+1}b_i...b_n}\,.
\end{align}
Here, $|b_i|$ stands for the degree of coordinate $x^{b_i}$. For nilpotency of $Q$, we have
\begin{align}
    \sum_{l+m=n+1}\sum_{p=1}^l\sum_{\substack{i_1,...,i_l \\ j_1,...,j_m}}\pm f^{a|\boldsymbol{l}}_{i_1...i_p...i_l}f^{i_p|\boldsymbol{m}}_{j_1...j_m}=0\,.
\end{align}
where the sign depends on the particular permutation. Assuming $f^{\bullet|\boldsymbol{0}}=0$, the first few relations read
\begin{subequations}
\begin{align}
  n=2&: \qquad  f^{a|\boldsymbol{1}}_{i_1}f^{i_1|\boldsymbol{1}}_{j_1}=0\,,\\
    n=3&: \qquad f^{a|\boldsymbol{2}}_{i_1i_2}f^{i_1|\boldsymbol{1}}_{j_1}+(-1)^{|i_1||j_1|}f^{a|\boldsymbol{2}}_{i_1j_1}f^{|\boldsymbol{1}}_{i_2}+f^{a|\boldsymbol{1}}_{i_1}f^{i_1|\boldsymbol{2}}_{j_1j_2}=0\,,\\
    n=4&:\qquad f^{a|\boldsymbol{2}}_{i_1i_2}f^{i_1|\boldsymbol{2}}_{j_1j_2}+(-)^{(|i_2|+|j_2|)|j_1|}f^{a|\boldsymbol{2}}_{i_1j_1}f^{|\boldsymbol{2}}_{j_2i_2}+(-)^{(|j_1|+|j_2|)|i_2|}f^{a|\boldsymbol{2}}_{i_1j_2}f^{i_1|\boldsymbol{2}}_{i_2j_1}+...=0\,,
\end{align}
\end{subequations}
where the ellipses contain terms with $f^{\bullet|\boldsymbol{1}}$ and $f^{\bullet|\boldsymbol{3}}$. We can identify $f^{\bullet|\boldsymbol{1}}$ with a differential $d$ since it squares to zero. Moreover, we can say that the coefficients $f^{\bullet|\boldsymbol{n}}$ determine an odd linear map of n-th tensor power of $T_p(\mathcal{M})$ into $T_p(\mathcal{M})$, where $T_p(\mathcal{M})$ is the tangent space of $\mathcal{M}$ at the stationary point $p$. This map induces a map $\boldsymbol{\ell}_k : V^{\otimes k} \rightarrow V$, where $V=\Pi T_p(\mathcal{M})$ (the space of tangent bundle to $\mathcal{M}$) is some graded vector space. The map $\boldsymbol{\ell}_1$ then determines a differential in $V$ and $\boldsymbol{\ell}_2$ determines a binary operation and so on.

\paragraph{Strong Homotopy Algebra} A linear space equipped with multilinear maps $\ell_k$ satisfying 
\begin{align}
    \llbracket \boldsymbol{\ell},\boldsymbol{\ell}\rrbracket=0\,,\qquad \boldsymbol{\ell}=\boldsymbol{\ell}_1+\boldsymbol{\ell}_2+...\,, \quad \boldsymbol{\ell}\in \Hom(TV,V) \ \ \text{and} \ \ \boldsymbol{\ell}_k\in \Hom(T^kV,V)
\end{align}
is called a $L_{\infty}$-algebra or strong homotopy Lie algebra. If we remove the condition \eqref{eq:commutativity}, then we have $A_{\infty}$-algbera or strong homotopy associative algebra. It is easy to see that at any stationary point $p$ on a $Q$-manifold we have strong homotopy algebras as local structures.



\end{appendix}

  \backmatter
  \markboth{}{}

  \addcontentsline{toc}{chapter}{\protect }
\chapter*{Acknowledgement}
This PhD thesis would not have been possible without my supervisor Prof. Ivo Sachs who gave me the opportunity to do my PhD in his group. I am really thankful and blessed for his support and guidance during my time in Munich. It was always a pleasure to discuss physics with Ivo in his office where conversation goes endlessly. 

I would like to specially address Dr. Evgeny Skvortsov for taking me up as a student and teaching me most of what I know about higher-spin theory. Under his wise guidance and advice either through conversation or emails, I have become a true higher-spin \textit{comrade}. I am deeply grateful for his continuous support during my PhD time. 

I am also very much indebted to Prof. Stefan Theisen for being my second official supervisor and letting me complete the second half of my PhD. I enjoyed Stefan's kindness and humour whenever we chat in the corridor or his office. Thank you also for refereeing this thesis. I further want to thank Prof. Herman Nicolai for kindly accepting me to AEI and for his generosity.  

My extended gratitude goes to Tomáš
Procházka, Sebastian Konopka and Jan Gerken for many fruitful and enlightening discussions. I am grateful to Maor Ben-Shahar for spending his time to correct the English in this thesis. Special thanks also go to all the members of the 'Theoretical Astroparticle Physics
and Cosmology' chair and beyond in Munich and to all the members or the 'Quantum Gravity and Unified Theories' division in Potsdam. In particular, I would like to thank Ottavia Balducci, Igor Bertan, Federico Gnesotto, Katrin Hammer, Till Heckelbacher, Frederik Lauf, Adiel Meyer, Luca Mattiello, Allison Pinto; Matteo Broccoli, Lorenzo Casarin, Hugo Camargo, Franz Ciceri, Caroline Jonas, Johannes Knaute, Lars Kreutzer and Hannes Malcha for making each day at the office entertaining and fun. 

I have also benefited from discussions and correspondence with: Antal Jevicki, Per Sundell, Igor Klebanov, Simone Giombi, Murat Gunaydin, Roberto Bonezzi, Gleb Arutyunov, Karapet Mkrtchyan, Livia Ferro, Xavier Bekaert, Maxim Grigoriev, Arthur Lipstein, Lionel Mason, Tristan McLaughlin, Mirian Tsulaia, Axel Kleinschmidt, Oliver Schlotterer, Matin Mojaza, Rakibur Rahman, Hadi Godazgar, André Coimbra, Alexey Sharapov, Ergin Sezgin, Thomas Basile, Lorenz Eberhardt, Olaf Hohm, Lance Dixon, Kostas Skenderis, Paolo Benincasa, Claudio Corianò, Maor Ben-Shahar, Francesco Serra, Justin Vines, Oleg Evnin, Euihun Joung, Jarah Evslin, Andrea Campoleoni and Nicolas Boulanger.

I thank the Max Planck Institut f\"ur Physik, the Ludwigs Maximilians-Universit\"at M\"unchen and the IMPRS program for providing me a stimulating scientific environment. I am particularly grateful to Mrs. Herta Wiesbeck-Yonis, Ms Darya Niakhaichyk for helping me go through all administrative and bureaucratic tangle.

I am greatly indebted to my wife Iroda Sodikova for her love, patience and support during all these years. I also thankful to my family for their endless support and unconditional love. I would never make it without them. This thesis is dedicated to them. 

Finally, I would like to thank all of my other teachers, colleagues and friends in Japan, America, Germany and somewhere around the world, who have helped me along the way.

\bibliographystyle{JHEP}
\bibliography{literatur.bib}

\end{document}